%% file: jqt-phys-rep.tex
\documentclass[a4paper,12pt,twoside,openright]{report}

\usepackage{a4wide}
\usepackage{xspace}
\usepackage{amsmath}
\usepackage{amssymb}
\usepackage{booktabs}
\usepackage{graphicx}
\usepackage{url}
\usepackage[breaklinks=true]{hyperref}
\usepackage{subfig}
\usepackage[countmax]{subfloat}
\usepackage{epigraph}
\usepackage{slashed}
\usepackage{cite}
\usepackage{mathtools}

\include{defs}

\usepackage{fancyhdr}
\newcommand{\helv}{\fontfamily{phv} \fontseries{b}\fontsize{10}{12}\selectfont}

\makeatletter
\renewcommand{\chaptermark}[1]{%
  \markboth{\MakeUppercase{\ifnum\c@secnumdepth>\m@ne \@chapapp \ \thechapter. \ \fi #1}}%
           {\MakeUppercase{\ifnum\c@secnumdepth>\m@ne \@chapapp \ \thechapter. \ \fi #1}}%
}
\makeatother

\makeatletter
\def\cleardoublepage{\clearpage\if@twoside \ifodd\c@page\else
\hbox{}
\vspace*{\fill}
\thispagestyle{empty}
\newpage
\if@twocolumn\hbox{}\newpage\fi\fi\fi}
\makeatother

\fancypagestyle{plain}{%
\fancyhf{} 
\fancyfoot[C]{\helv \thepage} 
}

\begin{document}

\begin{titlepage}
  
  \begin{center}
    \vspace*{5.0cm}
    \Huge \textsf{The Physics of Jet Quenching \\in Perturbative QCD}
    
    \vspace*{0.7cm}
    
  \end{center}

  \vspace*{2.0cm}
  
  \begin{center}
    {\large \textsf{Yacine Mehtar-Tani}}\\
    \vspace*{0.7cm}
  Physics Department, Brookhaven National Laboratory, \\
  Upton, NY 11973, USA\\
    \vspace*{0.3cm}
    email: \href{mailto:mehtartani@bnl.gov}{mehtartani@bnl.gov}
  \end{center}


\end{titlepage}


\begin{abstract}

Hard processes in collider experiments typically produce QCD jets, which have long served as precision tests of QCD in the vacuum. More recently, heavy-ion programs at RHIC and the LHC have offered a novel perspective on jets, establishing them as unique probes of strongly interacting matter. Experimental observations, including the suppression of high-$p_T$ hadrons and jets, provide compelling evidence for the formation of a new state of matter and its strong coupling to energetic partons.

These advances have motivated new theoretical approaches to jet quenching that go beyond standard perturbative techniques, aiming to elucidate the mechanisms of energy dissipation and thermalization of energetic partons in the quark-gluon plasma. This review highlights recent progress, beginning with a unified description of medium-induced radiation across the Landau-Pomeranchuk-Migdal regime and its role in turbulent gluon cascades. We then examine radiative corrections that renormalize the transport coefficient~$\hat q$, the mechanism of color decoherence in multi-parton systems, and nonlinear QCD evolution equations for jet energy loss. Finally, we confront this framework with experimental measurements, underscoring the need for precision phenomenology to fully exploit the rich data sets from RHIC and the LHC.

\end{abstract}

\tableofcontents

\include{intro}

\part{Partonic radiative energy loss}
\include{medinteract}
\include{formalism}
\part{Emergent phenomena: turbulence and superdiffusion}
\include{cascade}

\include{qhat-renormalization}

\part{Jet quenching weights}
\include{decoherence}
\include{jeteloss}

\include{phenomenology}

\include{conclusion}

\begin{acknowledgements}
I am deeply grateful to my collaborators Néstor Armesto, Jean-Paul Blaizot, Jorge Casalderrey-Solana, Fabio Dominguez, Edmond Iancu, Mauricio Martínez-Guerrero, Guilherme Milhano, Felix Ringer, Carlos Salgado, Konrad Tywoniuk, and Varun Vaidya, who deserve much of the credit for the work summarized in this report. I am especially indebted to my younger collaborators João Barata, Paul Caucal, Leonard Fister, Hao Ma, Daniel Pablos, Balbeer Singh, Alba Soto-Ontoso, and Marcus Torres for their invaluable contributions. I also wish to acknowledge Rudolf Baier, Alfred Mueller, and Dominique Schiff, who first introduced me to this subject early in my career. This work was supported by the U.S. Department of Energy under Contract No. DE-SC0012704.

\end{acknowledgements}

\appendix

\bibliographystyle{JHEP}
\bibliography{paper-jet-eloss.bib}

\end{document}

%% file: defs.tex
\def\del{\partial}
\def\bdel{\bs\partial}

\newcommand{\eqn}[1]{Eq.~\eqref{#1}}
\newcommand{\fign}[1]{Fig.~\ref{#1}}
\def\eqs#1{Eqs.~\eqref{#1}}

\def\lg{{\langle}}
\def\rg{{\rangle}}

\def\0{{\boldsymbol 0}}
\def\k{{\boldsymbol k}}
\def\p{{\boldsymbol p}}
\def\q{{\boldsymbol q}}
\def\r{{\boldsymbol r}}
\def\u{{\boldsymbol u}}
\def\v{{\boldsymbol v}}
\def\x{{\boldsymbol x}}
\def\y{{\boldsymbol y}}
\def\z{{\boldsymbol z}}

\newcommand{\xt}{\boldsymbol x}
\newcommand{\kt}{\boldsymbol k}

\newcommand{\qhat}{\hat{q}}

\def\bk{{ k}}

\def\bell{{\boldsymbol \ell}}
\def\bkg{{0}}

\def\A{{\boldsymbol A}}
\def\D{{\boldsymbol D}}
\def\l{{\boldsymbol l}}

\def\Q{{\boldsymbol Q}}

\def\btheta{{\boldsymbol \theta}}

\def\sst{\scriptscriptstyle}
\def\rme{{\rm e}}
\def\rmd{{\rm d}}

\def\and{ \quad\text{and}\quad}

\def\scal{\text{scal}}
\def\mfp{\text{mfp}}

\def\sM{\text{med}}

\def\scal{\text{scal}}

\def\tform{{t_\text{f}}}
\def\tdecoh{t_\text{d}}

\newcommand{\tr}{{\rm tr}}
\newcommand{\Tr}{{\rm Tr}}
\def\rmR{{\rm Re}}

\def\abar{{\rm \bar\alpha}}

\def\cC{{\cal C}}
\def\cG{{\cal G}}
\def\cP{{\cal P}}
\def\cK{{\cal K}}
\def\cM{{\cal M}}
\def\cD{{\cal D}}

\def\cU{{\cal U}}

\def\cH{{\cal H}}

\def\cO{{\cal O}}

\def\scal{{\rm scal}}

\def\Sii{{S^{(2)}}}

\def\Siii{{S^{(3)}}}

\def\tSiii{{\tilde S^{(3)}}}
\def\Siv{{S^{(4)}}}
\def\bdel{\boldsymbol \del}

\newcommand{\dif}{{\rm d}}

\newcommand{\rmTr}{{\rm Tr}}

\def\sM{\text{med}}

\def\ProbOne{P_{\scriptscriptstyle 1}}

\def\ProbTwo{ P_{\scriptscriptstyle 2} }
\def\ProbSing{ P_\text{sing}}

\def\br{\text{br}}

\def\med{\text{med}}
\def\Tr{\text{Tr}}

\def\sM{\text{med}}

\def\ProbOne{P_{\scriptscriptstyle 1}}

\def\ProbTwo{ P_{\scriptscriptstyle 2} }
\def\ProbSing{ P_\text{sing}}

\def\n{{\boldsymbol n}}

\long\def\comment#1{ }

\newcommand{\nn}{\nonumber\\ }

\def\be{\begin{eqnarray*}}
\def\ee{\end{eqnarray*}}
\def\beq{\begin{eqnarray}}
\def\eeq{\end{eqnarray}}

\usepackage{color}
\definecolor{darkgreen}{rgb}{0,0.5,0}
\definecolor{comment}{rgb}{0,0.3,0}
\definecolor{identifier}{rgb}{0.0,0,0.3}
\usepackage{listings}

\makeatletter
\newenvironment{acknowledgements}{%
  \if@twocolumn
    \section*{Acknowledgements}%
  \else
    \small
    \begin{center}%
      {\bfseries Acknowledgements\vspace{-.5em}\vspace{\z@}}%
    \end{center}%
    \quotation
    \fi}
  {\if@twocolumn\else\endquotation\fi}
\makeatother

\hypersetup{colorlinks=true,linkcolor=blue,urlcolor=blue}

%% file: intro.tex
\chapter{Introduction} \label{chap:intro}

Following the discovery of asymptotic freedom in non-Abelian gauge theories \cite{Gross:1973id,Politzer:1973fx}, Quantum Chromodynamics (QCD) \cite{Fritzsch:1973pi} was quickly recognized as the appropriate microscopic theory of strong interactions. This conceptual breakthrough was soon supported by growing experimental evidence from collider experiments, such as electron-proton collisions at SLAC and electron-positron collisions at DESY and LEP.  

A significant milestone was reached at DESY's PETRA collider in 1979 with the observation of jets, which provided direct experimental confirmation of gluons as the carriers of the strong interaction \cite{Brandelik:1979bd}. Jets, characterized by energetic sprays of particles, had been theoretically predicted by QCD just a few years earlier \cite{Sterman:1977wj}. Fundamentally, they represent the observable imprints of short-distance partonic dynamics during hard scattering events, making them a powerful tool for testing and validating the theory.

Since those early days, the field has advanced significantly, with higher-order theoretical computations of multijet events not only enhancing the precision of QCD tests but also playing a pivotal role in new physics searches at the Large Hadron Collider (LHC). Notably, these developments were instrumental in the discovery of the Higgs boson \cite{ATLAS:2012yve,CMS:2012qbp}. 

These advancements pertain to the low-density regime of QCD, where standard perturbation theory is applicable, albeit with the inclusion of resummation techniques. These techniques allow for the summation of logarithmically enhanced Feynman diagrams to all orders in the coupling constant $\alpha_s$. Such methods are typically encapsulated in renormalization group equations, such as the well-known Dokshitzer-Gribov-Altarelli-Parisi (DGLAP) equation \cite{Gribov:1972ri,Altarelli:1977zs,Dokshitzer:1977sg}, which describes the scale evolution of parton distribution functions at short distances and the collinear fragmentation of jets into hadrons.

At high density and high temperature, novel phenomena are predicted to emerge from QCD dynamics. One of the most notable is the phase transition that cold nuclear matter, described by hadrons, undergoes when a critical temperature of approximately $T_c \sim 150$–$160 \, \mathrm{MeV}$ is reached \cite{Aoki:2006we,Bazavov:2014pvz}. Beyond this point, a deconfined state of matter, known as the quark-gluon plasma (QGP), is created, where the elementary constituents of hadrons—quarks and gluons—are freed from their strong mutual attraction. This occurs under conditions of extremely high energy density and temperature, a result initially supported by early lattice gauge theory numerical simulations \cite{Kuti:1980gh}. This new state of matter, believed to have existed in the first moments of the universe, has been successfully recreated in laboratory settings through ultrarelativistic heavy-ion collisions conducted at RHIC and the LHC.

In his seminal work \cite{Bjorken:1982tu}, J. D. Bjorken hypothesized that at high energy a transient QGP could be produced in proton-proton ($pp$) collisions suppressing the yield of high-energy quark and gluon jets. The underlying mechanism is the following: as they traverse the plasma, jets lose a fraction of their energy to the hot medium causing a shift of the jet spectrum to lower $p_T$ values -- a phenomenon now known as ``jet quenching.'' This phenomenon was first observed at RHIC through the suppression of high-$p_T$ hadrons in Au-Au collisions at $\sqrt{s_{NN}}=200$ GeV energies \cite{Arsene:2004fa,Back:2004je,Adams:2005dq,Adcox:2004mh}, providing strong evidence for the creation of this new state of matter. Further support came from the observation of collective flow-like behavior in the large number of final-state particles produced in these collisions \cite{Shuryak:2004cy}.

A few years later, the LHC measured jet quenching in fully reconstructed jets \cite{Chatrchyan:2011sx,Aad:2010bu,CMS:2021vui,ALICE:2023waz,Connors:2017ptx}. These measurements spurred substantial experimental and theoretical efforts to exploit the perturbative nature of QCD jets and to adapt sophisticated tools, originally developed for jet studies in proton-proton collisions, to probe the properties of the QGP. In particular, the rapid development of jet substructure observables has enabled more detailed and differential studies of the modifications jets experience as they traverse the plasma. Beyond the well-established Jet nuclear modification factor \cite{Mehtar-Tani:2021fud}, a wide range of jet substructure analyses \cite{Larkoski:2014wba,CMS:2017qlm,Chien:2016led,Mehtar-Tani:2016aco,Caucal:2021bae,Caucal:2019uvr,Barata:2023bhh}, performed both analytically and through Monte Carlo event generators, have unveiled intriguing qualitative features of medium-modified jets \cite{ALICE:2021aqk,ALICE:2022hyz,Song:2023sxb}.

Before jet quenching was observed experimentally, early theoretical efforts focused on calculating the average properties of parton energy loss in QCD matter. The first estimate was for collisional energy loss -- analogous to energy loss in QED plasmas—the collisional mechanism scales linearly with the length of the medium, $L$ \cite{Braaten:1991we}. Specifically, the elastic energy loss can be expressed as $E_{\rm el} \sim \alpha_s^2 T^2 L \ln \sqrt{E T / m_D^2}$, where $T$ is the plasma temperature, $E$ is the energy of the hard parton undergoing energy loss, and $m_D \sim gT$ is the Debye screening mass of the plasma. However, it was soon recognized that, radiative processes cannot be neglected \cite{Gyulassy:1994ew,Baier:1994bd}. The energy loss of high energy parton in the plasma due to gluon radiation turns out to scale quadratically with medium length, i.e., $E_{\rm rad} \sim \alpha_s^3 T^3 L^2 $ \cite{Baier:1996kr}.

The medium-induced radiative spectrum lies at the heart of calculations for radiative energy loss in the quark-gluon plasma \cite{Wang:1991xy,Gyulassy:1990ye,Wang:1992qdg,Gyulassy:2000fs,Baier:1994bd,Baier:1996kr,Zakharov:1996fv,Zakharov:1997uu,Zakharov:1996fv,Gyulassy:2000er,Wiedemann:1999fq,Baier:2001yt,Wiedemann:2000za,Arnold:2002ja,Salgado:2003gb}. While initially computed to leading order in the coupling constant associated with the radiated gluons, the spectrum incorporates contributions from an arbitrary number of scatterings in the large and dense QGP created in heavy-ion collisions at RHIC and LHC. These multiple scatterings act coherently during the quantum mechanical formation time of the radiated gluon. 

This coherence effect, known as the Landau-Pomeranchuk-Migdal (LPM) effect \cite{Landau:1953gr,Migdal:1956tc}, suppresses gluon radiation with long formation times. In QCD, this suppression contrasts sharply with the incoherent sum of Bethe-Heitler (BH)-type radiation off individual scattering centers. It becomes significant at high gluon energies, where the gluon formation time exceeds the characteristic BH frequency \cite{Baier:1996kr,Zakharov:1996fv,Gyulassy:2000er,Wiedemann:2000za}. This highlights the role of quantum coherence in the physics of radiative energy loss.

In the last decade, the theory of jet quenching has developed rapidly due to the need to accurately describe the wealth of jet data produced at RHIC and LHC. However, in the absence of a systematic approach to higher-order computations of jet observables, early phenomenological models implemented in Monte Carlo event generators varied significantly in their treatment of how the parton shower is modified by interactions with the plasma \cite{Zapp:2013vla,Lokhtin:2008xi,Caucal:2018dla,Apolinario:2019amp,JETSCAPE:2017eso,He:2015pra,Cao:2017zih,Cao:2016gvr,Park:2018fuo,Schenke:2009gb,Zapp:2013vla,Armesto:2009fj,Gossiaux:2018}. The challenge lies in understanding how jets, as spatially extended and coherent multi-parton systems, interact with the dense QCD medium. Addressing this problem requires novel theoretical tools capable of treating jet evolution in heavy-ion collisions within the framework of open quantum systems.

In QCD, jets arise from the decay of highly virtual partons through a cascading process of successive collinear splittings. During this parton cascade, the virtuality $ Q $ decreases, and as long as $ Q $ remains larger than $ \Lambda_{\rm QCD} $, the jet fragmentation can be reliably computed using perturbation theory. Jets are typically defined as collections of high-energy hadrons clustered within a cone-shaped area of opening angle $R $, centered around the momentum direction of the total jet. The transverse momentum $ p_T$ of a jet approximately corresponds to the energy flow within this pencil-like structure. In proton-proton collisions, $ p_T $ refers to the projection of the jet momentum onto the plane perpendicular to the proton beam axis.

Operationally, jet algorithms \cite{Catani:1991hj,Dokshitzer:1997in,Cacciari:2008gp}, such as the anti-$ k_t $ algorithm \cite{Cacciari:2008gp}, cluster particles or calorimeter deposits into jets and assign a $ p_T $ based on the transverse momenta of their constituent particles. The primary goal of any jet definition is to reconstruct as accurately as possible the $ p_T $ of the original parton that initiated the jet. This enables access to the underlying microscopic partonic processes while suppressing the non-perturbative effects of hadronization, which occur at long distances and time scales of order $ \Lambda_{\rm QCD}/Q^2 $.

In heavy-ion collisions, the mapping between the final-state jet and its originating parton is significantly obscured due to the interaction of the jet with the surrounding hot and dense medium. Nevertheless, jets are a salient feature due to the large separation of scales: their transverse momentum, $p_T \sim 100 - 1000\,$  GeV their virtuality, $Q \sim p_T R \sim 30 - 300\, \text{GeV}$ for $R = 0.3$, and the medium temperature, $T$, which is typically less than $1\, \text{GeV}$. This pronounced separation makes jets unique and powerful probes of the quark-gluon plasma (QGP), offering valuable insights into the medium's properties and dynamics. The QGP is a product of ultra-relativistic heavy-ion collisions, while jets originate from rare hard scattering events, underscoring their significance as tools for studying the medium.

Jet quenching is a multi-scale problem. In addition to the aforementioned hard jet scales and local medium scales, such as the temperature and the Debye screening mass, $m_D$, the interactions between the jet's partonic fragments and the plasma constituents generate intermediate scales associated with emergent non-equilibrium phenomena. Elastic processes, which cause the broadening of the transverse momentum of high-energy partons relative to their direction of motion, are characterized by a global scale $Q_\mathrm{med} \sim (\hat{q}L)^{1/2}$, where $\hat{q} \sim m_D^2 / \ell_{\mathrm{mfp}}$ is the transverse momentum diffusion coefficient.  Due to its enhancement with the system size, this scale may lie in the perturbative regime, justifying the use of weak-coupling techniques to capture the leading effects of jet quenching. To estimate the magnitude of this scale, taking $L= 3 - 4$ fm and $\hat q = 2 $ GeV$^2$/fm and we arrive at $Q_\med\sim 2 - 3$ GeV.  At higher orders in perturbation theory, the transport coefficient $\hat{q}$ receives large logarithmic corrections, indicative of a renormalization group structure akin to the DGLAP evolution of structure functions. This renormalization group framework will be discussed in  Chap.~\ref{chap:qhat-rg}. 

The objective of this manuscript is to present a {\bf comprehensive theoretical framework for jet quenching phenomenology} rooted in first principles, enabling a unified and systematic description of jet evolution in a quark-gluon plasma. Our focus will be on the inclusive jet spectrum as a key application to phenomenology. This will serve as a basis for computing jet observables in the future. The overarching picture encompasses the transition from the hard, short-distance sector—spanning scales from $p_T$ down to medium scales such as $Q_\mathrm{med}$—to the medium dynamics at softer scales. As we will demonstrate, this transition is characterized by medium-induced gluon cascades, described by kinetic-like equations, in contrast to the virtuality- or angular-based evolution observed in vacuum.

A central challenge in achieving a coherent description of jet evolution lies in understanding the effects of color coherence among multiple color charges within the jet. The radiation pattern of these charges can interfere destructively, significantly altering the dynamics of energy loss. This phenomenon is well-established in vacuum parton cascades, where large-angle gluon bremsstrahlung fails to resolve the internal structure of the multi-parton emitting system. As a result, the radiation intensity is proportional to the total color charge rather than the incoherent sum of emissions from individual partons.  

In the case of medium-induced radiation, however, the determining factor for coherence or incoherence is not the transverse wavelength of the radiated gluon but the medium's resolution power, which is governed by the inverse of the medium scale, $(Q_\mathrm{med})^{-1}$. As discussed in Chap.~\ref{chap:decoherence}, this resolution scale translates into a coherence angle, $\theta_c$, below which jet fluctuations are unresolved and are perceived as a single effective color charge by the medium \cite{Mehtar-Tani:2011hma,Mehtar-Tani:2012mfa,Mehtar-Tani:2017ypq,Casalderrey-Solana:2011ule,Mehtar-Tani:2017ypq}.  Similar to the Landau-Pomeranchuk-Migdal (LPM) effect, the {\bf phenomenon of  color decoherence} plays a crucial role in determining the effective number of jet color charges that the medium resolves. This directly impacts the energy loss process and underscores the necessity of accurately accounting for the medium's interaction with the jet's color structure.

In addition to the conceptual aspects, several practical challenges must be addressed to achieve a quantitative phenomenology. These include accurately accounting for the dynamics of the quark-gluon plasma (QGP) and the geometry of the collision. While theoretical discussions often assume an idealized plasma as a static brick of fixed length $L$, the reality is far more complex. In actual collisions, the plasma is out of equilibrium, expanding and cooling as a function of time. The QGP phase, before transitioning back to hadronic degrees of freedom, typically lasts only a few femtometers.  This dynamical evolution is often encoded in a time-dependent transport coefficient, $\hat{q}(t)$ (see \cite{Mehtar-Tani:2021fud} and references therein). We will briefly address this issue in the final chapter, which focuses on phenomenology. However, the primary emphasis of this report will be on the QCD dynamics of the parton shower within the simplified ``brick" model.

In Chap.~\ref{chap:med-partons}, we provide a heuristic discussion of the Baier-Dokshitzer-Mueller-Peigné-Schiff-Zakharov-Wiedemann (BDMPS-ZW) approach to the medium-induced radiative spectrum \cite{Baier:1996kr,Zakharov:1996fv,Gyulassy:2000er,Wiedemann:2000za}, which serves as the central mechanism for parton radiative energy loss. This chapter also introduces key notations and concepts, avoiding a heavy reliance on formalism to ensure clarity. Chap.~\ref{chap:formalism} presents the key elements of the formalism underlying modern calculations of medium-induced radiation, building on U. Wiedemann's formulation of radiative energy loss \cite{Wiedemann:2000za,Kovner:2003zj}. This approach leverages the background field method, a powerful technique also applied effectively in other areas, such as the study of gluon saturation in protons and nuclei at high energies \cite{Balitsky:2001gj}.

In Chap.~\ref{chap:cascade}, we explore the formation of the medium-induced gluon cascade \cite{Blaizot:2012fh,Blaizot:2013vha,Blaizot:2013hx,Blaizot:2014rla}, which can be viewed as a classical branching process in time. The branching rate is described by quasi-instantaneous medium-induced gluon splittings, corresponding to the large $L$ limit of the BDMPS-ZW spectrum. This approach applies to large media and resums contributions enhanced by the medium's length, i.e., $\alpha_s L$. Recent studies of higher-order corrections to this asymptotic limit have shown that they are generally negligible \cite{Arnold:2023qwi}, except for large double logarithms, which can be absorbed into a {\bf renormalization of the jet quenching parameter $\hat{q}$}, preserving the classical picture established at leading order (LO).  

This framework can be applied to study the inclusive energy distribution of fragments as a function of time or, equivalently, system size. A striking feature of the resulting cascade is the transport of energy from the leading parton to the infrared (IR) scale in a local, self-similar manner, resembling wave turbulence. During this process, energy flows locally in frequency space without accumulation, within the so-called inertial window. This dynamic is characterized by power-law scaling across a broad range of scales.

While no measurable observable is directly discussed here, the rate equation we present will reappear as a crucial component in the broader context of computing jet energy loss distributions in Chap.~\ref{chap:jeteloss}. The inclusive properties of the turbulent gluon cascade have been extensively studied over the past decade \cite{Blaizot:2014ula,Blaizot:2014rla,Caucal:2019uvr,Mehtar-Tani:2018zba,Schlichting:2020lef,Mehtar-Tani:2022zwf,Soudi:2024yfy}, with significant progress at next-to-leading order (NLO) \cite{Arnold:2020uzm,Arnold:2021pin,Arnold:2022fku,Arnold:2022mby,Arnold:2023qwi,Arnold:2024whj,Arnold:2024bph}. This includes the renormalization of the jet quenching parameter that will be discussed in Chap.~\ref{chap:qhat-rg}\cite{Blaizot:2014bha,Caucal:2022mpp,Ghiglieri:2022gyv,Caucal:2022fhc,Arnold:2021pin,Arnold:2021mow,Caucal:2021lgf,Blaizot:2019muz,Iancu:2018trm,Mehtar-Tani:2017ypq,Wu:2014nca,Iancu:2014sha,Liou:2013qya,Iancu:2014kga}.

The final piece of the puzzle is addressing the interference and color decoherence of multiple radiating color charges within the jet \cite{Mehtar-Tani:2011hma,Mehtar-Tani:2012mfa,Mehtar-Tani:2017ypq,Casalderrey-Solana:2011ule,Mehtar-Tani:2017ypq}. This topic will be explored in detail in Chap.~\ref{chap:decoherence}. In that chapter, we extend the analysis beyond single-parton energy loss to compute the energy loss of a pair of collinear partons originating from the same vertex \cite{Mehtar-Tani:2017ypq}. In such cases, medium-induced radiation from one parton can interfere destructively with radiation from the other, suppressing the overall radiation intensity. This suppression occurs when the pair is collinear enough that the medium cannot resolve their internal structure.

A complete understanding of jet energy loss requires addressing how energy flows out of the jet cone via parton cascades, given the jet definition, where partons are reconstructed within a specific solid angle \cite{Mehtar-Tani:2024mvl}. This presents a significant challenge, as gluons radiated outside the cone can subsequently radiate back inside, leading to a complex interplay between radiation inside and outside the jet region. This intricate behavior, which will be explored in detail in Chap.~\ref{chap:jeteloss}, is reminiscent of the mechanisms that generate large logarithms in non-global observables \cite{Banfi:2002hw}. Here, we derive two coupled non-linear evolution equations governing the jet energy loss distribution. The first equation pertains to the medium-induced gluon cascade, establishing the initial condition for a {\bf non-linear DGLAP evolution equation}. This second equation describes the energy loss of resolved collinear color charges, which are produced early in the process through vacuum-like splittings. 

In the final chapter, Chap.~\ref{chap:phenomenology}, we apply the framework developed in the previous chapters to analyze data from the LHC. Beyond validating the agreement with experimental results as a function of $p_T$, centrality, and jet opening angle $R$, we conduct a thorough assessment of the theoretical and model uncertainties. These include approximations such as the leading logarithmic approximation in the perturbative regime and the effects of medium response in the soft sector, which necessitate additional modeling.

Before we delve into the discussions outlined above, a final note to the reader: this manuscript serves as a report on recent theoretical advancements toward a first-principles predictive theory of jet quenching. It builds upon a series of articles that have addressed various aspects of this problem, which we aim to integrate into a cohesive overview for the first time. Our goal is to provide a useful and comprehensive reference for future work. While many facets of this rich and complex physics remain beyond the scope of this work, we hope that the overarching framework presented here will inspire further theoretical and phenomenological developments. In particular, we will not cover here the extensive recent progress in jet substructure, such as groomed observables in heavy-ion collisions~\cite{Mehtar-Tani:2016aco,Chien:2016led,Caucal:2019uvr,Caucal:2021cfb} and the study of energy-energy correlators (EECs) in heavy-ion collisions~\cite{Andres:2022ovj,Andres:2023ymw,Andres:2023xwr,Andres:2024xvk,Barata:2025fzd,Singh:2024vwb,Barata:2023bhh}, nor the rapidly growing body of work on medium response and jet–medium interactions~\cite{Li:2010ts,Casalderrey-Solana:2016jvj,KunnawalkamElayavalli:2017hxo,Milhano:2017nzm,Tachibana:2017syd,Yang:2021qtl}. These important developments, while highly complementary to the present review, deserve dedicated discussions of their own.

%% file: medinteract.tex
\chapter{Basics of radiative parton energy loss }\label{chap:med-partons}

\section{The BDMPS-ZW mechanism for gluon radiation}\label{sec:rad-spect-regimes}
In a large and dense QCD medium of size $L$, such as created in ultra-relativistic heavy-ion collisions, an energetic parton produced alongside the plasma, undergoes multiple collisions with the medium's constituents. Each collision can, in principle, knock the gluon off-shell, enabling it to emit radiation after a characteristic formation time, 
\beq\label{eq:formation-time}
t_f \sim \omega/k_\perp^2\,,
\eeq
where $\omega$ and $k_\perp$ denote the energy and transverse momentum of the radiated gluon, respectively. However, coherence effects from multiple scatterings can significantly alter this simple picture.

These coherence effects, first explored in QED by Landau, Pomeranchuk, and Migdal (LPM) \cite{Landau:1953gr,Migdal:1956tc}, are collectively known as the \emph{Landau-Pomeranchuk-Migdal (LPM) effect}. The QCD analog of the LPM effect was initially studied by Gyulassy and Wang (GW)~\cite{Gyulassy:1993hr,Wang:1994fx} and later elaborated by Baier, Dokshitzer, Mueller, Peigné, Schiff,  Zakharov and Wiedemann (BDMPS-ZW)~\cite{Baier:1994bd,Baier:1996kr,Baier:1996sk,Zakharov:1997uu,Zakharov:1996fv,Wiedemann:1999fq,Wiedemann:2000za}. Early investigations primarily focused on radiative energy loss, which, for large media, scales quadratically with the medium length $L$, unlike collisional energy loss, which scales linearly~\cite{Braaten:1991we}, as alluded to in the introduction. 

During this time, multiple scatterings increase the gluon's transverse momentum by $k_\perp^2 \sim \hat{q} t_f$, where $\hat{q}$ denotes the jet quenching parameter, which quantifies transverse momentum diffusion \cite{Baier:1996sk}: 
\beq\label{eq:qhat-general}
\hat q \equiv \frac{\rmd \langle k_\perp^2\rangle }{\rmd t} \,.
\eeq
Combining this estimate for transverse momentum square with \eqn{eq:formation-time} gives the typical formation time for medium-induced gluons :
\begin{equation}\label{eq:formation-time-med}
t_f= \sqrt{\frac{\omega}{\hat{q}}}.
\end{equation}
In the regime of soft collisions, where transverse momentum broadening occurs via many small kicks, a diffusion approximation is appropriate.  To leading order in the strong coupling constant, the quenching parameter $\hat{q}$ is related to the $2 \to 2$ QCD matrix element $\mathrm{d} \sigma_\mathrm{el}/\mathrm{d}^2 q_\perp \simeq g^4 n/ {q_\perp^4}$, where $n$ is the density of medium color charges, as follows:
\begin{equation}
\hat{q}(Q) \simeq C_R \int  \, \frac{\mathrm{d}^2 q_\perp}{(2\pi)^2}\,q_\perp^2\, \frac{\mathrm{d} \sigma_\mathrm{el}}{\mathrm{d}^2 q_\perp} \simeq 4 \pi \alpha^2_s C_R n \ln\frac{Q^2}{\mu^2}.
\end{equation}
Here, $\alpha_s \equiv g^2/(4\pi)$ is the coupling constant, $C_R = C_F = \frac{N_c^2 - 1}{2N_c}$ ($C_R = C_A = N_c$) is the color charge of a quark (gluon), and $\ln(Q^2/\mu^2)$ is the Coulomb logarithm. The parameter $\mu$ represents an infrared cutoff related to the Debye screening mass in the QGP, while $Q^2$ is a process-dependent hard scale. Finally, $n$ denotes the density of scattering centers in the medium.

For a medium characterized by the Debye screening mass $m_D$, the transverse momentum squared acquired over a distance $L$ is $\hat{q} L$, and the elastic mean free path $\ell_\mathrm{mfp}$ is related to $m_D^2$ via $m_D^2 \simeq \hat{q} \ell_\mathrm{mfp}$. It is instructive to quote known parametric estimates for these local medium scales for a weakly coupled thermal plasma as a function of the temperature $T$ \cite{Arnold:2002ja}:
\begin{equation}
m_D \sim gT\,, \quad \ell_\mathrm{mfp} \sim (g^2 T)^{-1}\,, \quad \hat{q} \sim \frac{m_D^2}{\ell_\mathrm{mfp}} \sim g^4 T^3\,.
\end{equation}
A hierarchy of length scales emerges when $g \ll 1$, namely,
\begin{equation}
\xi_D \ll \ell_\mathrm{mfp} \ll L\,,
\end{equation}
where $\xi_D = m_D^{-1}$ is the Debye length. This separation of scales is fundamental to the independent scattering approximation, which will be assumed to compute analytically the expectation values in the plasma in the next chapter.

Depending on the relationship between the gluon formation time and the characteristic medium scales $L$ and $\ell_\mfp$, the radiative spectrum can be classified into three distinct regimes:

\begin{itemize} 

\item{{\bf Bethe-Heitler regime} ($t_f  \lesssim \ell_\mfp$)}

In this regime, collisions act incoherently, producing the radiation spectrum:
\begin{equation}
\omega \frac{\mathrm{d} I}{\mathrm{d} \omega} \simeq \bar{\alpha} \frac{L}{\ell_\mfp}\,
\end{equation}
where $\bar{\alpha} = \alpha_s N_c/\pi$.
\item{{\bf Coherent regime} ($L \gg t_f \gg \ell_\mfp$)}

In this regime many collisions act coherently during the formation time of the gluon, resulting in the radiation spectrum becomes:
\begin{equation}\label{eq:bdmps-coh}
\omega \frac{\mathrm{d} I}{\mathrm{d} \omega} \simeq \bar{\alpha} \sqrt{\frac{\omega_c}{\omega}}\,,
\end{equation}
where 
\beq \label{eq:omegac}
\omega_c= \hat{q} L^2\,,
\eeq
is the critical frequency at which $t_f(\omega_c) = L$.

\item{ {\bf Large formation times: }$t_f \gg L$, or equivalently $\omega \gg \omega_c$}

 In this regime, the gluon cannot distinguish the hard scattering event from the multiple scatterings within the medium. As a result, the radiation spectrum is strongly suppressed, 
 \beq\label{eq:bdmps-hard}
 \omega \frac{\mathrm{d} I}{\mathrm{d} \omega} \simeq \frac{ \pi \alpha^3_s C_R N_c n  }{ \omega}\, L^2
 \eeq
 Notably, in this case, the radiation is predominantly governed by rare, hard scatterings in the medium.
\end{itemize}

In the LPM regime, the transverse momentum acquired by the gluon during its formation can be estimated as
\beq\label{eq:formation-kt}
k_f \sim \hat{q} t_f \sim (\hat{q} \omega)^{1/4},
\eeq
where $\hat{q}$ is the jet quenching parameter, and $\omega$ is the gluon energy. 

It is useful to discuss angular scales, as they naturally arise in the description of jet energy loss outside a jet cone. The angular scale corresponding to the transverse momentum in \eqn{eq:formation-kt} is given by

\begin{equation}\label{eq:formation-angle}
\theta_f = \frac{k_f}{\omega} = \left(\frac{\hat{q}}{\omega^3}\right)^{1/4}\,.
\end{equation}
However, after the gluon is produced, it will continue to undergo transverse momentum broadening due to elastic scattering, resulting in a typical transverse momentum of order 
\begin{equation}
\langle k_\perp^2 \rangle \sim \hat{q} L\,,
\end{equation}
where $L$ is the path length through the medium.

Note that the formation angle reaches a minimum value corresponding to the transition to the third regime of strong LPM suppression, which is associated with long formation times $t_f \gg L$. Inserting $\omega_c$ into \eqn{eq:formation-angle}, we obtain the characteristic angle 
\begin{equation}\label{eq:angle-c}
\theta_c = \frac{1}{\sqrt{\hat{q} L^3}} \,,
\end{equation}
below which radiation is strongly suppressed. This corresponds to rare emissions that are put on-shell by medium hard modes. We will recover this characteristic angular scale when discussing the physics of color decoherence in Chap.~\ref{chap:decoherence}.

\section{Mean and typical radiative energy loss }\label{sec:mean-eloss}

Upon closely examining \eqs{eq:bdmps-coh} and (\ref{eq:bdmps-hard}), we observe that the mean energy loss is dominated by the hard regime, $E \gg \omega \gg \omega_c$, and exhibits a logarithmic divergence. This divergence indicates a sensitivity to the emitter's energy, $E$, which effectively sets an upper cutoff for the $\omega$-integral. Integrating the spectrum over the full $\omega$ range yields:

\begin{equation} \langle \epsilon\rangle  = \int_0^{E} \mathrm{d} \omega \, \omega \frac{\mathrm{d} I}{\mathrm{d} \omega} \simeq \pi \alpha^3_s C_R N_c n \,  L^2 \ln \frac{E}{\omega_c}\,. \end{equation}

The contribution from the region $\omega < \omega_c$ is subleading, adding only a constant. Here, we observe the characteristic quadratic dependence on $L^2$ in the radiative energy loss, which compensates for the smallness of the coupling constant $\alpha_s$. This $L^2$ scaling underscores the dominance of importance of radiative processes in parton energy loss. 

Due to the probabilistic nature of radiation, the effect of parton energy loss on the observed jet spectrum is inherently sensitive to the full energy loss distribution. As we will discuss later, the steeply falling jet spectrum, when convolved with the energy loss distribution, skews the observed energy loss toward values smaller than the mean. While the mean energy loss is dominated by rare hard emissions, the typical energy loss, corresponding to the peak of the energy loss probability distribution, is primarily governed by the cumulative effect of multiple soft gluon emissions. 

A parametric estimate of this soft energy scale $\omega_s$, can be obtained by integrating \eqn{eq:bdmps-coh} dropping the factor $\omega$, to compute the number of radiated gluons above a given frequency $\omega$,
\begin{equation}
N(\omega) = \int_\omega^{+\infty} \rmd \omega'\frac{\rmd I}{\rmd \omega'}\sim \abar \sqrt{\frac{\omega_c}{\omega}}, 
\end{equation}
When $N(\omega) \gtrsim 1$, multiple gluon emissions become highly probable. This transition occurs at the characteristic soft scale:
\begin{equation}\label{eq:omega_s}
\omega_s \equiv \bar{\alpha}^2 \hat{q} L^2\,.
\end{equation}
This scale serves as a critical threshold, delineating the perturbative regime, in which gluon radiation is suppressed by the coupling constant, from the regime where a resummation of all orders in powers of 
\beq \label{eq:multi-rad-cond}
\bar{\alpha} \, \frac{L }{t_f }\sim 1\,,
\eeq
 becomes essential to achieve meaningful results. The soft scale $\omega_s$ plays a crucial role in defining the statistical properties of gluon branching within the medium. 

\section{Parton energy loss distribution }

\begin{figure}[t!]
\begin{center}
\includegraphics[width=0.6\textwidth]{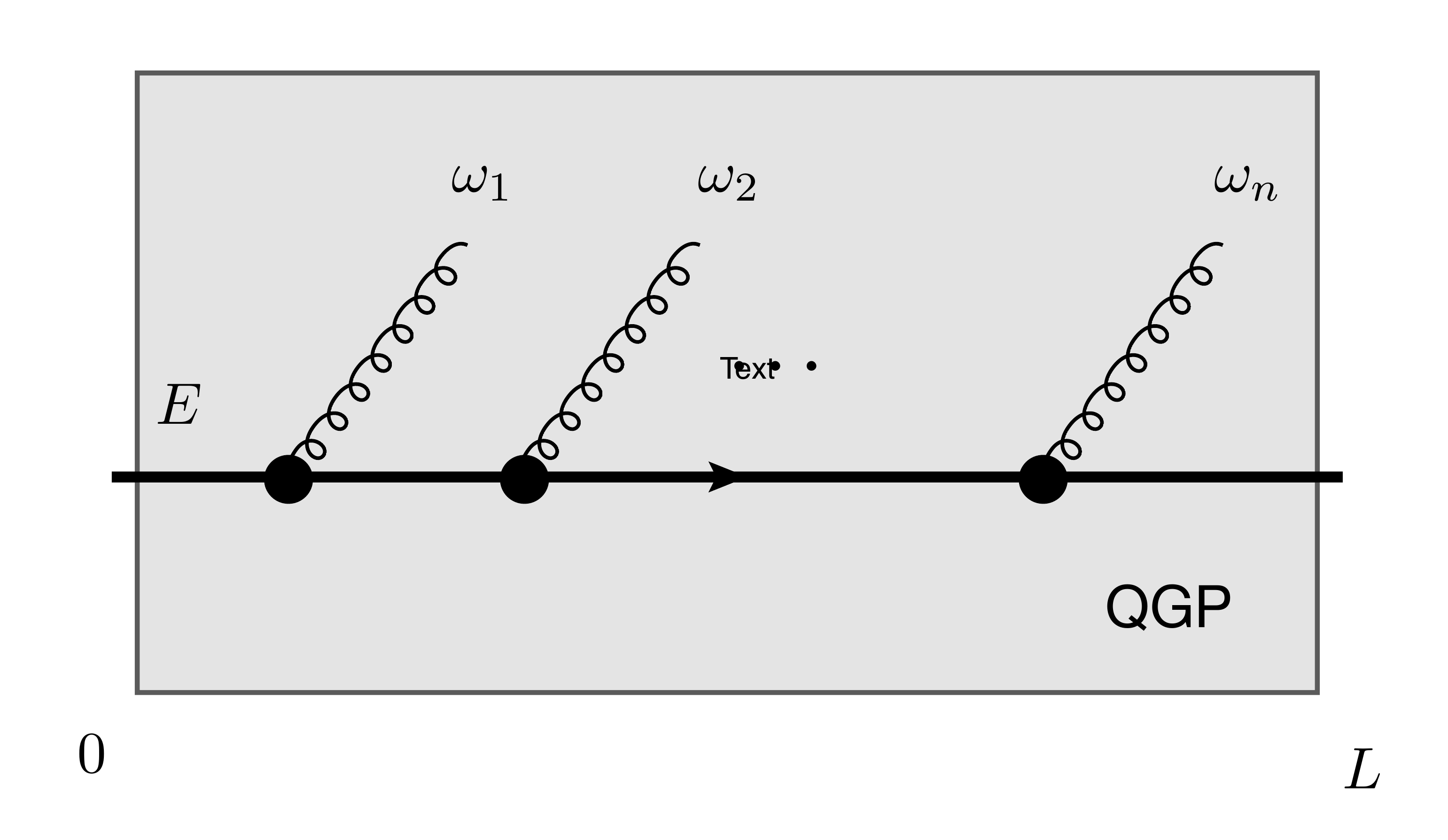}
\caption{Illustration of a high-energy parton undergoing energy loss through multiple quasi-instantaneous soft gluon emissions, characterized by the Poisson-like distribution described in \eqn{eq:poisson}, in a plasma of length $L$.  }
\label{fig:poisson-rad}
\end{center}
\end{figure}

Due to the fact that the gluon coherence time is much shorter than the medium length, i.e., $t_f\ll L$, in the regime of multiple emissions determined by the condition  \eqn{eq:multi-rad-cond}, and the overlap between successive emissions is suppressed by powers of $\bar{\alpha}$ \cite{Blaizot:2014ula,Arnold:2022mby}, the probability for a high-energy quark to lose energy $\epsilon$ can be approximated by a Poisson-like distribution, where successive soft gluon emissions are treated as independent. An illustration is given in Fig.~\ref{fig:poisson-rad}. This allows the energy loss distribution to be constructed order by order in the coupling constant.
At leading order in $\alpha_s$, the energy loss vanishes, corresponding to no gluon radiation. The probability distribution in this case is given by: 
\begin{equation} P_0(\epsilon) = \delta(\epsilon)\,.
 \end{equation}
At the next order, the energy loss distribution is: 
\begin{equation} 
P_1(\epsilon,L) = \int_0^L \rmd t\int_0^\infty \mathrm{d}\omega\,\frac{\mathrm{d}I}{\mathrm{d}\omega \rmd t}\, \Big[ \delta(\epsilon - \omega) - \delta(\epsilon) \Big] \,,
\end{equation} 
where the first term represents the probability of losing energy $\epsilon = \omega$ through the emission of a single gluon with energy $\omega$, while the second term accounts for the corresponding reduction in the probability of zero energy loss, which arises from virtual corrections. Here, $\rmd I / \rmd \omega \rmd t$ denotes the radiation rate, obtained by differentiating the spectrum with respect to $L = t$. In the coherent regime, assuming a uniform medium, this rate is approximately constant. However, in practice, the rate depends on the time $t$ at which the emission occurs, as it is influenced by the time-dependent quenching parameter $\hat{q}(t)$. 

Before discussing the next order, it is useful to note that the virtual contributions exponentiate to all orders.
\begin{align} \label{eq:no-rad-prob}
&P_{\rm no-rad} (\epsilon) =\nn
&  \delta(\epsilon) \left(1- \int_0^L \rmd t_1\int_0^\infty \mathrm{d}\omega\,\frac{\mathrm{d}I}{\mathrm{d}\omega \rmd t_1}\,+  \int_0^L \rmd t_2 \int_0^{t_2}  \rmd t_1\int_0^\infty \mathrm{d}\omega_1\int_0^\infty \mathrm{d}\omega_2\,\frac{\mathrm{d}I}{\mathrm{d}\omega \rmd t_2}\frac{\mathrm{d}I}{\mathrm{d}\omega \rmd t_1}+...\right) \nn&\quad \quad \quad=\delta(\epsilon) \exp\left[ - \int_0^L \rmd t\int_0^\infty \mathrm{d}\omega\,\frac{\mathrm{d}I}{\mathrm{d}\omega \rmd t} \right]\,\nn
&\quad \quad \quad=\delta(\epsilon) \Delta(L)\,.
\end{align} 
At $\mathcal{O}(\bar{\alpha}^2)$, excluding the virtual contributions that exponentiate to all orders, as discussed earlier, the emission of two successive gluons with energies $\omega_1$ and $\omega_2$, occurring at times $t_1$ and $t_2$, respectively, results in:
\beq 
P_{2,\rm real} (\epsilon) = \int_0^L \rmd t_2 \int_0^{t_2}  \rmd t_1\int_0^\infty \mathrm{d}\omega_1\int_0^\infty \mathrm{d}\omega_1\,\frac{\mathrm{d}I}{\mathrm{d}\omega \rmd t_2}\frac{\mathrm{d}I}{\mathrm{d}\omega \rmd t_1} \, \delta(\epsilon-\omega_1-\omega_2)\,.
\eeq
The time integrations decouple, but this comes at the cost of a symmetry factor of $1/2!$. Multiplying $P_{2,\mathrm{real}}(\epsilon)$ by the exponential factor in \eqn{eq:no-rad-prob} yields the exclusive probability of energy loss due to the radiation of exactly two gluons. 

The generalization to arbitrary numbers of gluons is straightforward and consists in summing over all exclusive probabilities as follows \cite{Baier:2001yt,Salgado:2003gb}
\begin{align}\label{eq:poisson}
P(\epsilon,L) = \delta(\epsilon)\,\Delta(L)  \left[ \delta(\epsilon) +  \sum_{n=1}^\infty \frac{1}{n!} \prod_{i=1}^n\int_0^{L} \rmd t_i\int_0^{+\infty} \rmd \omega_i \frac{\mathrm{d}I}{\mathrm{d}\omega_i \rmd t_i} \, \delta(\epsilon-\sum_{i=1}^n)  \right]\,,
\end{align}
where $\Delta(L)$, defined in \eqn{eq:no-rad-prob}, denotes the probability to radiate between $0$ and $L$. 

Note that, although the integral of the spectrum over the gluon frequency diverges in the infrared (IR), this divergence is canceled by the corresponding virtual contributions at each order in perturbation theory, ensuring that the probability distribution remains IR finite.

In order to make contact with the rate equation that will be discussed in the following chapters, we observe that  \eqn{eq:poisson} is solution of the evolution equation 
\beq\label{eq:eloss-rate-eq}
 \frac{\del}{\del t }P(\epsilon,t) = \int_0^{+\infty} \rmd \omega \frac{\rmd I}{\rmd \omega \rmd t }  \left[P(\epsilon-\omega,t) -P(\epsilon,t) \right]\,.
\eeq 
Introducing the Laplace transform of the energy loss distribution (and its inverse): 
\beq
 Q_\nu (t)=\int_0^{+\infty}\rmd \epsilon  \, P(\epsilon,t) \,\rme^{-\epsilon \nu }\,, \quad \text{and} \quad  P(\epsilon,t)  = \int_{c-i\infty}^{c+i\infty} \frac{\rmd \nu}{2 \pi i } \, Q_\nu (t)\,\rme^{\nu \epsilon}\,, 
\eeq
where the $\nu$ integral runs vertically to the real axis. The value $c$  (the real part of $\nu$) is chosen such that it is greater than the real parts of all singularities of the integrand. 

We can readily show that the solution takes the exponential form in Laplace space
\beq \label{eq:LT-LO}
 Q_\nu (L)\,= \exp\left[ \int_0^t \rmd L\int_0^{+\infty} \rmd \omega \frac{\rmd I}{\rmd \omega \rmd t } (1-\rme^{-\nu \omega})\right]\, \,,
\eeq

where the cancellation of the IR divergence, $\omega\to 0$, is explicit. 

Using the simple form of the spectrum in the LPM regime given by \eqn{eq:bdmps-coh}, we can easily integrate over $\omega$, using the standard integral
\beq
 I(a)=\int_0^{+\infty} \rmd x \, x^a \, (1-\rme^{-x})=-\Gamma(1 + a)\, ,
\eeq
for $-2 < {\rm Re}(a) <1$, where $\Gamma(1 + a)$ is the gamma function.  In our case $a=1/2$ and thus,  $I(-1/2)= -\sqrt{\pi}$. As a result, we obtain for the \eqn{eq:LT-LO} ,
\beq\label{eq:qw-lo}
 Q_\nu = \exp\left( -\sqrt{\pi \omega_s \nu }\right)\,,
\eeq
whose inverse transform yields the following energy loss distribution: 
\beq\label{eq:qw-approx}
P(\epsilon) = \sqrt{\frac{\omega_s}{\epsilon^3}}  \exp\left(-\frac{\pi \omega_s}{\epsilon} \right)\,
\eeq
This is the approximation used in \cite{Baier:2001yt} and \cite{Salgado:2003gb}, where the concept of quenching weights was introduced. It explicitly demonstrates the heuristic argument presented in Sec.~\ref{sec:mean-eloss}, showing that the scale governing the typical energy loss, where $P(\epsilon)$ peaks, is significantly softer than the critical scale $\omega_c$. 

Of course, this approximation breaks down when $\epsilon \sim \omega_c$, i.e., in the tail of the distribution. At first glance, this might seem concerning, given that we have shown the mean energy loss is sensitive to the ultraviolet tail of the distribution. However, in the context of jet energy loss, only radiation emitted at angles larger than the jet angle $R$ contributes to the energy loss. Such gluons are typically much softer than $\omega_c$. Indeed, hard gluons, which are predominantly collinear, are radiated within the jet at angles $\theta_c < R$ and do not contribute to the energy loss. Furthermore, these are rare events and can be treated perturbatively, order by order in the coupling constant.

\section{The quenching of hadron spectra}\label{sec:hadron-quenching}
We proceed with a review of the BDMPS-ZW approach to ``jet quenching," \cite{Baier:2001yt} which focuses exclusively on parton energy loss as a simplified approximation for jet, or more precisely, hadron energy loss. This relatively straightforward framework, which neglects the effects of collinear splittings arising from the evolving virtuality of the initial parton, provides an introductory foundation for understanding the core elements of jet quenching physics. In Chapter~\ref{chap:jeteloss}, the problem of the energy lost by a fully developed parton shower will be addressed. 

The jet spectrum in vacuum, or more specifically in proton-proton (pp) collisions, decreases steeply with increasing jet energy or transverse momentum ($p_T$). This behavior is well approximated by a power-law function:
\beq
\frac{\rmd \sigma_{\rm pp}}{ \rmd p_T}  \sim \frac{1}{p_T^n}\,.
\eeq 
The power-law exponent $n \gg 1$ is typically on the order of $5$–$6$ for jet production and even larger for hadrons at RHIC and LHC energies. Consequently, $n$ serves as a large parameter, enabling a series of well-justified and controlled approximations. 

Since the physics governing the spectrum is dominated by short-distance processes, the resulting cross-section in heavy-ion collisions factorizes into two components: a short-distance hard part, which—aside from potential modifications due to nuclear PDFs—is identical to that in proton-proton (pp) collisions, and an energy loss probability distribution that encapsulates the medium's effects, 
\beq\label{eq:AA-xsec-LO}
\frac{\rmd \sigma_{_{\rm AA}}}{ \rmd p_T}  \sim \int \rmd E  \int \rmd \epsilon \, \delta(E- p_T-\epsilon) \,P(\epsilon) \, \frac{\rmd \sigma_{\rm pp}}{ \rmd E} \,,
\eeq 
where the nucleus-nucleus (AA) cross-section is appropriately normalized to account for the scaling w.r.t. nucleon-nucleon collisions.  Now, we can take advantage of the fact that $n$ is large to derive a closed analytic form. The initial transverse momentum, $p_T$, in the vacuum spectrum is reduced by an energy loss $\epsilon$, resulting in a lower final transverse momentum. Assuming that 
\beq 
n\gg1 \quad \text{ and } \quad \frac{n \epsilon }{p_T}\sim 1 \,,
\eeq
that is, $\epsilon \ll p_T$, we can make the following approximation
\begin{align}
\frac{1}{(p_T+\epsilon)^n} &=  \frac{1}{p_T^n} \exp\left[ -n \ln\left(1+ \frac{\epsilon}{p_T}\right)\right]  \nn
& \simeq \frac{1}{p_T^n} \exp\left( -\frac{n\epsilon}{p_T}\right)(1+\cO(n \epsilon^2/p_T))\,.
\end{align}

The resulting exponential dependence on $\epsilon$ allows us to use the simple form of the Laplace transform in \eqn{eq:AA-xsec-LO} as follows
\beq\label{eq:aa-xsec-lo}
 \frac{\rmd \sigma_{_{\rm AA}}}{ \rmd p_T}  \simeq  \frac{\rmd \sigma_{\rm pp}}{ \rmd p_T} \int_0^\infty \rmd \epsilon \exp\left( -\frac{n\epsilon}{p_T}\right) P(\epsilon) = \frac{\rmd \sigma_{\rm pp}}{ \rmd p_T} \, Q_{\nu=n/p_T}(L)\,.
\eeq 
Introducing the nuclear modification factor as the ratio of the $AA$ to the $pp$ cross-section and inserting \eqn{eq:qw-lo}, letting $\nu=n/p_T$,  in \eqn{eq:aa-xsec-lo}, we obtain the well known BDSM approximate result 
\beq\label{eq:raa-approx}
R_{_{\rm AA}}  \approx \exp\left( -\sqrt{\frac{\pi n \abar^2\hat q L^2 }{ p_T} }\right)\,.
\eeq
Let us summarize the key insights from this result, which will be crucial when generalizing our approach to fully reconstructed jets. First, we note that while the mean parton energy loss may provide a reasonable approximation when $R_{_{AA}} \sim 1 + \langle \epsilon \rangle / p_T \sim 1$, it is expected to become less accurate when $R{_{AA}} \ll 1$. The transition occurs when the exponent in \eqn{eq:raa-approx} becomes of order unity, specifically when $\omega_s = \bar{\alpha}^2 \hat{q} L^2 \sim p_T / n$. This indicates that the scale governing the strong suppression of leading partons can be an order of magnitude smaller than the jet $p_T$, primarily due to the bias introduced by the steeply falling spectrum.

This concludes our brief overview of radiative parton energy loss. In the remainder of this report, we will extend this framework to address jet energy loss, which involves answering two key questions:
\begin{itemize} 
\item 
What is the effect of further branchings of the primary gluon emissions discussed above?
\item What is the energy loss distribution of multiple collinear partons that constitute the energetic core of the jet?
\end{itemize}

%% file: formalism.tex
\chapter{Elements of the formalism}  \label{chap:formalism}

In this chapter, we introduce the field-theory formalism that underpins the computations presented in subsequent chapters. The foundation of our approach draws inspiration from the light-front formalism of high-energy scattering, where the light-cone gauge plays a central role. In this gauge, gluon emissions from charges moving in the direction opposite to the jet are absent, significantly simplifying the calculations. Another key aspect of our methodology is the use of perturbation theory in the presence of a background field—representing the QGP—which is treated to all orders, at least in the initial stages.

\section{Collinear approximation}

We aim to describe the propagation and fragmentation of high-energy partons, produced early in a hard collision on time scales of order the inverse of their energy, $t_0 \sim 1/E$. As they traverse the quark-gluon plasma (QGP) over a time of typically a few fermis, they undergo multiple interactions with the soft partons that make up the transient QGP formed in the same heavy-ion collision. The typical momentum exchange between the hard parton and a plasma constituent is governed by the screening mass, which for a weakly coupled plasma scales as $m_D \sim gT \ll E$. This large separation of scales, a central theme of this work, enables the interaction to be systematically described to all orders in perturbation theory using a strong background field, $A_\text{bkg}$.

The direction of propagation of the initial hard parton, which closely matches the
final jet direction, is usually chosen along the $z$-axis. Because jets are boosted and collimated objects, the transverse scales involved in their dynamics are much smaller than the jet energy, which is the largest scale in the problem:
\beq
p_\perp \ll E\,,
\eeq
where the transverse momentum $p_\perp$ is defined with respect to the jet axis. This transverse momentum can be generated by collinear splitting or elastic interactions with the medium.

This large separation of scales allows for the construction of an effective field theory to describe jet evolution in the plasma.

It is customary to use light-cone variables, where the 4-momentum $p$ of a parton is expressed as:
\beq
p^+ = \frac{1}{2}(E + p_z), \qquad p^- = E - p_z, \qquad \p \equiv (p_x, p_y),
\eeq
where $E$ is the energy, $p_z$ the longitudinal momentum, and $\p$ the transverse momentum. With this convention, the 4-coordinate vector becomes:
\beq \label{eq:lc-var}
x \equiv (x^+, x^-, \x)\,,
\eeq
where $x^+ = (t + z)/2$, $x^- = t - z$, and $\x \equiv (x, y)$.

The angle $\theta$ formed by the 3-momentum $\vec{p}$ and the jet axis is related to the transverse component by $p_\perp \equiv |\p| \sin \theta$. Because jets are collimated objects, their angular broadening is small. Thus, we assume $\theta \ll 1$ in all calculations, leading to the approximations:
\beq
p^+ \simeq E, \qquad |\p| \equiv p_\perp \simeq E \theta,
\eeq
and the light-cone time can be identified with the real time:
\beq
x^+ \simeq t \simeq z.
\eeq
Corrections to this approximation are power-suppressed in $\theta \sim p_\perp / E$.

We work in the light-cone gauge $A^+ = 0$, a convenient choice in this context. In this gauge, the current $J^-$, which describes the medium through which the jet propagates, couples to the $+$ component of the gauge field, which vanishes. This suppresses gluon radiation off particles propagating in the $-z$ direction, so only the jet constituents propagating in the $+z$ direction radiate. Additionally, the light-cone gauge facilitates a partonic picture of radiation dynamics.

The choice of light-cone gauge, combined with the special role played by the time variable labeling interactions with the medium, naturally motivates a hybrid representation involving both momenta and time, instead of the standard covariant formulation used in vacuum. Fourier transforming with respect to $p^-$ results in the following change of representation:
\beq
(p^+, p^-, p_\perp) \quad \rightarrow \quad (p^+, x^+, p_\perp).
\eeq

Although not evident at first glance, this approach shares many similarities with light-front perturbation theory (LFPT), where quantization is performed on the light-cone \cite{Carbonell:1998rj,Burkardt:1995ct,Brodsky:1997de}. However, the approach adopted here begins with a fully covariant description, followed by Fourier transformation with respect to the $-$ component of momenta. This reduction simplifies the dynamics to a 2+1-dimensional non-relativistic quantum mechanics framework, where $p^+$ acts as the mass and $x^+$ as the time variable.

\section{Background field method}\label{sec:bkg-field}

The interaction term in the QCD Lagrangian that encodes the interaction of a fast moving quark with the background field greatly simplifies in the limit where $p^+ \gg k^+$ where, $k^+$ is the longitudinal momentum transfer from to the plasma.  Performing a multipole expansion as $k^+\to 0$: 
\beq
 g  \int \rmd k^+ \bar \psi(-p) \slashed{A}_\bkg(k) \psi(p-k) \simeq g  \bar \psi(-p)\left( \int \rmd k^+ \ \slashed{A}_\bkg (k)\right)\psi(p-k)\Big|_{k^+=0} + \cO(k^+)\,,
\eeq
enables us to apply the integral over $k^+$ solely on the gauge field leading to 
\beq
\int \rmd k^+ A_\bkg^\mu(k) = A_\bkg^\mu(x^-=0)\,.
\eeq
To leading power, $A^+ \sim k^+ \ll p^+$ can be neglected in a general gauge. In light-cone gauge, it is zero by choice. The leading interaction term involves the component $A^-$, as it is integrated over the length of the jet trajectory, i.e., $\int \rmd x^+ A^- \sim L A^-$, while $\int \rmd \x \cdot \A_\perp$ can be neglected since it has support along a transverse segment of size $x_\perp \sim \theta L \ll L$. As noted earlier, the angle explored by the high-energy parton is small, $\theta \ll 1$.

Consider for instance a quark field moving in the QGP background field. As it is customary in LFPT, we can integrate out the subleading  component of the Dirac spinor, known as the ``bad'' component,  $\chi = P_- \psi$, by using the constraint that relates $\chi$ to $\phi$, and we left with the dominant component $ \phi = P_+ \psi $. Here the projectors are given by:$P_+=\gamma^+\gamma^-/2$ and $P_-=\gamma^-\gamma^+/2$, with $P_++P_-=1$.

Integrating the bad spinor component, which is done by using the constraint that relates $\chi$ to $\phi$, one obtains the leading power Lagrangian: 
$\phi^\dag(x) \left[D^--\D^2/ 2\del^++ (F^{12}/\del^+) \Sigma^3 \right] \phi(x) $ 
where $\Sigma^3= \frac{i}{2}[\gamma^1,\gamma^2] $. Here, $D^\mu \equiv \del^\mu-ig A^\mu$ and $F^{\mu\nu}\equiv \del^\mu A^\nu -\del^\nu A^\mu-ig [A^\mu,A^\nu]$ stand for the covariant derivation and the field strength tensor. 
Th term proportional to the magnetic field along the jet direction of motion, i.e., $F^{12} = \vec{v}\cdot \vec{B}$ is responsible for spin flip, which is neglected in the high energy limit. This Lagrangian describes the motion of non-relativistic spin $1/2$ particle in a background field. Besides the two-component spinor there is no dependence in $x^-$. Furthermore, because the non-local operator $1/\del^+$ takes a simple form is momentum space we can trade the integral over $x^-$ in the action with an integral over $k^+=-i\del^+$. 

There are two potential expansion parameters: $g$ and $1/k^+$. To $\mathcal{O}(1/(k^+)^2)$ and $\mathcal{O}(A/k^+)$ accuracy, one can neglect the transverse field. Moreover, terms involving the transverse component are suppressed compared to the $A^-$ term by a factor of $k_\perp / k^+ \ll 1$. However, the kinetic term $k_\perp^2 / k^+$ will not be neglected. This allows for the resummation of sub-eikonal power corrections of the form \cite{Kovner:2003zj}:
\beq\label{eq:q-phase}
 \frac{k_\perp^2}{2 k^+} L\sim1\,,
\eeq
With $x^+$ playing the role of time variable, under the above approximation we have the following hierarchy of time scales:
\beq
 \frac{1}{gA^-} \, \ll \, \frac{k^+}{k_\perp^2} \, \ll\, \frac{k^+ }{k_\perp (g A_\perp) }\,.
\eeq
The first time, $t\sim 1/gA^-$ is the time it takes for a scattering to happen with order 1 probability. This interaction has mostly the effect of rotating the color of the quark and simply amounts to multiplying the wave function by a phase in QED. 
The second time, $k^+/k_\perp^2$,  is the time scale (or phase) associated with quantum diffusion. The last time scale that shall be assumed to be infinite is related to the interactions with the magnetic field of the plasma. 

In the previous section, we used the notation $\omega$ to represent the gluon frequency. Moving forward, we will identify $\omega$ with the longitudinal component of the gluon momentum, such that $k^+ \sim \omega$.

\subsection{Eikonal approximation and Wilson lines  }\label{eq:Wilson-line}
For sufficiently high energies, in the strict $k^+ \to +\infty$ limit, the quantum phase $k_\perp^2 / k^+$ can be neglected, and the propagation of the energetic quark is simply described by a path-ordered exponential:
\begin{equation}\label{eq:Wilson-line}
U(n) \equiv \mathcal{P} \exp\left[ ig \int_0^\infty \rmd s \, \bar{n} \cdot A_\text{bkg}(x(s)) \right],
\end{equation}
where, for the following choices, $n \equiv (n^+, n^-, n_\perp) = (1, 0, 0)$ and $n \equiv (n^+, n^-, n_\perp) = (0, 1, 0)$, two null vectors define the jet direction and its opposite. Here, $A \equiv A^a t^a$ are matrices in color space, with $t^a$ being the generators of SU(3).

Assuming the quark starts at an initial transverse position $x_\perp$, we have
\begin{equation}
x^\mu(s) \equiv s n^\mu + x_\perp^\mu \, .
\end{equation}
In this case, the transverse position is frozen, and the light-cone time $x^+$ is the only variable:
\begin{equation}
\bar{n} \cdot A_\text{bkg}(n s) = A_\text{bkg}^-(x^+, x_\perp, x^- = 0).
\end{equation}

The Wilson line in \eqn{eq:Wilson-line} is then the solution of the following equation:
\begin{equation}
\left( \partial^- - ig A_\bkg^-(x^+, x_\perp) \right) U_\x(x^+,0) = 0\,,
\end{equation}
where we have used a slightly different notation to make the dependence on the coordinate explicit: $U(n) \to U_\x(x^+,0)$. 

\begin{figure}[t!]
\begin{center}
\includegraphics[width=0.5\textwidth]{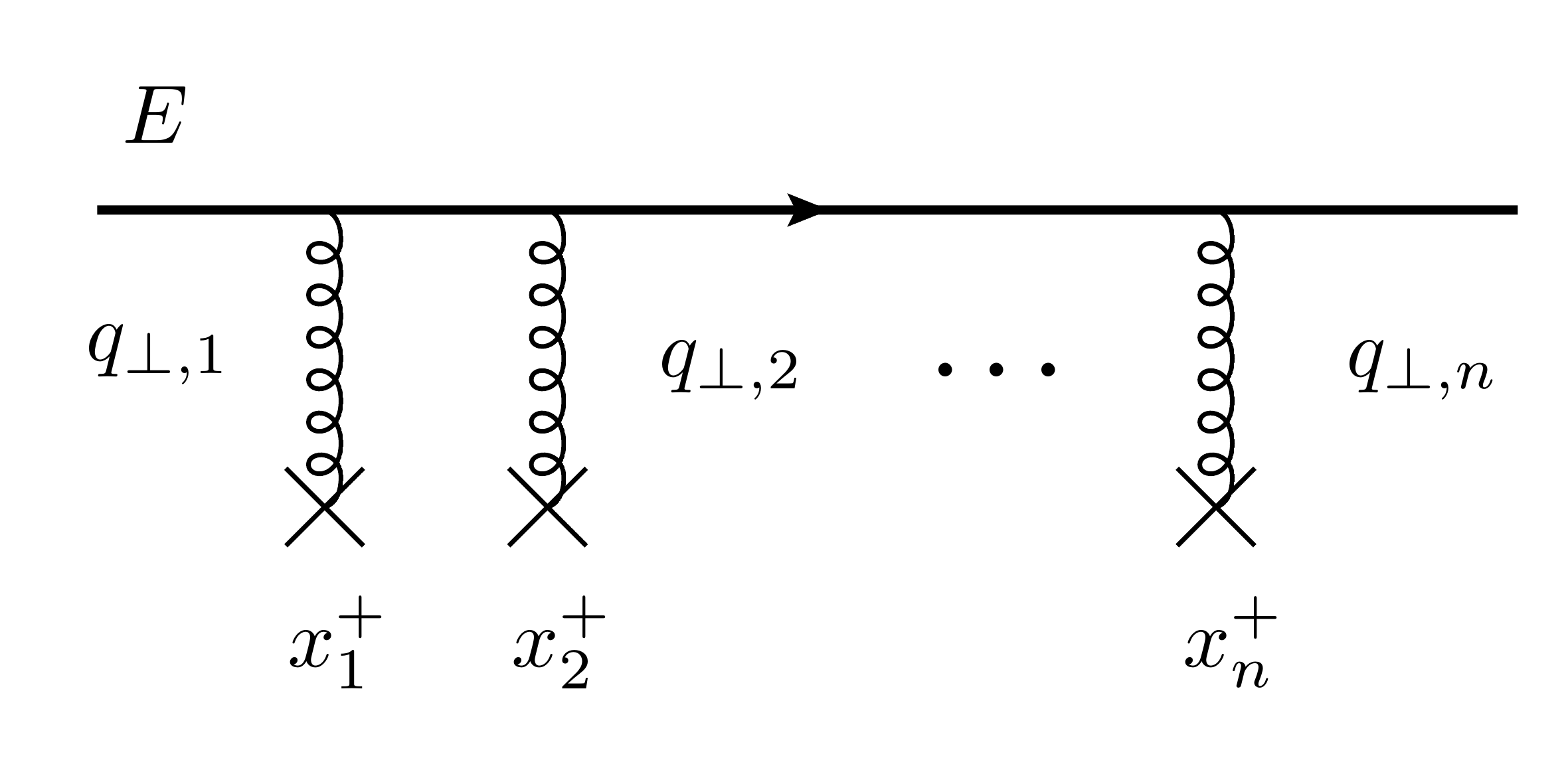}
\caption{Illustration of a eikonal propagation encoded in Wilson lines. The longitudinal momentum $p^+= E$ of the energetic parton is much larger than the momentum transfers to the medium. The background field is depicted vertical gluon lines. }
\label{fig:wilson-line}
\end{center}
\end{figure}

Wilson lines are natural objects in factorization approaches to QCD \cite{Becher:2014oda,Collins:2011zzd,Balitsky:2001gj}. They encode the physics of soft radiation and soft interactions with the medium to all orders. An illustration is given in FIg.~\ref{fig:wilson-line}.
In momentum space we can represent the Wilson line as an infinite series in the gauge field
\beq
U_\x(x^+,0) &= &1 + ig \int_0^{x^+} \rmd x_1^+ \int_{\q_1}A_\bkg^-(x_1^+,\q_1) \nn 
&+& (ig)^2 \int_0^{x^+}\rmd x_1^+ \int_{\q_1}A_\bkg^-(x_2^+,\q_2) \int_0^{x^+_2} \rmd x_1^+ \int_{\q_1}A_\bkg^-(x_1^+,\q_1) +...
\eeq

In contrast to the hard collinear partons, whose interaction is well described by Wilson lines, the propagation of semi-hard gluons—such as those radiated by fast color charges—through the medium is not eikonal. To see this, consider the quantum phase for transverse momenta of the order of the medium scale $k_\perp^2\sim Q_\text{med}^2 \sim \hat{q} L$:
\begin{equation}
\frac{k_\perp^2}{2k^+} L \sim \frac{\omega_c}{k^+},
\end{equation}
where we observe that the eikonal approximation, which leads to the Wilson line operator, is no longer valid for frequencies in the LPM regime:
\begin{equation}
k^+ \equiv \omega \ll \omega_c.
\end{equation}
We will discuss the non-eikonal propagator in the next section.

\subsection{Propagators in the background field  } \label{eq:propagator}

The hard gluons that compose a collimated jet travel along  nearly straight-line trajectories, and interact with the color field of the medium. Only the $-$ component of this field couples to the high energy gluon with the helicity conserving vertex, $V^{ij+}A_\bkg^-\sim p^+ A_\bkg^- \delta^{ij}$, where $i,j = 1,2$  depict the two transverse components of the physical polarization vector $\epsilon_\lambda^i$. The transverse component of the field can be removed by making use of the additional degree of freedom of the light-cone gauge \cite{Blaizot:2008yb}. Moreover, the gluon with very large longitudinal momentum $p^+$, propagates near the light-cone branch, $x^- = 0$, so that, finally, the only component of the medium field that matters is $A_\bkg(x^+,\x)\equiv \left. A_\bkg^-(x)\right|_{x^-=0}$.

The choice of the light-cone gauge along with the eikonal approximation for the vertex imply that the gluon propagator in the medium background field is diagonal in helicity space, i.e., $G^{ij}\sim \delta^{ij} G$, with  $G$ obeying the following equation, 
\beq\label{eq:backprop}
\left[\Box-2 ig \del^+ A_\bkg^-(x)\cdot T\right] G(x, x')=\delta^{(4)}(x-x'),
\eeq
where $g$ is the gauge  coupling constant.  

\begin{figure}[t!]
\begin{center}
\includegraphics[width=0.5\textwidth]{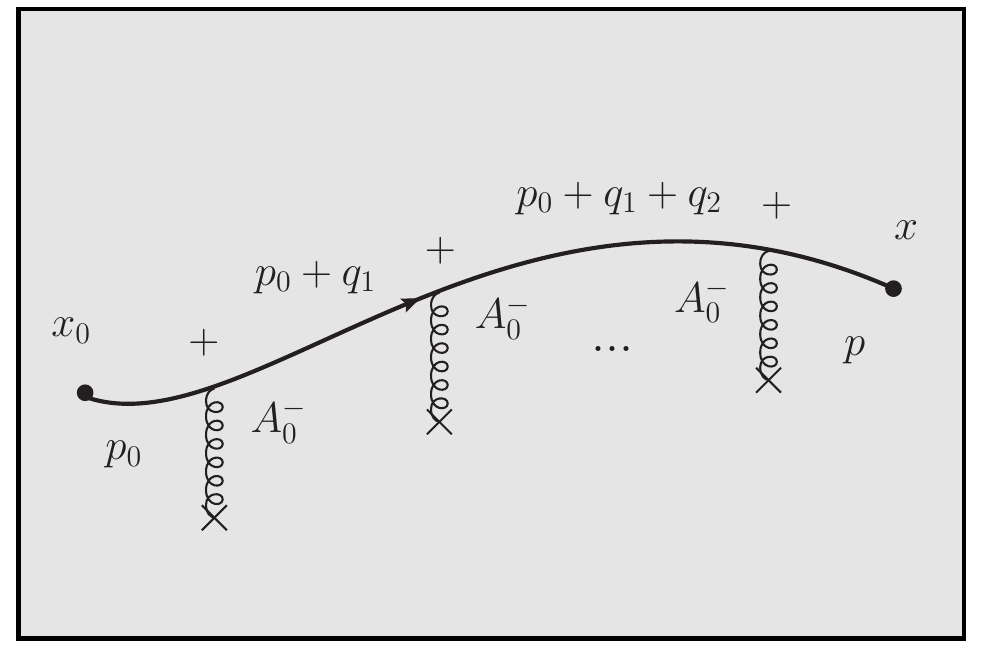}
\caption{Illustration of a non-eikonal propagation in the background field of the plasma. Even though the longitudinal momentum $p^+= E$ of the energetic parton is much larger than the momentum transfers to the medium the large extent of the medium results in non-eikonal effects encoded in the free propagation between interactions.  }
\label{fig:noneik-prop}
\end{center}
\end{figure}
This propagator encodes non-eikonal propagation in the plasma via quantum diffusion in transverse space between two interactions, while the interaction with the field remains strictly eikona (cf. Fig.~\ref{fig:noneik-prop}). reduces to a Wilson line in the limit when $\del^+ \sim  p^+\to \infty$. This approximation amounts to reduce the D'Alembertian  $\Box= 2\del^+\del^- -\del_\perp^2\sim 2\del^+\del^-$, where the transverse part responsible for brownian motion in transverse plane, is neglected. As a result \eqn{eq:backprop} becomes: 
\beq\label{eq:prop-wl}
2 \del^+\left[ \del^- -ig A_\bkg^-(x)\cdot T\right] G(x, x')=\delta(x-x')\,,
\eeq
whose solution is given by the eikonal form
\beq
G(x, x') = G_{\rm eik}(x, x') \equiv \theta(x^--x'^-)\theta(x^+-x'^+) \delta^{(2)}(\x-\x')U_\x(x^+,x'^+)\,
\eeq
where $ U_\x(x^+,x'^+)$ is a Wilson-line in the fundamental representation of the gauge group.

For future reference we present here the quark and gluon propagator  propagators in momentum space: 
\beq\label{eq:quark-prop}
D(p,p_0)  =  (2\pi)^4\delta^{(4)}(p-p_0)D_0(p) + \frac{\slashed{p}  \gamma^+  \slashed{p_0} }{2 p^+} \, \left[ G_\scal(p,p_0) - G^0_\scal(p)\delta(p-p_0)\right], 
\eeq
for the quark and 
\beq\label{eqntransfermatrix3}
G^{\mu\nu}(p,p_0)= -   (2\pi)^4 \delta^{(4)}(p-p_0)  \, G_0(p)\delta^{\mu -} \delta^{\nu -}\,\frac{1}{(p^+)^2} +d^{\mu i}(p)d^{i\nu}(p_0) G_\scal(p,p_0).
\eeq
A direct calculation, using the explicit expression of $d^{\mu\nu}$ given in Eq.~(\ref{freepropagator}) reveals that
\beq
G_0(p)\left[- d^{\mu\nu}(p) -d^{\mu i}(p)d^{i\nu}(p)\right]=- \delta^{\mu -}\delta^{\nu -}\,\frac{1}{(p^+)^2}.
\eeq
for the gluon, where
\beq\label{freepropagator}
d^{\mu\nu}(p)\equiv -i\left[   g^{\mu\nu}-\frac{p^\mu n^\nu+p^\nu n^\mu}{n\cdot p} \right], \qquad G_0(p)\equiv \frac{1}{p^2+i0},
\eeq
and 
\beq
D_0(p) = \frac{i\slashed{p}}{p^2+i0}\,.
\eeq

Moving forward, drop the light-cone time notation for standard time to alleviate them: 
\beq
x^+ \to t\,. 
\eeq
We have already mentioned that the jet probes the medium gauge field only near the light cone, i.e.  $x^-\approx 0$. We may then assume that over such small distance, the medium gauge field is constant, and perform a Fourier transform with respect to $x^-$.\footnote{Since $x^-$ is the conjugate variable to $p^+\sim E$, assuming that the field $A$ does not depend on $x^-$ leads to the conservation of $p^+$ at each vertex coupling the field to the parton. Alternatively, this conservation law could be obtained by observing that the transfer of $+$ momentum from the plasma to the parton is negligible compared to the large energy $E$ of the parton.} We define
\beq
(\x|  \cG(t,t')|\x') =\frac{i}{2E} \int \frac{dx^-}{2\pi }  \rme^{-iE (x-x')^-} G(x-x'), 
\eeq
and from Eq.~(\ref{eq:backprop}) above, one finds that $\cG(t,t')$ obeys the following
 2-dimensional Schr\"odinger equation
\beq\label{SchroG}
\left[i\frac{\del}{\del t} + \frac{\del^2_\perp}{2 E}+gA(t,\x) \right] (\x|\cG(t-t')|\x')=i\delta(t-t')\delta(\x-\x'),
\eeq
where the background field, $A\equiv A_a T^a$, with $T^a$ the generators of $SU$(3) in the adjoint representation, is a matrix in color space, and so is the propagator $\cG$. In the high energy limit, the interesting dynamics is therefore localized in the transverse plane. It is the 2-D equivalent   of a non-relativistic particle in a background potential, where the invariant $p^+= E$ component of the momentum plays the role of the mass.

The propagation of the gluon from $(t_0,\x_0)$ to $(t,\x)$ is described by the Green's function $(\x|\cG(t,t_0)|\x_0)$ that  solves the  2-dimensional Schr\"odinger equation (\ref{SchroG}) (see e.g. Ref.~\cite{Kovner:2003zj}).
Its solution can be formally represented by a path integral in transverse coordinate space:
\beq\label{prop-G}
(\x|\cG(t,t_0)|\x_0)=\Theta(t-t_0)\, \int^{\x}_{\x_0} \cD\r \exp\left[i\frac{E}{2}\int_{t_0}^t \rmd t' \dot \r^2(t') \right]\, U_\r(t,t_0),
\eeq
where retarded conditions are used and
\beq
U_\r(t,t_0)=\cP\exp\left[ig \int_{t_0}^t \rmd t' A(t',\r(t'))\cdot T\right]
\eeq
is a Wilson-line along the path $\r(t')$, with $\r(t_0)\equiv \x_0$ and $\r(t)\equiv \x$. 
Between two scatterings, the gluon propagation is free, 
\beq\label{free-prop-G}
(\x|\cG_0(t,t_0)|\x_0)=\Theta(t-t_0)\,  \frac{E}{2\pi i (t-t_0)} \exp\left[\frac{iE(\x-\x_0)^2}{2(t-t_0)}\right].
\eeq
His Fourier transform yields the well-known kinetic phase
\beq
 (\k|\cG_0(t,t_0)|\k_0) = (2\pi)^2\delta^{(2)}(\k-\k_0) \, \rme^{-i\frac{\p^2}{2E}(t-t_0)}\,.
\eeq
It is sometimes convenient to work with the integral form of \eqn{SchroG}, 
 \beq\label{eq:cG-prop-2}
(\z|\cG(t,t_0)|\z_0)=(\z|\cG_0(t,t_0)|\z_0)+ig (\z_1|\cG_0(t,t_1)|\z_1)A_\bkg(t_1,\z_1)(\z_1|\cG(t_1,t_0)|\z_0) \,.
\eeq
The Green's function $\cG(t,t')$ plays the role of a unitary time evolution operator for a free particle evolving from the position state $|\x)$ to the state $|\x')$.
One easily deduces the following composition property (using the matrix notation)
\beq\label{convprop}
\cG (t,t_0)=\cG (t,t_1) \,  \cG(t_1,t_0)\, 
\eeq
for any $t_0 < t_1 < t $, which in coordinate space yields
\beq\label{convprop-x}
(\x| \cG(t,t_0)|\x_0)=\int \rmd \x_1 (\x|\cG(t,t_1)|\x_1)(\x_1|\cG(t_1,t_0)|\x_0)\,. 
\eeq

 In momentum representation,
\beq
|\p) = \int \rmd^2 \x\, \rme^{-i\p\cdot\x}\,|\x), \qquad \int \frac{\rmd^2\p}{(2\pi)^2}\, |\p)(\p| = 1,
\eeq
we define the  amplitude for a gluon of (large) momentum $p_0\equiv(p^+_0,\p_0=0)$, 
present in the system at time $t_0$, to evolve in the medium into a gluon with momentum $p_1=(p^+_1,\p_1)$, where $p^+_0=p^+_1\equiv E$ (to within an irrelevant phase factor),
\beq\label{eq:amp-prop}
{\cal M}(p_1|p_0)=  
(\p_1|\,{\cal
G}(t_1,t_0)\,|\p_0).
\eeq
 The amplitude ${\cal M}\equiv {\cal M}_{ab}$ is a matrix in color space, propagating a gluon with color $b$ to color $a$. It is also a (diagonal) matrix in the helicity space. We do not write explicitly color and helicity indices to alleviate the notation. 

The free Hamiltonian in 2-dimensional space, reads
\beq
\cH_0=\frac{\p^2}{2E}.
\eeq
The interaction part of the hamiltonian is the sum of two components, $\cH_{\rm int}= \cH_1+ \cH_2$,
\beq\label{int-H}
(\q|  \cH_1|\p)=g A(\q-\p,t), \quad\text{and}\quad (\q; \q'|  \cH_2|\p)=g(\q; \q'|V|\p).
\eeq
where $\delta \cH_1$ represents the elastic scattering of a particle off the potential $A(\q,t)$, while  $\delta \cH_2$ represents the splitting of a particle into two from LFPT (see \eqn{vertex} for the gluon vertex). Note that the latter process does not conserve the non relativistic energy $\p^2/2E$, but it conserves the ``mass''  $E$. When using the propagator (\ref{prop-G}) we effectively treat $\cH_1$ to all orders. Later, we shall calculate the branching probability of a gluon to first order in   $ \cH_2$. Multiple branchings will be treated via probabilistic cascades.
\subsection{Independent scattering approximation}\label{sec:field-correlator}
Jet observables can generally be expressed as functions of background field configurations, which must be integrated over. This integration requires knowledge of the statistical properties of the hot medium. These observables are, in principle, non-perturbative quantities that can, for instance, be evaluated on the lattice. However, analytical formulas can be derived by leveraging the large separation between the medium size and the QGP correlation length:

\beq\label{eq:local-int}
\xi_D \ll L \,,
\eeq  
which permits treating medium interactions at distances exceeding $\xi_D$ as independent. These independent interactions are enhanced by the medium size, such that $g A^\bkg L \sim 1$, and thus, need to be resummed to all orders, while interactions over distances of order $\xi_D$ can be computed perturbatively for a weakly coupled medium. This allows neglecting higher-order connected contributions, such as $g \langle A A A \rangle \ll \langle A A \rangle$. This is equivalent to neglecting the nonlinear self-interaction of the background and neglecting the interactions between different scatterers. Alternatively, one may define local non-perturbative operators, a possibility we will revisit when discussing the renormalization of the jet quenching parameter. Neglecting non-Gaussian correlations in the background gauge field would not be allowable if the processes under consideration were sensitive to infrared physics. However, we can argue that jet-medium interactions are dominated by Coulomb interactions, which produce large logarithms arising from perturbative interactions between $m_D$ and the emergent medium scale: $Q_\med \gg \Lambda_\text{QCD}$. 

Under these assumption (\ref{eq:local-int}), the medium correlations can be approximated to obey Gaussian statistics. To leading order, these correlations factorize into a product of two-point correlation functions:

\beq
\frac{1}{d_R} \mathrm{Tr} \langle A^{\ast \mu}(x) A^\nu(y) \rangle =  \int \frac{\rmd^4 q}{(2\pi)^4} \, \rho^{\mu\nu}(q) \rme^{-i (x-y)\cdot q},
\eeq
with the one-point correlation function vanishing, $\langle A^- \rangle = 0$. Here, $\rho^{\mu\nu}(q)$ is the spectral density of gauge field fluctuations  in a homogeneous medium  \cite{Arnold:2002ja}, which can be computed for a plasma in thermal equilibrium in the framework of hard-thermal loop (HTL). However, as alluded to earlier, it is sufficient to confine our discussion the perturbative regime where the momentum transfer is larger than the plasma temperature, $q_\perp \gg T$. 

Performing the Fourier transform w.r.t. to $\x$ and $\y$ fro the case of interest, i.e., $\mu=\nu=-$ and $x^-=y^-=0$: 
\beq
 \frac{1}{d_R} {\rm Tr}\langle A^{\ast -}(x^+,\q') A^{ -}(y^+,\q) \rangle =(2\pi)^2 \delta^{(2)}(\q-\q')\int \frac{\rmd q^-\rmd q^+}{(2\pi)^2} \rme^{-iq^-(x^+-y^+)} \rho^{--}(q)\,.
 \eeq
Now, recall that the time difference $\tau = x^+-y^+$ is bounded by the in-medium correlation length $\xi_D$ and thus, is parametrically much smaller then $x^+\sim L \gg \tau$. In this limit the scattering process in the medium is quasi-instantaneous when compared with the size of the target, resulting in a Markovian dynamics. In particular, since the scattering time is well localized, one can neglect $\tau$ dependence outside the correlator and extend the $\tau$ integral to infinity yielding a delta function $\delta(q^-)$. This simple argument justifies the usual assumption of time locality of the medium averages.

Consequently, integrating over $\tau$ we obtain the elastic rate 
\beq\label{eq:el-rate}
 \,   \gamma(\q)  \equiv \int \frac{\rmd q^+\rmd q^-}{(2\pi)^2} \, (2\pi)\delta(q^-)\, \rho^{--}(q)\,
 \eeq
where we have factored out the density of scattering center $n$ that may depend on time and $\gamma(\q)$ is related to the $t$-channel of the 2 to 2 scattering matrix element at leading ordder:
\beq
\gamma(\q) \equiv \frac{g^2 n}{\q^4} \,.
\eeq

Defining the Fourier transform of $\gamma$: 
\beq
\gamma(\x-\y)  \equiv  \int_\q \,\gamma(\q) \, \rme^{i \q\cdot (\x-\y)}\,.
\eeq
where we have used the shorthand notation for simplicity
\beq
  \int_\q \equiv \int \frac{\rmd^2\q}{(2\pi)^2}   \,.
\eeq
We are presently equipped to write down the effective two-point correlation function at leading power $\xi_D/L$ that will be used to perform systematic computations of observables 
 \beq\label{eq:2-point-corr}
 \langle A_\bkg^{a,\ast \mu} (x^+,\x)  A_\bkg^{b,\nu} (y^+,\y)  \rangle =  \delta^{ab} \delta(x^+-y^+) \, \gamma(\x-\y)\,. 
\eeq  
where the delta function $\delta(x^+-y^+)$ is an effective way to implement the strong separation between the relevant time scales for jet evolution and the medium correlation time. 

The brackets $\lg ...\rg$ stand for the ensemble average. By the delta function in the r.h.s. of Eq.~(\ref{eq:2-point-corr}), we have formally set $\xi_D\to 0$. 

This formulation is reminiscent of the McLerran-Venugopalan model \cite{McLerran:1993ni,McLerran:1993ka} that deals with cold nuclear matter.  Also, Eq.~(\ref{eq:2-point-corr}) is equivalent to the Gyulassy-Wang model in which the medium is modeled as a collection of static scattering centers \cite{Wang:1991xy,Gyulassy:1993hr,Wang:1994fx} .

\section{Transverse momentum broadening}\label{sec:TMB}
In the previous section, we argued that the propagation of an energetic gluon is only minimally affected by multiple interactions within the plasma. Its longitudinal momentum, $p^+ \propto E$, remains conserved throughout its propagation and is significantly larger than its transverse momentum, $p_\perp$, which is gradually acquired through multiple scatterings. Additionally, its helicity remains unchanged during this process.

In this section, we focus on the elastic processes responsible for the transverse momentum broadening of high-energy partons. Our goal is to derive the probability for a high-energy parton to acquire a transverse momentum $p_\perp$ through multiple independent scatterings in the plasma. This derivation represents the first building block of jet quenching calculations, through which we will introduce the jet quenching parameter, $\hat q$.

The eikonal propagation of the gluon from $(t_0,\x_0)$ to $(t,\x)$ is described by the Green's function $ (\x,t|\x_0,t_0)\equiv (\x|\cG(t,t_0)|\x_0)$ that  solves the  2-dimensional Schr\"odinger equation (\ref{SchroG}) (see Ref.~\cite{Kovner:2003zj} for an extended discussion of in-medium gluon propagation).

Let $\cM(\p,t)\equiv (\p,t|\cM)$ be  the amplitude that relates to the gluon production cross-section as follows,
\beq \label{eq:cross-sec}
\frac{\rmd \sigma}{\rmd \Omega_p}\equiv \lg\cM(\p,t)\cM^\ast(\p,t)\rg\,\quad  \text{with} \quad \rmd\Omega_p\equiv\frac{\rmd E\rmd^2\p}{2E(2\pi)^3}
\eeq
The ensemble average $\lg...\rg$ is to be carried out with the help of Eq.~(\ref{eq:2-point-corr}) at the level of the observable (amplitude squared) whose measure is performed over a large number of realizations. 

Using Eq.~(\ref{convprop-x}) we can relate the observable (\ref{eq:cross-sec}) at $t$ to that at an initial time $t_0$, 
\beq\label{2p-fct}
\frac{\rmd \sigma}{\rmd \Omega_p}&=&\lg\cM(\p,t)\cM^\ast(\p,t)\rg \nn
&=&\frac{1}{N_c^2-1} \int_{\p_0\bar\p_0}\lg \rmTr (\p|\cG(t,t_0)|\p_0)(\bar\p_0|\cG^\dag(t_0,t)|\p)\rg \lg\cM(\p_0,t_0)\cM^\ast(\p_0,t_0)\rg\nn
&=&\int_{\p_0\bar\p_0} (\p;\p|S^{(2)}(t,t_0)|\p_0;\bar\p_0)  \lg\cM(\p_0,t_0)\cM^\ast(\p_0,t_0)\rg\nn
&=& \int_{\p_0 } \cP(\p,t|\p_0,t_0) \frac{\rmd \sigma}{\rmd \Omega_{p_0}}, 
\eeq
where the trace in the second line is over the color indices. The quantity $\cP(\p,t|\p_0,t_0)$ has a natural interpretation as a $p_\perp$-broadening probability and it can be easily checked that it is properly normalized, i.e., $\int_\p \cP(\p,t|\p_0,t_0)=1$.  

Due to the instantaneous field correlations, the color structure is simplified, with each pair of connected fields contributing a color factor $C_R$. 

In Eq.~(\ref{2p-fct}), the Fourier transform of a fundamental quantity appears: the two-point function correlator $(X_1|\Sii(t,t_0)|X_0)$. This correlator describes the evolution of a dipole consisting of two gluons with opposite energies, $E$ and $-E$, where $E$ denotes the absolute value. The evolution proceeds from the initial coordinates $|X_0) \equiv |\x_0, \bar{\x}_0)$ to the final coordinates $|X_1) \equiv |\x_1, \bar{\x}_1)$, where $\x_i$ and $\bar{\x}_i$ represent the transverse positions of the gluon and anti-gluon, respectively.

Translational invariance in the transverse coordinate space, encoded in the two-field correlators [cf. Eq.~(\ref{eq:2-point-corr})], manifests in momentum space as an exact balance of transverse momentum between the amplitude and its complex conjugate. Consequently, the correlator satisfies (see \cite{Blaizot:2012fh} for a detailed discussion)
\beq 
(\p; \p|S^{(2)}(t, t_0)|\p_0; \bar{\p}_0) \equiv \delta(\p_0 - \bar{\p}_0) \cP(\p, t | \p_0, t_0),
\eeq
where $\cP(\p, t | \p_0, t_0)$ is the corresponding evolution kernel.

The average of 2-point function over the field configurations using the field correlator (\ref{eq:2-point-corr}), yields the following rate equation for the time evolution of the broadening probability ($\cP(\p,t|\p_0,t_0)\equiv\cP(\p,t) $)\cite{Blaizot:2013vha}
\beq\label{ME-P}
\frac{\del}{\del t} \cP(\p,t) = N_c \, \int_\q \gamma(\q)\,\left[  \cP(\p-\q,t) - \cP(\p,t)\right] \,.
\eeq
The factor $N_c$ follows from the color trace $\Tr T^a T^b = N_c \delta^{ab}$ contracted with \eqn{eq:2-point-corr}. 
The quantity $ \gamma(\q)=g^4/\q^4$ is the rate of elastic scattering encoded in the correlator \eqn{eq:2-point-corr}. Eq.~(\ref{ME-P}) can be solved by Fourier transforming to coordinate space. The r.h.s of Eq.~(\ref{ME-P}) becomes local in $\x$. Hence, it is straightforward to get (for $t\equiv L$)
\beq\label{GM-P}
\cP(\p,L) = \int \rmd^2 \x\, \exp\left[-\frac{N_c}{2} n\sigma(\x) L-i\p\cdot\x \right],
\eeq
which is the Glauber-Mueller formula\cite{Kovchegov:1998bi,Mueller:2012bn}.

Note that the quantum phases in the propagator completely cancel out due to translational invariance. In this cse the broadening probability reduces to the correlation of Wilson lines with relates to the gluon scattering cross-section at high energy 
\beq\label{eq:S-dipole}
S(\x)\equiv \frac{1}{N_c^2-1}\lg \Tr \, U_\x U_\0\,  \rg =\exp\left[-\frac{N_cn}{2} \sigma(\x) L \right] \,,
\eeq
that has the meaning of a forward dipole scattering amplitude \cite{Nikolaev:1990ja}, as it involves the scattering of a pair of gluons  in a color singlet state. Here, the Wilson lines are evaluated between $U_\x \equiv U_\x(+\infty,0)$. 

In the diffusion limit, namely when the final momentum $\p$ is acquired by a large number of small momentum transfers, i.e., $\q\ll \p$, Eq.~(\ref{ME-P})  takes the form of a Fokker-Planck equation
\beq\label{diff-P}
\frac{\del}{\del t} \cP(\p,t) = \frac{1}{4}\left(\frac{\del}{\del \p}\right)^2 \left[\hat q (\p^2) \cP(\p,t)\right],
\eeq
where the momentum dependent diffusion coefficient reads,
\beq\label{qhat}
\hat q (\p^2) = N_c \int_\q \q^2 \gamma(\q)\simeq \frac{g^4N_cn}{4\pi }\ln \frac{\p^2}{m_D^2},
\eeq
to logarithmic accuracy: the lower cut-off of the logarithmic transverse momentum integration is given by the  Debye screening mass $m_D$.
The transport coefficient (\ref{qhat}) is the fundamental parameter that characterizes the properties of the medium which the jet is sensitive to \cite{Baier:1996sk}. It is often referred to in the literature as the quenching parameter.  

The dipole cross-section can also be directly related to the quenching parameter in coordinate space, for small dipole seizes $\x^2\ll  \q^{-2}\ll m_D^{-2}$. Using $ \gamma(\q)=g^2/\q^4$  and expanding the oscillating phase to quadratic order in $\q$, we get
 \beq\label{qhat-r2}
\frac{N_c}{2}n\sigma(\x)=g^4N_cn\int_\q \frac{1-\rme^{i\q\cdot\x}}{\q^4}\approx  \frac{g^4N_cn}{16 \pi} \x^2  \ln \frac{1}{\x^2 m_D^2}\, =  \frac{1}{4 } \x^2 \hat q (\x^{-2}) \,.
\eeq\,
Large transverse momenta are dominated by a single scattering, and  
\beq \label{eq:hard-scat}
\cP_{\rm hard }(\p,L)\propto  \frac{1}{\p^4}\,.
\eeq
It follows that the broadening probability does not admit moments at leading order. 
Nevertheless, in the diffusion approximation, where we can neglect the slow variation of $\hat q$, by taking it to be constant, we can define a typical transverse momentum squared,
\beq\label{kt2-typ}
\lg p^2_\perp\rg_\text{typ}= \hat q L.
\eeq
And so long as  $\p^2 \lesssim \hat q L$,
\beq
\cP(\p,L)\approx \frac{4\pi}{\hat q L} \exp\left(-\frac{\p^2}{\hat q L}\right), 
\eeq
is an approximate solution of Eq.~(\ref{diff-P}) and Eq.~(\ref{ME-P}). 

This approximation also known as the harmonic-oscillator since by neglects the Coulomb logarithm we have a quadratic form for the dipole cross-section:
\beq\label{eq:sigma-ho}
\frac{N_c n}{2}\sigma_{\rm HO}(\x)  \sim   \frac{1}{4} \hat q \, \x^2 \,,
\eeq 
and treating $\hat q$ as a constant. However, at this order we have no control over the scale in the logarithm and therefore $\hat q$ is not properly determined. We will solve this issue below.

A closed analytic form can be obtained using the  Moli\`ere's theory~\cite{Moliere:1948zz,Bethe:1953va} of multiple scattering
 which splits the Coulomb log in \eqn{qhat-r2} into a large constant part and a perturbation that can be treated order by order \cite{Barata:2020rdn}
\beq\label{eq:sigma-split}
\sigma(\x)  &\sim & \x^2 \left[  \ln \frac{Q^2_\med }{m_D^2} +\ln \frac{1}{Q^2_\med \x^2 }  \right]\,\nn
 &=& \sigma^{(0)}(\x)+ \delta \sigma(\x) \,. \nn
\eeq 
At all order in the quadratic part $\sigma^{(0)}(\x)$  and leading order in $\delta \sigma(\x) $ we obtain \cite{Barata:2020rdn}
\begin{align}
\label{eq:golden}
    \cP^{(0)+(1)}(\p,L)=\frac{4\pi}{Q_\med^2} \rme^{-x}\Big[ 1 -\lambda \left(\rme^{x}-2+\left(1-x\right)\,\big({\rm Ei}\left(x\right)-\log (4x\,a)\right) \big)\Big] \, ,
 \end{align}
where 
\beq
x\equiv\frac{\p^2}{Q_\med^2} \,,
\eeq
and lambda is the expansion parameter:
\beq
\lambda \equiv \frac{1}{\ln (Q^2_{\med} /m_D^2)}  \ll 1 \,,
\eeq
where the emergent scale $Q_\med$  solves the transcendental equation, that follows from \eqn{qhat}
\beq
 Q^2 _\med = \frac{g^4N_c\,n\,L}{4\pi }\, \ln (Q^2_{\med} /m_D^2)\sim \hat q(Q^2_{\med}) L\,.
\eeq
The resulting curves  are plotted in Fig.~\ref{fig:theory-plot}. 

We shall see in Chap. \ref{chap:qhat-rg} that the transverse momentum distribution receives large radiative corrections that can be absorbed in a redefinition of the quenching parameter. 

\begin{figure}
\centering
\includegraphics[width=9cm]{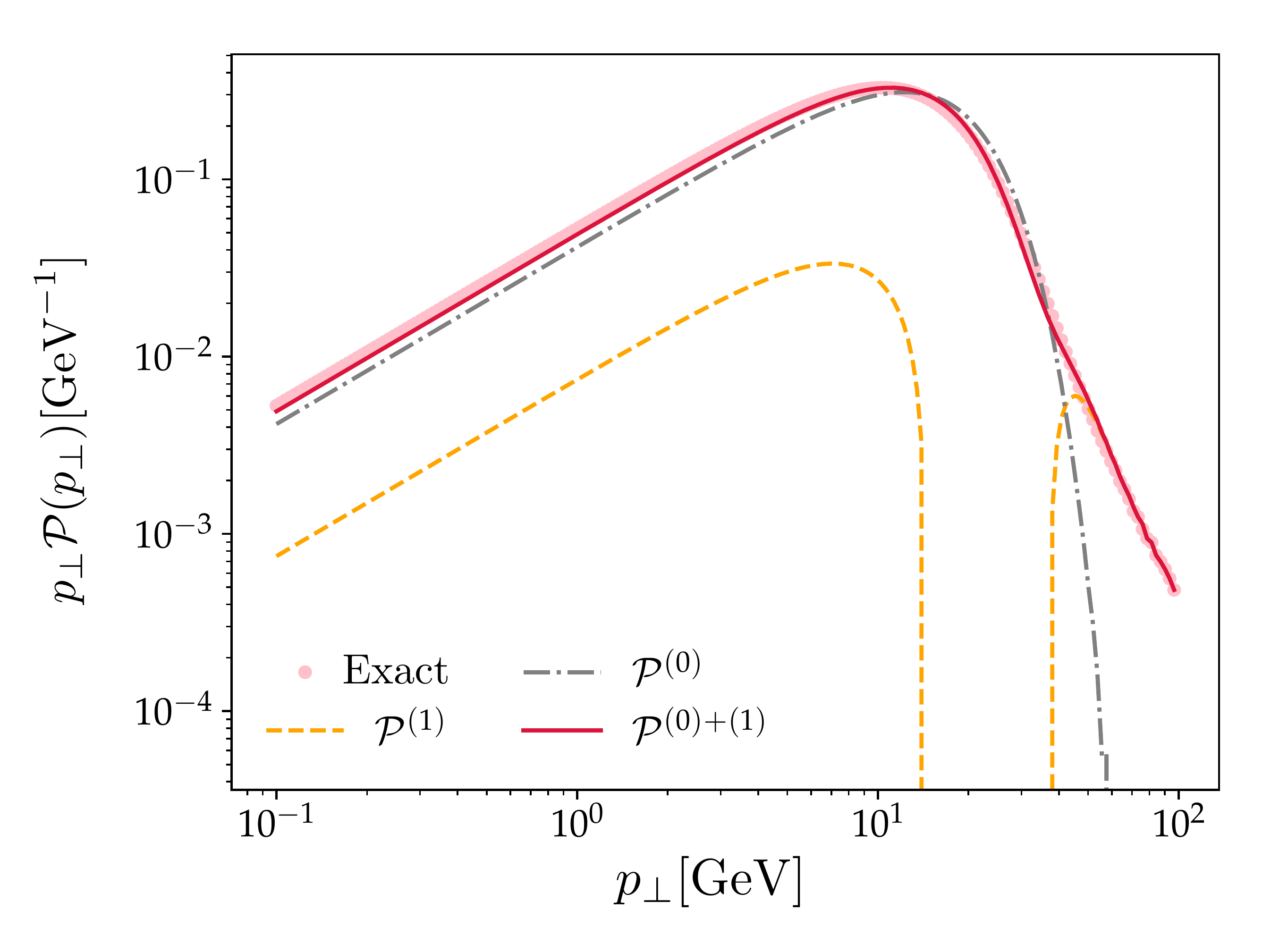}
\caption{Momentum broadening probability distribution at first two orders in in Moli\`ere's-expansion  compared to the exact solution of the Fourier transform \eqn{GM-P} with $\lambda\!=\!0.1$ corresponding to ($Q^2_{\med}\!=\!30$~GeV$^2$, $m^2_D\!=\!0.13$~GeV$^2$). In this and following figures $p_\perp\equiv |\p|$. $\cP^{(0)}$ stands for the Gaussian form order $\lambda^0$ in \eqn{eq:golden}, while the next order  $\cP^{(1)}$ contains the Coulomb tail $1/p_\perp^4$. }
\label{fig:theory-plot}
\end{figure}

\section{Derivation of medium-induced spectrum }\label{sec:soft-rad}

Now that we have derived the transverse momentum broadening probability within our formalism, we turn to the derivation of the soft medium-induced radiative spectrum. Rather than directly computing the spectrum using standard Feynman rules, we use this opportunity to frame the calculation in terms of the parton energy loss probability distribution. This distribution is defined through Wilson line correlation functions, which will also be discussed in detail in Chap.~\ref{chap:decoherence} (cf. Refs.~\cite{Mehtar-Tani:2024smp,Mehtar-Tani:2025xxd} for an all order formulation based on effective field theory techniques).

\beq\label{eq:P1-operator-0}
P(\epsilon) = \frac{1}{d_R} \sum_X \delta(\epsilon - \bar{n} \cdot k_{\rm loss}) \, \tr_c \left[ \langle \med | U(n) | X \rangle \langle X | U^\dagger(n) | \med \rangle \right],
\eeq
where multiple soft-gluon radiation and interactions with the plasma background field are encoded in the semi-infinite Wilson line:
\begin{align}
U(n) &\equiv  \exp\left[ ig \int_0^\infty \rmd x^+ \, A^-(n s) \right],
\end{align}
where, the 4-vector $n = (1, 0, 0, 1)$ points in the jet direction  of propagation, i.e., along the $z$-axis. Here, $d_R$ represents the dimension of the emitter's color representation: $d_R = N_c$ for a quark or $d_R = N_c^2 - 1$ for a gluon. The delta function acts as a measure function. 

Here 
\beq 
\bar{n} \cdot k_{\rm loss}= \sum_{i=1}^\infty \omega_{i,\rm loss}\,
\eeq
denotes the sum over all the radiated gluon energies ($\omega_i\equiv k^+_i$), or more precisely their longitudinal momenta. The main difference with the equation introduced later, \eqn{eq:P1-operator}, is the absence of the jet cone condition $k_\perp > R$ that ensures that only gluon radiated outside the jet contribute to energy loss $\epsilon$ . 

The gauge field $ A^-(n s) $ encodes both the background field $A_\bkg$, and the quantum fluctuations that gives rise to gluon radiation $a^-$\, 
\beq
A^-(n s)=A^-(x^+,x^-=0,\x) \equiv  A_\bkg^-(x^+,\x=0) + a^-(x^+,x^-=\x=0)\,.
\eeq

As a proof of concept, let us compute the leading radiative correction to the soft function $\cU_m$ for a single collinear mode. Namely, we shall consider, for the quark case 
\beq \label{eq:soft-function-1}
&&P  (\epsilon) \equiv \frac{1}{N_c} {\rm tr}_c \sum_{X}   \, \delta(\epsilon - \bar n \cdot k_{\rm loss} ) \,  \langle \med|  U^\dag(n) | X \rangle  \langle X| U(n)| \med \rangle  \,
 \eeq
 
 To zeroth order in the soft-collinear field and all order in the background field we have 
 \beq 
&&P_0 (\epsilon) \equiv \frac{1}{N_c} {\rm tr}_c \sum_{X}   \, \delta(\epsilon - \bar n \cdot k_{\rm loss} ) \,  \langle \med |  U_{\bkg }^\dag(n) | X \rangle  \langle X| U_{\rm \bkg }(n)| \med  \rangle  \,. 
 \eeq
Here, the background field does not contribute to energy loss since $A_\bkg(x^-) \to \delta(k^+)$. This approximation can be relaxed easily leading to collisional energy loss which will be neglected here.  In this situation we can write 
\beq
P_0 (\epsilon)   &=&  \frac{1}{N_c} {\rm tr}_c \sum_{X}   \, \delta(\epsilon  ) \,  \langle \med |  U_\bkg^\dag(n) | X \rangle  \langle X| U_\bkg (n)| \med \rangle  \,\nn
  &=& \delta(\epsilon  )  \frac{1}{N_c} {\rm tr}_c \langle \med|  U_\bkg^\dag(n) U_\bkg(n)| \med  \rangle \nn
    &=& \delta(\epsilon  ) 
\eeq
where we have used the completeness relation   $\sum_{X}  | X \rangle  \langle X|  =1$.

At the next order we expand the Wilson line to linear  order in $a^-$ (the  quadratic order will be necessary for the virtual contribution). That is, 
\beq
 U_1(+\infty, 0) &\equiv& \cP \exp\left[ ig \int_0^{+\infty} \rmd x^+ \left( A^- _\bkg(x^+,0^-,0_\perp) +a^-(x^+,0^-,0_\perp)\right)\right]\nn &\simeq& 1 +i g \int_0^\infty \rmd z^+ U_{\bkg , 0_\perp} (+\infty, z^+) a^-(z^+,0^-,0_\perp)U_{\bkg , 0_\perp} (z^+,0) \,.\nn
\eeq
This formula describes the color precession of the fast-moving gluon between $0^+$ and $z^+$ during the emission of a soft gluon. After the radiation, the fast gluon resumes its color precession in the plasma from $z^+$ to $+\infty$.

The real contribution at leading order yields 
\beq
P_1 (\epsilon,R)  &\approx &   \frac{g^2 }{N_c} {\rm tr}_c \sum_{X}   \,  \,  \langle \med |  \int_0^{+\infty} \rmd z_2^+  U^\dag _\bkg (z_2^+,0)  a^-(z_2^+) U^\dag_\bkg (+\infty, z_2^+)   | X \rangle  \delta(\epsilon -\hat\omega   ) \nn
 &\times&  \langle X|   \int_0^{+\infty}\rmd z_1^+ U_\bkg(+\infty, z^+)   a^-(z_1^+)U_\bkg (z^+,0) )| \med \rangle \Big|_{x_\perp=0_\perp}  \,\nn
 &= &   \frac{g^2 }{N_c}   \sum_{X}  \int_0^{+\infty} \rmd z_2^+  \int_0^{+\infty} \rmd z_1^+   \delta(\epsilon -\hat\omega  ) \nn
 &\times&\langle \med| a^{b,-}(z_1^+) \delta(\epsilon -\hat\omega  )  |X\rangle \langle X| {\rm tr}_c  \left(t^a[z_1^+, z_2^+] t^b[z_2^+, z_1^+]\right) a^{a,-}(z_1^+)| \med \rangle  \,\nn
\eeq
This has the form of an amplitude squared. Assuming that the background field gluons do not contribute to the final state the corresponding Wilson lines cancel out outside the time interval $[z^+_1,z^+_2]$. Here we have used the symbolic notation $\hat \omega$ that denotes an operator that measures the energy radiated out of the cone. 

Using the following algebraic identity we may further simplify by combining the two fundamental Wilson lines into an adjoint one:
\beq
 [z_2^+, z_1^+] t^a[z_1^+, z_2^+] = t^b \cU^{ba} (z_2^+, z_1^+) \,.
\eeq
As a result we have 
\beq
 {\rm tr}_c  \left(t^a[z_1^+, z_2^+] t^b[z_2^+, z_1^+]\right) =\frac{1}{2} \cU^{ba} (z_2^+, z_1^+) 
\eeq
To compute the the correlator of the cs gauge field  we need their propagator in the presence of the medium, i.e., \eqn{eqntransfermatrix3} ,
\beq
&&A^{b,-}_{\rm cs}(z_1^+) \delta(\epsilon - k^+_{\rm loss}  )  |k_{\rm loss}\rangle \langle k_{\rm loss}| A^{a,-}_{\rm cs}(z_2^+) 
\nn &&\to \int \frac{\rmd^{d} k}{(2\pi)^4} (2\pi) \delta(k^2)\theta(k^+) \delta(\epsilon - k^+ ) \nn
&&  k^4\int \rmd^d y\,\rme^{-i k\cdot y } \, G^{\dag, b -}_{\mu}( y, z_2^+)    \int \rmd^d x\, \rme^{i k\cdot x } \,G^{ca,\mu -}( x, z_1^+) \nn
\eeq
Applying the following reduction formula for dressed external propagators, we obtain
\beq
 \lim_{k^2\to 0 } k^2 \int_{k_0} G_{\rm scal}(k,k_0) J(k_0) \to   \int \rmd^{d-2} \k_0  \int \rmd z^+_0 (\k| \cG(+\infty, z^+_0) |\x_0)  J(\x_0,k^+, z^+_0)\nn
 \eeq 
 where $J(k_0)$ is some current, allowing us to connect to our mixed representation where the non-eikonal propagation is encoded in the 2+1D propagators \eqn{SchroG}. Using the polarization decomposition in \eqn{eqntransfermatrix3} yields
 
 \beq
&& P_1 (\epsilon)  \approx    \frac{g^2}{2 N_c}  \int \frac{\rmd^{4} k}{(2\pi)^4}(2\pi) \delta(k^2)\theta(k^+) \delta(\epsilon - k^+ ) \nn
 &&   \int_{\k_1,\k_2}\frac{(\k_1\cdot \k_2)}{(k^+)^2} \,\cU^{ba} (z_1^+, z_2^+)  (\k_2| \cG^{bc}(z^+_2, {+\infty}) |\k)    (\k| \cG^{ca}({+\infty}, z^+_1) |\k_1)  \nn 
 && =   \frac{g^2}{4 N_c}  \int \frac{\rmd  k^+}{ (k^+)^3}  \delta(\epsilon - k^+ )  \int_{\k_1,\k_2}(\k_1\cdot \k_2)   {\rm Tr}_c \left[\cU^\dag (z_1^+, z_2^+)  (\k_2| \cG^\dag(z^+_2, z^+_1) |\k_1) \right] + (z^+_2\leftrightarrow z^+_1)\nn\,, 
 \eeq
where in the last line we have integrated over unmeasured final gluon transverse momentum $\k$, using the completeness relation $|\k)(\k|=1$. Note that the gluon propagation after the emission in the complex conjugate amplitude at $z^+_2$ cancels out owing to the relation $\cG(z^+_2, {+\infty}) \cG^\dag(z^+_2, {+\infty}) =1$.   Medium average is omitted to alleviate the notations. 
 
   Since we are working at leading order the real part of $P(\epsilon_1)$ is nothing else but the radiation spectrum, ($k^+=\omega$)
   \beq
    P_1 (\epsilon) = \int_0^{+\infty} \rmd  \omega\,  \delta(\epsilon - \omega )  \,  \frac{\rmd I}{\rmd \omega}\,.
   \eeq 
   that takes this general operatorial  form 
   \beq
 \omega \frac{\rmd I}{\rmd \omega} &=&     \frac{\alpha_s}{2(\omega^2) N_c} 2 {\rm Re} \int_{\k_1,\k_2}(\k_1\cdot \k_2)  \lg \med|   {\rm Tr}_c \left[\cU^\dag (z_1^+, z_2^+)  (\k_2| \cG^\dag(z^+_2, z^+_1) |\k_1) \right] |\med\rg \nn
 &=&     \frac{\alpha_s C_F}{\omega^2} \,2 {\rm Re}\,  \bdel_1\cdot \bdel_2\cK(\z_2,z^+_2,\z_1, z_1^+)\Big|_{\z_1=\z_2=0}  \nn
   \eeq
   where the kernel $\cK$  is given by 
   \beq \label{eq:K-NP}
   \cK(\z_2,z^+_2,\z_1, z_1^+)  = \frac{1}{N_c^2-1}   \lg \med| {\rm Tr}_c \left[\cU_{0_\perp}^\dag (z_1^+, z_2^+)  (\z_2|\cG^\dag(z^+_2, z^+_1) |\z_1) \right] |\med\rg\,.
   \eeq
   
This result encompasses both medium-induced radiation and vacuum radiation, both of which contribute to energy loss. The vacuum part diverges in the ultraviolet (UV) and therefore requires regularization, typically performed using dimensional regularization. This issue is discussed in \cite{Mehtar-Tani:2024smp}, where Wilson line operators were first introduced in this context. 

The infrared (IR) divergence of the real emission, however, cancels against the corresponding virtual contribution. Here, we will simply subtract the vacuum contribution to focus exclusively on the medium-induced component of the energy loss. From an operator standpoint, this procedure amounts to dividing the Wilson-line correlator in \eqn{eq:P1-operator-0} by its vacuum limit.

\section{Improved harmonic oscillator approximation }

At this point, a few remarks are in order. Thus far, we have only performed exact algebraic manipulations in physical and color space. As a result, the spectrum in \eqn{eq:K-NP} is a non-perturbative quantity that, in principle, can be evaluated on the lattice. To obtain closed analytic expressions, we now evaluate the radiation spectrum under the independent scattering approximation, utilizing the Gaussian statistics encoded in the field correlator \eqn{eq:2-point-corr}. 

This approach assumes that gluon exchanges with the medium are quasi-instantaneous compared to the time scale of collinear-soft radiation. Similarly to the broadening probability in \eqn{2p-fct}, the radiation kernel $\cK$, which describes the propagation of a two-body system in the transverse plane, involves the correlator of two Wilson lines. However, while one of these Wilson lines is described by a path integral \eqn{prop-G}, it can be shown that it reduces to the field correlator in \eqn{eq:2-point-corr}.

 We shall now change slightly notations to better reflect the analogy with quantum mechanics. Setting: $z^+_1=t_1$ and $z^+_2=t_2$.  The emission kernel $\cK$ written in coordinate space obeys the Sch\"dinger equation: 

\begin{equation}\label{eq:cK_Sch_app}
  \left[i\frac{\del}{\del t}+\frac{\bdel^2_\x}{2\omega}+i\frac{N_c n}{2}\sigma(\x)\right]\cK(\x,t;\y,t_1)=i\delta^{(2)}(\x-\y)\delta(t-t_1) \, .   
  \end{equation}
In integral form it reduces to a Dyson-like relation reading
\begin{equation}\label{eq:Dyson_GLV_app}
\cK(\x,t;\y,t_1)=\cK_0(\x,t;\y,t_1)- \frac{N_c n}{2}\int_\u \int_{t_1}^{t_2}  \rmd s \, \cK_0(\x,t_2;\z,s) \sigma(\u,s)\cK(\z,s;\y,t_1) \, ,
  \end{equation}
where $\cK_0$ corresponds to vacuum solution to Eq.~\eqref{eq:cK_Sch_app} with $\sigma=0$:
\begin{align}\label{eq:cK_vac_app}
\cK_0(\x,t_2;\y,t_1)=   \frac{\omega }{2\pi i (t_2-t_1) } \exp\left(i\frac{\omega(\x-\y)^2}{2(t_2-t_1)}\right) \, .
\end{align}
The solution to \eqn{eq:Dyson_GLV_app} can be expressed formally as a path integral:
\beq\label{S2harm}
&&\cK(\x,t_2;\y,t_1)=\int_{\x_1}^{\x_2}\cD \u\,  \exp\left \{\frac{i\omega}{2}\int_{t_1}^{t_2}\rmd t \,\dot\u^2-\frac{N_cn}{2}\int_{t_1}^{t_2}\rmd t \,\sigma(\u)\right\},\nn
\eeq
However, to make progress in phenomenology we want to obtain accurate analytical expressions. 

Following \cite{Wiedemann:2000za,Kovner:2003zj}, the problem can be fully solved in the harmonic oscillator (HOA) approximation using the quadratic approximation \eqn{eq:sigma-ho}. However, the HOA misses the physics at large frequencies that dominate the mean energy loss. Not only that, but it  also result in a scale ambiguity in $\hat q $. This limitation is resolved by expanding around the HOA, similarly to what was done for the broadening distribution. Splitting introducing a factorization scale $Q^2_\med$ we split the dipole cross-section into a constant part that corresponds to the harmonic oscillator potential and a logarithmic term that is assumed to be small if $Q_\med$ is fixed to the typical transverse momentum generated in the radiative process. 
\beq\label{eq:sigma-split-2}
\sigma(\x)  &\sim & \x^2 \left[  \ln \frac{Q^2_\med }{m_D^2} +\ln \frac{1}{Q^2_\med \x^2 }  \right]\,\nn
 &=& \sigma^{(0)}(\x)+ \delta \sigma(\x) \,. \nn
\eeq 
The kernel can then be expressed as a series in the parameter $\lambda\sim 1/\ln(Q_\med^2/m_D^2)$,
\beq
 \cK(\x,t_2;\y,t_1) =  \cK^{(0)}(\x,t_2;\y,t_1)  + \cK^{(1)}(\x,t_2;\y,t_1)  + \cK^{(2)}(\x,t_2;\y,t_1)+...
\eeq
where $\cK^{(0)}(\x,t_2;\y,t_1)$ is solution in the HOA: 
\beq\label{eq:green-fct-sol}
&& \cK^{(0)}(\x,t;\y,t_1)=  \frac{i\omega \Omega}{2 \pi \sin\left(\Omega (t-t_1)\right)} \nn
&\times&\exp\left[  i  \frac{\omega \Omega}{2 \sin\left(\Omega (t-t_1)\right)} \left[\cos\left(\Omega (t-t_1)\right)\,( \x^2 + \y^2) -2 \, \,\x\cdot \y\right]\right]\,.\nn
\eeq 
where 
\beq\label{eq:im-omega}
\Omega = \frac{1-i}{2} \sqrt{ \frac{\hat q(Q_\med)  }{\omega}}.
\eeq

We shall refer the reader to these papers that detail the derivations that are rather technical \cite{Mehtar-Tani:2019tvy,Mehtar-Tani:2019ygg,Barata:2020sav,Barata:2021wuf}, for exact calculation up to the next-to-next-to-leading order $\cK^{(2)}$.  We simple present here the results for the first two orders: 
\beq\label{eq:ioe-0}
 \omega\frac{\rmd I^{(0)}}{\rmd \omega} &=& 2 \abar  \ln \left|\cos\left(\frac{1-i}{2}\sqrt{\frac{2 \omega_c}{\omega}}\right)\right|\simeq  2 \abar
 \begin{dcases}
\,\,\sqrt{\frac{\omega_c}{2\omega}}   \quad\qquad\text{for}\quad \omega \ll \omega_c\\
\,\,\frac{1}{12} \left(\frac{\omega_c}{\omega} \right)^2\quad\text{for} \quad \omega \gg \omega_c\\
\end{dcases}
\eeq
where 
\beq
\omega_c\equiv \frac{1}{2}\hat q(Q_\med) L^2\,. 
\eeq
The NLO contribution reads
\beq\label{eq:nlo}
\omega\frac{\rmd I^{(1)}}{\rmd\omega}&& = \frac{1}{2} \, \abar \, {\hat q_0}  \,{\rm Re} \int_0^L \rmd s \,  
\frac{1}{k^2(s)} \left[ \ln\frac{k^2(s)}{Q^2} +\gamma_E\right]\,,
\eeq
where $\gamma_E=0.5772...$ is the Euler constant and $\hat q_0=(4\pi) \alpha^2_s N_cn $ the prefactor in front of the Coulomb logarithm in the definiton of $\hat q$  \eqn{qhat}.
\begin{figure}
\centering
\includegraphics[width=1\textwidth]{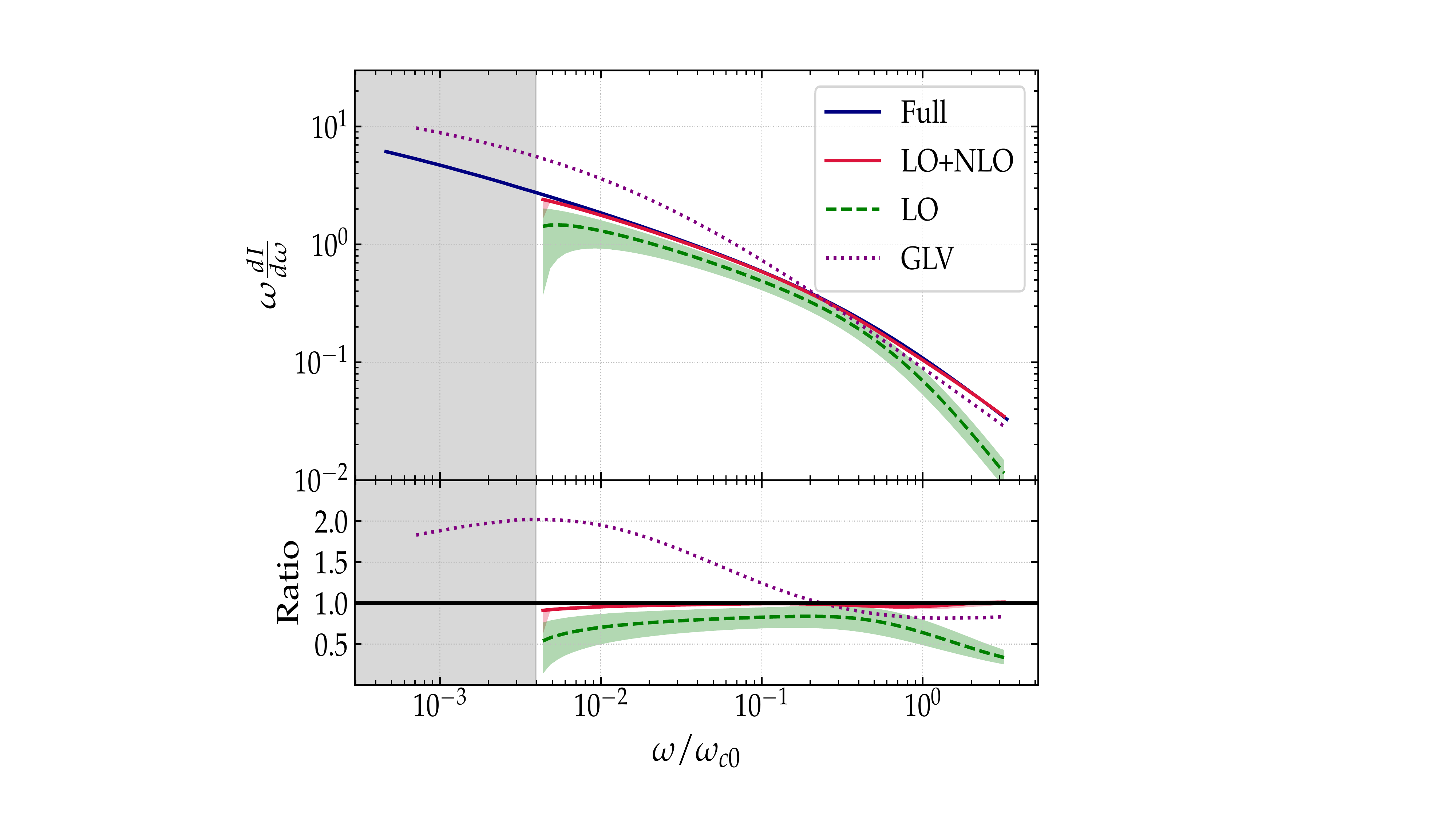}
\caption{Comparison between the energy spectrum computed with GLV (dotted, purple), the Improved Harmonic oscillator (or improved opacity expansion) at LO (dashed, green), LO+NLO (solid, red), and the all-order spectrum (solid, navy) as computed in \cite{Andres:2020vxs}. The ratio to the full solution is presented in the bottom panels. The uncertainty band arises from variations in the matching scale, and the gray region indicates the soft BH regime where our approach breaks down. The parameters used $\hat q_0 = 0.16$ GeV$^3$, $L = 6$ fm and $m_D = 0.355 $ GeV, with $\omega_0 \equiv \hat{q}_0 L^2$.
}
\label{fig:theory-plot}
\end{figure}

For completeness, let us show that we recover the Gyulassy-Levai-Vitev (GLV) result at large frequencies, $\omega \gg \omega_c$. When $\omega \to \infty$, we have $\tan(\Omega(L-s)) \to 0$ and $\cot(\Omega s) \to (\Omega s)^{-1}$. It follows that $k^2(s) \approx i \omega / 2s$. Hence,
\begin{equation}
\frac{1}{2} \, \mathrm{Re} \int_0^L \rmd s \,  
\frac{1}{k^2(s)} \left[ \ln\frac{k^2(s)}{Q_\mathrm{med}^2} + \gamma_E\right]  
\,\rightarrow \,\frac{\pi}{4} \frac{L^2}{\omega},
\end{equation}
which yields the UV limit of \eqn{eq:bdmps-hard} for $\omega \gg \omega_c$. 

It is not surprising that the physical choice for the matching scale, which ensures stability of the solution in the IR, is the transverse momentum acquired during the formation of the gluon. This scale is a function of $\omega$ and is given by:
\begin{equation}
Q^2_\mathrm{med}(\omega) = \sqrt{\hat{q}_0 \omega \ln\left(\frac{Q_\mathrm{med}^2(\omega)}{m_D^2}\right)}\,.
\end{equation}
In Fig.~\ref{fig:theory-plot}, we present numerical results, including the leading-order calculation based on the HOA, which does not compare well with the exact numerical solutions from \cite{Andres:2020vxs}. The GLV spectrum, representing the leading order in opacity, breaks down below $\omega_c$. In contrast, our approach, which expands around the HOA, shows excellent agreement with the exact result and unifies the various approximations for energy loss that have been used in phenomenology to date. This constitutes a significant advancement toward the precise computation of jet observables in the future.

:

%% file: cascade.tex
\chapter{Medium-induced gluon cascade } \label{chap:cascade}

We have demonstrated that a leading hard parton emits multiple successive soft gluons with high probability, $P\sim \abar L/t_f(\omega)$, which are characterized by formation times much shorter than the medium length, $t_f(\omega) \ll L$ . These gluons collectively contribute to the total radiative energy loss. While this picture is reasonable when focusing solely on the leading parton—such as a heavy flavor quark—it cannot be directly applied to jet energy loss. In jets, the primary radiated gluons themselves can undergo further radiation, redistributing energy in a non-linear manner between the interior and exterior of the jet.

To address this, it is necessary to develop an approach that tracks the evolution of these primary gluon emissions. These gluons tend to undergo rapid branching, as we shall see, forming a gluon cascade. This cascade is the key mechanism driving the transport of energy to large angles and the soft sector.

\section{Quasi-instantaneous gluon branching}\label{br-prob}

The framework we will develop in this section closely resembles kinetic theory, where the gluon cascade is governed by quasi-instantaneous branchings interspersed with transverse momentum broadening between successive branchings. Our approach follows the derivation outlined in Ref.~\cite{Blaizot:2012fh}.

To build the gluon cascade to all orders in powers of the medium length,  we need to compute the leading order transition probability $\cP_2(\k_a,\k_b,L| \p_0,t_0)$, that relates the 2-gluon production probability to the single gluon spectrum  
\beq\label{br-amp}
\frac{\rmd N}{d\Omega_{k_a}d\Omega_{k_b}}= \int_{\p_0} \cP_2(\k_a,\k_b,L| \p_0,t_0) \frac{\rmd N}{d\Omega_{p_0}}.
\eeq
The transition probability is derived from the amplitude describing the evolution of a gluon with momentum $p_0 \equiv (E, \p_0)$ at time $t_0$. The gluon propagates through the medium until time $t_1$, where it acquires transverse momentum $\p_1$ and branches into two offspring gluons with momenta $q_1 \equiv (zE, \q_1)$ and $q_1' \equiv ((1-z)E, \q_1')$, respectively. These offspring gluons continue to propagate through the medium until they exit at $L$ with final momenta $k_a \equiv (zE, \k_a)$ and $k_b \equiv ((1-z)E, \k_b)$.

\begin{figure}[]
\centering
\includegraphics[width=8cm]{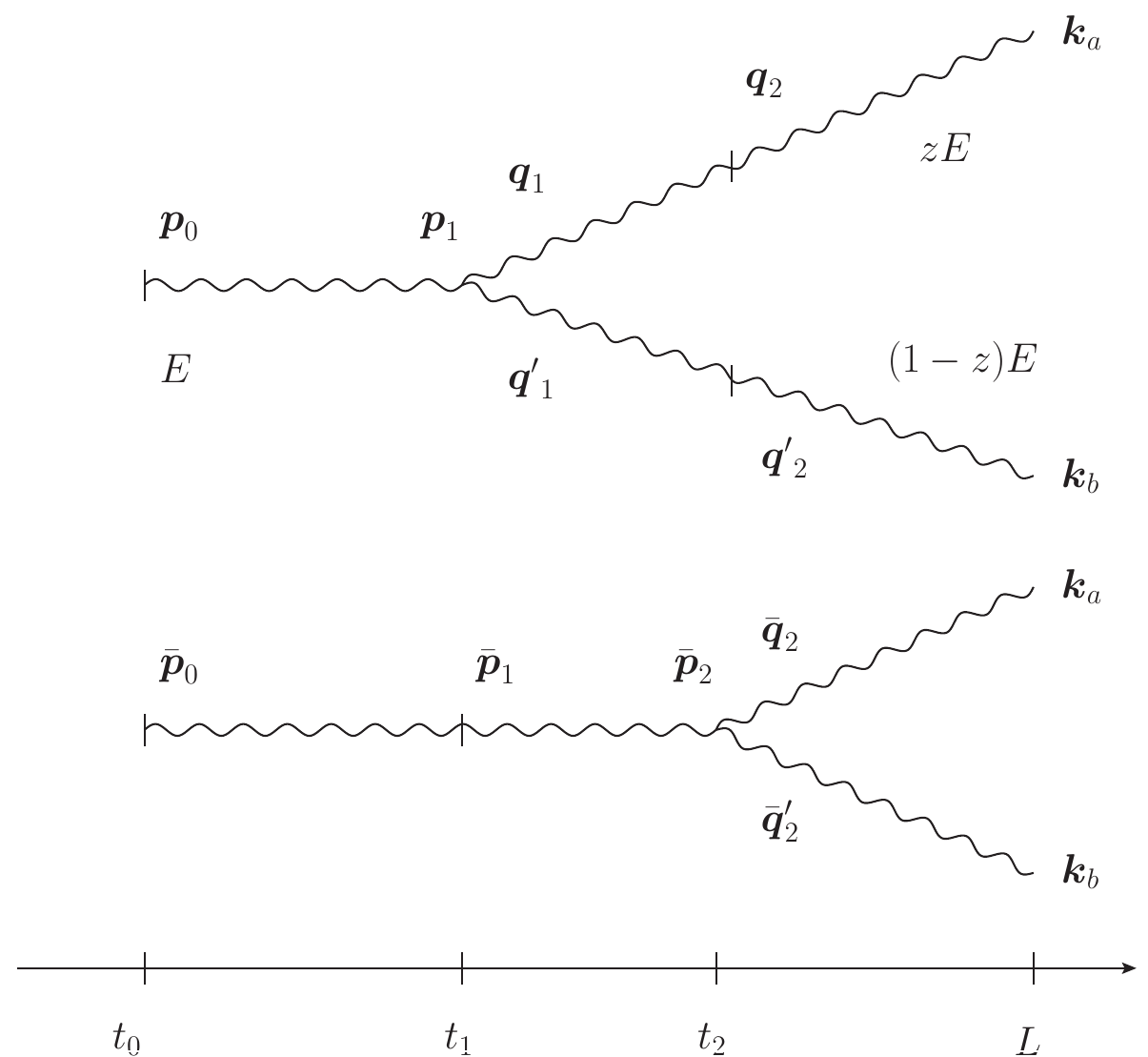} 
\caption{A diagrammatic representation of the time evolution of the amplitude for gluon branching, as described by Eq.(\ref{br-amp}) (upper diagram), and its complex conjugate amplitude (lower diagram), is provided for the calculation of the branching probability in Eq.(\ref{g-br-1}). The evolution spans from $t_0$ to $L$, with the branching occurring at $t_1$ in the amplitude and at $t_2$ in the complex conjugate amplitude. Refer to the text for further details and explanations. }\label{fig:gluon-br-amp}
\end{figure}

The amplitude, with color indices omitted for simplicity, is expressed as:
\beq\label{br-amp}
 \frac{g}{2E}\int \rmd t_1\int_{\q_1\q'_1\p_1}\,(\k_a|\cG(L,t_1)|\q_1) \, (\k_b|\cG(L,t_1)|\q_1') \, (\q_1;\q'_1|  V|\p_1)(\p_1|\cG(t_1,t_0)|\p_0)\,,\nn
\eeq 
and is depicted in the upper diagram of Fig.~\ref{fig:gluon-br-amp}. Here, the 3-gluon vertex takes its well-known form:
\beq\label{vertex}
(\q a;\q' b | V |\p c)=\delta(\q-\q'-\p)\, 2 f^{abc}\epsilon^i (\q) \epsilon^j(\q')\epsilon^k(\p)
 \Gamma^{ijk}(\q-z\p,z),\nn
\eeq
with 
\beq
\Gamma^{ijk}(\Q,z) = \frac{1}{1-z} Q^k\delta^{ij}- Q^i\delta^{jk}+\frac{1}{z}Q^j\delta^{ik}.
\eeq 
and $z=\omega_a/E$ is the energy fraction of the parent gluon carried by gluon $a$. In \eqn{vertex}, 
$f^{abc}$ are the SU(3) structure constants, and $\epsilon^i$ represents the transverse polarization vectors of the radiated gluons, corresponding to their physical degrees of freedom.
The specific combination of momenta, $\Q = \q - z\p$, reflects a Galilean invariance in the effective 2D dynamics. A detailed derivation of this formalism, starting from Feynman diagrams in the light-cone gauge ($A^+ = 0$), can be found in Ref.~\cite{Blaizot:2012fh}. Equation~(\ref{br-amp}) represents the $\mathcal{O}(g)$ correction to the evolution operator. 

By squaring the amplitude (\ref{br-amp}) to obtain the two-gluon production distribution, we are left with an average over the gauge field configurations. Since the correlations between two $A$ fields are instantaneous, this operation can be conveniently visualized by plotting the amplitude and its complex conjugate, as shown in Fig.~\ref{fig:gluon-br-amp}).

To perform the average over the field configurations, we use Eq.~(\ref{eq:2-point-corr}), identifying three distinct factors corresponding to three different time intervals, involving two, three, and four propagators, respectively. The two-point function, $\Sii$, associated with the first interval $[t_0, t_1]$, is related to the broadening probability, as described in Eq.(\ref{2p-fct}). The three-point function, $\Siii$, associated with the second interval $[t_1, t_2]$, determines the branching rate, as we will discuss shortly. Finally, the last interval involves a four-point function, $\Siv$, which factorizes into two two-point functions in the limit of a short formation time \cite{Blaizot:2012fh}.

\beq
(\k_a\k_b;\k_a\k_b|\Siv (L,t_2)|\bar\q_2 \bar\q'_2 ; \q_2 \q'_2)\simeq \delta(\q_2-\q'_2)\, \delta(\bar\q_2-\bar\q'_2)\, \cP(\k_a,L| \q_2,t_2) \cP(\k_b,t|\q'_2,t_2),\nn
\eeq
up to terms of the order of $t_f/L$ that we neglect. One can show that the color correlations between the offspring gluons are exponentially suppressed after a time of the order of the branching time, $t_f$, as they move apart. This behavior is intimately connected to the phenomenon of color decoherence, which will be discussed in more detail in next section. 

This factorization property is crucial for achieving a closed system of equations for high-point correlators. It enables the formulation of exclusive probabilities for producing any number of gluons in the final state, expressed solely in terms of the broadening probability, $\cP$, and the branching rate, $\cK \sim \tSiii$.

By integrating over the intermediate states, summing over the final polarization vectors and averaging over the initial ones using the completeness relation $\sum_{\lambda=1,2} \epsilon_\lambda^i(\k)\epsilon^{\ast j}_\lambda(\k)=\delta^{ij}$, one finds that the amplitude squared depends on two quantities, the broadening probability and the reduced 3-point function as follows  
\beq\label{g-br-1}
&& \cP_2(\k_a,\k_b,L| \p_0,t_0) =\frac{g^2}{2E^2} 2 \rmR \int_0^L \rmd t_1 \int_0^{t_1}   \rmd t_2 \int_{\Q_1, \Q_2, \l}  \Gamma^{ijk}(\Q_1) \Gamma^{ijk}(\Q_2)\nn 
&& \cP(\k_a,t| \q_2,t_2)\cP(\k_b,t| \q'_2,t_2)  \, \tSiii(\Q_2,\Q_1\l;t_1,t_2) \, \cP(\p_1,t| \p_0,t_2), \nn
\eeq  
where the 3-point function $\tSiii$, representing the correlator of 3 scalar propagators, reads (cf. Ref.~\cite{Blaizot:2013vha}). It obeys the Sch\"odinger equation 
\begin{align}
&\left[ i \frac{\del }{\del t} - \frac{\bdel^2}{2z(1-z)E} - \frac{N_cn}{4} \left(\sigma(\u)+\sigma(\v-z\u)+\sigma(\v+(1-z)\u)\right)\right] \tSiii(\u,\u_1,\v; t,t_1) \nn
&= \delta^{(2)}(\u-\u_1) \delta(t-t_1)\,.
\end{align}
whose solution can be formally expressed as a path integral in two space dimensions,
\beq\label{S3harm}
&&\tSiii= \int  \rmd \u_2 \rmd \u_1 \rmd \v\, \rme^{i\u_1\cdot\Q_1-i\u_2\cdot\Q_2-i\v\cdot\l}\int_{\u_1}^{\u_2}\cD \u\nn
&& \times  \exp\left \{\frac{iz(1-z)E}{2}\int_{t_1}^{t_2}\rmd t \dot\u^2-\frac{N_cn}{4}\int_{t_1}^{t_2}\rmd t \left[\sigma(\u)+\sigma(\v-z\u)+\sigma(\v+(1-z)\u)\right]\right\},\nn
\eeq
with $\Q_1=\q_1-z\p_1$ and $\Q_2=\q_2-z\p_2$ the transverse momentum of the splitting at $t_1$ and $t_2$, respectively, and $\l=\bar\p_2-\bar\p_1$, is transverse momentum transfer to the medium during the branching. These momenta are conjugate variables to the transverse sizes $u_1$, $\u_2$ and $\v$.  

Momentum conservation at each of the two vertices implies that $\p_1=\q_1+\q_1'$ and $\bar\p_2=\bar\q_2+\bar\q_2'$.
The function $\tSiii$ is an important quantity that encodes the evolution, from $t_1$ to $t_2$, of the the system of two gluons with energies $zE$ and $(1-z)E$ respectively, and one gluon with energy $E$.  The transverse vector $\u$ corresponds to the transverse distance between the offspring gluons a and b, while $\v$ corresponds to the coordinate of the center of mass of the system. 

Furthermore, the sum of gluon momenta in the amplitude and the complex conjugate amplitude vanishes at any given time, $\q_2+\q_2'=\bar\q_2+\bar\q_2'$, $\p_1=\bar\p_2$ and at each vertex we have $\q_1'=\p_1-\q_1$ and $\bar\q_2'=\bar\p_2-\bar \q_2$. The contraction of  the three gluon vertices in the amplitude and complex conjugate amplitude yields
\beq
\frac{1}{2} \Gamma^{ijk}(\Q_1)\Gamma^{ijk}(\Q_2)=\frac{2}{z(1-z)}p_{gg}(z) (\Q_1\cdot\Q_2),
\eeq
where 
\beq
p_{gg}(z)=N_c\left[\frac{z}{1-z}+ \frac{1-z}{z}+z(1-z)\right], 
\eeq
is the the unregularized Altarelli-Parisi gluon to gluon splitting function \cite{Altarelli:1977zs}.
 
In the short formation time approximation, Eq.~(\ref{g-br-1}) can be further simplified.  The difference between $t_1$ in the amplitude and $t_2$ in the complex conjugate amplitude, $\tau=\Delta t =t_2-t_1$ is at most of the order of $t_f$. The 3-point function decays exponentially for $\tau > t_f$. It follows that the upper bound of the integral over $\tau$ can be send to infinity, 
\beq\label{t-integrals}
\int_0^L \rmd t_2 \int_0^{t_2} \rmd t_1=  \int_0^L \rmd t_2 \int_0^{t_2} \rmd \tau\approx  \int_0^L \rmd t_2 \int_0^{\infty} \rmd \tau.  
\eeq
Similarly, the relative transverse momenta generated in the branching, $\Q_1$, $\Q_2$, $\l$ are typically of the order of $k_\br\sim \sqrt{\hat q t_f}$  and thus, can be neglected compared to the external momenta. This is the collinear branchings approximation. Hence, one can integrate over these internal transverse momenta and $\tau$, and this allows us to define the transverse momentum integrated kernel \cite{Zakharov:1996fv,Baier:1998kq}, \footnote{The vacuum part of the 3-point point function, $\tSiii_0$,  is implicitly subtracted.}
\beq\label{kernel}
{\cal K}(z,E)&= & \frac{p_{gg}(z)}{[z(1-z)E]^2} \rmR\int_0^\infty \rmd \tau \int_{\Q_1\Q_2\l} (\Q_1\cdot\Q_2) \tSiii(\Q_1,\Q_2,\l,t_1,\tau), \nn
\eeq
In general, exact analytic solutions for the kernel (\ref{kernel}) are not available. Numerical approaches have been extensively explored \cite{Caron-Huot:2010qjx,Andres:2020vxs}. More recently, a novel expansion technique inspired by Molière's theory of scattering \cite{Moliere:1948zz} has been developed. This method allows for a unified treatment of both the soft scattering regime, using the harmonic oscillator (HO) approximation, and the hard tail of the distribution, which is treated as a perturbation around the HO approximation \cite{Mehtar-Tani:2019tvy,Barata:2021wuf}. This approach is dubbed the Improved-Opacity-Expansion (IOE). 

For simplicity, although it is not strictly necessary, we shall restrict our discussion to the harmonic approximation. In this approximation, we neglect the logarithmic dependence of $\hat{q}$ that appears in the dipole cross-section (\ref{qhat}), allowing the reduced three-point function (\ref{S3harm}) to be explicitly calculated. This yields for the kernel (\ref{kernel}),
\beq\label{Kdef}
{\cal K}(z,E)=\frac{(1-z+z^2)^{5/2}}{(z(1-z))^{3/2}}\,\sqrt{\frac{\hat q }{E}}.\,
\eeq
Finally, the transition probability (\ref{g-br-1}) takes the compact form\cite{Blaizot:2012fh} (letting $t_1\equiv t$) 
\beq\label{P2}
 &&\cP_2(\k_a,\k_b,z; L| \p_0,t_0) \nn
 &&=  2g^2z(1-z)\int_0^L \rmd t \, \cP(\k_a,L| z\p,t)\cP(\k_b,L| (1-z)\p,t)  \, {\cal K}(z,E)\, \cP(\p,t| \p_0,t_0).\nn
\eeq  

When the branching time is comparable to the size of the system, $t_f \sim L$, finite-size effects become significant. In this regime, only a single branching typically occurs \cite{Blaizot:2012fh, Apolinario:2014csa}. The corresponding branching probability, integrated over transverse momentum, generalizes Eq.(\ref{S2harm}) by including finite energy corrections where $z \sim 1-z$.

This result is leading-order in the coupling and describes the instantaneous branching of a semi-hard gluon inside a medium of size $L$. The probability distributions (\ref{GM-P}) and (\ref{P2}) can be used as fundamental building blocks for constructing the in-medium gluon cascade, which will be discussed in Section \ref{se:multi-br-approx}.

\section{Independent multiple branchings approximation}\label{se:multi-br-approx}
The probability distribution (\ref{P2}) takes the parametric form
\beq
\cP_2 \sim\abar \frac{L}{t_f} \,,
\eeq
where $\bar\alpha / t_f$ represents the rate of (quasi-instantaneous) branching, and the factor $L$ accounts for the branching occurring anywhere within the medium, i.e., $0 < t < L$.

Now consider the branching probability $\cP_3$, where two successive branchings produce three gluons in the final state. These branchings occur at times $t_1$ and $t_2$ in the amplitude, and at times $t'_1$ and $t'_2$ in the complex conjugate amplitude. When the time intervals $[t_1, t'_1]$ and $[t_2, t'_2]$ do not overlap, i.e., $t_1 < t'_1 < t_2 < t'_2$, the successive integrations over these times yield $L^2 \, t_f^2$, assuming $t_f \sim \tau_{1\br} \sim \tau_{2\br} \ll L$. In this regime, the branchings are independent, and the 3-gluon probability factorizes as
\beq\label{fact-br}
\cP^{\text{indep}}_3 \sim (\cP_2)^2 \sim 
\left(\bar\alpha \frac{L}{t_f} \right)^2.
\eeq

However, interferences between subsequent branchings can arise when the time intervals $[t_1, t'_1]$ and $[t_2, t'_2]$ overlap. For instance, this happens when $t_1 < t_2 < t'_1 < t'_2$ with $\tau_1, \tau_2 < t_f \ll L$. The corresponding time integrals contribute a factor $L \, t_f^3$. In this case, the interference contribution to $\cP_3$, denoted $\cP_3^\text{inter}$, is suppressed by a factor $t_f / L$ compared to the factorized result (\ref{fact-br}). This reasoning can be generalized to all possible time configurations. 

Higher-order corrections to this approach have been extensively studied \cite{Arnold:2020uzm,Arnold:2021pin,Arnold:2022fku,Arnold:2022mby,Arnold:2023qwi,Arnold:2024whj,Arnold:2024bph}, including their impact on the renormalization of the jet quenching parameter \cite{Blaizot:2014bha,Caucal:2022mpp,Ghiglieri:2022gyv,Caucal:2022fhc,Arnold:2021pin,Arnold:2021mow,Caucal:2021lgf,Blaizot:2019muz,Iancu:2018trm,Mehtar-Tani:2017ypq,Wu:2014nca,Iancu:2014sha,Liou:2013qya,Iancu:2014kga}. The main conclusion from these studies is that corrections arising from overlapping formation times in large media are numerically negligible, except for those associated with large logarithms. These logarithmic corrections can be absorbed into the renormalization of the jet quenching parameter, thereby preserving the classical branching process established at leading order.

For consistency, and to validate the independent multiple branching approximation, the time between successive branchings must be much larger than the branching time. For inclusive quantities at weak coupling, this condition is naturally satisfied because $t_\ast(\omega) = t_f(\omega) / \bar\alpha \gg t_f(\omega)$. 

Thus, as long as the medium length $L$ is much larger than the branching time $t_f$, successive gluon branchings in a large medium can be approximated as a probabilistic cascade. The properties of this cascade can be derived from a hierarchy of $n$-point functions, for which a generating functional was constructed in \cite{Blaizot:2013vha}.

The essential building blocks of this construction are the elastic and collinear branching rates, $\cP$ and $\cK$. The complete dynamics of the medium-induced cascade can be systematically described using the generating functional formalism \cite{Blaizot:2013vha}. Alternatively, a numerical approach based on Monte Carlo simulations of the medium-induced cascade provides a powerful framework for the systematic computation of jet observables \cite{Caucal:2019uvr, Caucal:2020zcz}.

In this report, we shall focus solely on the lowest of the aforementioned $n$-point functions, namely the single-inclusive gluon distribution  
\beq\label{D-def}
D(x,\k)=x\frac{\rmd N }{\rmd x \rmd^2\k}&=& \frac{1}{2(2\pi)^3}  \sum_{n=1}^\infty\frac{1}{n!}\int \prod_{i=1}^{n} \,\rmd\Omega_i\, n\, P_n(\bk, \bk_1,\cdots,\bk_{n-1}),
\eeq
where 
\beq
 x =\frac{\omega}{E}\,,
\eeq
is the energy fraction of parent gluon carried by the gluon fragment $\omega$. 
$D(x,\k)$ can be viewed as a fragmentation function of gluons in the medium. 

It obeys the evolution rate equation 
\begin{align}\label{Dk}
\frac{\partial}{\partial t}D(x,\k,t)&=N_c \int_{\q}\,
\gamma(\bell,t)\left[D\left(x,\k-\bell,t\right)-D\left(x,\k,t\right)\right]\nn
&+\alpha_s\int_0^1\rmd
z \bigg[\frac{2}{z^2}{\cal K}\left(z,\frac{x}{z}E\right)
D\left(\frac{x}{z},\frac{\k}{z},t\right)-{\cal K}\left(z,xE\right)D\left(x,\k,t\right)\bigg].
\end{align}
The first and second terms in the r.h.s. correspond, respectively, to  the elastic and the branching rates  with in the latter case $\bell=0$.

This equation describes the average properties of energy distribution in frequency and angle around the initial parton direction. 

\section{Energy distribution and wave turbulence}  \label{sec:energy-dist-WT}
We now investigate some characteristic features of Eq.~(\ref{evol-eq-ang0}), focusing particularly on the energy distribution of the emitted gluons as a function of time $t$, or equivalently, the system size $L$.

In many cases, it is more convenient to switch to angular variables and work in angle space, as this is more natural when discussing jets defined as energy flowing within a cone of opening angle $R$. In terms of angular variables, the rate equation becomes
\begin{eqnarray} 
\frac{\del}{\del t}D(x,\btheta,t) &=& \int \rmd z\, {\cal K}(z)\left[\frac{D\left(x/z,\btheta,t \right)}{t_\ast(x/z)} - z\frac{D\left(x,\btheta,t\right)}{t_\ast(x)} \right] \nn
&&+ \int \frac{\rmd^2\btheta'}{(2\pi)^2}C(\btheta',x) D(x,\btheta-\btheta',t)\,,
\label{evol-eq-ang0}
\end{eqnarray}
where
\beq
x = \frac{\omega}{E}, \qquad \text{and} \qquad \btheta \equiv \frac{\k}{\omega} = \frac{\k}{xE}\,,
\eeq
with (see Eq.~(\ref{ME-P}))
\beq\label{calCx}
C(x,\btheta) = (xE)^2 \sigma(\q) = \frac{g^4 N_c}{(xE)^2} \left[\frac{1}{\btheta^4} - \delta^{(2)}(\btheta)\int \frac{\rmd^2\btheta'}{\btheta'^4} \right]\,.
\eeq
Here, $\btheta$ is a 2-dimensional vector collinear to $\k$, with a (small) magnitude equal to the polar angle of the emitted gluon relative to the initial direction of the leading particle.

\begin{figure}[htbp]
\centering
		\includegraphics[width=9cm]{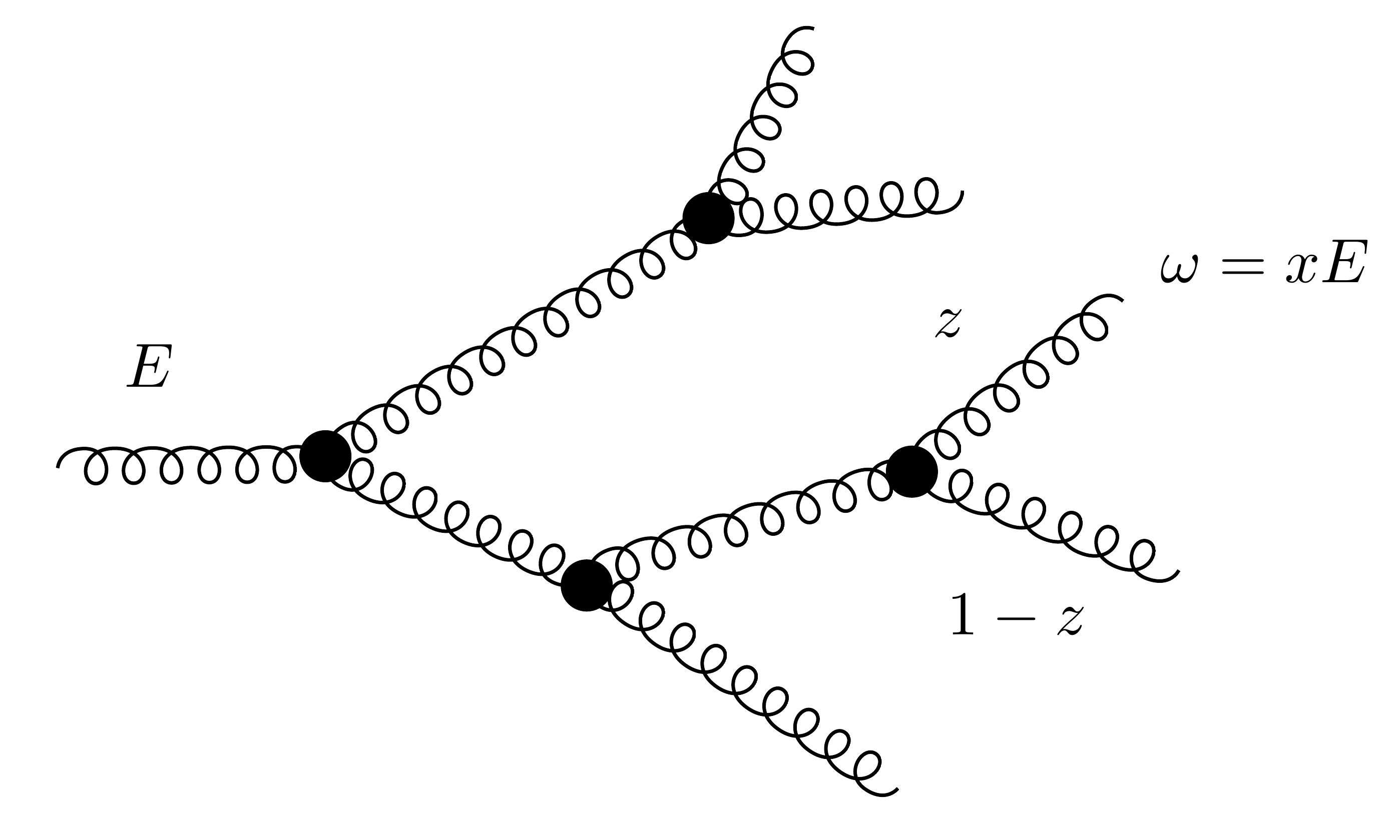}
\caption{A depiction of a gluon cascade that is initiated by a gluon with energy $E$. The blobs illustrate the quasi-instantaneous splitting time $t_f \ll L$. The inclusive distribution $D(x,\btheta)$ measures the probability to find in the cascade, at time $t$,  a gluon with energy $xE$.  The rate equation (\ref{evol-eq-ang0}) describes how this distribution evolves with time $t$. }
	\label{fig:shower-2}
\end{figure}

An illustration of the gluon cascade described by Eq.~(\ref{evol-eq-ang0}) is provided in Fig.~\ref{fig:shower-2}. The locality (in angle) of the splitting term reflects the effective collinearity of the splitting process.

In addition to switching to angular variables, we have explicitly factorized the kernel (\ref{kernel}) in Eq.~(\ref{evol-eq-ang0}) as
\beq
2\alpha_s \cK(z,xE) = \frac{\cK(z)}{t_\ast(x)},     
\eeq
where
\beq\label{factorizedkernel}
\cK(z) = \frac{[1-z(1-z)]^{5/2}}{z^{3/2}(1-z)^{3/2}}, \quad \text{and} \quad t_\ast(x) = \frac{1}{\bar\alpha}\sqrt{\frac{xE}{\hat q}}.
\eeq
We refer to $\cK(z)$ as the \emph{reduced kernel}. The quantity
\beq\label{stop-time}
t_\ast(E) \equiv \frac{1}{\bar\alpha}\sqrt{\frac{E}{\hat q}}
\eeq
is a characteristic time scale of the BDMPS-Z cascade. After a time of order $t_\ast(E)$, most of the initial energy is radiated into soft gluons (assuming $L > t_\ast(E)$). This defines the so-called \emph{stopping time} or \emph{stopping distance} \cite{Arnold:2009ik}. For simplicity, we often denote $t_\ast(E)$ as $t_\ast$, except when clarity requires otherwise. Note that $t_\ast(x) = t_\ast \sqrt{x}$.

As written, the kernel (\ref{factorizedkernel}) does not account for finite-size effects, which become significant when $t_\ast(x) > L$, corresponding to $\omega_c > E$. In the following, we neglect these corrections, implicitly restricting the initial gluon energy to $E \lesssim \omega_c$ (see \cite{Blaizot:2014rla, Fister:2014zxa} for detailed discussions). This is a reasonable assumption, as the medium-induced cascade is triggered by primary gluon emissions discussed in Chap.~\ref{chap:med-partons}.

We focus on the energy distribution, obtained by integrating $D(x,\btheta,t)$ over the angle. Specifically, we define
\beq
D(x,t) = \int \frac{\rmd^2 \btheta}{(2\pi)^2} D(x,\btheta, t),
\eeq
which satisfies the following evolution equation \cite{Baier:2000sb, Arnold:2002zm, Jeon:2003gi, Blaizot:2013hx}:
\begin{eqnarray}\label{evol-eq}
\frac{\del}{\del t}D(x,t) &=& \int \rmd z\, {\cal K}(z) \left[\frac{D\left(x/z,t \right)}{t_\ast(x/z)} - z\frac{D\left(x,t\right)}{t_\ast(x)} \right] 
\end{eqnarray}

Since initially the total energy is carried by a single gluon, the initial condition for Eq.~(\ref{evol-eq}) is simply $D_0(x)= \delta(1-x)$. In the regime where the length of the medium is so short that at most a single branching can be induced,  $L\lesssim t_\ast(x(1-x))$,  Eq.~(\ref{evol-eq}) can be solved by iterations, with the first  one given by 
\beq
D_1(x,L)= \frac{L}{t_\ast} x\cK(x),
\eeq
which coincides with  the BDMPS-ZW  spectrum (\ref{eq:bdmps-coh}) at small $x$. In the opposite regime where $L \gg t_\ast(x)$, which corresponds to gluon energies $\omega\ll \omega_s\equiv \bar\alpha^2 \,\hat q \,L^2$, multiple branchings are important. A non perturbative solution becomes mandatory.
 
Eq.~(\ref{evol-eq}) can be solved exactly for a simplified version of the reduced kernel (\ref{factorizedkernel}) in which one  neglects the $z$ dependence of the numerator, that is
\beq\label{reduced-kernel}
\cK(z)\approx \frac{1}{z^{3/2}(1-z)^{3/2}}. 
\eeq
This turns out to be an excellent approximation. In fact, we shall argue later that the exact form of the kernel plays a minor role in the determination of the general features of the cascade. 
The solution of Eq.~(\ref{evol-eq}) for the simplified kernel reads \cite{Blaizot:2013hx,Blaizot:2015jea} 
\beq\label{Gsol2Ap}
D(x,\tau)=\frac{\tau}{\sqrt{x}\,(1-x)^{3/2}}\, \exp\left(-\pi\frac{\tau^2}{1-x}\right)\,, \quad\text{with}\quad  \tau=\frac{L}{t_\ast}.
\eeq
This solution exhibits two remarkable features: a peak near $x=1$ associated with the leading particle, and a scaling behavior in $1/\sqrt{x}$ at small $x$ where the $x$ dependence factorizes from the time dependence, i.e. 
\beq\label{Gsol2-soft}
D(x,\tau)\approx \frac{\tau}{\sqrt{x}}\, \rme^{-\pi \tau^2 }.
\eeq

\begin{figure}[h]
	\centering
	\includegraphics[width=9cm]{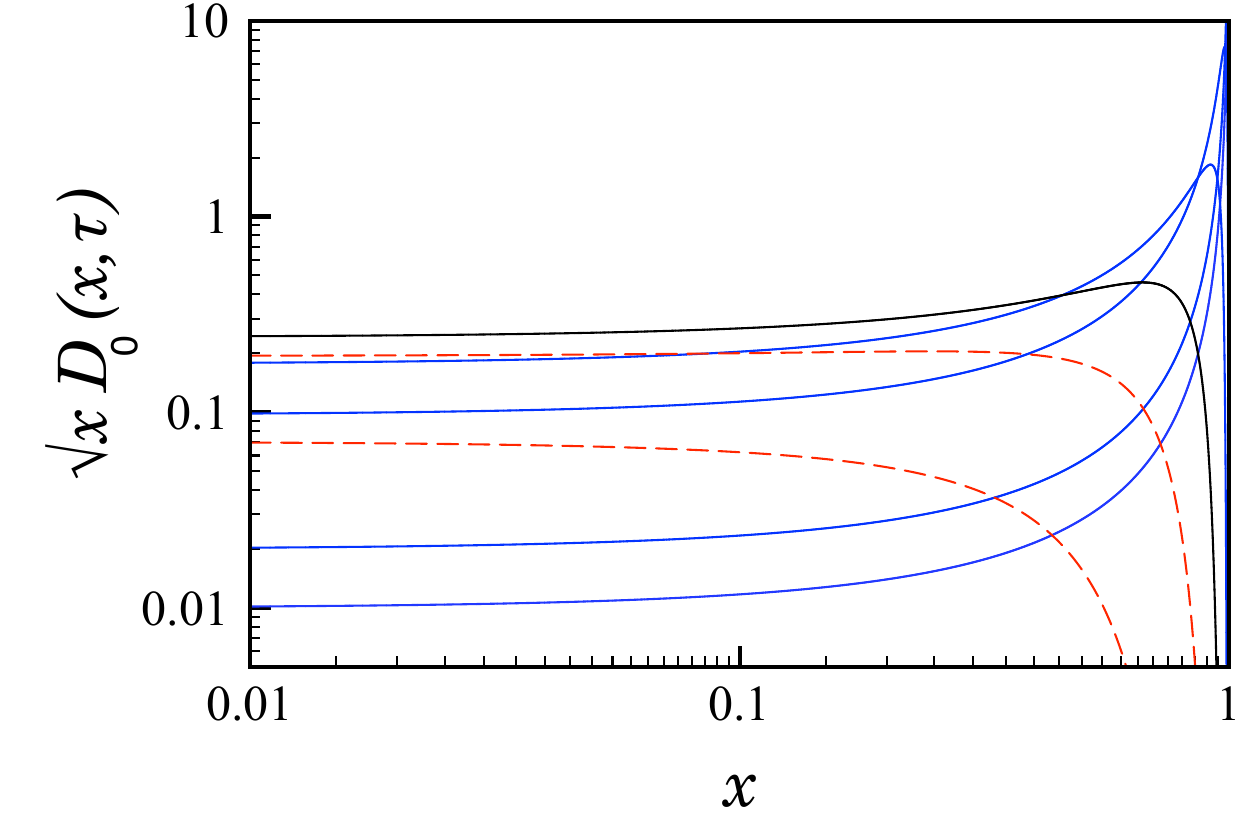}
		\caption{Plot (in log-log scale) of $\sqrt{x} D(x,\tau)$, given by Eq.~(\ref{Gsol2Ap}), as a function of $x$ for various values of $\tau$ (full lines from bottom to top, $\tau=0.01,0.02,0.1,0.2,0.4$; dashed lines from the top down: $\tau=0.6,0.9$).}
		\label{fig2}

\end{figure}

 An illustration of this solution is given in Fig.~\ref{fig2}, left panel. The energy of the leading particle, initially concentrated in the peak at $x\lesssim 1$, gradually disappears into radiated soft gluons,  and after a time $t\sim t_\ast$ (i.e. $\tau\sim 1/\sqrt{\pi}\approx 0.5$)  most of the energy is to be found in the form of radiated soft ($x\lesssim 0.1$) gluons. This is also the time at which the peak corresponding to the leading particle disappears (see Fig.~\ref{fig2}). At the same time the occupation of the small $x$ modes increases (linearly) with time, keeping the characteristic form of the scaling spectrum. When the peak has disappeared, the cascade continues to lower $x$, causing a uniform, shape conserving, decrease of the occupations of the modes, and a flow of energy towards small $x$.

We define 
\beq\label{flowdef}
{\cal F}(x_0,\tau)=-\frac{\del {\cal E}(x_0,\tau)}{\del \tau}, \qquad {\cal E}(x_0,\tau)\equiv \int_{x_0}^1 \rmd x D(x,\tau), 
\eeq
where  ${\cal E}(x_0,\tau)$ is the amount of energy contained in the modes with $x>x_0$, and ${\cal F}(x_0,\tau)$ is the corresponding flux of energy,  counted positively for energy moving to values of $x$ smaller than $x_0$. 
These quantities can be calculated explicitly. We have for instance
\beq
{\cal E}(x_0,\tau)=\int_{x_0}^1 \rmd x\, D(x,\tau) ={\rm e}^{-\pi \tau^2}\,{\rm erfc}\left(  \sqrt{\frac{\pi x_0}{1-x_0}}\,\tau\right),
\eeq
with  ${\rm erfc}(x)$ the complementary error function. 
We note that the fraction of the total energy ``stored in the spectrum'', namely
\beq
\lim_{x_0\to 0} {\cal E}(x_0,\tau)={\rm e}^{-\pi \tau^2},
\eeq
 decreases with time, and accordingly there is a non vanishing flux of energy reaching $x=0$
\beq\label{flowexactwo}
{\cal F}(0,\tau)=2\pi \tau \, \rme^{-\pi\,\tau^2}\,.
\eeq
 It follows that the complete, energy conserving, solution involves a contribution 
 \beq\label{condensate}
 D_c(x)=n_c(\tau)\delta(x)\quad \text{with}  \quad n_c(\tau)=1-{\rm e}^{-\pi\tau^2},
 \eeq
somewhat analogous to a condensate where the radiated energy accumulates. Note that when $\tau\sim 1/\sqrt{\pi}$, corresponding to the disappearance of the leading particle into soft radiation,  about 60\% of the initial energy has flown into the condensate.
    \begin{figure}[h]
	\centering
	\includegraphics[width=9cm]{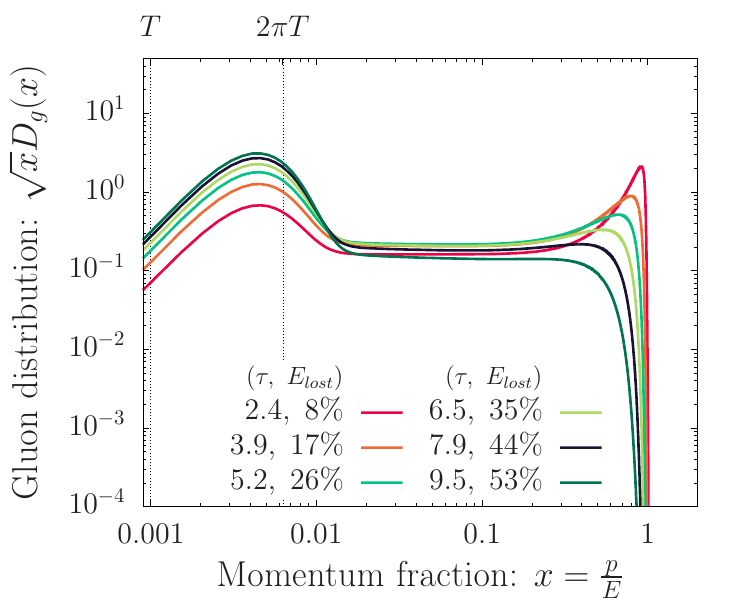}
		\caption{Evolution of the energy distribution at intermediate times for a gluon jet from kinetic theory \cite{Schlichting:2020lef}. One clearly observes the Kolmogorov scaling in \eqn{eq:kolmogorov} at intermediate energies $T/E \ll  x  \ll 1$. We also observe the thermalization in the soft sector around the temperature scale. 
}
		\label{fig3}

\end{figure}

 At this point, it is useful to  comment briefly on a peculiarity of the BDMPS-Z  cascade, that makes the transition between the dilute, single-branching regime, and the the multiple branching regime completely smooth, with no sign of a change of regime as  $x$ crosses the value $x_s$. Indeed, the perturbative solution, valid at $\tau\ll 1$), is proportional to the BDMPS-Z  spectrum 
\beq\label{eq:kolmogorov}
D(x,\tau)\simeq \frac{\tau}{\sqrt{x}},
\eeq  
and this is already in the scaling form. We shall see that the scale $x_s$ becomes visible in the angular distribution, to which we now turn. 

The properties of the cascades depend also on the splitting kernel, that is, on the way the energy is distributed between the offsprings during a splitting. However, this turns out to have a minor effect on the main characteristics of the cascade, as compared to that of the transport time scale just mentioned, at least as long as the splitting kernel is not too singular. In that case the cascades develop as if the branching were completely democratic, with the two offsprings taking each half the energy of the parent gluon. The interactions responsible for the splittings can then be considered as local (in energy space). As already mentioned, these properties of the BDMPS-Z  cascade that we have briefly listed, are typical of wave turbulence \cite{Nazarenko:2011zz}.   

The existence of the inertial window, which characterizes a constant energy flow without accumulation, was confirmed through simulations of the full kinetic equations that account for detailed balance and, consequently, thermalization in the plasma, as shown in Fig.~\ref{fig3}. In addition to the leading parton peak and the scaling regime (inertial range), we observe an accumulation and thermalization of the lost energy around the plasma temperature.

%% file: qhat-renormalization.tex
\chapter{Renormalization of the jet quenching parameter $\hat q$ }\label{chap:qhat-rg}

\section{Radiative transverse momentum broadening }\label{gluon-dist}

In constructing the probabilistic picture in the previous section, we explicitly assumed that the dominant contribution to transverse momentum broadening along the cascade arises from elastic scatterings. While some increase in transverse momentum can also occur due to branching processes, it is suppressed by a factor of $t_f / L$. This suppression forms the foundation of our kinetic description. However, this parametric power counting does not exclude the presence of logarithmically enhanced terms, which, as we shall see, indeed exist.

A simple way to diagnose such contributions is to estimate the radiative contribution to transverse momentum broadening. Recall that the typical transverse momentum generated by the radiation of a gluon with frequency $\omega$ is $k_f(\omega) \sim (\omega \hat{q})^{1/4}$. Combining this with the radiative spectrum, $\omega \, \rmd I / \rmd \omega \sim \bar{\alpha} \sqrt{\omega_c/\omega}$ (cf. Eq.~(\ref{eq:bdmps-coh})), we obtain the following parametric estimate for the total squared transverse momentum generated by splittings:
\begin{equation}\label{kt2-rad-estimate}
\langle k_\perp^2 \rangle_\text{rad} \sim \bar{\alpha} \int^{\omega_c}_{\omega_\mathrm{BH}} \rmd \omega\, \frac{\rmd I}{\rmd \omega}\,   k_f^2(\omega) \,
\sim \bar{\alpha} \ln\frac{L}{\ell_\mathrm{mfp}} \, \langle k_\perp^2 \rangle_\text{el},
\end{equation}
where we have integrated over the relevant range of gluon energies, from $\omega_\mathrm{BH}$ to $\omega_c$, and used the relation $\omega_\mathrm{BH} = \hat{q} \ell_\mathrm{mfp}$.

Compared to the typical transverse momentum acquired through elastic scatterings, $\langle k_\perp^2 \rangle_\text{el} \sim \hat{q} L$, the radiative contribution is suppressed by a factor $\bar{\alpha}$ but enhanced by a potentially large logarithm, $\ln (L / \ell_\mathrm{mfp})$ and therefore, cannot be neglected. However, this is not the dominant contribution. A more detailed analysis reveals an additional double logarithmic enhancement \cite{Liou:2013qya,Blaizot:2013vha,Blaizot:2014bha,Iancu:2014kga}, which is missed by the estimate above. This necessitates going beyond the collinear approximation in the branching rate and performing a calculation whose validity is not immediately apparent but will be justified later.

We note that the 3-point function defined earlier contains the information about the  transverse momentum broadening accompanying a single branching. Hence, instead of integrating over $\bell$ and $\Q$, as we did for Eq.~(\ref{kernel}), we define the unintegrated kernel 
\beq\label{kernel-unint}
{\cal K}(\Q,\bell,z,E)&= & \frac{p_{gg}(z)}{[z(1-z)E]^2} \rmR\int_0^\infty \rmd \tau \int_{\Q_1} (\Q\cdot\Q_1) \tSiii(\Q,\Q_1,\bell,t_1,\tau).\nn
\eeq
One can thus generalize the radiative part of Eq.~(\ref{Dk}),
 \begin{align}\label{Dkt-2}
\frac{\partial}{\partial t}D(x,\k,t)\Big|_{\rm rad}&=\alpha_s\int_0^1\rmd
z\int_{\Q,\bell}\,\bigg[\frac{2}{z^2}\,
{\cal K}\left(\Q,\bell,z,\frac{x}{z}E\right)
D\left(\frac{x}{z},\q,t\right)\nn
&-{\cal K}\left(\Q,\bell,z,xE\right)D\left(x,\k-\bell,t\right)\bigg]\,.
\end{align}
where $\Q\equiv\k - z(\q+\bell)$. This combination of momenta is a consequence of Galilean symmetry in transverse plane resulting itself from the rotational invariance of the splitting. The vector $\Q$ denotes the transverse momentum generated in the splitting from a parent gluon with momentum $\q$ that splits into two daughters with momenta $\k$ and $\q-\k-\bell$, respectively.   

We shall now follow the strategy that we adopted in order to reduce Eq.~(\ref{ME-P})  to the  diffusion equation (\ref{diff-P}). This involves an expansion around the large momentum $\k$ of the followed gluon. The momenta $\Q$ and $\l$, which are at most of the order of $k_f \equiv \sqrt{z(1-z)\hat q E}$, are small compared to $\k$ which is typically of the order of $\hat q L$. The expansion of the gluon distributions around $\k$ yields for the first term of Eq.~(\ref{Dkt-2})
\beq\label{D-exp1}
&&D\left(\frac{x}{z},\frac{\k-\delta \k}{z}\right)\nn
&&=  D\left(\frac{x}{z},\frac{\k}{z}\right)-\delta \k\cdot\frac{\del}{\del \k}
D\left(\frac{x}{z},\frac{\k}{z}\right)
+\frac{1}{2!}\, \delta k^i \delta k^j\frac{\del}{\del k_i}\frac{\del}{\del k_j}D\left(\frac{x}{z},\frac{\k}{z}\right)+\cdots\nn
 \eeq 
where we have set $\delta \k\equiv \Q +z\bell$. One expands similarly $D\left(x, \k-\bell\right)$. 
It is easy to see that the leading terms reproduce Eq.~(\ref{Dk}). The linear terms vanish upon angular integration. Remain the quadratic terms, whose contribution can be cast in the form of the diffusion term, thereby exhibiting a correction $\Delta\hat q$ to the jet quenching parameter.  
For consistency, we shall also simplify the elastic term by using the diffusion approximation.
At this point, we anticipate that evaluation of the correction $\Delta\hat q$ meets with logarithmic divergences. These arise from the region $z\sim 1$. To the leading-logarithmic accuracy, we can set $z=1$ everywhere, except in the dominant singularity. The dominant contribution to $\Delta\hat q$ can be then written as
 \begin{align}\label{qhat-rad2}
\Delta\hat q (\k^2)\,=\, 2\alpha_s \int_{x}^1 \rmd z \int_{\Q,\l}\,
 \big[(\Q+\l)^2 - \l^2\big]\,
{\cal K}\left(\Q,\l,z,xE\right),
 \end{align}
where the $\k^2$ dependence arises from the integration boundary $\Q^2,\l^2\ll \k^2$. 

The explicit calculation of $\Delta\hat q$, using the expression (\ref{S3harm}) of the reduced 3-point function in the harmonic oscillator approximation, yields 
\beq\label{dsig-0}
\Delta \hat q\,&=&
\frac{\alpha_s \,N_c}{\pi}\, 2\text{Re }
\int \rmd\omega \,\int \rmd \tau  \,\frac{i \Omega^3}{\sinh(\Omega \tau)}\left[1+\frac{4}{\sinh^2(\Omega  \tau)}\right]\,,\nn
&\simeq & \frac{\alpha_s \,N_c }{\pi}\, \,
\int  \frac{\rmd\omega}{\omega}\int^{\Omega^{-1}}
\frac{\rmd\tau}{\tau}\,\hat q\,,\nn
\eeq
where $\omega=zE$ and $\Omega$ is given by \eqn{eq:im-omega}:
\beq\label{eq:im-omega-2}
\Omega = \frac{1-i}{2} \sqrt{ \frac{\hat q }{\omega}}\,.
\eeq
As anticipated the radiative correction to $\hat q$ exhibits a double logarithmic divergence, when $\tau=\omega^2/k_\perp^2\to 0$, that represents the formation time of the gluon and $\omega\to 0$.   The complete calculation of the integral (\ref{qhat-rad2}) is presented in \cite{Blaizot:2013vha,Blaizot:2014bha}.  One gets 
 \begin{align}\label{qhat-right}
\Delta\hat q (\k^2)=\frac{\alpha_s \,N_c }{2\pi}\, \hat q \,
\ln^2\frac{\k^2}{\hat q \tau_{0}}, 
\end{align}
where $\tau_{0}^{-1}$ is the maximum energy that can be extracted from the medium in a single scattering (e.g. $\tau_{0} = 1/T$ for a weakly coupled plasma with temperature $T$). 
We return to this result in the next section. Finally, Eq.~(\ref{Dkt-2}) can be recast in a form that is similar to that of Eq.~(\ref{Dk}) (after performing the diffusion approximation)\cite{Blaizot:2013vha}, 
\begin{align}\label{Ddiff}
\frac{\partial}{\partial t}D(x,\k,t)&=\frac{1}{4}\left(\frac{\del}{\del \k}\right)^2 \left[(\hat q(\k^2) + \Delta \hat q (\k^2) )D\left(x,\k-\q,t\right)\right]\nn
&+\alpha_s\int_0^1\rmd
z \bigg[\frac{2}{z^2}{\cal K}\left(z,\frac{x}{z}E\right)
D\left(\frac{x}{z},\frac{\k}{z},t\right)-{\cal K}\left(z,xE\right)D\left(x,\k,t\right)\bigg],
\end{align}
where $p_\perp$-broadening due to soft gluon radiation is now taken into account effectively via a redefinition of the quenching parameter  $\hat q \to \hat q+\Delta \hat q$.

As clear from Eq.~(\ref{dsig-0}) the $\tau$ integral is bounded at the upper end by  $|\Omega|^{-1}\sim t_f (\omega)=\sqrt{\omega/\hat q}$ (cf. Eq.~(\ref{eq:im-omega-2})) corresponding to the onset of the multiple scattering regime and the LPM effect: the relevant gluon fluctuation experiences a single scattering with the medium constituents. 
In order to systematically account for the boundaries of the double integral,\footnote{For a detailed discussion on the boundaries of the logarithmic integrals see Ref.~\cite{Liou:2013qya,Iancu:2014kga}}  it is in fact more convenient to change variables, from $(\omega,\tau)$ to $ (\q, \tau)$,  
with $\tau\equiv \omega/\q^2$ the formation time of the radiated gluon, and $\q$ its transverse momentum which can run up to $\k^2\equiv \v^{-2}$ (the logarithmic transverse momentum integration can be rephrased in terms of the transverse coordinate of the radiated gluon $\q^{-1}\sim \u \gg \v$, see Fig.~\ref{fig5}). We obtain then
\beq\label{qhat1}
\Delta \hat q (\tau_\text{max}, \k^2)\equiv \frac{\alpha_sN_c}{\pi} \, \int^{\tau_\text{max}}_{\tau_0} \frac{\rmd\tau}{\tau}\int^{\k^2}_{\hat q\tau}\frac{\rmd\q^2}{\q^2}\, \hat q (\q^2)\,,
\eeq
where we have explicitly indicated the scale dependence of $\hat q$. The boundary corresponding to the region of multiple scattering now appears as the lower bound of the $\q$ integration,  $ \q^2 \gg  k_f^2\equiv \hat q\tau $. 
 Since medium-induced gluons forms inside the medium, the largest value for $\tau$ is the length of the medium $L\sim t-t_0$. As for the lowest value $\tau_0$, it can be interpreted as the inverse of the largest energy that can be extracted from the medium through a single scattering. 

For a constant $\hat q $ in the integral, one can easily perform the integrations, and by keeping the leading contributions, we recover the result first derived in \cite{Liou:2013qya,Blaizot:2013vha}, 
\begin{align}\label{deltaqhat}
 \Delta \hat q \simeq  \frac{ \alpha_s C_A}{2\pi}\, \hat q \ln^2\left(\frac{L}{\tau_0}\right),
\end{align}
where we have used the fact that $\k^2\sim \hat q L$.

We are now ready to revisit the standard estimate of the typical momentum broadening Eq.~(\ref{kt2-typ}), making use of the redefinition of the quenching parameter above, substituting $\k^2 \sim \hat q L $ in Eq.~(\ref{qhat-right}),  we obtain 
\beq
\lg k^2_\perp\rg_\text{typ}  \simeq \hat q L \left[1+\frac{\alpha_s N_c}{\pi}\ln^2\frac{L}{\tau_0}\right].
\eeq
This result agrees with that obtained in Ref.~\cite{Liou:2013qya} using a different approach.\footnote{Note that NLO corrections to $p_\perp$-broadening have been also investigated in Deep Inelastic Scattering in the High-Twist approach \cite{Kang:2013raa}.}
Note that we obtain a double logarithm of the medium length, as opposed to the heuristic estimate (\ref{kt2-rad-estimate}), which involves a single logarithm. Although, the soft (short time) logarithmic enhancement is correctly captured in this estimate, the single scattering that contributes to the branching probability that in turn leads to the second logarithm, is missing in the BDMPS-Z spectrum that is used Eq.~(\ref{kt2-rad-estimate}).  Single-logarithmic corrections have also been computed in~\cite{Liou:2013qya,Arnold:2021mow,Ghiglieri:2022gyv}

\section{Locality and universality of radiative corrections}\label{locality-RC}

The correction to the  jet quenching parameter discussed in the previous section points to the existence of potentially important radiative corrections. The calculation that we just presented focussed on a correction that is singular, and this is what allowed us to retain  in the branching kernel contributions that should not be kept a priori since, if they were not divergent,  they would be of the same order of magnitude as terms that  have been consistently neglected in arriving at Eq.~(\ref{Dk}). Now that we identify the singular origin of the leading logs we can use more systematic approaches based on radiative corrections of Wilson line operators to compute directly the soft limit of radiative corrections.

So far we have discussed the calculation only for the leading singular part. Furthermore, it is only this singular part that can be interpreted as a correction to the jet quenching parameter; less singular corrections would presumably be non local, thereby spoiling their interpretation in terms of a correction to the jet quenching parameter. However, they can be computed order by order in perturbation theory.  A lot of work has been done to answer these questions and extend the early result to better understand the effect of overlapping radiation in higher order computations, \cite{Liou:2013qya,Blaizot:2013vha,Blaizot:2014bha,Iancu:2014kga,Wu:2014nca,Caucal:2022mpp,Ghiglieri:2022gyv,Caucal:2022fhc,Arnold:2021pin,Arnold:2021mow,Caucal:2021lgf,Blaizot:2019muz,Iancu:2018trm,Mehtar-Tani:2017ypq,Iancu:2014sha}.

The key to the locality of radiative corrections lies in the nature of the logarithmic integral whose boundaries do not affect the overall multiplicative constant. Thus, to the extent that one restricts oneself to a leading order calculation, one can proceed as if the lifetimes of the fluctuations involved in the radiative corrections were small, and treat them as local. The coefficient of the leading double logarithm is calculated correctly, and corrections coming from overlapping contributions will be subleading. In short, the leading corrections  can be treated as being effectively local in time and thus independent \cite{Blaizot:2014bha}. 

Another fundamental question is that of the universality of these corrections, that is, whether they can be absorbed in a redefinition of $\hat q$ for other observables than the broadening probability. This has important implications as it would affect the branching rate in Eq.~(\ref{Ddiff}), that controls the number of final state gluons. 

\begin{figure}[htbp]
\centering{
\includegraphics[width=8cm]{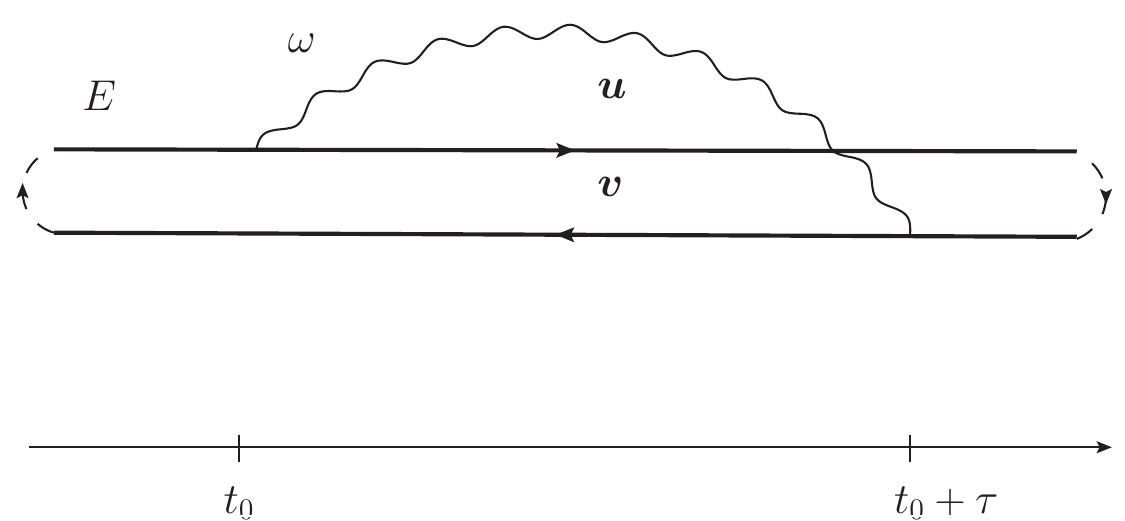} 
\caption{Diagrammatic illustration of the BDMPS-Z spectrum (\ref{eq:K-NP}) that corresponds to the radiation of a soft gluon with energy $\omega$. We denote by $\u$ the transverse position of the radiated gluon with respect its radiator, the energetic dipole whose size is denoted by  $\v$.  }\label{fig:S3}}
\end{figure}

\begin{figure}[htbp]
\centering{
\includegraphics[width=8cm]{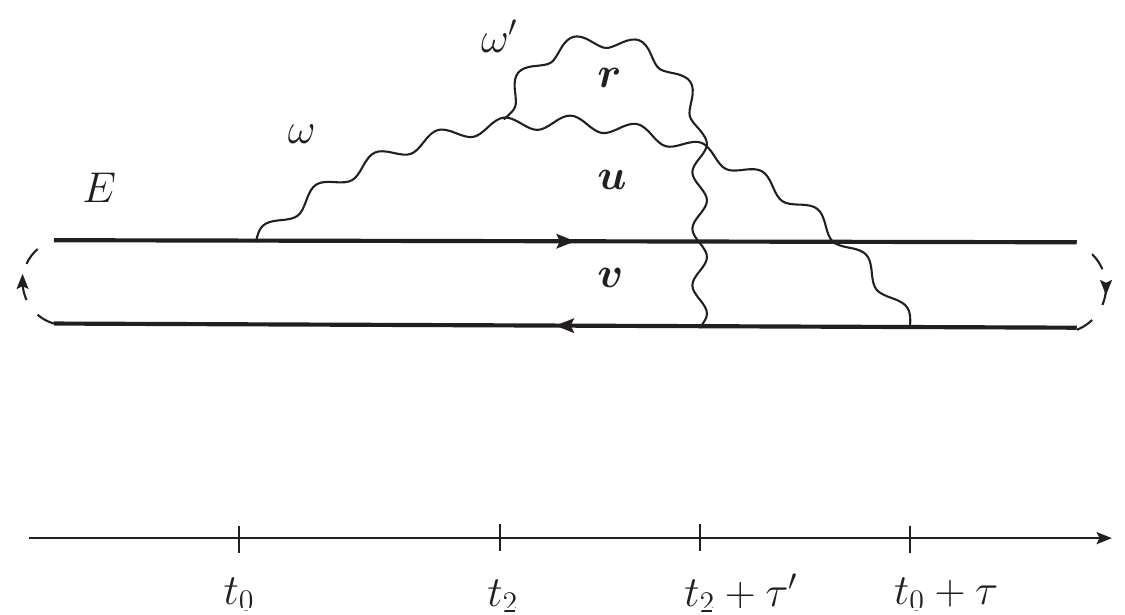} 
\caption{ Diagrammatic illustration of the radiative correction to the BDMPS-Z spectrum (\ref{eq:K-NP}). The gluon $\omega'$ is integrated over and account for real and virtual contributions. Here one topology among 9 is displayed corresponding to the radiation of the gluon $\omega'$ from gluon $\omega$ and its absorption by the gluon $E$ in the complex conjugate amplitude. The labeling is as in the previous figure, with here, in addition, $\r$ denoting the transverse coordinate of the gluon fluctuation $\omega'$. The integration over $\r \sim \q^{-1}$ in the region $ \r \gg \u$ yields the second logarithmic enhancement of the radiative correction to $\hat q$ (cf. Eq.~(\ref{dsig-0})).  }\label{fig:S3-RC}}
\end{figure}

To address this issue, radiative corrections to the BDMPS-Z spectrum (\ref{eq:bdmps-coh}) were computed \cite{Blaizot:2014bha}, where the radiative correction to the soft radiation kernel \eqn{eq:K-NP}, given explicitly at leading order by Eq.~(\ref{S2harm}) in the harmonic approximation, as depicted in Fig. \ref{fig:S3}. The BDMPS-Z spectrum of Eq.~(\ref{eq:bdmps-coh}), that is computed from  \eqn{eq:K-NP}, is easily recovered from this expression by performing the integrations over the transverse momenta and over $\tau$ \cite{Mehtar-Tani:2013pia,Blaizot:2012fh}.  One can show that the leading radiative corrections do not modify $\cK$ defined in \eqn{eq:K-NP}, that follow from computing a diagram of the type shown in Fig.~\ref{fig:S3-RC}, except for a change in the value of the parameter $\hat q$, the correction to $\hat q$ being the same as that calculated previously for momentum broadening. 

 The  diagram displayed there corresponds typically to a branching process where a gluon with initial energy $E$ (represented by the lower two thick lines in the amplitude and the conjugate amplitude, respectively), emits a  gluon at $t_0$ in the amplitude and reabsorbs it in the complex conjugate amplitude  at $t_1=t_0+\tau$, with energy $\omega=zE$.

We find that, to double logarithmic accuracy, the radiative corrections to the radiation kernel are accounted for by correcting the dipole cross-section and thus the jet quenching parameter.  This result applies to the particular 2-point function involved in the calculation of the gluon spectrum (\ref{eq:K-NP}). In order to evaluate the corrected spectrum, we need to perform the integration over the times in  Eq.~(\ref{eq:K-NP}), replacing $\hat q$ by $\hat q +\Delta\hat q (\tau_\text{max})$. However, since the correction to $\hat q$ is computed to double logarithmic accuracy one can simply replace the variable $\tau_\text{max}$ in Eq.~(\ref{qhat1}) by its typical value in the radiation process, i.e., $\tau_\text{max}\simeq t_f\equiv \sqrt{\omega / \hat q}$. The integral over $\tau$ in the BDMPS-Z spectrum (\ref{eq:K-NP}) can then be performed as for the case with no radiative correction. Doing so, we obtain \cite{Blaizot:2014bha} 
\beq\label{BDMPS-Z -rad}
\frac{\rmd I}{\rmd\omega \rmd t}\equiv \frac{\alpha_s N_c}{\pi } \sqrt{\frac{\hat q+\Delta \hat q}{\omega}},
\eeq
where for a constant $\hat q$ one gets, from Eq.~(\ref{qhat1}) letting $\k^2\simeq k^2_f(\omega)\equiv \sqrt{\omega\hat q} $,
\beq
\hat q+\Delta \hat q \approx \hat q\left[1+\frac{\alpha_sN_c}{2\pi}\ln^2\sqrt{\frac{\omega}{\hat q \tau^2_0}}\right]. 
\eeq
It can be shown that this result extends to the full kernel (\ref{kernel}) in the large $N_c$ limit \cite{Blaizot:2014bha}.  These results were recently confirmed by 
an alternative calculation of the radiative corrections to the mean-energy loss to double logarithmic accuracy\cite{Wu:2014nca}.
 
\section{Non-linear evolution of $\hat q$ }

\begin{figure}[htbp]
\centering
\includegraphics[width=7.cm]{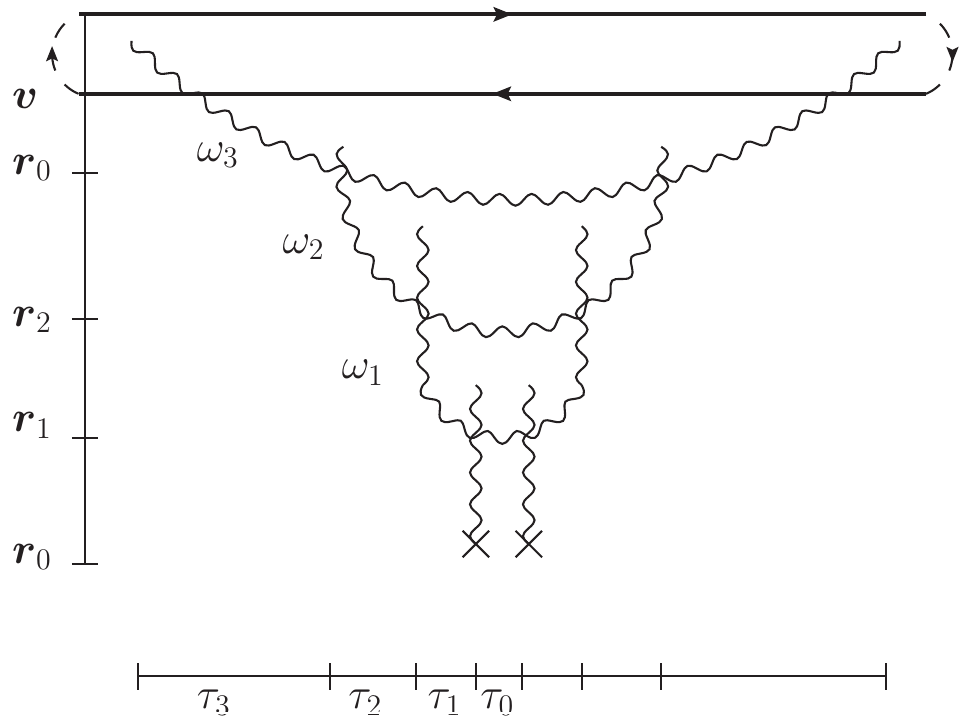}	
\caption{A diagrammatic representation of the resummation of double logarithmic contributions for a single scattering. Here three gluon emissions are depicted.  In the double logarithmic approximation successive radiations are strongly ordered in formation time, i.e. $\tau_0 \ll \tau_1\ll \tau_2\ll \tau_3\ll L$, and in transverse sizes, i.e., $1/m_D \sim \r_0 \gg \r_1\gg \r_2\gg \r_3\gg \v $. For the original dipole the subsequent radiations look {\it as if } they were instantaneous.  }		\label{fig:RG-qhat} 
\end{figure}

For large media, as soon as  $\bar\alpha\ln^2(L/\tau_0)\sim1$, one has to resum the double logarithmic power corrections. Unlike the previous resummation of independent multiple radiative corrections, this now involves radiative corrections that are correlated to each other. To understand how this resummation proceeds, we denote the standard leading order definition of the jet-quenching parameter by $\hat q_0$ and we note that the first correction to the jet-quenching parameter, $\hat q_1(\tau,\k^2)\equiv \Delta \hat q(\tau,\k^2)$ is proportional to the 3-point function, $S^{(3)}[\hat q_0]$ which is itself  a function of the leading order $\hat q_0$.  As we have shown, under radiative corrections,  the 3-point function gets renormalized by a simple modification of $\hat q$, that is,  $S^{(3)}[\hat q_0]\to S^{(3)}[\hat q_0+\hat q_1] $. This allows us to compute the second correction from Eq.~(\ref{qhat1}),
\beq
\hat q_2(\tau,\k^2)= \bar\alpha \, \int^{\tau }_{\tau_0} \frac{\rmd\tau'}{\tau'}\int^{1/\x^2}_{\hat q_0\tau'}\frac{\rmd\q^2}{\q^2}\,\hat q_1(\tau',\q^2).
\eeq
One sees  emerging a self-similarity that results from the separation of time scales involved in the computation of the leading logarithms.
The structure of the first double logarithmic corrections being set, the next corrections that yield double logarithms will follow the same systematics, with successive gluonic fluctuations ordered in formation time $\tau_0 \ll \tau _1\ll ...\ll\tau_n\equiv \tau_{\text{max}} $, or in transverse size $\r_0 \gg \r _1\gg ...\gg \r_n\equiv \r_{\text{max}} $, or in transverse momentum $m_D  \ll \q _1\ll ...\ll\q_n\equiv \k $. A diagrammatic illustration is given in Fig.~\ref{fig:RG-qhat}. The difference with the standard Double-Logarithmic Approximation (DLA) is the limits of logarithmic phase-space set by the LPM effect since, i.e., multiple-scatterings since in the DLA only a single scattering contributes, which imposes that the formation time of a fluctuation to be smaller than the BDMPS-Z formation time, or in terms of our transverse momentum variables,  $\q^2 \gg  \hat q_0\tau $ \footnote{The resummation of double logarithms in the quenching parameter was postulated earlier in Ref.~\cite{CasalderreySolana:2007sw} where the LPM suppression was not taken into account.}. The following equation resums the  double logarithmic corrections to all orders
\beq\label{qhat-evol}
\frac{\del \hat q (\tau,\k^2)}{\del \ln \tau} = \, \int_{\hat q  \tau }^{\k^2}\frac{d\q^2}{\q^2}\, \bar \alpha(\q)\, \hat q (\tau,\q^2).
\eeq
with some initial condition $ \hat q (\tau_0,\k)$. We have let the coupling running at the transverse scale $\q$.
The important feature of this equation is that it predicts the evolution of the jet-quenching parameter from an initial condition $\hat q_0$ (which can be computed e.g. on the lattice, or  to leading order in $\alpha_s$, which implies, $\hat q(\tau_0) \equiv\hat q_0  $ as given by the leading order result (\ref{qhat}). The $\tau_0$ cut-off that was introduced  to cut the logarithmic divergence in the radiative corrections, can be seen as a factorization scale.  
The solution to this equation was derived in \cite{Liou:2013qya} for the pt-broadening in the case where $\hat q_0=\hat q(\tau_0)$ is constant and for a final $\tau=L$ and $\k^2=\hat q_0L$, merging the 2 independent variables at the end of the evolution. The solution reads 
\beq\label{qhat-L}
\hat q(L)=  \frac{1}{\sqrt{\bar\alpha}}\,I_1\left(2\sqrt{\bar\alpha}\ln\frac{L}{\tau_0}\right)\,\hat q(\tau_0).
\eeq
A semi-analytic analysis of this equation including the running of the coupling has been performed in Ref.~\cite{Iancu:2014sha}.
For large $L$, the quenching parameter scales like $\hat q(L)\sim L^\gamma $, with the anomalous 
\beq\label{anomalous-d}
\gamma=2\sqrt{\bar\alpha}\,.
\eeq
Interestingly, the resummation of large double logarithms modifies the scaling of the energy loss with $L$,
 \beq
\Delta E \sim L^{2+\gamma},
\eeq
a correction that  seems to fall between the standard small coupling result, $\Delta E  \sim L^2 $ and the strong coupling result obtained with the help of the AdS-CFT correspondence in ${\cal N}=4$ SYM theory\cite{Hatta:2007cs,Chesler:2008uy}, $\Delta E\sim L^3 $.

Finally, let us make now a rough estimate of the renormalized quenching parameter. The standard perturbative estimate yields a value of about $\hat q _0\sim 1$ GeV$^2$/fm \cite{Baier:1996sk}.  For $\alpha_s\sim0.5$, $L\sim5$ fm and $T=\tau_0^{-1}\sim 500$ MeV, Eq.~(\ref{qhat-L}) yields a sizable increase by a factor 2, $\hat q \sim 2$ GeV$^2$/fm. This result is in the ballpark of the jet quenching parameter values extracted from the data \cite{JET:2013cls}.

\section{Geometric scaling and traveling waves}
\label{sec:geoscal}
In this section, we extend the analysis by exploring the consequences of the quasi-local nature of quantum corrections on the transverse momentum broadening (TMB) distribution. Specifically, we find that the TMB distribution reaches a universal scaling solution at late times (large $L$), which we compute analytically along with its sub-asymptotic deviations. This analysis leverages a formal analogy with traveling wave solutions in reaction-diffusion processes \cite{Munier:2003vc}. The self-similarity of the anomalous random walk results in a TMB distribution of L\'{e}vy type, characterized by a heavy power-law tail that describes rare, long steps spanning a wide range of transverse momenta beyond the typical scale.
L\'{e}vy flights, known for their ubiquity in nature, are observed across various stochastic processes, including biological systems, molecular chemistry, optical lattices, turbulent diffusion, and polymer transport. 

As already mentioned in Sec.~\ref{sec:TMB} (cf.~\eqn{eq:S-dipole}), the TMB distribution is related to the forward scattering amplitude $S(\xt)$ of an effective dipole in color representation $R=A,F$ with transverse size $\xt$ (see e.g. \cite{Blaizot:2013vha,Kovchegov:2012mbw}) via a Fourier transform,
\begin{equation}
 \mathcal{P}(\kt)=\int\dif^2\xt \,\rme^{-i\kt\cdot\xt} S(\xt)=\exp\left(-\frac{1}{4}\frac{C_R}{N_c}\, \xt^2\,\qhat(1/\xt^2) L\, \right) \,.\label{eq:scatt-ampl}
\end{equation}
Considering the dipole formulation in position space allows for a straightforward resummation of multiple scattering by exponentiating the single scattering cross-section, so long as the interactions between the dipole and the medium are local and instantaneous. The latter relation defines the quenching parameter in the adjoint representation which is assumed to be a slowly varying function of $\xt$. At tree-level it reads
\begin{equation}
 \qhat^{(0)}(1/\xt^2)=\qhat_0\ln\left(\frac{1}{\xt^2m_D^2}\right)\,,\label{eq:qhat-tree}
\end{equation}
up to powers of $x_\perp m_D$ suppressed terms. For a weakly coupled QGP the bare quenching parameter $\qhat_0$ and the infrared transverse scale $m_D^2$.

As discussed in sec{sec:TMB}, the resummation of multiple-scattering generates a scale $Q_\med\equiv Q_s(L)$, reminiscent of the saturation scale in the Color Glass Condensate framework \cite{}, via the relation $S(\xt^2=1/Q_s^2(L))\equiv \mathrm{e}^{-1/4}$, or equivalently, $\qhat(Q_s^2(L))L\equiv Q_s^2(L)$. This definition, standard in small-$x$ physics \cite{Kowalski:2003hm,Lappi:2011ju}, is  motivated by Moli\`ere theory of multiple scattering \cite{Moliere:1948zz,Barata:2020rdn} in which $Q_s$ is the transverse scale that controls 
the transition between the multiple soft scattering and the single hard scattering regimes.

In the first form of the evolution equation for $\hat q$ the lower bound of the $\q$ integral depends on the $\hat q_0$. However, this lower bound should be replace by the resummed saturation scale is obtained by solving the evolution equation for $\hat q$ itself. This yields the follow system of non-linear coupled equations: 
$\tau$~\cite{Liou:2013qya,Blaizot:2014bha,Iancu:2014kga}: 
\begin{align}
\qhat(\tau,\kt^2)&=\qhat^{(0)}+\int_{\tau_0}^{\tau}\frac{\dif\tau'}{\tau'}\int_{Q^2_{ s}(\tau')}^{\kt^2}\frac{\dif \q^2}{\q^2} \ \abar(\q^2)\ \qhat(\tau',\q^2)\,,\nn
Q^2_{s}(\tau)&=\qhat(\tau,Q_{ s}^2(\tau))\tau\,,\label{eq:Qsat}
\end{align}
where $\abar=\alpha_sN_c/\pi$ and $\qhat^{(0)}\equiv \hat q (\tau_0,\kt)$ .
The condition $\kt^2\ge Q^2_s(\tau)$ in Eq.\,\eqref{eq:Qsat} enforces the gluon fluctuations to be triggered by a single scattering with plasma constituents whose contribution is logarithmically enhanced  compared to multiple scattering. 

One may go beyond the DLA approximation \cite{Caucal:2021lgf,Caucal:2022mpp} by solving the Balitsky-Fadin-Kuraeev-Lipatov (BFKL) equation augmented with a saturation boundary \cite{Munier:2003vc}. In \cite{Caucal:2022mpp}, analytic results were obained to NNLL accuracy for $\hat q$.

The final value of $\tau$ is fixed at the largest time allowed by the saturation condition, i.e.\ at the time scale $\tau$ such that $Q_s^2(\tau)=\kt^2$, as long as $\tau<L$ and $\tau =L$ otherwise. The latter case corresponds to the dilute regime \cite{Blaizot:2019muz}, that is when $\kt^2 \gg Q_s^2(L)$.

In what follows, we address both analytically and numerically the non-linear system \eqref{eq:Qsat}.  Analytic solutions are in general difficult to obtain, however, a solution for the linearized problem \eqn{qhat-evol} that consists in approximating $Q_s^2(\tau)\simeq \qhat_0\tau$ for the emission phase space can be found in \cite{Iancu:2014sha}. Formally, this linearization is valid in DLA.

The TMB distribution is said to obey geometric scaling if it is only a function of $\kt^2/Q^2_s(L)$ as a result of scaling invariance of the radiative process at late times. More precisely, we would have 
\begin{equation}
 \lim\limits_{L\to\infty}\qhat(\kt^2)L=Q_s^2(L)f\left(\frac{\kt^2}{Q_s^2(L)}\right)\,,
\end{equation}
for some scaling function $f$ to be determined.

Geometric scaling was extensively studied in the context of deep inelastic scattering, where it has been shown that the gluon distribution $g(x,Q^2)$ at small $x$ satisfies this property over a broad region of photon virtuality $-Q^2$ \cite{Stasto:2000er,Iancu:2002tr,Kwiecinski:2002ep}. 
We shall demonstrate that TMB exhibits similar properties.

Remarkably, it is possible to find the scaling function $f$ for the non-linear problem defined by Eqs.\,\eqref{eq:Qsat}. In terms of the variables 
\beq
Y=\ln(\tau/\tau_0) \,, \qquad \rho=\ln(\kt^2/(\qhat_0\tau_0)) \,, 
\eeq
 the differential equation satisfied by $\qhat$ reads
\begin{equation}
  \frac{\partial \qhat(Y,\rho)}{\partial Y}=\abar \int_{\rho_{s}(Y)}^{\rho}\rmd\rho' \ \qhat(Y,\rho')\,,\label{eq:eqndiff_Yrho}
\end{equation}
with $\rho_s=\ln(Q_s^2/(\qhat_0\tau_0))$.

In addition to scaling solution, sub-asymptotic corrections that violate geometric scaling can also be computed near the wave front, typically for $x\gg 1$. The solution takes the form \cite{Brunet:1997zz,Munier:2003sj}

\begin{align}
& \qhat(Y,\rho)=\qhat_0\, \rme^{\rho_s(Y)-Y}\exp\left(\beta  x-\frac{\beta  x^2}{4cY}\right)\nonumber\\
 &\times\left[1+\beta  x+\frac{bx}{c^2Y}\left(1+\frac{\beta (c+4)x}{6}\right)+\mathcal{O}\left(\frac{1}{Y^2}\right)\right]\,,\label{eq:qhat-final}
\end{align}
with 
\beq 
 \rho_s(Y)=cY-\frac{3}{2(1-\beta )}\ln(Y)+\rm const.\label{eq:Qs-Y}
\eeq
with $c=1+2\sqrt{\abar+\abar^2}+2\abar\simeq 1+2\sqrt{\abar}$ and $\beta=(c-1)/2c$.  This is reminiscent of the TW solutions \cite{Munier:2003vc} to the Balitsky-Kovchekov (BK) equation \cite{Balitsky:1995ub, Kovchegov:1999yj}.

The TW solution \eqref{eq:qhat-final} provides the functional form of $\qhat(Y,\rho)$ near the wave front, i.e.\ for $x=\rho-\rho_s(Y)\sim \sqrt{Y}\gg1$ and 
is independent of the initial condition (for physically relevant ones). 
The resummed TMB distribution displays a universal behavior independent of the non-perturbative modeling of the tree-level distribution.

%

%
\section{Super-diffusion and modification of Rutherford scattering}
\label{sec:supdiff}

We now discuss the physical consequences of the asymptotic solution \eqref{eq:qhat-final} for $\hat{q}$ on the TMB distribution given by Eq.\,\eqref{eq:scatt-ampl}. Our focus is on the large $k_\perp$ region, where the distribution is characterized by rare events that are sensitive to the point-like nature of the medium scattering centers \cite{DEramo:2012uzl}. As previously argued, in the large $L$ limit, the TMB distribution $\mathcal{P}(k_\perp)$ depends only on the scaling variable $k_\perp/Q_s(L)$. 

\begin{figure}[t]
 \centerline{\includegraphics[width=0.45\columnwidth]{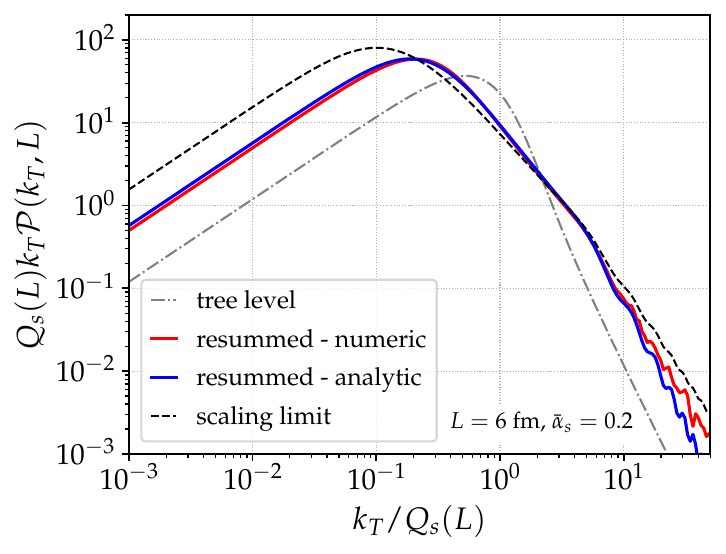} \includegraphics[width=0.45\columnwidth]{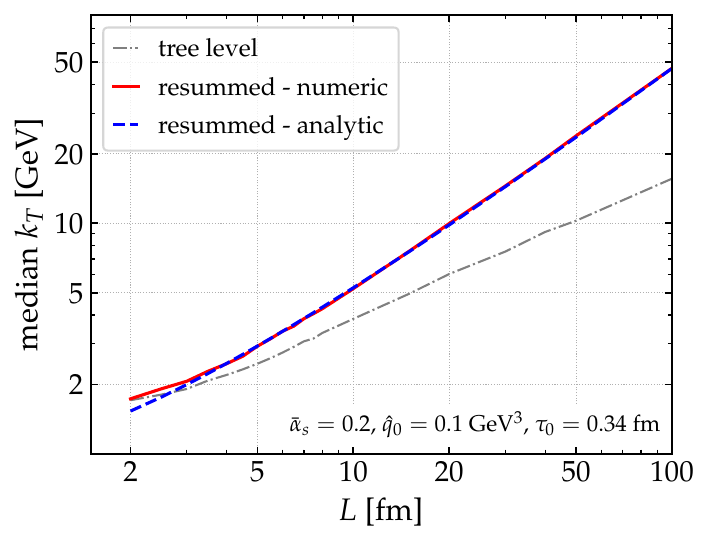}}
 \caption{\textbf{Left:} The TMB distribution of a high-energy gluon propagating through a dense medium of size $L$ is shown at tree level (dash-dotted line) and after the resummation of leading radiative corrections (solid red). The dotted black line represents the scaling limit as $L \to \infty$, while the blue curve corresponds to our analytic result, given by Eq.\,\eqref{eq:qhat-final}, which includes sub-asymptotic corrections (color online). 
\textbf{Right:} The system size dependence of the median of the TMB distribution is shown at tree level (grey line) and after the numerical resummation of radiative corrections (red line). The dashed blue line represents our analytic prediction, given by Eq.\,\eqref{eq:qhat-final}.
}
  \label{fig:pt-broad-2}
\end{figure}

In Fig.\,\ref{fig:pt-broad-2}, we test this asymptotic scaling solution against exact numerical solutions of the non-linear evolution equation for $\hat{q}$. The distribution is displayed as a function of $k_\perp/Q_s(L)$ with $Y = \ln(L/\tau_0) = 4$, for the following cases: 
(i) in dash-dotted grey, the tree-level result using Eq.\,\eqref{qhat}, 
(ii) in red, after quantum evolution obtained by numerically solving Eqs.\,\eqref{eq:Qsat}, 
(iii) in blue, using the expression in Eq.\,\eqref{eq:qhat-final}, which includes sub-asymptotic corrections to the scaling limit, and 
(iv) in dashed black, the scaling limit $Y \to \infty$ from Eq.\,\eqref{eq:qhat-final}. Interestingly, the sub-asymptotic corrections account for the relatively large deviations between the asymptotic curve and the exact numerical result at the moderate value of $L = 6$ fm.

The $k_\perp$ distribution exhibits two distinct regimes: the region around the peak near $Q_s(L)$ and the large $k_\perp$ tail, where $k_\perp \gg Q_s(L)$. These results can be interpreted in terms of a special type of random walk in momentum space, known as a L\'{e}vy flight. This remarkable connection with statistical physics allows us to highlight several interesting features: 
(i) self-similar dynamics, 
(ii) super-diffusion, and 
(iii) a power-law tail with a slower decay compared to the Rutherford behavior, $\kt^{-4}$.

Because of its heavy tail, the mean $k_\perp^2$ of the TMB distribution is not defined. Nevertheless, it is possible to introduce a measure of the characteristic width of the $k_\perp$ distribution, and study its behavior as a function of the medium size $L$. In what follows, we shall use the median value $\langle k_\perp\rangle_{\rm med}$ of $k_\perp\mathcal{P}(k_\perp)$  which is shown in Fig.~\ref{fig:pt-broad-2} for three different scenarios. In grey, we plot the tree-level resulting from Eq.~\eqref{eq:scatt-ampl} and \eqref{eq:qhat-tree}. The median scales approximately like $(L \ln L)^{1/2}$, which up to the logarithmic factor resulting from the Coulomb logarithm in \eqn{eq:qhat-tree},  exhibits the standard diffusion scaling.
 The red line is the median of the $k_\perp$ distribution obtained using the resummed value of $\qhat$ with fixed coupling, after numerical resolution of Eqs.\,\eqref{eq:Qs-Y}. We then compare this result with our analytic prediction \eqref{eq:qhat-final} (assuming $\langle k_\perp\rangle_{\rm median}\propto Q_s$ \footnote{the unknown pre-factor is determined by a fit to the numerical result at large $L$}), $\langle k_\perp\rangle_{\rm median}\propto L^{^{\frac{c}{2}+\frac{b}{2}\frac{\ln(Y)}{Y}}}$,
which is represented in blue in Fig.\,\ref{fig:pt-broad-2} (right panel) .
Remarkably, the agreement is excellent down to rather small values of $L\sim 3$ fm.
Since $c/2>1/2$, the median grows faster than $\sqrt{L}$ at large $L$, illustrating the super-diffusive behavior of TMB beyond leading order, with a deviation to the standard diffusion of order $\sqrt{\abar}$.

Another important aspect of L\'{e}vy flights is
the power law decay of the step length distribution for a L\'{e}vy walker. This reflects the fact that long jumps with arbitrary length may occur with non-negligible probability. In the problem at hand, this power-law tail can also be understood as a consequence of the self-similar nature of overlapping successive gluon fluctuations.

The tail of the TMB distribution is controlled by the large $\kt^2$ behavior of $\qhat(\kt^2)$, and consequently, by the exponential in  \eqref{eq:qhat-final}.  
Without loss of generality, one can derive the leading behavior of $\mathcal{P}(\kt)$ at large $\kt$ by expanding the dipole S-matrix for small dipole sizes, as a result the Fourier transform can be approximated by:
\begin{align}
 \mathcal{P}(\kt) \sim \vec{\nabla}_{\kt}^2\frac{\pi}{\kt^2}\frac{\dif \qhat (\kt^2)L }{\dif \ln \kt^2 }\,,\label{eq:P1-largekt}
\end{align}
up to logarithmically suppressed terms. 
This formula quantifies the deviations from the Rutherford scattering cross-section that are induced by radiative corrections.
Applying Eq.~\eqref{eq:P1-largekt} to scaling part of our solution \eqref{eq:qhat-final}, one finds the tail
 \begin{equation}
  \mathcal{P}(\kt)\propto\frac{1}{Q_s^2(L)}\left(\frac{Q_s^2(L)}{\kt^2}\right)^{\nu}\,,
 \end{equation}
 with $\nu=2-\beta +\mathcal{O}(\ln(x)/x)$ and $x=\ln(\kt^2/Q_s^2(L))$.  The power of the tail deviates from the tree level Rutherford behavior by $\sim -2\sqrt{\abar}$. The form of $\nu$ is correct in the strict scaling limit $L\to \infty$. For finite $L$ values, the $1/\kt^4$ tail is recovered at very large $k_\perp$, since  $\nu=2-2\sqrt{\abar Y/x}$ (when $x\gg Y$). The fact that geometric scaling extends in the tail region is known in the context of saturation physics as the ``extended geometric scaling window" corresponding to $Q_s\ll k_\perp \ll Q_s^2/\mu$.  
 
Finally, note that this analysis is valid so long as $E/k_\perp^2 \gg L$, where $E$ is the energy of the fast parton. This allows us to neglect the quantum diffusion of the dipole in the medium. In the opposite case the quantum phase is suppressed when $E/k_\perp^2  \ll \tau \ll L$ and the evolution is expected to be of DGLAP type with the substitution $\ln L/\tau_0 \to \ln  E/(k_\perp^2\tau_0)$.

As a final remark, we stress that this analysis can be extended to including the running of the coupling and most importantly can be pushed to NNLL accuracy using the leading log equations such as BFKL or DGLAP with a saturation boundary that implement the non-linear evolution \cite{Caucal:2022fhc,Caucal:2022mpp}.

%% file: decoherence.tex
\chapter{Color Decoherence }\label{chap:decoherence}

\section{Parametric estimates}
So far, we have discussed the multiple interactions of a leading hard parton traversing the plasma, without addressing the early collinear and hard branching processes initiated by the creation of a highly virtual hard parton, regardless of the presence of the medium. The goal was to first focus on the ``longitudinal" coherence phenomena associated with the Landau-Pomeranchuk-Migdal (LPM) effect, which is responsible for coherent medium-induced radiation, where multiple scattering centers act collectively during the formation of the radiated gluon. We then demonstrated that, in the gluon cascade regime, this coherence time is negligible compared to the length of the medium, enabling us to adopt a classical description of the medium-induced shower in terms of kinetic equations.

Now, consider the interaction of the plasma with two or more collinear charges entangled in color space—such as a virtual photon splitting into a quark-antiquark pair in a color-singlet state. In this case, the interaction may not factorize into incoherent collinear partons that can be individually treated as single partons in the plasma, thereby reducing the problem to multiple copies of the single-parton case. 

 Consider this simple argument: due to the property of color transparency, a sufficiently narrow quark-antiquark pair in a singlet state is not resolved by the medium color charges, up to power-suppressed terms of order $r_\perp^2 \sim 1/Q^2$, where $r_\perp$ is the pair's transverse size in the plasma. This negligible contribution is associated with hard modes in the medium that are strongly power suppressed. As a result, the pair would traverse the plasma without interacting or losing energy to the medium \cite{Mehtar-Tani:2010ebp}.

However, we should naturally expect the opposite limiting case, where the pair is well-separated in transverse space, or equivalently in angle. At any given time $t$, parametrically of order $L$, the transverse separation is given by $r_\perp \sim \theta_{12} L$. In this regime, the medium resolves the individual color charges, and the two jets can be treated as independent with respect to their interactions with the medium (see Fig.~\ref{2parton-eloss} for an illustration)

In contrast, the transverse size of a medium-induced splitting matches the medium's resolution scale at the time of its formation, leading to an incoherent subsequent evolution.

The purpose of this chapter is to present the QCD framework for computing the energy-loss distribution of such an entangled pair in the plasma, as the next non-trivial step beyond the single-parton energy loss discussed in Chap.~\ref{chap:med-partons}. We expect interference effects to play a fundamental role in the transition between the coherent and decoherent regimes.

A simple parametric estimate can help identify the scales governing the process of color decoherence in the plasma. We have already identified a transverse scale associated with the pair, $r_\perp \sim \theta L$. The medium transverse scale is related to the typical transverse momentum exchange with the plasma, namely, $Q_{\rm med} \equiv \sqrt{\hat{q} L}$, the transverse momentum broadening scale. Using these scales, we can distinguish two regimes \cite{Mehtar-Tani:2010ebp,Mehtar-Tani:2011hma,Mehtar-Tani:2012mfa,Casalderrey-Solana:2011ule,Mehtar-Tani:2017ypq,Mehtar-Tani:2017ypq}:

\begin{itemize}
    \item {\it Coherence regime:} When $r_\perp \ll Q_{\rm med}^{-1}$, the medium does not resolve the inner structure of the collinear splittings and interacts with the total charge, i.e., that of the parent parton.
    \item {\it Decoherence regime:} When $r_\perp \gg Q_{\rm med}^{-1}$, the medium resolves the individual color charges, which can be treated as independent emitters in the plasma.
\end{itemize}

An angular scale naturally emerges from this analysis, called the coherence angle:
\beq\label{eq:coh-angle}
\theta_c \equiv \frac{1}{Q_{\rm med} L} \sim \frac{1}{(\hat{q} L^3)^{1/2}}\,,
\eeq
which also corresponds to the minimum medium-induced radiation angle below which the spectrum is strongly suppressed. The physics of transverse coherence is closely related to the LPM effect due to the connection between the gluon formation time and the transverse medium scale.

\begin{figure}
\centering
\includegraphics[width=0.6\textwidth]{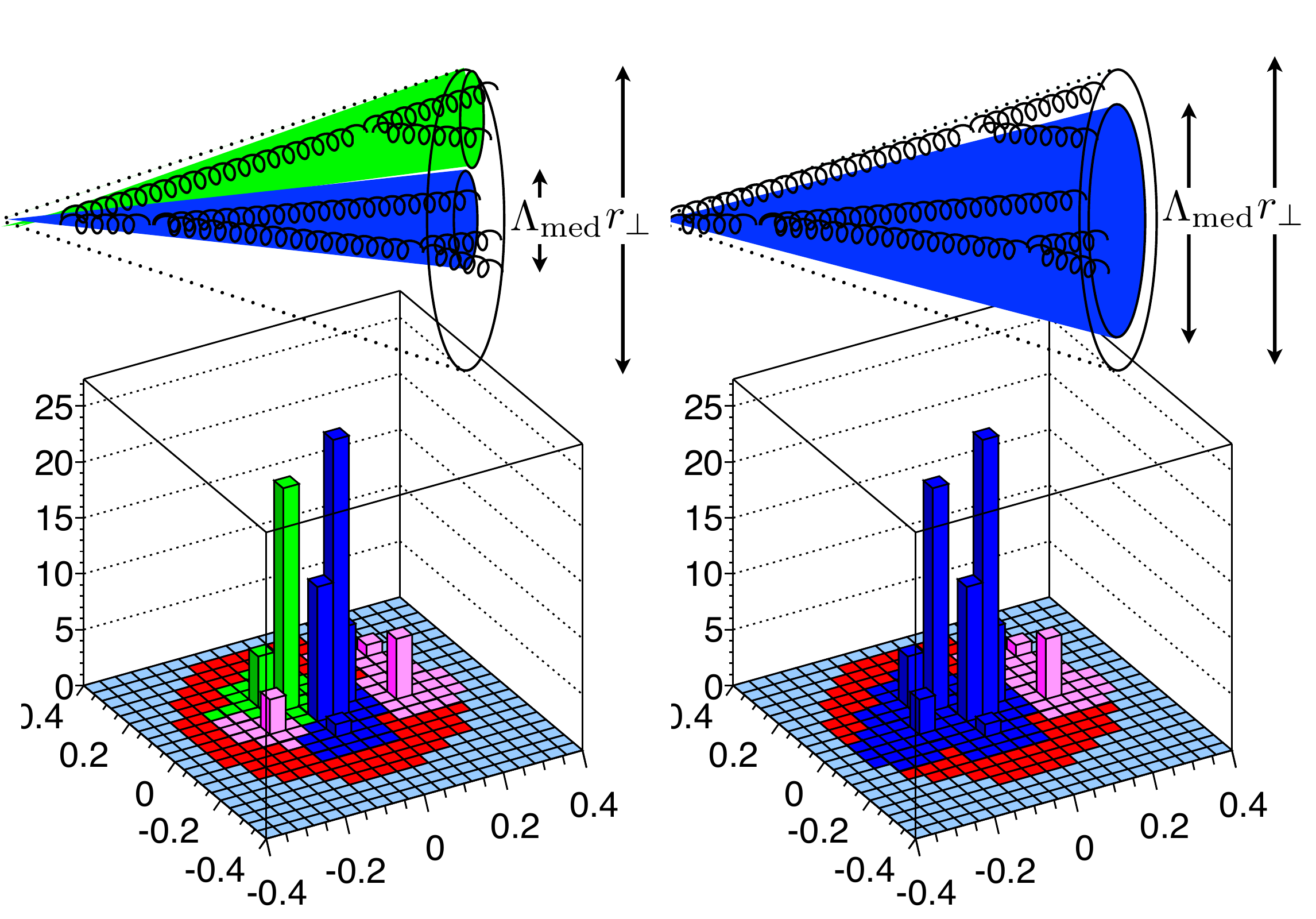}
\caption{Illustration of a sample jet event resolved with angular resolution scales $\theta_c = 0.1$ (left) and $\theta_c = 0.15$ (right). The medium resoltution scale $\Lambda_{\rm med}\equiv Q^{-1}_{\rm med}$ is compared to the jet transverse size $r_\perp$. The blue histogram corresponds to the hardest resolved sub-jet, the green histogram to the next-to-hardest, while the pink histogram represents the distribution of soft fragments. Adapted from Ref.~\cite{Casalderrey-Solana:2012evi}.}
\label{fig:legoplot}
\end{figure}

\begin{figure}[t!]
\begin{center}
\includegraphics[width=0.7\textwidth]{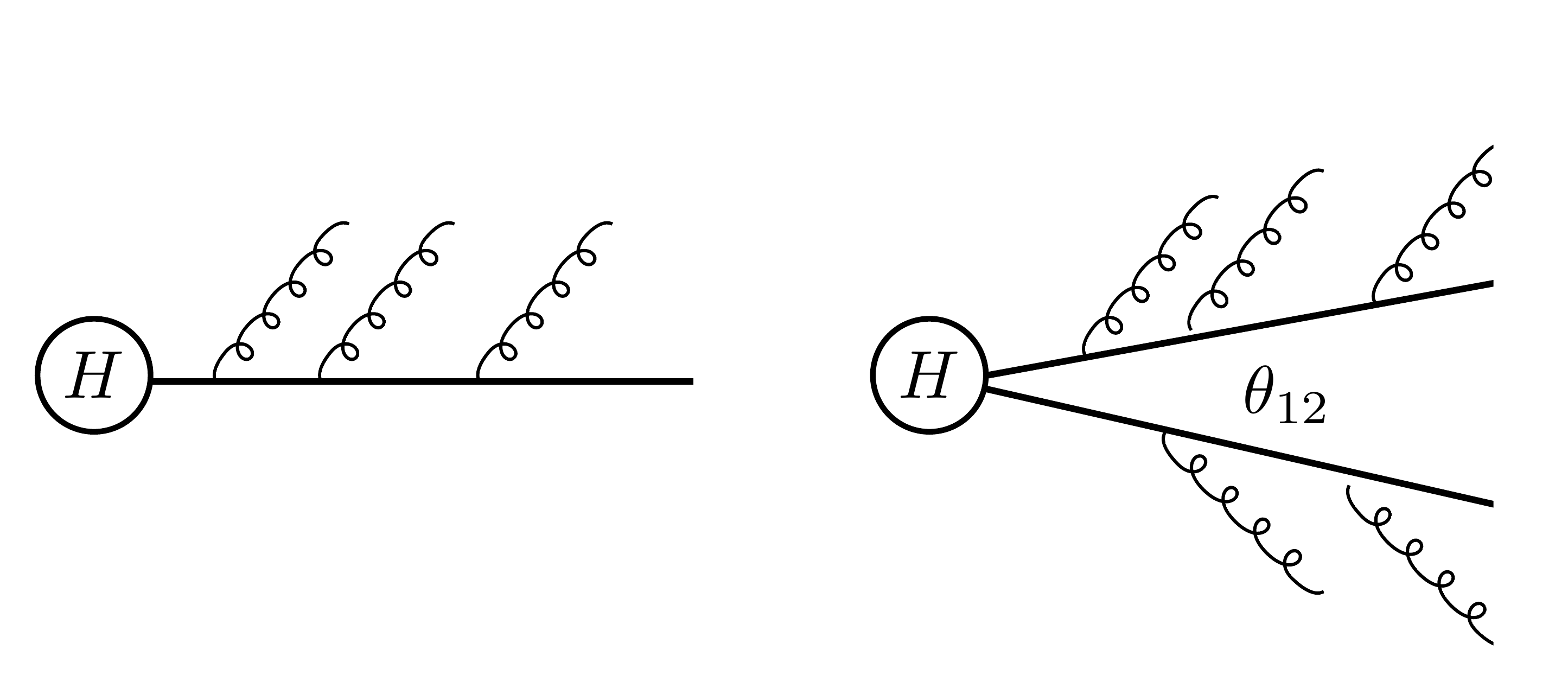}
\caption{Depiction of the amplitude for radiative energy loss for a single particle (left) and a two-pronged color object (right). Multiple soft gluon emissions are represented through Wilson lines, while elastic processes are omitted for simplicity. }
\label{2parton-eloss}
\end{center}
\end{figure}

An illustrative representation of color decoherence in a medium is shown in Fig.~\ref{fig:legoplot}, from Ref.~\cite{Casalderrey-Solana:2012evi}. The figure highlights how the resolution power of the medium determines the number of effective color charges that can be resolved within a multi-prong jet system, thereby governing the onset of decoherence.

Similarly, by fixing the opening angle of the pair, we can derive a time scale for the transition from a color-coherent pair—whose transverse distance grows linearly with time and is initially unresolved—until the decoherence time is reached:
\beq\label{eq:decoh-time}
t_{\rm d} \equiv \frac{1}{(\hat{q} \theta_{12}^2)^{1/3}} \,.
\eeq
The generalization of this discussion beyond the singlet case is straightforward and is most easily understood in the large-$N_c$ limit. For example, the energy loss of a quark-antiquark pair in a color octet state can be factorized into the energy loss of a singlet antenna and the energy loss associated with the total charge of the parent gluon.

There is well known feature that reflects the coherence of the pair, active near the origin of the splitting, before interactions with the medium occur. In vacuum, a dipole can only radiate gluons that can resolve its size, i.e., those with a transverse wavelength $\lambda_\perp \sim k_\perp^{-1} < r_\perp$, where for a time-like dipole the size grows linearly with time, $r_\perp \sim \theta t$. However, at a given time $t$, only certain gluons have had sufficient time to develop quantum-mechanically. Substituting $t$ with the typical formation time of a splitting, $t_{\rm form} \sim \omega/k_\perp^2$, transforms this condition into the well-known angular ordering constraint $\theta < \theta_{12}$. This constraint reflects the fact that large-angle radiation is strongly suppressed due to color coherence \cite{Dokshitzer:1991wu,Mueller:1981ex,Bassetto:1982ma,Dokshitzer:1982ia}.

The effect of in-medium color decoherence is cumulative, with multiple soft gluon exchanges with the medium causing rapid color precession of the antenna's color charges. This, in turn, leads to the decoupling of the pair. After the decoherence time, the pair of hard partons can be treated as independent, fast-moving color charges, to which the single-parton energy loss framework can be applied. In the next chapter, we will extend this picture to a large number of collinear color charges within the so-called leading logarithmic approximation.

\section{Factorization and energy loss}

Anticipating the study of jets characterized by an opening angle $R$, we distinguish between two angular scales: collinear vacuum radiation with angles $\theta_{12} < R$, confined within the jet, and medium-induced radiation that escapes outside the jet at angles larger than $R$. In this discussion, we focus exclusively on medium-induced radiation from the antenna, postponing the treatment of vacuum radiation to the next chapter.

As discussed in the previous chapters, soft, medium-induced radiation can be initiated at any point within the medium, with $\tform \ll t < L$, unlike vacuum radiation, where $t \sim \tform$. In the regime of interest, $\tdecoh \gg \tform$, the decoherence time $\tdecoh$ marks the transition from coherent radiation ($t \ll \tdecoh$), where the system radiates as a single charge, to independent radiation ($t \gg \tdecoh$), where the two charges radiate independently.

The sequence of events unfolds as follows: collinear splitting occurs first, after a time $\tform \sim E / p_\perp^2 \sim 1 / (E \theta_{12}^2)$, where $E$ is the energy of the parent parton and $p_\perp$ the transverse momentum of the splitting. Initially, the pair remains close in spatial separation and loses energy collectively as a single charge. Over time, as the pair separates and becomes increasingly susceptible to in-medium color precession, it eventually decoheres after a time $\tdecoh$, transitioning to independent energy loss.

Let us, for the moment, consider the initial parton to be a quark. In this scenario, medium scales represent only a small correction, so $\omega \ll E$, where $E$ is the energy of the projectile parton, and $\omega$ corresponds to the medium-induced gluons. Medium interactions that lead to energy loss outside the jet cone with opening angle $R$ set an upper energy scale, $E_{\rm med} \sim k_\perp / R \sim \sqrt{\hat{q} L} / R < \omega_c \ll E$ for medium-induced radiation relevant for energy loss. This scale justifies the high-energy limit assumption. Note that the inequality $\sqrt{\hat{q} L} / R \ll \omega_c$, which restricts the soft gluon frequency to the short formation time regime ($\tform \ll L$), also implies that the jet opening angle satisfies $R \gg \theta_c$.

In this limit the emissions spectrum does not depend explicitly on the parent energy,
and the cross-section factorizes as follows:
\beq
\frac{\rmd \sigma }{\rmd \omega \rmd p_T} \simeq   \frac{\rmd I}{\rmd \omega} \, \frac{\rmd \sigma_\text{vac}}{\rmd E}  \,,
\eeq
Here, $\sigma_\text{vac}$ represents the quark cross-section, and $\rmd I / \rmd \omega$ denotes the medium-induced bremsstrahlung spectrum, as discussed in Chap.~\ref{chap:med-partons} (see \eqn{eq:bdmps-coh}). The quark energy before radiation is given by $E = p_T + \omega$. The gluon frequency $\omega$ is neglected everywhere except in $\sigma_\text{vac}$, as the steeply falling spectrum amplifies the impact of even a small energy loss on the final spectrum. Typically, the vacuum cross-section scales as $\rmd \sigma_\text{vac}/\rmd E \propto E^{-n}$, where $n \gg 1$ \cite{Baier:2001yt}. We refer the reader to the discussion in Chap.~\ref{chap:med-partons} on the details of the energy loss of single hard parton.  

To determine the inclusive spectrum of single quarks in heavy-ion collisions, one must integrate over the gluon frequency and include virtual corrections. The final-state spectrum of quarks is then expressed as a convolution of the vacuum production spectrum, $\rmd \sigma_\text{vac} / \rmd E$, with the probability of energy loss:

\beq \label{eq:one-prong-factorization} \frac{\rmd \sigma_\text{med}}{\rmd p_T} = \int_0^{\infty} \rmd \epsilon, \ProbOne(\epsilon) \frac{\rmd \sigma_\text{vac}(p_T + \epsilon)}{\rmd E} ,, \eeq

where $\ProbOne(\epsilon)$ represents the energy-loss probability distribution, as defined in \eqn{eq:poisson}. We have previously demonstrated that the single-parton energy-loss probability can be expressed as a product of independent medium-induced emissions (see Fig.~\ref{2parton-eloss}).

The probability distribution in \eqn{eq:poisson} is usually referred to as the quenching weight \cite{Salgado:2003gb} and forms the basis of most theoretical studies of jet quenching. The main underlying assumption of these studies is that the initial projectile does not split inside the medium with a vacuum probability. For high-energy jets, this requirement neglects the contributions coming from hard radiation that can be formed early in the medium. Calculating how energy-loss affects such processes is the topic of this section.

Let us now turn to the question of the energy loss probability for a neighboring pair of hard partons, separated by an angle $\theta_{12} < R$, originating from the same hard vertex, as depicted  in Fig.~\ref{2parton-eloss}. We will summarize the main results here and refer the reader to Ref.~\cite{Mehtar-Tani:2017ypq} for a detailed derivation.

The pair is produced in a hard event at scales much larger than the medium scale, with $Q = \theta_{12} E \gg Q_{\rm med}$. Consequently, the splitting occurs quasi-instantaneously, allowing the Born-level process to be factored out from subsequent interactions with the medium. This setup extends the concept of quenching weights, as described in the previous subsection, and serves as a stepping stone toward understanding the energy loss of full-fledged jets. The two partons are color-connected and can be produced in an arbitrary color representation of SU(3), such as singlet, triplet, or octet.

As for single-parton quenching, we shall compute the medium modification of the two-parton system spectrum. In the collinear limit, $p_\perp \ll E$, the quark-gluon spectrum in vacuum is given by 
\beq\label{eq:vac-2prong-spect}
\frac{\rmd\sigma_\text{vac}}{\rmd z \rmd E \,\rmd \p^2} \simeq \frac{\alpha_s C_F}{2\pi} \frac{p_{gq}(z)}{\p^2} \frac{\rmd \sigma_\text{vac}}{\rmd E} \,,
\eeq
to leading order in perturbation theory and at leading logarithm, where we denote $p_\perp \equiv |\p|$ throughout the paper.  In \eqn{eq:vac-2prong-spect}, $p_{gq}(z)$ stands for the quark-gluon  Altarelli-Parisi splitting function and $\rmd \sigma_\text{vac} \big/\rmd E$ is the Born-level quark spectrum.

For a hard ``instantaneous'' splitting the two-parton distribution in the presence of a medium  reads 
\beq
\label{eq:two-prong-spectrum}
\frac{\rmd \sigma_\text{med}}{ \rmd z \rmd p_T \,\rmd \p^2} = \int_0^\infty \rmd \epsilon \, \ProbTwo(\epsilon) \, \frac{\rmd\sigma_\text{vac}}{\rmd z \, \rmd E \,\rmd \p^2} \,,
\eeq
analogously to Eq.~(\ref{eq:poisson}), here $\ProbTwo(\epsilon)\equiv \ProbTwo^{\sst R}(\epsilon,\theta_{12},L)$ stands for the energy loss probability distribution for a two-prong structure in the color representation $R$. In this case, the opening angle is $\theta_{12} = p_\perp/(z(1-z)E)$.
This approximation holds when there is large separation between the time-scales of production of the hard pair and the time-scale of medium modifications. In particular, this involves the case $\tform \ll \tdecoh$, which contains the leading logarithmic phase space that will be discussed in the next chapter, for a further discussion see also \cite{Mehtar-Tani:2017web}.\eqn{eq:two-prong-spectrum} is valid at leading power and leading logarithmic accuracy. Recent efforts to extend this factorized description beyond the leading approximation have been pursued in~\cite{Casalderrey-Solana:2015bww,Dominguez:2019ges,Abreu:2024wka}.

The main dependence on $\epsilon$ in the spectrum arises from the parent parton energy, where $E = p_T + \epsilon$. The splitting function $P(z)$ depends also on $\epsilon$, but such dependence can be neglected in a first approximation, as long as the energies of the daughter partons satisfy $zE \gg \epsilon$. Indeed, the steeply falling spectrum, assumed to follow a power-law behavior as discussed earlier, makes the ratio of medium-modified to vacuum spectra particularly sensitive to small energy shift in the spectrum as discussed in Sec.~\ref{sec:hadron-quenching}.

The singlet antenna represents the fundamental color configuration to consider in the large-$N_c$ limit for constructing general results applicable to arbitrary color states. Therefore, we focus on the decay of a boosted massive object in a singlet color state, such as a virtual photon decaying into a quark-antiquark ($q\bar q$) pair, while neglecting the quark masses for simplicity.

We adopt the same set of approximations as in the single-quark case discussed previously. Specifically, in the treatment of radiation, we work in the soft gluon approximation for medium-induced gluons, assuming quasi-instantaneous emissions and neglecting overlapping formation times (see Chap.~\ref{chap:cascade} for details). As a final step, we will generalize this description to include the contributions from a generic total charge of the parent parton.

\section{Operator definition }
It is possible to construct an operator definition for the $n$-parton energy loss probabilities as a function of Wilson lines, which are the fundamental objects encoding the soft limit of QCD. The single-parton energy loss takes the following form \cite{Mehtar-Tani:2024smp}:

\beq\label{eq:P1-operator}
P_1(\epsilon) = \frac{1}{d_R} \sum_X \delta(\epsilon - \bar{n} \cdot k_{\rm loss}) \, \tr_c \left[ \langle \med | U(n) | X \rangle \langle X | U^\dagger(n) | \med \rangle \right],
\eeq
where $\bar{n} \cdot k_{\rm loss} = k^+_{\rm loss} = \sum_i k_{i,s}^+$ represents the sum over the longitudinal momenta of the radiated soft gluons in the final state $|X\rangle$.

Multiple soft-gluon radiation and interactions with the plasma background field are encoded in the semi-infinite Wilson line:
\begin{align}
U(n) &\equiv \exp \left[ ig \int_0^\infty \rmd s \, \bar{n} \cdot A(n s) \right]\equiv \exp\left[ ig \int_0^\infty \rmd x^+ \, A^-(n s) \right],
\end{align}

where, in the last line, we assume the 4-vector $n = (1, 0, 0, 1)$ to point along the $z$-axis. Here, $d_R$ represents the dimension of the emitter's color representation: $d_R = N_c$ for a quark or $d_R = N_c^2 - 1$ for a gluon. The delta function acts as a measure function. 

In general, this definition encompasses both medium-induced and vacuum radiation. However, vacuum radiation can be subtracted by normalizing the right-hand side of \eqn{eq:P1-operator} to its vacuum version, where the medium state $|\med\rangle$ is replaced by the vacuum state $|0\rangle$.

For two-parton energy loss, the distribution involves two directions, $n_1$ and $n_2$, and two Wilson lines on each side of the cut \cite{Mehtar-Tani:2024smp}:
\beq\label{eq:P2-operator}
P_{2(\rm sing)}(\epsilon) = \frac{1}{d_R} \sum_X \delta(\epsilon - \bar{n} \cdot k_{\rm loss}) \, \tr_c \left[ \langle \med | U(n_1) U(n_2) | X \rangle \langle X | U^\dagger(n_2) U^\dagger(n_1) | \med \rangle \right].
\eeq

The gauge field $A^\mu$ can be split into a background field $A_\bkg^-(x^+,x^-=0,\x)$ and a quantum flied, $a^\mu(x)$ that depicts radiative corrections:
\beq
 A^\mu(x) \equiv g^{\mu -} A_\bkg ^-(x^+,x^-=0,\x) + a^\mu(x)\,.
\eeq
The propagation of the quantum takes place in the presence of the background field and is described by the non-eikonal propagator introduced in Chap.~\ref{chap:formalism}. 

For simplicity, we have expressed the two-parton energy loss in the case of the singlet color configuration. For  general color representations $R$ we have to include a Wilson line in the opposite directed $\bar n$ with representation $\bar R$ to ensure gauge invariance. However, it drops out in light-cone gauge $A^+$. 

The generalization to arbitrary number of emitters is straightforward, however, it lies beyond the scope of the present report.

\begin{figure}[t!]
\begin{center}
\includegraphics[width=0.6\textwidth]{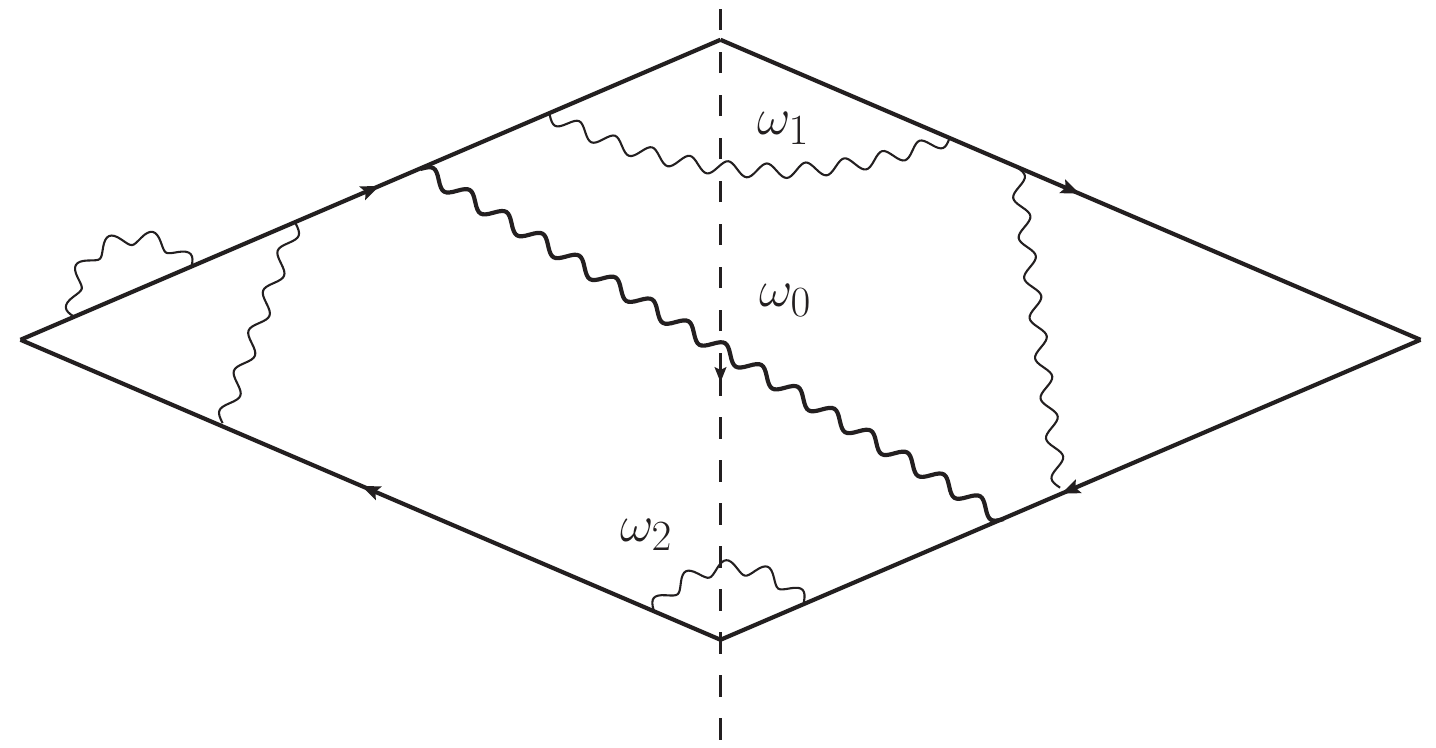}
\caption{Feynman diagram depiction of the process of decoherence and subsequent energy loss of two partons created early in the medium. In the presence of the two-prong structure, we integrate over all medium-induced emissions, depicted by wavy lines, that can occur as direct contributions (e.g., such as the gluons labeled by $\omega_1$ and $\omega_2$), interferences (e.g., the gluon labeled by $\omega_0$) and virtual contributions. Only radiative processes are explicitly represented, while elastic interactions with the medium are implicit and omitted for simplicity.}
\label{fig5}
\end{center}
\end{figure}

\section{Two-parton energy loss probability at large-$N_c$}\label{sec:two-prong-solution}

The generic contributions to the Two-parton energy loss (\ref{eq:P2-operator}), we are interested in is depicted in the standard form in \fign{fig5}, where the amplitude (and its complex conjugate) is depicted on the left (right) side of the cut, represented by the dashed line, and we integrate over all medium-induced gluons, depicted by wavy lines.
The crucial observation that reduces the complexity of the task at hand consists in realizing that there can only be one gluon connecting the quark (antiquark) in the amplitude and the antiquark (quark) in the complex conjugate amplitude in the large-$N_c$ limit  \cite{Dominguez:2012ad}, as represented by a thick wavy line labeled by $\omega_0$ in \fign{fig5}. This emission constitutes at the same time a modification of the color structure of the antenna, which is why we refer to it as a ``flip''. Hence, to the left of the flip in the amplitude, 
only virtual diagrams contribute will lead to the renormalization of the quenching parameter $\hat q$ inside the decoherence parameter.  To the right of the flip radiation off the quark and the antiquark factorize, resulting in independent radiative energy loss, as denoted by the thin, gluon lines labeled $\omega_1$ and $\omega_2$ in \fign{fig5}. 

\begin{figure}[t!]
\begin{center}
\includegraphics[width=0.9\textwidth]{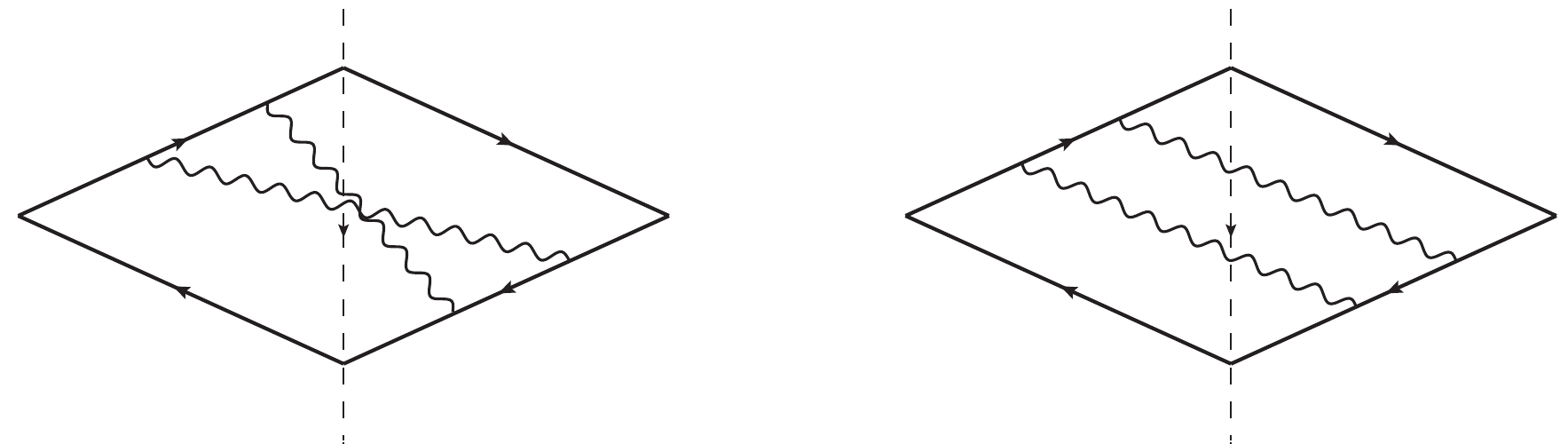}
\caption{Example diagrams of sub-leading contributions that are neglected in our treatment. The left diagram is non-planar and, thus, suppressed in the large-$N_c$ limit, and the right diagram is suppressed because of overlapping formation times. }
\label{fig7}
\end{center}
\end{figure}

This intuitive picture arises from our set of approximations, which simplify the analysis by neglecting more complex situations and sub-leading topologies. Two examples of such interference contributions are illustrated in \fign{fig7}. The diagram on the left represents a non-planar contribution, which is explicitly suppressed in the large-$N_c$ limit. The diagram on the right, while leading in $N_c$, corresponds to the emission of two gluons with overlapping formation times, where the order of radiation times is reversed between the amplitude and the complex conjugate amplitude. This contribution is sub-leading in the soft gluon emission limit due to the restricted phase space available for overlapping formation times in a large medium (see, e.g., \cite{Blaizot:2013vha}).

We define a regularized splitting rate, acting on the propagating quark-antiquark system as
\beq
\label{eq:gamma-ij}
\Gamma_{ij}(\omega, t) \equiv \frac{\rmd I_{ij}}{\rmd \omega\, \rmd t} -\delta(\omega)\int_0^\infty  \rmd \omega' \frac{\rmd I_{ij}}{\rmd \omega'\, \rmd t} \,,
\eeq
where the index $i$ ($j$) denotes the leg that is emitting (absorbing) the medium-induced gluon (regardless whether it is in the amplitude or complex conjugate amplitude). 
In the following, the index ``1'' (``2'') will refer to the quark (antiquark). For example, $\n_1 = \p_1/E_1$ ($\n_2 = \p_2/E_2$) refers to the direction of propagation of the quark (antiquark), and so on. We find that, 
\begin{align}
\label{eq:bdmps-direct}
\frac{\rmd I_{11}}{\rmd \omega \rmd t} = \frac{\rmd I_{22}}{\rmd \omega \rmd t}& = \abar \sqrt{\frac{\hat q}{\omega^3}} \,,\\
\label{eq:bdmps-interference}
\frac{\rmd I_{12}}{\rmd \omega \rmd t} = \frac{\rmd I_{21}}{\rmd \omega \rmd t}& = - \abar \sqrt{\frac{\hat q}{\omega^3}} \,\textsl{F}\left(t/t_\text{quant}\right) \,,
\end{align}
for the direct and interference spectra, respectively. The function $\textsl{F}\left(x\right) \sim\Theta(1-x)$, (for details see discussion in \cite{Mehtar-Tani:2017ypq}), incorporates the effect of quantum decoherence which suppresses the interference term for hard vacuum emissions that resolve the dipole when they are formed at times $t  > t_\text{quant} \sim (\theta_{12}^2 \omega)^{-1} \sim (\theta_\text{f}/\theta_{12})^{4/3} \,\tdecoh $. In the regime of interest, that is for medium-induced radiation at angles larger than the opening angle $\theta_\text{f} \gg \theta_{12}$, we have $\tdecoh  \ll t_\text{quant} $.\footnote{In order to quantify this statement further, we point out that radiation inside the cone occurs with probability $\mathcal{O}(\alpha_s)$ as long as $\omega_s \ll (\hat q/\theta_{12}^4)^{1/3} < \omega_c$.} Therefore, the mechanism of color decoherence is active before quantum decoherence. One can thus neglect the latter by letting $F(x) \sim 1$. 
As a result, the interference spectra (stripped of the decoherence parameter (\ref{eq:decoh-parameter}))  are approximately equal to the direct emission spectra, 
\beq
\frac{\rmd I_{11}}{\rmd \omega \rmd t} \simeq \frac{\rmd I_{22}}{\rmd \omega \rmd t}\simeq -\frac{\rmd I_{12}}{\rmd \omega \rmd t}\simeq -\frac{\rmd I_{21}}{\rmd \omega \rmd t}\simeq \frac{\rmd I}{\rmd \omega \rmd t}.
\eeq
This is a remarkable observation. Unlike vacuum emissions, which are angularly ordered and suppressed at large angles, medium-induced radiation is incoherent and independent of the emitter's angle $\theta_{12}$. As a result, medium-induced radiation can populate the entire angular range. 

Putting all the pieces together,  the two-prong energy loss probability of a color singlet dipole in the large-$N_c$ limit reads
\begin{align}
\label{eq:main-result}
\ProbSing(\epsilon) &= \int_0^\infty\rmd \epsilon_1 \int_0^\infty \rmd \epsilon_2 \,  \ProbOne(\epsilon_1, L) \, \ProbOne(\epsilon_2, L)\, \delta(\epsilon-\epsilon_1-\epsilon_2) \nn 
& - 2 \int_0^L \rmd t  \int_0^\infty\rmd \epsilon_1 \int_0^\infty \rmd \epsilon_2 \, \ProbOne(\epsilon_1, L-t) \, \ProbOne(\epsilon_2, L-t) \nn
& \times \,  \Big[ 1- \Delta_\sM(t) \Big] \int_0^\infty \rmd \omega  \, \Gamma (\omega, t) \, \delta(\epsilon-\epsilon_1-\epsilon_2-\omega) \,,
\end{align}
where $\ProbSing(\epsilon) \equiv \ProbSing(\epsilon,\theta_{12},L)$. The factor $-2$ in the last line results from the sum over the two contributions to the interference spectrum $\sum_{i \neq j}\Gamma_{ij} (\omega, t) \approx -2 \Gamma (\omega, t)$.
This is one of the main results of this paper.
Here, $\ProbOne(\epsilon,L)$ describes the independent energy loss of the antenna legs, and $\Delta_\text{med}(t)$ is the so-called decoherence parameter \cite{Mehtar-Tani:2011hma,Mehtar-Tani:2011vlz,Mehtar-Tani:2012mfa}
that incorporates the effect of color decoherence, and
reads (for a homogeneous medium) 
\begin{align}
\label{eq:decoh-parameter}
\Delta_\text{med}(t) &= \frac{1}{N_c^2-1} \tr_c \left[ \langle \med|  U_\bkg(n_1) U_\bkg^\dag(n_2) | \med\rangle \right]\nn
&\approx 1- \exp \left[ - \frac{1}{12} \hat q \, \theta^2_{12}\,  t^3 \right]  \,,
\end{align}
where the Wilson lines $U_\bkg$ are only function of the background field $A_\bkg$. The last line is obtained in the harmonic oscillator approximation. \eqn{eq:decoh-parameter} can be interpreted as the probability for the medium to resolve the color structure of the pair after traversing a distance $t$ in the medium and is sensitive to characteristic angle $ \sim (\hat q t^3)^{-1/2}$, above which color coherence is wiped out.
One can check that \eqn{eq:main-result} contains two limiting cases, corresponding to coherent and decoherent antennas. In order to illustrate this point, let us for the moment focus exclusively on the effect of color decoherence, contained in the decoherence parameter \eqn{eq:decoh-parameter}. First, we deal with the incoherent case. Letting $\Delta_\sM(L) =1$, i.e. $\theta_{12} \gg \theta_c$ where $\theta_c \sim (\hat q L^3)^{-1/2}$, suppresses the second term and therefore the total energy loss probability of the singlet two-prong structure is given by the product of one-prong energy loss probabilities,
\beq\label{eq:main-result-decoh}
\ProbSing(\epsilon) \simeq \int_0^\infty\rmd \epsilon_1 \int_0^\infty \rmd \epsilon_2 \,  \ProbOne(\epsilon_1) \, \ProbOne(\epsilon_2)\, \delta(\epsilon-\epsilon_1-\epsilon_2) \,.
\eeq 
In the opposite case $\theta_{12}=0$, so that  $\Delta_\sM(t) =0$. In this limit the interferences cancel the direct contributions, $\rmd I_{12} \simeq -\, \rmd I_{11}$.\footnote{Strictly speaking, this is true for soft, medium-induced emissions that are not affected by angular ordering. } In the interference term, proportional to $\Gamma( \omega,t)$, we can shift $\epsilon_1\to \epsilon_1-\omega$ to allow the integral over $\omega$ to act on $\ProbOne(\epsilon_1-\omega, L-t)$.
Then one can use that, cf. \eqn{eq:eloss-rate-eq},
\beq
\frac{\del }{\del t } \ProbOne(\epsilon_1, L-t) = -  \int \rmd \omega\,  \Gamma (\omega,t)    \ProbOne (\epsilon_1-\omega, L-t) \,,
\eeq
and similarly with $\ProbOne(\epsilon_2, L-t)$. We reconstruct in this way the total derivative acting on the product of energy-loss probabilities, such that the interference term gives rise to 
\beq
\int_0^L  \rmd t\, \frac{\del }{\del t }  \big[ \ProbOne (\epsilon_1, L-t) \, \ProbOne(\epsilon_2, L-t)\big] = \delta(\epsilon_1) \delta(\epsilon_2)-  \ProbOne(\epsilon_1, L) \, \ProbOne(\epsilon_2, L),
\eeq
where the second term cancels exactly the first term in \eqn{eq:main-result-decoh}. Therefore, the energy loss probability of the infinitely narrow singlet antenna vanishes, that is when $\theta_{12} \ll \theta_c$, reads 
\beq
\ProbSing(\epsilon) \simeq  \delta(\epsilon) \,,
\eeq
 as expected.

The generalization to arbitrary color representation of the parent parton is straightforward in the large-$N_c$ limit \cite{Mehtar-Tani:2017ypq}. Let us for the moment consider parton splittings via a gluon emission.\footnote{This does not directly apply to $g \to q+\bar q$ splittings which do not involve any interferences in the large-$N_c$ limit and which are anyway suppressed in the leading-logarithmic approximation.}
Hence, for a parent parton with color representation $R$, the two-pronged energy loss probability reads,
\beq
\label{eq:main-result-colored}
\ProbTwo^{\sst R}(\epsilon) = \int_0^\infty \rmd \epsilon_1 \int_0^\infty \rmd \epsilon_2 \,\ProbOne^{\sst R}(\epsilon_1) \ProbSing(\epsilon_2) \delta(\epsilon - \epsilon_1 - \epsilon_2) \,,
\eeq
where $\ProbSing(\epsilon)$ denotes the color-singlet two-prong quenching weight in \eqn{eq:main-result}.
Keep however in mind that the color factor in the interference spectrum entering $\ProbSing(\epsilon)$ has to be replaced by $C_F \approx N_c/2$ and the one-prong quenching weight $\ProbOne^{\sst R}(\epsilon)$ becomes sensitive to the relevant color charge through the generalized coupling $\bar \alpha_{\sst R}= \alpha_s C_R/\pi$. This is the second main result of the paper.

\begin{figure}[t!]
\begin{center}
\includegraphics[width=0.47\textwidth]{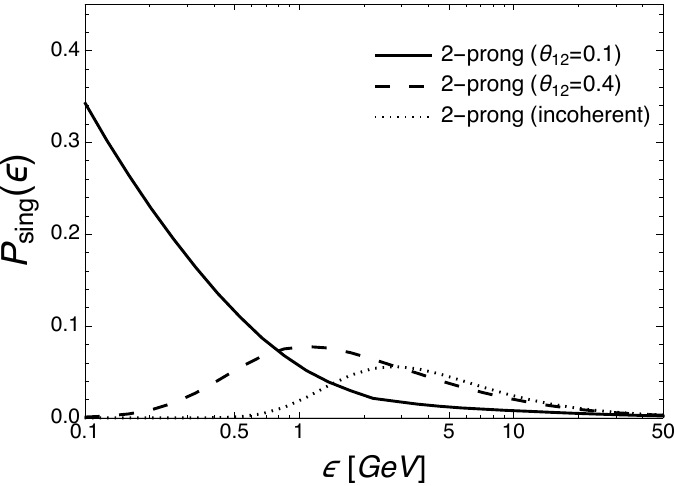}
\includegraphics[width=0.5\textwidth]{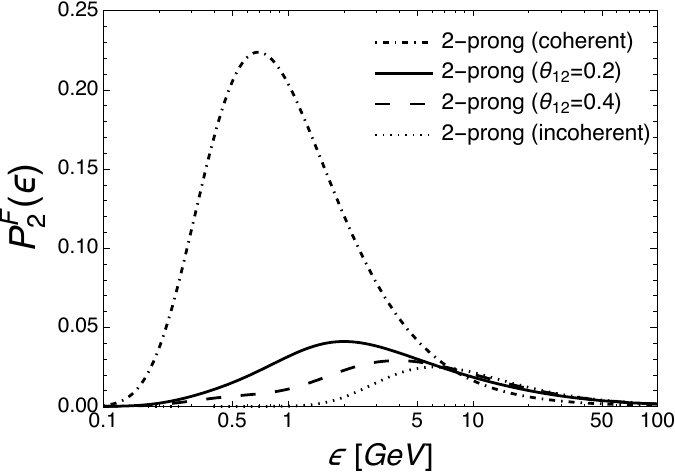} 
\caption{The two-prong energy loss distribution for a color singlet antenna (left) and for a quark splitting (right). Medium parameters were chosen to be $\hat q = 1$ GeV$^2$/fm and $L=2$ fm, and $\alpha_s=0.3$. We plot the corresponding distributions for two opening angles, $\theta_{12} = 0.2$ (solid curves) and $\theta_{12}=0.4$ (dashed curves) in addition to the completely incoherent case (dotted curves) and, for the quark, the one-prong energy loss distribution corresponding to a completely coherent splitting (dash-dotted curve).}
\label{figEloss}
\end{center}
\end{figure}

We plot the two-prong energy loss distribution for a color singlet antenna, evaluated according to \eqn{eq:main-result}, and for a quark splitting, evaluated according to \eqn{eq:main-result-colored}, in \fign{figEloss} for two different opening angles, see figure caption for details.
For comparison, we also evaluate a two-pronged incoherent energy loss distribution, which simply is defined by the first line in \eqn{eq:main-result}. Finally, for the quark we also plot the single-prong energy loss distributions that corresponds to a completely coherent splitting.
The color singlet distribution interpolates between a distribution peaked around small values of $\epsilon$ for small opening angles to the incoherent case at large angles. We notice immediately the sensitivity to the opening angle. At sufficiently high energy, all distributions fall off as $\epsilon^{-1/3}$, due to the incoherent nature of hard emissions.

To summarize, for arbitrary color representation in the large-$N_c$ limit, the two-pronged energy loss distribution is therefore a convolution of the quenching weight of the total charge along the whole length of the medium with the two-pronged color singlet distribution. This result is the basis for multi-parton energy loss that is the basis for jet energy loss. Recently a factorization approach to this problem has been proposed \cite{Mehtar-Tani:2024smp,Mehtar-Tani:2025xxd}. 

In the next section we will derive non-linear evolution equations that encode the physics of decoherence to leading collinear  logarithm accuracy. In this approximation, the energy two-prong energy loss distribution \eqn{eq:main-result}  is approximated by 
\beq \label{eq:2prong-eloss-approx}
P^R_2(\epsilon) &\approx& \Theta(\theta_{12}-\theta_c)\int \rmd \epsilon_1\int \rmd \epsilon_2 \delta(\epsilon- \epsilon_1-\epsilon_2)  \ProbOne^{R_1}(\epsilon_1, L) \, \ProbOne^{R_2}(\epsilon_2, L) \nn
&+& \Theta(\theta_c-\theta_{12}) \ProbOne^R(\epsilon, L)\,.
\eeq
where $R_1$ and $R_2$ are the color representation of the final state daughters that lose energy independently if $\theta_{12}>\theta_c$. Conversely, when  $\theta_{12}<\theta_c$ energy loss is coherent and the two parton system loses energy as a single color charge in representation $R$ of the parent. This approximation is valid to leading power in $\theta_c$ and enables a resummation of collinear logs $\ln (R/\theta_c)$ resulting from the angular integral of the collinear hard splitting:
\beq
\int_0^R\frac{\rmd \theta_{12}}{\theta_{12}} \Theta(\theta_{12}-\theta_c)=  \ln\frac{R}{\theta_c}\,.
\eeq
At angles smaller than the critical angle $\theta_c$, real and virtual collinear splitting cancel out since they are multiplied by the same total charge energy loss distribution.

%% file: jeteloss.tex
\chapter{Jet energy loss }\label{chap:jeteloss}
We are now prepared to address the central problem of this work: applying the theoretical framework to jet observables to confront our predictions with data. Our focus will be on inclusive jet production in heavy-ion collisions, a key testing ground for our approach. For simplicity, we will continue to restrict our analysis to purely gluonic parton showers, with straightforward generalization to arbitrary flavors left for future work.

In this chapter, we develop an all-orders analytic description of energy flow outside the jet region, defined by a cone of opening angle $R \ll 1 $ centered around the parent parton. We derive a non-linear evolution equation that resums medium-length-dependent powers of $ \alpha_s L $, arising from the medium-induced cascade discussed in Chap.~\ref{chap:cascade}. This is then extended to include the collinear, virtuality-driven vacuum cascade, resumming logarithmic powers of $ \ln(1/R) $ (out-of-cone DGLAP evolution) and $ \ln(R/\theta_c) $ (intra-jet color charges resolved by the medium). Here, $ \theta_c \sim (\hat{q} L^3)^{-1/2} \ll 1 $ is the resolution angle below which splittings are suppressed due to destructive interference, as explained in Chap.~\ref{chap:decoherence}~\cite{Mehtar-Tani:2011hma,Mehtar-Tani:2012mfa,Mehtar-Tani:2017ypq,Casalderrey-Solana:2011ule,Mehtar-Tani:2017ypq}. The vacuum cascade, occurring prior to medium-induced processes, sets the initial condition for the non-linear DGLAP evolution.

The goal of this section is to generalize the concept of ``quenching weights''~\cite{Salgado:2003gb} to fully developed jets. This chapter builds upon~\cite{Mehtar-Tani:2024mvl} and~\cite{Mehtar-Tani:2017web}.

\begin{figure}
\centering
\includegraphics[width=12cm]{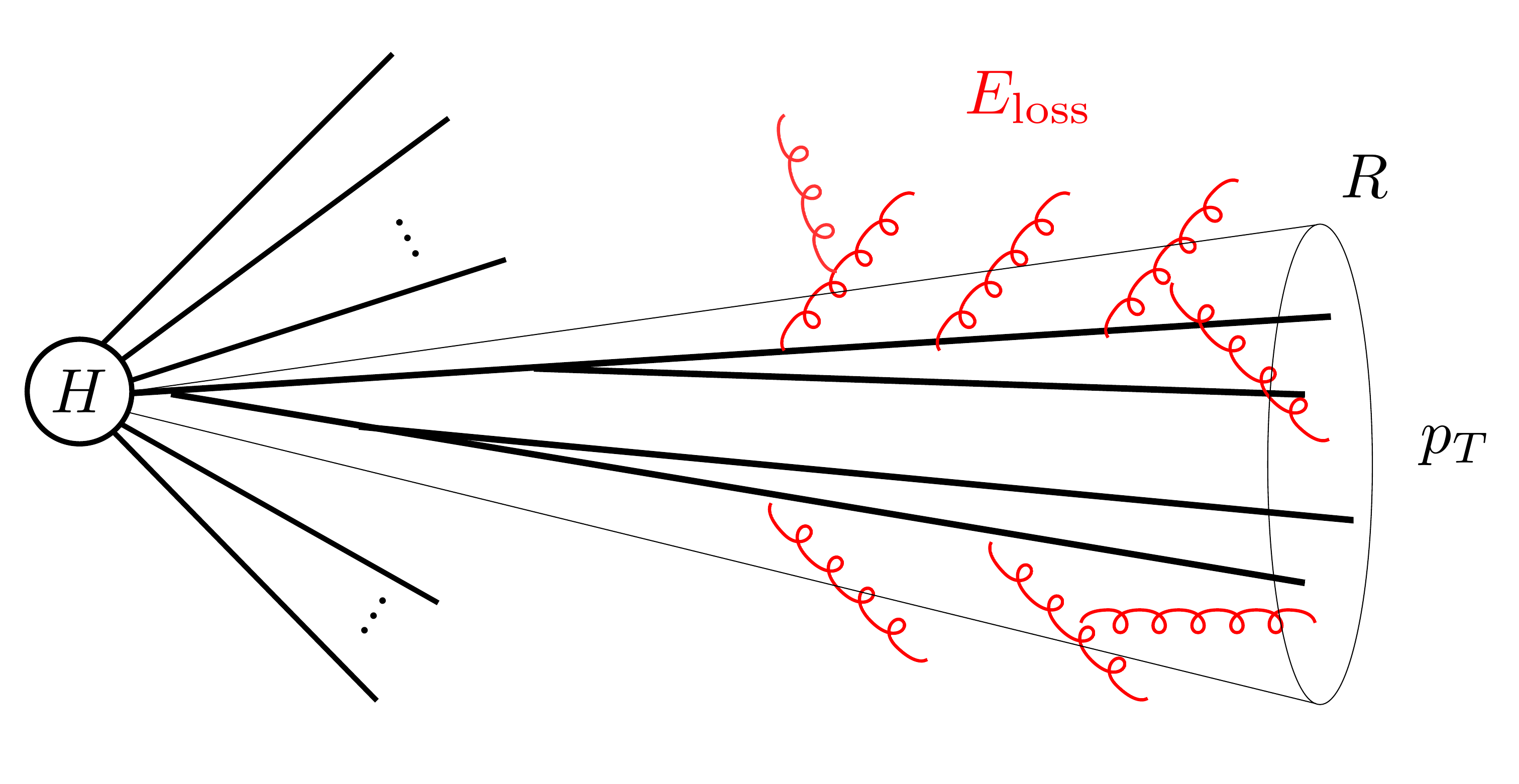}
\caption{An illustration of a high-energy jet produced by a hard scattering and losing energy to the plasma via medium-induced gluon radiation.   Note that gluons that have been radiated out side the cone can radiate back inside the cone reducing the total energy loss. .} \label{fig:jet-eloss}
\end{figure}

\section{The inclusive jet spectrum}\label{sec:eloss-dist} 

We start we a high energy gluon an energetic gluon with energy (longitudinal momentum) $E\equiv p^+ $ produced intially by a hard collision. This highly virtual gluon, assumed to be the jet initiator, will undergo to type of branching processes. It will first branch via a collinear DGLAP cascade which then will be followed by a medium induced cascade that is characterized by random angular distributions.

The jet is defined by all particles that fall within the domain $\cC_{\rm in}$ in solid angle, typically characterized by an opening angle $R$. The jet region is thus defined as $\cC_{\rm in} = \{ \btheta \in \mathbb{R}^2 \, | \, |\btheta| < R \}$.

We want to compute the energy loss distribution $S_{\rm loss}(\epsilon)$  of the initial gluon longitudinal momentum $E$ that loses an amount $\epsilon$ out of the jet cone,
\beq\label{eq:excl-amp}
S_{\rm loss}(\epsilon, t, t_0, E,\btheta) = \sum_{m=1}^\infty   \prod_{i=1}^m  \int\frac{\rmd k^+_i \rmd^2 \k_i}{2(2\pi)^3 k_i^+ } |A_m(k_1,...,k_m,p;t,t_0)|^2 \delta(\epsilon-  \bar n \cdot k_{\rm out} ) \Theta_{alg} \,.
\eeq
Here, $\bar n \cdot k_{\rm out}= k^+_{\rm out}=\sum\limits_{i=1}^{m} k^+_{i,\rm out}$ represents the total longitudinal momentum of the collinear partons located in the `out' region at the time $t$, with $\bar n\equiv(1,0,0,-1)/2$, a null vector that defines the light cone direction opposite to the jet direction of propagation, and $A_m$ is the exclusive amplitude for an initial parton $p$ at $t_0$ fragmenting into $m$ collinear partons at time $t$. The delta function represents the measurement of energy lost by particles with momentum that are measured outside the jet  region $\cC_{\rm in}$ and $\Theta_{alg} $ denotes the jet clustering algorithm such as the anti-$k_t$ algorithm \cite{Cacciari:2008gp}.  In \eqn{eq:excl-amp}, momentum conservation is implicit.

By construction, the energy loss distribution is normalized to 1 as a consequence of unitarity,
\beq \label{eq:norm}
\int_0^{+\infty } \rmd \epsilon \, S_{\rm loss}(\epsilon) =1\,.
\eeq

To compute the inclusive cross-section we need the function $S_{\rm loss}$ evaluated at $t_0=0$, $t=+\infty$ and $\btheta=0$, i.e.,   $S_{\rm loss}(\epsilon, E) \equiv  S_{\rm loss}(\epsilon,+\infty, 0, E,\btheta=0)$, which we then convolve with the LO inclusive jet cross-section as an illustration (we shall generalize to all orders shortly):
\beq\label{eq:factorization}
\frac{\rmd \sigma_{\rm incl}}{\rmd p_T}  = \int_0^{+\infty}\rmd E\,\int_0^{+\infty} \rmd \, \epsilon\,\,  \delta(E-\epsilon- p_T)\, S_{\rm loss}(\epsilon, E)\,  \frac{\rmd \sigma^{\rm LO}_{\rm incl}}{\rmd E}  \,.
\eeq
Here, we implicitly assume that the parent leading parton defines the jet direction, though this is not strictly accurate, as jet definition algorithms depend on the final state. However, in the high-energy limit where the jet energy far exceeds the lost energy, the effects of the jet algorithms are power suppressed. Under this assumption, the jet algorithm effectively acts on the initial collinear parton cascade. A detailed discussion of relaxing this approximation is deferred to future work.

Introducing the quenching weight as a Laplace Transform of the energy loss distribution,
\beq  \label{eq:eloss-LT}
Q_\nu(E) \equiv \int^{+\infty}_0 \rmd \epsilon \,  S_{\rm loss}(\epsilon, E)\,\rme^{-\nu \epsilon  } \,,
\eeq
with the inverse transform,
\beq
S_{\rm loss}(\epsilon, E) = \int_{c-i\infty}^{c+i\infty}\frac{\rmd \nu}{2\pi i}\, Q_\nu(E)\, \rme^{\nu \epsilon}\,,
\eeq
where the $\nu$ integral runs vertically to the real axis. The value $c$  (the real part of $\nu$) is chosen such that it is greater than the real parts of all singularities of the integrand. Note that, we may equally use the Mellin representation by transforming w.r.t. to the rescaled variable $y=\epsilon/E$. The reason the Laplace transform is more convenient stems from the fact that the medium-induced part of energy loss distribution is independent of the jet energy for large values of $E$. Moreover, in the non-linear regime discussed in the next section, the evolution equation is local in $\nu$.

\eqn{eq:factorization} can be rewritten as
\beq
\frac{\rmd \sigma_{\rm incl}}{\rmd p_T}  = \int_0^{+\infty}\rmd E\, \int \frac{\rmd \nu}{2\pi i} \, \rme^{ (E-p_T) \nu}Q_\nu(E) \,  \frac{\rmd \sigma^{\rm LO}_{\rm incl}}{\rmd E} \,.
\eeq
We observe that in the case where the jet loses all of its energy to the plasma we have the limiting result:
\beq
\frac{\rmd \sigma_{\rm incl}}{\rmd p_T}  \to \delta(p_T)  \int_0^{+\infty}\rmd E \frac{\rmd \sigma^{\rm LO}_{\rm incl}}{\rmd E} \,.
\eeq
In order to connect our formulation with the so-called jet function that appears in the factorization of the jet cross-section in vacuum \cite{Kang:2016mcy,Dasgupta:2014yra}, we make the change of variables
\beq
p_T = x E\,.
\eeq
We then obtain
\beq\label{eq:factorization-LO}
\frac{\rmd \sigma_{\rm incl}}{\rmd p_T}  =p_T \int_0^{1}\frac{ \rmd x}{x^2}\, \int \frac{\rmd \nu}{2\pi i} \, \rme^{ \frac{(1-x)}{x}p_T \nu}Q_{\nu}(p_T/x) \,  \frac{\rmd \sigma^{\rm LO}_{\rm incl}(E=p_T/x)}{\rmd E} \,.
\eeq
In terms of the jet function \cite{Kang:2016mcy}, we have
\beq\label{eq:jet-function}
J\left(x,E\right) =E\int \frac{\rmd \nu}{2\pi i} \, \rme^{  (1-x)E \nu}\, Q_{\nu}\left(E\right)  \, .
\eeq
Then, introducing the factorization scale $\mu$, the general form of \eqn{eq:factorization-LO} reads
\beq\label{eq:factorization}
\frac{\rmd \sigma_{\rm incl}}{\rmd p_T}  = \int_0^{1}\frac{\rmd x}{x}\,J(x,E,R,\mu)  \,  H(E=p_T/x,\mu)\,.
\eeq
where $H(p_T,\mu)\sim  \rmd \sigma_{\rm incl}(\mu)/\rmd p_T $, the hard matrix element that is related to jet cross-section evaluation to a factorization scale $\mu$. This formula factorizes collinear jet dynamics from the hard scattering.

Assuming the presence of single medium scale, $E_{\rm med}$, the jet function can be written as a function of dimensionless variables:
\beq
J(x,E,R,E_{\rm med},\mu)\,\rightarrow   \,J\left (x, \frac{\mu}{E R},\frac{E_{\rm med}}{E},R \right)  \, \underset{ \rm vacuum}{\rightarrow}  \,  J\left (x, \frac{\mu}{E R}\right)\,,
\eeq
where the extra $R$ dependence is due to power corrections related to medium dynamics and can be dropped in a first approximation.

The operator definition of the jet function for quark initiated jet in vacuum can be found in Ref.~\cite{Kang:2016mcy}.

In this unified framework, the jet function can be interpreted as a jet energy loss distribution even in vacuum. However, a key distinction arises between the vacuum and medium cases. In vacuum, only collinear splittings at angles larger than $R \ll 1$ produce significant logarithmic contributions, i.e., $\alpha_s \ln (1/R) \sim 1$. Since only the total transverse momentum ($p_T$) of the jet is measured, the observable is insensitive to fluctuations in the jet substructure \cite{Kang:2016mcy,Dasgupta:2014yra}. As a result, splittings within the jet cone do not affect the inclusive observable and must cancel out due to unitarity.

In contrast, a QCD medium resolves different substructure fluctuations, leading to differential energy loss that impacts the observable. For example, a jet that splits into two subjets will experience more energy loss than one that does not split, causing an imbalance between real and virtual contributions. This effect manifests as a Sudakov-type suppression, first noted in \cite{Mehtar-Tani:2017web}, resumming terms of the form $(\alpha_s \ln (R/\theta_c))^m$ (with $m=0,1,...,+\infty$) where $\theta_c$ represents the medium's angular resolution scale. When $\theta_c > R$, the medium is unable to resolve the jet's finer color structure, and the result converges to the vacuum case, apart from an overall quenching factor corresponding to the total color charge. As we shall show, this mismatch will result in a non-linear evolution when collinearly enhanced contributions are resummed to all orders in perturbation theory.

When $\theta_c$ is small but not much smaller than $R$, it may become necessary to resum soft logarithms of the form $(\alpha_s \ln (p_T/E_{\rm med}))^m \sim (\alpha_s \ln (n))^m$, where $n \gg 1$ denotes the power-law index of the jet spectrum. In such scenarios, a recently developed effective field theory (EFT) based factorization approach offers a systematic framework for separating hard collinear modes associated with virtuality-induced splittings. These splittings lose energy in the medium through soft gluon emissions, which are encoded in multi-Wilson-line correlators \cite{Mehtar-Tani:2024smp}.

\begin{figure}
\centering
\includegraphics[width=9cm]{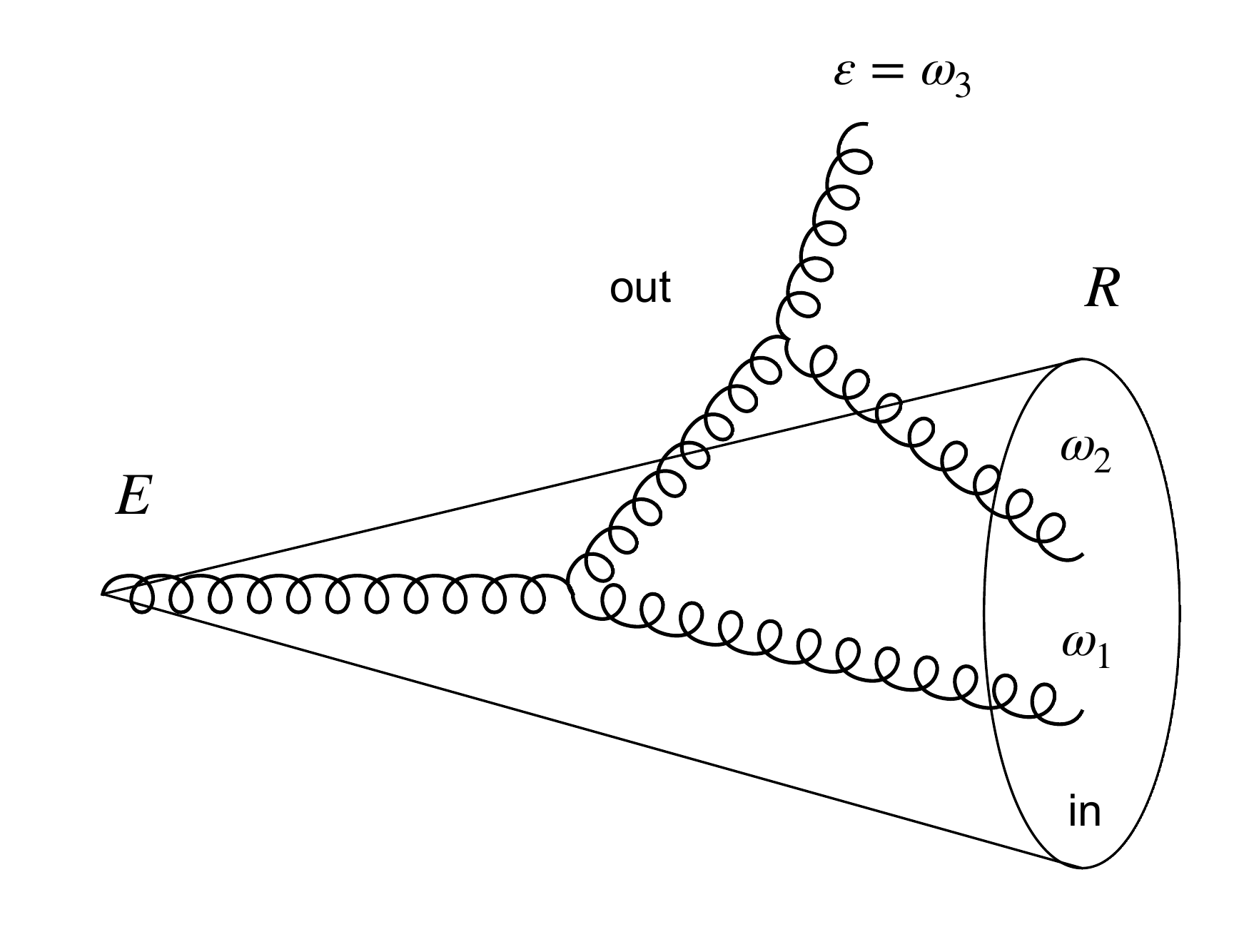}
\caption{An illustration of a high-energy gluon with energy $E$ splitting into two gluons: one that remains within the jet region $\mathcal{C}_{\rm in}$ with an opening angle $R$, and another that propagates outside the jet region before radiating back into it. This type of non-linear dynamics resembles QCD processes that generate large non-global logarithms in vacuum \cite{Dasgupta:2001sh,Banfi:2002hw}.} \label{fig:cascade}
\end{figure}

\section{non-linear evolution driven by medium-induced cascades}\label{sec:med-shower}

Using the parton shower language, we shall first discuss the effect of medium induced-cascade on jet energy loss, as it is the main aspect that sets the medium and vacuum cases apart. We will then discuss the case of the virtuality ordered  vacuum cascade in Section.~\ref{sec:med-shower}.
The question we want to address here is, given a gluon with energy $E$ located initially at an angle $\btheta$ at the time $t_0$, what is the energy distribution that flows out of a roughly circular region $\cC_{\rm in}$ of radius $R$ and centered around the origin $\btheta=0$ (see Fig.~\ref{fig:cascade}).

\begin{figure}
\centering
\includegraphics[width=10cm]{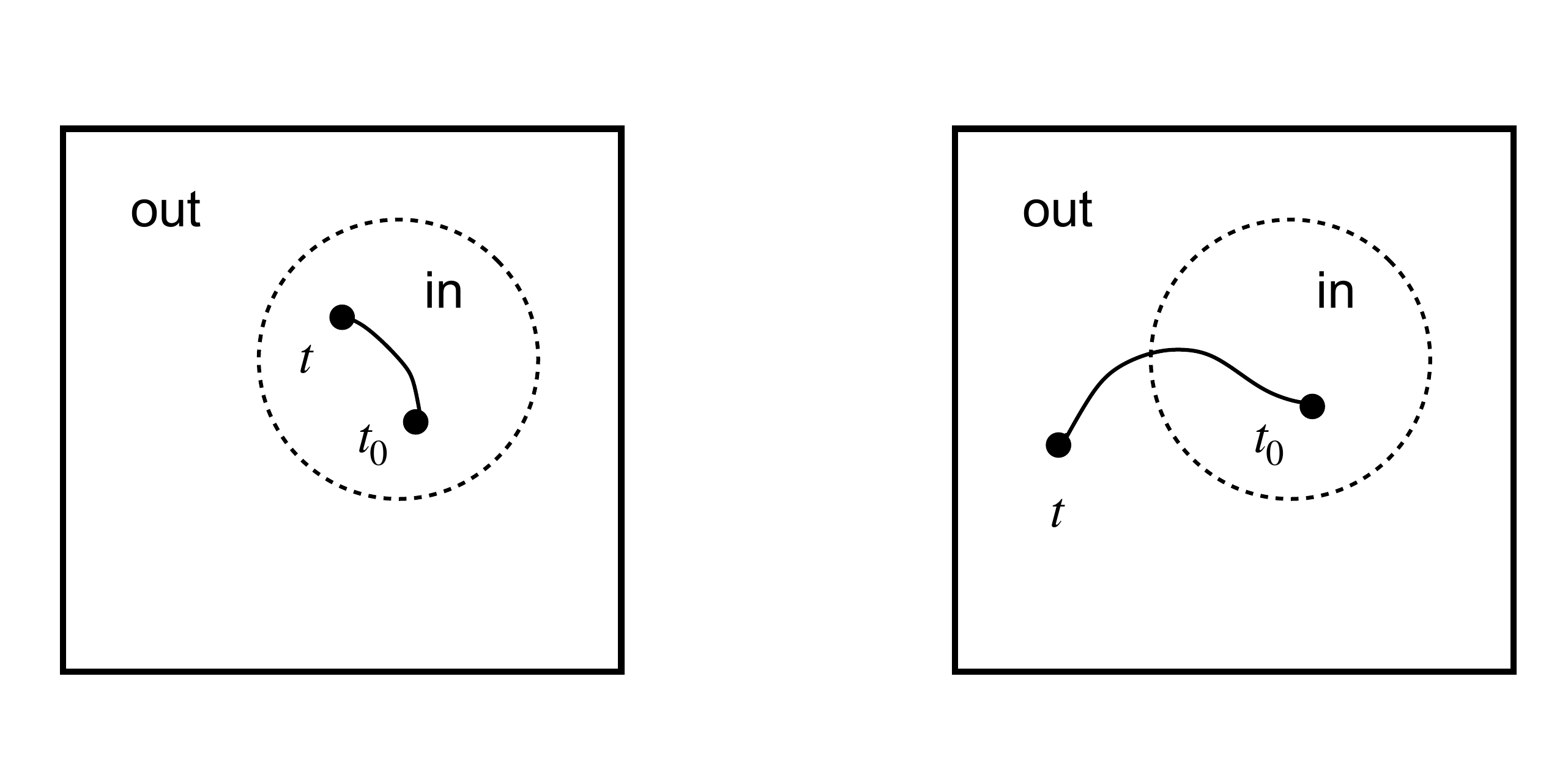}\caption{The dynamics of collinear parton evolution can be described on the transverse plane $\btheta=\q/E$. The two figures depict the two terms in the elastic ($p_\perp$-broadening) contribution to energy loss \eqn{eq:S-elastic}: the left figure corresponds to no energy loss at time $t$, since the gluon ends up inside the jet cone for which $|\btheta|<R$, while, the right figure corresponds to the case $|\btheta|<R$ where the gluon is registered outside the jet region $\cC_{\rm in}$.  }\label{fig:TMB}
\end{figure}

In the simpler case of a gluon that undergoes only elastic scattering the energy loss distribution reads
\beq\label{eq:S-elastic}
S_{\rm el}(\epsilon, t, t_0;E, \btheta)  = \int \frac{\rmd^2 \q}{(2\pi)^2}\, \cP(\q-E \btheta,t,t_0) \left[ \Theta(|\q|<RE)\, \delta(\epsilon)+\Theta(|\q|>RE)  \,\delta(E-\epsilon) \right]\,,\nn
\eeq
where $\q$ and  $\k=E\btheta$ are the final and initial gluon transverse momenta, respectively. A depiction of the two contributions is given in Fig.~\ref{fig:TMB}. The first term corresponds to the scenario where the gluon ends up inside the jet cone, resulting in no energy deposition outside the cone, i.e., $\epsilon = 0$. The second term accounts for the opposite case, where the gluon exits the jet region, leading to a total out of cone energy of $\epsilon = E$. 

The distribution $\cP(\q-\k,t,t_0)$,  introduced in Chap.~\ref{chap:formalism}, describes the probability for a gluon to acquire a transverse momentum kick of $\q - \k$ due to interactions with the medium between an initial time $t_0$ and final time $t$. In the independent multiple-scattering approximation it obeys the kinetic equation \eqn{ME-P} (cf.~ \cite{Blaizot:2013vha})

Evidently, in vacuum we have $\cP(\k-\bell,t,t_0)  \to (2\pi)^2\delta^{(2)}(\k-\bell)$, implying  $S_{\rm el}(\epsilon, t, t_0;E, \btheta=0)\to \delta(\epsilon)$.
\begin{figure}
\centering
\includegraphics[width=6cm]{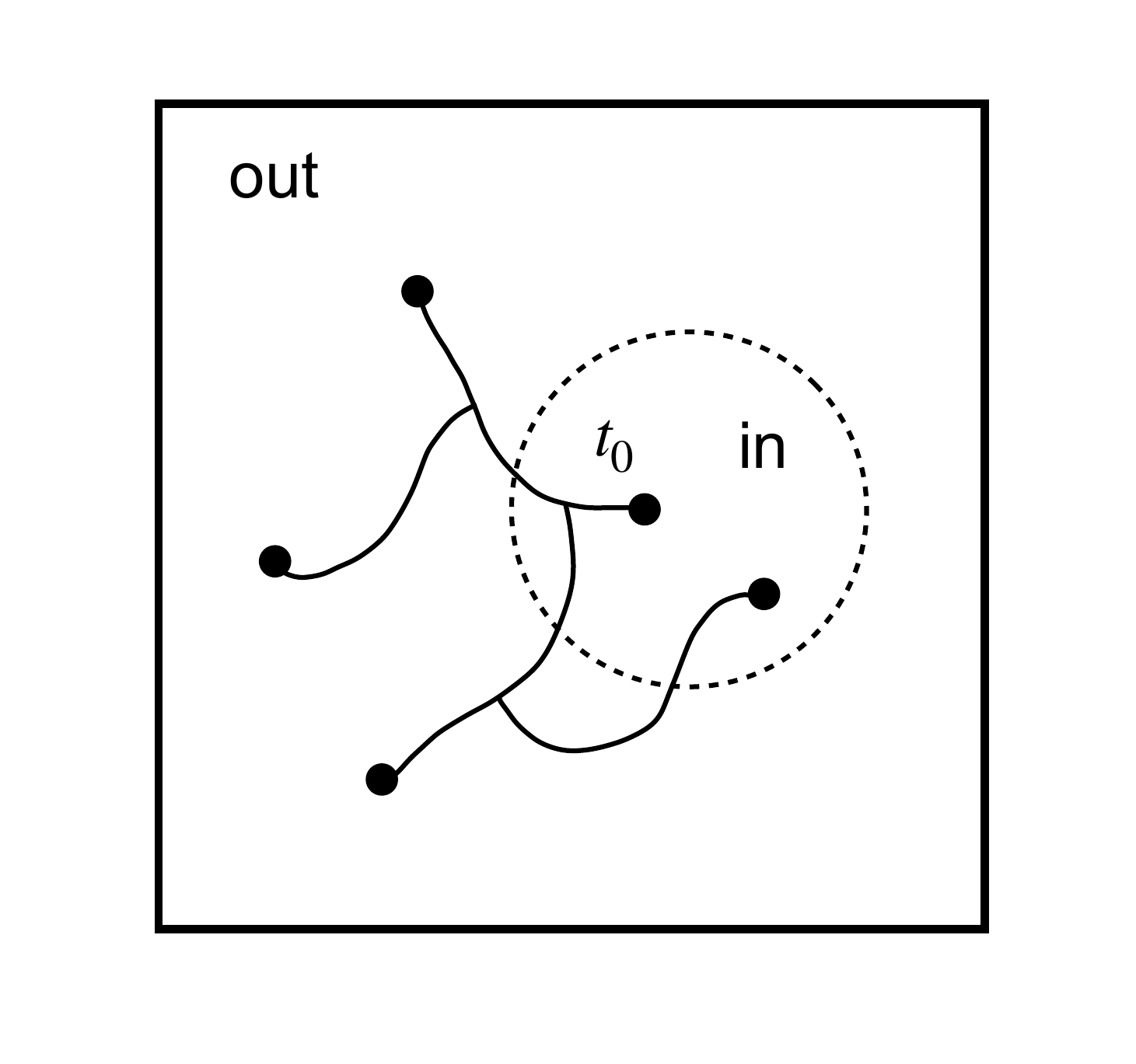}\caption{Illustration of the effect of medium-induced splitting on the evolution of energy flow out of the cone. Note that a gluon can flow out of the cone at intermediate time but then radiates a gluon back inside the cone leading to fluctuating energy loss as a consequence of the evolution of the parton cascade. In contrast to the vacuum cascade successive gluon branchings are not ordered in angles.  }\label{fig:cascade-2d}
\end{figure}

\subsection{The master equation }

Interactions with the QGP can cause the gluon to branch, initiating a gluon cascade that redistributes the initial energy among many quanta (see Fig.~\ref{fig:cascade-2d}). Therefore, in addition to elastic processes, we must account for multiple gluon branchings. During such a process, multiple scatterings cause a gluon with energy $E$ to split into two gluons with energies $zE$ and $(1-z)E$, respectively, at a rate given by $\cK(z,E)$, which, in the multiple soft-scattering approximation based on the harmonic approximation takes a simple analytic form given by \eqn{Kdef}) \cite{Blaizot:2013vha}. 

As discussed in Chap.~\ref{chap:cascade}, the inelastic rate describes the quasi-instantaneous collinear splitting of a gluon. Consequently, no transverse momentum is acquire during this process which is valid in the approximation where $t_f/t \to 0$.

Details on the computation of the rate, along with generalizations that incorporate the hard Coulomb tail and finite-size effects, can be found in the following references \cite{Mehtar-Tani:2019tvy,Mehtar-Tani:2019ygg,Barata:2020sav,Barata:2020rdn,Barata:2021wuf,Caron-Huot:2010qjx,Andres:2023jao,Andres:2020vxs}. Additional information on the non-perturbative component of the collision rate is available in \cite{Moore:2021jwe,Schlichting:2021idr}, and for an expanding medium in \cite{Adhya:2019qse,Caucal:2020uic,Soudi:2024yfy}. This medium-induced parton shower can be viewed as a limiting case of a kinetic theory description of the plasma dynamics \cite{Arnold:2002zm,Schlichting:2020lef,Mehtar-Tani:2022zwf}.

Since each splitting creates two branches that can independently contribute to the total energy loss, the evolution equation for the energy distribution outside the cone becomes non-linear, as depicted in Fig.~\ref{fig:jet-eloss}, and takes on a familiar form
\begin{align}\label{eq:non-lin-eq}
 & S_{\rm loss}(\epsilon, t, t_0 ;E,\btheta)  = S_{\rm el}(\epsilon, t, t_0;E,\btheta) \nn
& \qquad+ \alpha_s \int_{t_0}^t \rmd t_1   \int_0^1 \rmd z\, \cK(z,E)  \, \int \frac{\rmd^2 \btheta'}{(2\pi)^2 E^2}\, \cP(E(\btheta'-\btheta),t_1,t_0) \nn
& \times\left[ \int_{\epsilon_1,\epsilon_2}\,S_{\rm loss}(\epsilon_1, t, t_1 ;zE,\btheta') \,S_{\rm loss}(\epsilon_2, t, t_1 ;(1-z)E,\btheta') \, \delta(\epsilon-\epsilon_1-\epsilon_2) - S_{\rm loss}(\epsilon, t, t_1 ;E,\btheta')  \right]  \,,
\end{align}

The evolution halts at $t=L$, after which the produced gluons propagate in vacuum, resulting in no further medium-induced effects.
\begin{figure}
\centering
\includegraphics[width=14cm]{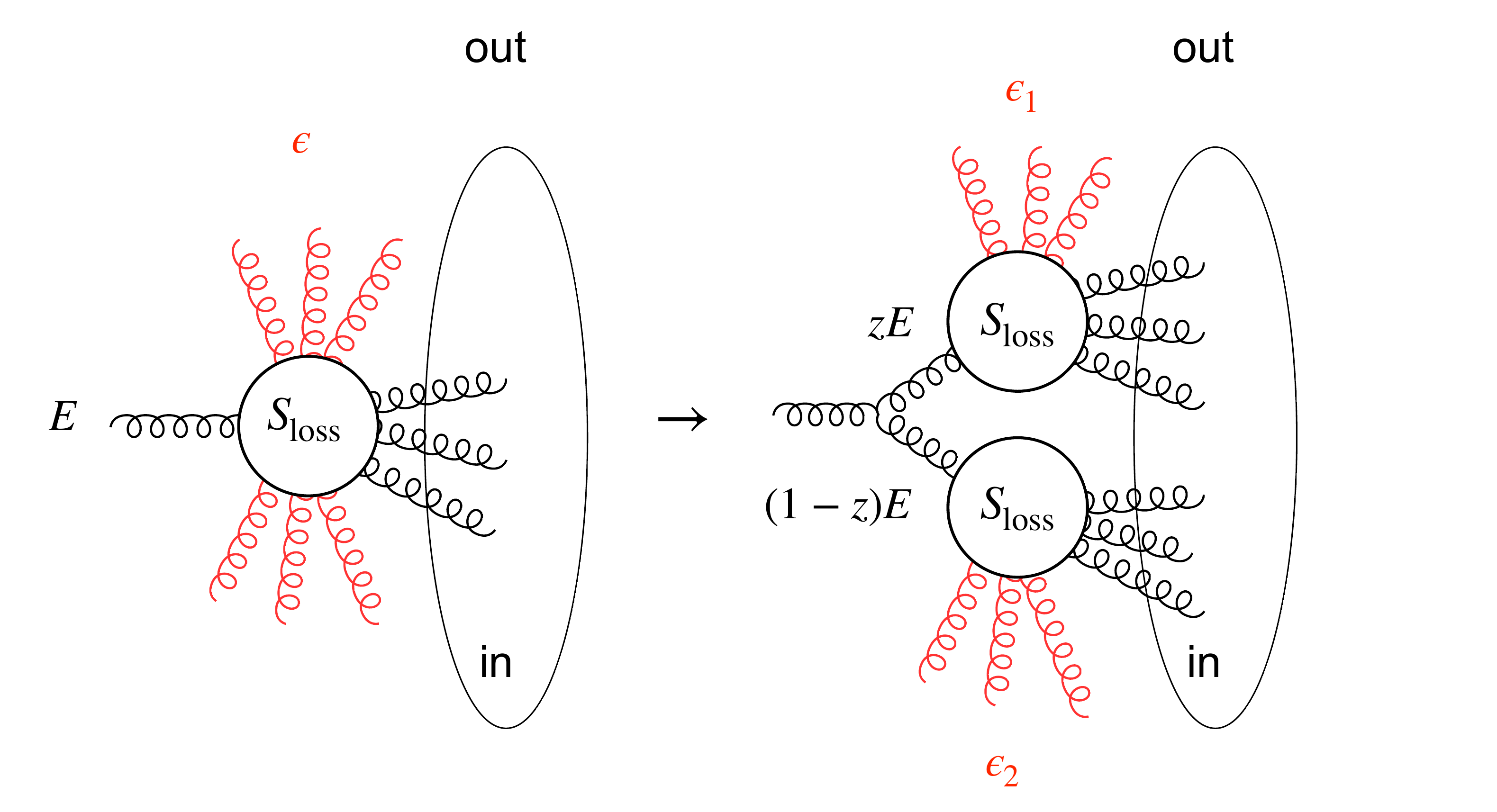}\caption{Illustration of \eqn{eq:non-lin-eq}. The left figure depicts the energy loss distribution from the initial time $t_0 $ to the final time $ t $. Between the instants $ t_0 $ and $ t_0 + \Delta t $, the initial gluon may split into two gluons. Due to rapid color decoherence, each gluon then evolves independently from $ t_0 + \Delta t $ to $ t $, contributing additively to the total energy loss, such that $ \epsilon = \epsilon_1+ \epsilon_2 $.
}\label{fig:jet-eloss}
\end{figure}

Note that when $E=0$, we deduce from \eqn{eq:S-elastic} that
\beq\label{eq:null-eloss}
S_{\rm loss}(\epsilon,E=0)= \delta(\epsilon)\,,
\eeq
which is just a statement that if the parent gluon energy vanishes so does its energy loss, hence, $\epsilon=0$. This property remains true in the presence of the medium-induced cascade. It is also crucial to ensure that, even though the splitting kernel is singular in the limit $z\to 1$ and $z\to 0$ the equation is finite. This can be checked by noting that the non-linear term in \eqn{eq:non-lin-eq} admits the limit
\beq
\lim_{z\to 0}S_{\rm loss}(\epsilon_1,zE) \, S_{\rm loss}(\epsilon_2(1-z),zE) =  \delta( \epsilon_1)\, S_{\rm loss}(\epsilon_2,E)\,.
\eeq
An analogous identity applies in the limit $z\to1$, precisely eliminating the linear term.

\eqn{eq:non-lin-eq}, a key result of this work, simplifies when expressed in Laplace space. To proceed, we introduce the Laplace transform of the energy loss distribution
\beq
Q_\nu (E)= \int_0^{+\infty}\rmd \epsilon \, \rme^{-\epsilon \nu}\,  S_{\rm loss}(\epsilon,E) \,.
\eeq
With this transformation, \eqn{eq:non-lin-eq} yields
\begin{align}\label{eq:non-lin-eq-LT}
& Q_\nu(t, t_0;E,\btheta)  =  Q^{\rm el}_\nu(t, t_0;E,\btheta)  + \alpha_s  \int_{t_0}^t  \rmd t_1  \int_0^1 \rmd z\, \cK(z,E)\, \int \frac{\rmd^2 \btheta'}{(2\pi)^2E^2 }\, \cP(E(\btheta'-\btheta),t_1,t_0) \nn
& \times\left[ Q_\nu( t, t_1;zE,\btheta')\,  Q_\nu( t, t_1;(1-z)E,\btheta') -  Q_\nu( t, t_1;E,\btheta')  \right]  \,.
\end{align}
and \eqn{eq:S-elastic} yields
\begin{align}
Q_\nu^{\rm el}(t, t_0;E,\btheta)  &= \int \frac{\rmd^2 \btheta'}{(2\pi)^2E^2}\, \cP(E(\btheta'-\btheta),t,t_0) \left[ \Theta(|\btheta'|<R)\, +\Theta(|\btheta'|>R)  \,\rme^{-\nu E} \right]\,. \nn
&=\rme^{-\nu E}+  \int \frac{\rmd^2 \btheta'}{(2\pi)^2E^2}\, \cP(E(\btheta'-\btheta),t,t_0) \,  \Theta(|\btheta'|<R)\, (1-\rme^{-\nu E}) \,. \nn
\end{align}
The normalization condition \eqn{eq:norm} implies
\beq
Q_0^{\rm el}(t, t_0;E,\btheta) =1\,.
\eeq
We also wish to verify that if the `in' region vanishes, then $Q_\nu(t, t_0, E) = \rme^{-\nu E}$. Conversely, if the `out' region vanishes, we must have $Q_\nu(t, t_0) = 1$. Additionally, we note that the evolution equation \eqn{eq:non-lin-eq-LT} admits a thermal fixed point:
\beq Q_\nu^{\rm th}(E) = \rme^{-\nu E}\,,\label{eq:thermal-fp}
\eeq
which corresponds to the complete decay of the parent gluon. Of course, this asymptotic fixed point is not strictly physical, as some residual energy will always remain within the `in' region after the gluon has dissipated its energy into the plasma. Nevertheless, we can interpret this by introducing an IR cutoff scale, $\omega_{\rm IR} \gtrsim T$, on the order of a few times the plasma temperature, such that only modes above this threshold are considered part of the jet.

\eqn{eq:non-lin-eq-LT} takes the same form as the collimator function, which describes the energy loss associated with the vacuum-like collinear cascade (cf. \eqn{eq:non-lin-eq-coll}) \cite{Mehtar-Tani:2017web,Mehtar-Tani:2021fud}. In Sec.~\ref{sec:vacuum}, we will discuss how the collimator function can be naturally extended to incorporate the medium-induced cascade.

\subsection{Single parton energy loss}

The most physically relevant case for phenomenology is when a produced jet emerges out of the plasma with sufficiently high $p_T$ to be reconstructed and identified as a jet originating from a hard event. In this scenario, the energy loss that the jet experiences in the plasma is small compared to its energy, namely, $\epsilon\sim E_{\rm med} \ll E$, where $E_{\rm med}\sim \alpha^2 \hat q L^2$ is the typical gluon energy below which the gluon is fully quenched \cite{Baier:2001yt,Blaizot:2013hx,Mehtar-Tani:2017web}. Moreover,   the parent parton's angular broadening is negligible compared to the jet cone size which allows us to approximate the first term in \eqn{eq:non-lin-eq-LT} as
\begin{align}
Q^{\rm el}_\nu(t, 0 ; E,\btheta=0)& =1 \,.
\end{align}
where we have used $\cP(\q-\k,t,t_0) \approx (2\pi)^2\delta^{(2)}(\q-\k)$.  Since the energy flow is dominated by the leading parton, it approximately defines the jet center around which the jet area $|\btheta| < R$ is determined.

The same approximation is applied to the second term in \eqn{eq:non-lin-eq-LT} and we distinguish the leading (hard) branch contribution from a subleading branch for which $z\ll 1$ (or $z\ll 0$ by symmetry of the gg splitting function).  In addition, we need to approximate the splitting function by its soft limit, namely the soft radiation rate. This allows us to factorize the hard part  to obtain
\begin{align}\label{eq:non-lin-eq-jet}
& Q^{\rm hard}_\nu(t,t_0;E)  = 1 + \int_{t_0}^t  \rmd t_1  \int_0^E \rmd \omega \frac{\rmd I}{\rmd \omega}\,\left[ Q^{\rm soft}_\nu( t,t_1, \omega) -  1  \right] Q^{\rm hard}_\nu( t,t_1;E) \,,
\end{align}
where we have introduced the LO soft radiation rate (cf.~\eqn{eq:bdmps-coh})
\beq
\omega \frac{\rmd I}{\rmd \omega} = 2 \lim_{z\to 0 } z\cK(z,E) = \frac{\alpha_s N_c}{\pi} \sqrt{\frac{\hat q}{\omega}}\,,
\eeq
with $\omega=zE$ and $Q^{\rm hard}_\nu(t,t_1;E) \equiv Q^{\rm hard}_\nu(t,t_1,E;0)$. To keep the discussion simple and focused on the core properties of our evolution, we have intentionally neglected the fact that the spectrum changes for $\omega \gtrsim \omega_c$, where finite-size effects become significant as $t_f \sim L$. However, we emphasize that our formulation remains valid over the entire range of gluon frequencies. Furthermore, quarks can be easily incorporated into this framework.

Assuming weak quenching, i.e., $Q^{\rm hard}_\nu(t,t_0;E)\sim 1$, to first non-trivial order \eqn{eq:non-lin-eq-jet} yields
\begin{align}\label{eq:non-lin-eq-jet}
& Q^{\rm hard}_\nu(t,t_0;E)  = 1 + \int_{t_0}^t  \rmd t_1  \int_0^E \rmd \omega \frac{\rmd I}{\rmd \omega}\,\left[ Q^{\rm soft}_\nu( t,t_1, \omega) -  1  \right]+\cO((1-Q^{\rm hard})^2) \,,
\end{align}

\eqn{eq:non-lin-eq-jet}  can be readily solved, yielding
\begin{align}\label{eq:lin-eq-jet-sol}
& Q^{\rm hard}_\nu(t,t_0;E)  = \exp\left\{ -  \int_{t_0}^t  \rmd t_1  \int_0^E \rmd \omega \frac{\rmd I}{\rmd \omega}\,\left[ 1-Q^{\rm soft}_\nu( t,t_1;\omega)   \right] \right\} \,.
\end{align}
where $Q^{\rm soft}$ obeys the full non-linear equation \eqn{eq:non-lin-eq-LT}. This formula is valid so long as $E\gg \alpha_s^2 \omega_c$. In the Sec..~\ref{sec:steep-spect} we shall discuss this approximation in the context of the fully developed DGLAP shower.

The inverse Laplace transform leads to the energy distribution 
\begin{align}
&S^{\rm hard}(\epsilon, t,t_0;E) = \exp\left[-\int_{t_0}^t \rmd t' \int \rmd \omega \frac{\rmd I}{\rmd \omega \rmd t'}\right]  \nn
&\times\left\{\delta(\epsilon)+  \sum_{n=1}^\infty \frac{1}{n!} \prod_{i=1}^n \left[\int_{t_0}^t\rmd t_i \int\rmd \epsilon_i\,\rmd \omega_i\frac{\rmd I}{\rmd \omega_i\rmd t_i}\,  S^{\rm soft }(\epsilon_i, t,t_i;\omega_i) \right]\delta(\epsilon-\sum_{i=1}^n\epsilon_i)\right\}\,,
\end{align}
which is a  generalization of the Poisson-like energy loss distribution, namely  \eqn{eq:poisson} presented in Chap.~\ref{chap:med-partons} \cite{Baier:2001yt}, that can be recovered by neglecting the soft gluon shower, or by setting $R = 0$, .i.e., 
\beq 
\lim_{R\to 0}S^{\rm soft }(\epsilon_i, t,t_i;\omega_i)  \to \delta(\epsilon_i-\omega_i)\,.
\eeq
or equivalently, $\lim_{R\to 0}Q^{\rm soft}_\nu( t,t_1; \omega) =\rme^{-\nu \omega}\,$. 
%

\section{Collinear vacuum cascade}\label{sec:vacuum}

Up to this point, we have neglected the collinear parton cascade that arises from the decay of the initial parton's virtuality, which is responsible for jet fragmentation in vacuum. This cascade occurs prior to the medium-induced cascade, making the solution to \eqn{eq:non-lin-eq} the initial condition for the early-time non-linear DGLAP evolution.

Following the same construction as for the medium-induced cascade, we write

\begin{align}\label{eq:non-lin-eq-coll}
&S_{\rm loss}(\epsilon,E,\k,\theta_{\rm max})  = S_{\rm loss}^{\rm med}(\epsilon,E,\k,R)  \nn
&+ \frac{\alpha_s}{2\pi^2}\int_0^{\theta_{\rm max}} \rmd \theta\int \frac{\rmd^2 \q}{\q^2}  \int_0^1 \rmd z p_{gg}(z)  \delta\left(\theta-\frac{|\q|}{z(1-z)}\right) \Theta_{\rm PS}(\theta,z) \nn
& \times\left[ \int_{\epsilon_1,\epsilon_2} S_{\rm loss}(\epsilon_1, zE,\q+z\k, \theta)    S_{\rm loss}(\epsilon_2,(1-z)E,-\q+(1-z)\k,\theta)   \delta(\epsilon-\epsilon_1-\epsilon_2)\right. \nn
&\left.-  S_{\rm loss}(\epsilon_1, E,\k,\theta)   \right]  \,,\,
\end{align}
where the non-linear term describes the splitting of a gluon with transverse momentum $\k$ into two gluons of momenta $-\q+z\k$ and $\q+(1-z)\k$, respectively \cite{Catani:1990rr}. The initial condition is $ S_{\rm loss}^{\rm med}$ encodes medium-induced gluon cascade and is solution of the master equation \eqn{eq:non-lin-eq}. The maximum angle $ \theta_{\rm max} $ serves as the evolution variable, running up to angles of order $R$, and is related to the virtuality of the decaying parton $\mu $ through $ \mu \sim E \theta_{\rm max} $. The small $R$ approximation will lead to a further factorization.  The angular constraint $\theta< \theta_{\rm max}$
arises from the leading infrared (IR) contributions in perturbative QCD. To correctly account for leading soft and collinear logarithms, the kinematically available phase space for parton emission must be restricted to the angular-ordered region, where the branching angles decrease progressively along the cascade from the hard vertex to the final state \cite{Mueller:1981ex,Bassetto:1982ma,Dokshitzer:1982ia}. Outside this angular-ordered region, different parton emitters act coherently, resulting in destructive interference, where the azimuthally integrated distribution vanishes at leading order. This phenomenon is known as color coherence \cite{Dokshitzer:1991wu}.

The angular ordered constraint $ \theta < \theta_{\rm max}$ ensures the proper accounting of leading soft and collinear logarithms and is a result of color coherence \cite{Dokshitzer:1991wu}. The kinematically available phase space for parton emission is limited to the angular-ordered region, where branching angles decrease sequentially along the cascade from the hard vertex to the final state \cite{Mueller:1981ex,Bassetto:1982ma,Dokshitzer:1982ia}. Destructive interference suppresses large-angle emissions outside this region, as multiple partons behave coherently as a single effective emitter.

Interaction with the medium color charges tends to alter the color coherence of the jet, thereby opening up the phase space for radiation outside the angular-ordered region. This anti-angular-ordered radiation \cite{Mehtar-Tani:2010ebp,Mehtar-Tani:2011hma}, occurs at medium scales and does not generate large collinear logarithms, such as $\ln (R/\theta_c)$. However, it can be enhanced by the length of the medium \cite{Mehtar-Tani:2017ypq}. Color decoherence radiation takes place when the medium resolves individual color charges within the jet.

In the simplest case of an antenna forming an angle $\theta $, if the medium's resolution scale, which is related to transverse momentum broadening as $(\hat{q} L)^{-1/2} $, is smaller than the typical in-medium antenna size, $ r_\perp \sim \theta L $, i.e., $ (\hat{q} L)^{-1/2} < \theta L $, the two emitters radiate independently \cite{Casalderrey-Solana:2011ule,Mehtar-Tani:2012mfa,Mehtar-Tani:2017ypq}. This inequality implies that for incoherent radiation to occur, the angle between the hard collinear partons must satisfy the condition
\beq
\theta > \theta_c \equiv \frac{1}{(\hat{q} L^3)^{1/2}}\,,
\eeq
where $ \theta_c $ is the in-medium coherence angle \cite{Mehtar-Tani:2011hma,Mehtar-Tani:2012mfa,Mehtar-Tani:2017ypq}, which also corresponds to the minimum angle for medium-induced radiation \cite{Baier:1996kr}. Collinear splittings at smaller angles act coherently as a single color charge, and therefore do not impact the observable.

Apart from the angular constraint, the transverse momentum of the collinear splitting must exceed the typical transverse momentum for medium-induced splittings, given by $ k_\perp \equiv \sqrt{z(1-z) \hat{q}} $ (see \cite{Blaizot:2015lma} and references therein).

\begin{figure}
\begin{center}
\includegraphics[width=0.6\textwidth]{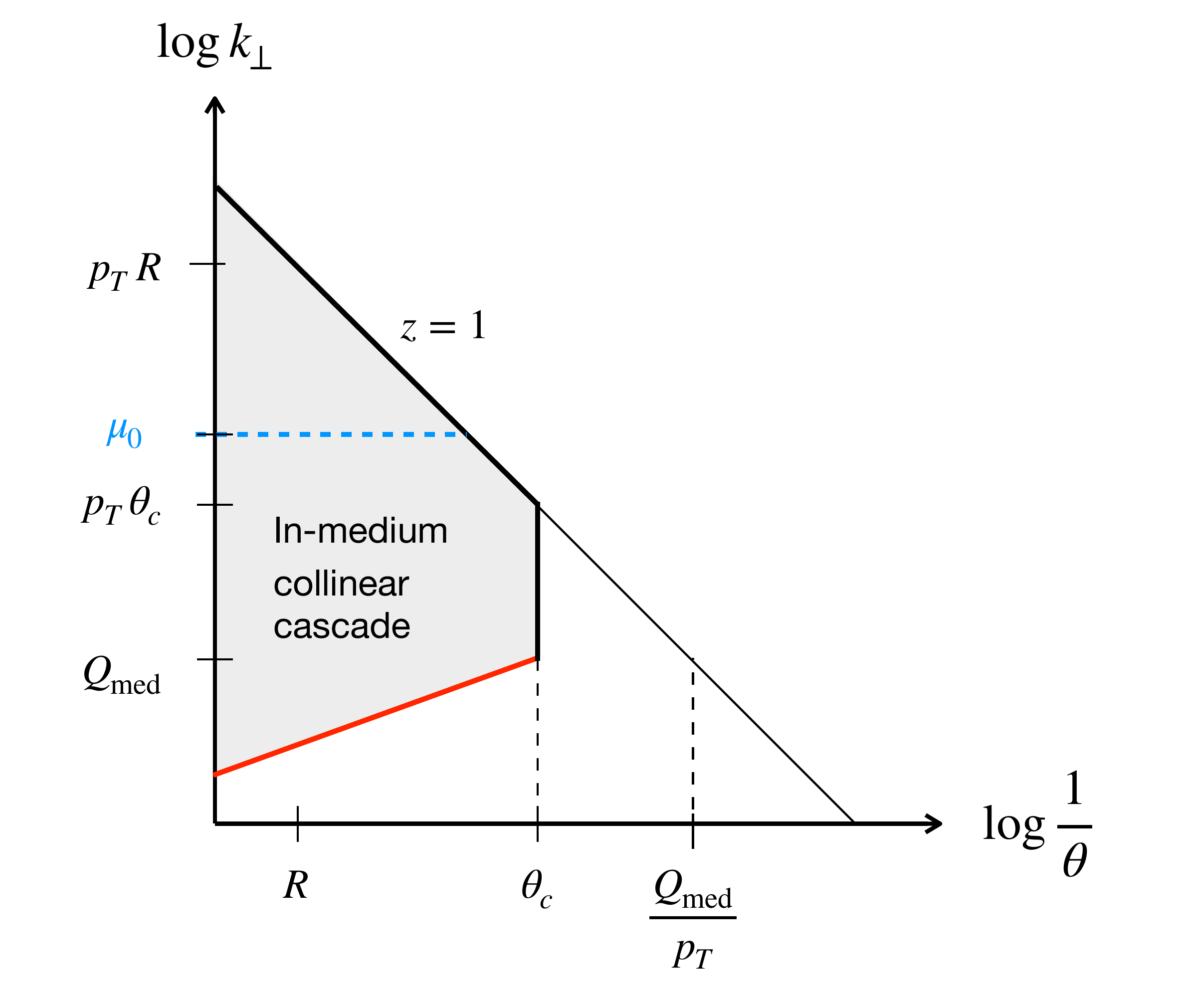}
\end{center}
\caption{ Lund-plane representation of the phase-space for in-medium vacuum shower. The dynamics near the boundaries $\theta=\theta_c$ and $k_\perp=(\omega \hat q)^{1/4}$, which correspond to the medium resolution angle and the medium-induced transverse momentum scale is driven by the details for the interactions with the medium. However, the physics in the hard sector, namely, the upper corner of the triangle is fully determined by DGLAP vacuum physics. The large separation of scales is illustrated by the intermediate factorization scale $\mu_0$. The medium  scale is typically given by $Q_{\rm med}\sim \sqrt{\hat q L}$. }
\label{fig:Lund-Plane}
\end{figure}

The angular and transverse momentum constraints described above define the phase space for resolved vacuum emissions, as illustrated in Fig.~\ref{fig:Lund-Plane}, and can be expressed as
\beq\label{eq:PS}
\Theta_{\rm PS}(\theta,z)  = \Theta(\theta>\theta_c) \Theta(|\q|> (z(1-z) E\hat q )^{1/4}) \,.
\eeq
Collinear splittings outside this phase space are either unresolved by the medium or occur outside the medium, and therefore do not contribute to inclusive jet production.

\subsection{Non-linear DGLAP evolution}
Again, it is convenient to express the above equation in Laplace space
\begin{align}\label{eq:non-lin-eq-coll}
& Q_\nu(E,\k,\theta_{\rm max})  = Q_\nu^{\rm med}(E,\k)  + \frac{\alpha_s}{2\pi^2}\int_0^{\theta_{\rm max}}\rmd \theta\int \frac{\rmd^2 \q}{\q^2}  \int_0^1 \rmd z \,p_{gg}(z)  \delta\left(\theta-\frac{|\q|}{z(1-z)}\right)\nn
& \times\left[ Q(zE,\q+z\k,\theta)    Q_\nu((1-z)E,-\q+(1-z)\k,\theta)  -  Q_\nu(E,\k,\theta)   \right]  \,,
\end{align}
To leading logarithmic accuracy, $\theta\ll \theta_{\rm max} \sim R$, therefore we may neglect variation of energy loss due to small deviations of collinear parton angles, i.e., $Q_\nu(E,\k,R)  \simeq Q_\nu(E,\k=0,R)\equiv Q_\nu(E,R)$, that are power suppressed. Upon this simplification and after integration over $\q$ then taking the derivative w.r.t. $\theta_{\rm max} \sim R$ \footnote{Throughout the evolution, we distinguish between $R$ and $\theta_{\rm max} $, the actual evolution variable, as the initial condition also depends on $R$. We set $\theta_{\rm max}=R$ only at the final stage.}, we recover the renormalization group equation (RG) for the quenching weight from \cite{Mehtar-Tani:2017web,Mehtar-Tani:2021fud}
\begin{align}\label{eq:non-lin-eq-coll}
& \theta\frac{\del }{\del \theta}Q_\nu(E,\theta)  =   \frac{\alpha_s}{\pi} \int_0^1 \rmd z \,  p_{gg}(z) \, \left[ Q_\nu(zE,\theta)    Q_\nu((1-z)E,\theta)  -  Q_\nu(E,\theta)   \right]  \,.
\end{align}
For the case of a running coupling, we should use $\alpha_s\equiv \alpha_s(z(1-z)\theta))$.  Here, we can ignore the phase space constraint since the equation is IR finite, i.e., when  $z\to 0$ or $z\to 1$,  and so long as $\theta>\theta_c$.  Indeed, the $z$ integral is cutoff dynamically in the hard collinear sector. A similar non-linear DGLAP evolution has been derived for track functions \cite{Chen:2022muj}, which measure a subset of hadrons, such as charged particles, within the jet \cite{Chang:2013rca}. The connection between the non-linear evolution of energy loss and track functions was noted recently \cite{Barata:2024bmx}. However, this straightforward extension of the track function approach to jet energy loss incorrectly assumes that the jet energy loss distribution is boost invariant. As discussed in this paper, the energy loss distribution depends not only on the energy fraction $ x $ and the factorization scale $ \mu $, but also on the initial energy $E $, which has motivated our (local) formulation in Laplace space.

This non-linear DGLAP equation is expressed as an evolution in angles to account for angular ordering along the cascade. We may also write a virtuality evolution, defining $\mu\equiv ER$ and $Q_\nu(E,R)\to Q_\nu(E,\mu)$, we obtain
\begin{align}\label{eq:non-lin-eq-coll-mu}
& \mu \frac{\del }{\del \mu}Q_\nu(E,\mu)  =  \frac{\alpha_s}{\pi} \int_0^1 \rmd z \,p_{gg}(z) \, \left[ Q_\nu(zE,z\mu)    Q_\nu((1-z)E,(1-z)\mu)  -  Q_\nu(E,\mu)   \right]  \,.
\end{align}
Neglecting angular ordering amounts to removing the factors $z$ and $(1-z)$ multiplying $\mu$ in the r.h.s. This is justified so long as the $z$ integral does not receive large contributions from the soft sector.

While the discussion above focused on collinear splittings within the jet ($ \theta < R $), \eqn{eq:non-lin-eq-coll-mu} remains applicable at larger angles to leading logarithmic (LL) accuracy. For $\theta \gg R $, one of the branches lies outside the jet cone, and, assuming negligible radiation back into the jet, all of its energy is lost, leading to $Q_\nu = 1 $. In this limit, the non-linear equation \eqn{eq:non-lin-eq-coll-mu} simplifies to a linear DGLAP-like equation. Thus, \eqn{eq:non-lin-eq-coll-mu} also resums large $ \ln R $ terms, corresponding to early out-of-cone vacuum radiation. In this case, we obtain
\beq
Q_\nu(zE,\theta>R) \simeq\rme^{-zE \nu} \,\quad \text{and} \quad  Q_\nu((1-z)zE,\theta>R) \simeq\rme^{-(1-z)E \nu}\,,
\eeq
for the $z$ and $1-z$ branches, respectively.  Hence, when  $ \theta \gg R$, or $\mu >Q\equiv  ER$, we obtain
\begin{align}\label{eq:lin-eq-coll}
& \mu\frac{\del }{\del {\mu}}Q_\nu(E,\mu) =\frac{\alpha_s}{\pi} \int_0^1 \rmd z \,  p_{gg}(z) \nn
&  \,\times \left[  \rme^{-(1-z) zE \nu}Q_\nu(zE,\mu)    +  \rme^{-zE \nu}Q_\nu((1-z)E,\mu)  -  Q_\nu(E,\mu)   \right]  \,,
\end{align}

Recalling the relationship with the jet function \eqn{eq:jet-function} to derive the corresponding evolution equation, we find, upon performing the inverse Laplace Transform of the first term
\begin{align}\label{eq:J-term-1}
\mu\frac{\del }{\del {\mu}}J(x,E,\mu) &=\frac{\alpha_s}{\pi} \int_0^1 \rmd z \,  p_{gg}(z) \int \frac{\rmd \nu}{2\pi i } \rme^{(1-x) p_T \nu - (1-z)E\nu} Q_{x\nu}\left(\frac{z E}{x},\mu\right)\nn
&=\frac{\alpha_s}{\pi} \int_0^1 \frac{\rmd z}{z} \,  p_{gg}(z) \int \frac{\rmd \nu'}{2\pi i } \rme^{(1-x) E \nu'} Q_{(x/z)\nu}\left(\frac{z E}{x},\mu\right)\nn
&=\frac{\alpha_s}{\pi} \int_0^1 \frac{\rmd z}{z} \,  p_{gg}(z) J\left(\frac{x}{z},zE,\mu\right)\,,
\end{align}
where in the second to last step we have changed variables, namely,  $\nu'=z\nu$.

Finally, using the symmetry $z \leftrightarrow 1-z$ of the equation and replacing the splitting function by the regularized Altarelli-Parisi splitting function \cite{Altarelli:1977zs}
\beq\label{eq:AP}
P_{gg}(z) =p_{gg}(z)_++p_{gg}(1-z) \,,
\eeq
we finally obtain the DGLAP-like equation
\begin{align}\label{eq:J-DGLAP-1}
& \mu\frac{\del }{\del {\mu}}J(x,E,\mu) =\frac{\alpha_s}{\pi} \int_0^1 \frac{\rmd z}{z} \,  P_{gg}(z)  J(x/z,zE,\mu) \,.
\end{align}
In the pure vacuum case there is no other scale other than $E$ and $\mu$, hence, owing to boost invariance we can combine the two scales to form dimensionless arguments, i.e., $J(x,E,R,\mu)\simeq J(x,\mu/ER)$, where we have restored  the implicit $R$ dependence. However, the presence of medium scales breaks this property and thus and extra $E$ dependence remains, as discussed in Sec..~\ref{sec:eloss-dist}.

Interestingly, with this observation, \eqn{eq:J-DGLAP-1} reduces to the recently proposed form of the jet function evolution equation valid at NLL \cite{Lee:2024tzc,vanBeekveld:2024jnx},
\begin{align}\label{eq:J-DGLAP-2}
& \mu\frac{\del }{\del {\mu}}J\left(x,\frac{\mu}{ER}\right) =\frac{\alpha_s}{\pi} \int_0^1 \frac{\rmd z}{z} \,  P_{gg}(z) \, J\left(\frac{x}{z},\frac{\mu}{zER}\right)  \,.
\end{align}
Extending beyond leading logarithmic accuracy requires matching coefficients at the transition scale $\mu = ER$, which connects the linear DGLAP-like renormalization group evolution for $\mu \gg ER$ with the non-linear evolution for $ \mu \ll ER$ \cite{Mehtar-Tani:2024smp}.

This leading logarithmic approach captures the essential features of the all-order structure of our formula at leading power in $R^2$, while retaining certain $R^2$ contributions associated with ene

At higher orders, it is necessary to compute matching coefficients that account for splittings at angles around $\theta \sim R$. A factorization formula for this case was recently derived in \cite{Mehtar-Tani:2024smp}, specifically for the regime $R \sim \theta_c$, enabling a detailed treatment of the infrared region in jet evolution. In the regime $R \gg \theta_c$, collinear logarithms $\ln(R/\theta_c)$ can be resummed, as discussed above, where interference contributions are suppressed. Additional matching is required at angles $\theta \sim \theta_c$, where the factorization framework developed in \cite{Mehtar-Tani:2024smp} is applicable.

\section{Non-perturbative effects and medium response }\label{sec:NP-effects}

Although our evolution equation is infrared finite the dynamics around the temperature scale should be described by kinetic theory. However, since soft modes relax quasi-instantaneously we can encode formally this physics by introducing a non-perturbative term that accounts for the thermalized energy whose distribution is known. By varying the infrared separation scale we can gauge to uncertainties associated with the non-perturbative regime. Such an approach allows us to account for the effect of medium response on energy loss. Similar strategy was previously implemented in Monte Carlos simulations such as CoLBT \cite{Chen:2017zte} and MARTINI \cite{Park:2018acg} where jet quanta that fall below an IR cutoff are modeled as sources for subsequent hydro evolution.

To illustrate our point, take for instance the expression given by \eqn{eq:lin-eq-jet-sol}. It would be modified as follows
\begin{align}\label{eq:lin-eq-jet-sol}
& Q^{\rm hard}_\nu(t,t_0;E)  \to  \exp\left\{ - \int_{t_0}^t  \rmd t_1  \int_{\omega_{\rm th}}^E \rmd \omega \frac{\rmd I}{\rmd \omega \rmd t_1}\,\left[ 1-Q^{\rm soft}_\nu( t,t_1;\omega)   \right] - F_{\rm NP}(\nu, \omega_{\rm th}) \right\} \,,
\end{align}
where here, the function $ F_{\rm NP}$ account for the energy redistribution of primary gluon radiation with energy below $\omega_{\rm th}$. It reads
\beq
F_{\rm NP}(\nu, \omega_{\rm th})= \int_{t_0}^t  \rmd t_1  \int^{\omega_{\rm th}}_0 \rmd \omega \frac{\rmd I}{\rmd \omega \rmd t_1}\,\left[ 1-Q^{\rm NP}_\nu(\omega)\right]\,.
\eeq
Now, assuming full thermalization of such soft modes their distribution should be thermal
\beq
Q^{\rm NP}_\nu(\omega) = \rme^{-\nu (\omega- E^{\rm th}(R,\omega)) }\,.
\eeq
This corresponds to the energy loss function
\beq
S_{\rm loss}^{\rm th}= \delta(\omega-E^{\rm th}(R,\omega))\,.
\eeq
This average energy loss approximation is based on the fact that  large number of soft gluons are produced below $\omega_{\rm th}\sim T$ which justifies neglecting fluctuations. The latter are accounted for in the semi-hard sector above $\omega_{\rm th}$.
Of course in the case were the soft gluon energy remaining inside the jet cone can be computed  \cite{Mehtar-Tani:2022zwf}
\beq
E^{\rm th}(R,\omega) =\omega  \left[1+\frac{3}{2}(1+\cos(R))\right]\sin^2\frac{R}{2}\, =\,  \omega R^2+\cO(R^4)\,.
\eeq
This result exhibits the expected scaling of isotropic energy with the jet area, $R^2$. To leading power in $R$, provided $R \ll 1$, medium response can be neglected. Moreover, we are equipped to systematically compute corrections associated with medium response as a power series in $R$.
\section{Sudakov suppression of jets }\label{sec:steep-spect}

The power exponent of the spectrum has long been identified as a important amplifying effect in jet energy loss \cite{Baier:2001yt}. Indeed, assuming a constant shift $\epsilon \ll p_T$ of the jet spectrum due to energy loss and expanding to second order we have
\beq
\frac{\rmd \sigma}{\rmd p_T} \sim \frac{1}{(p_T+\epsilon)^n} \simeq \frac{1}{p_T^n} \left(1-\frac{n \epsilon}{p_T} +\cO(n \epsilon^2)\right)\,,
\eeq
where we see that a small shift $\epsilon \sim p_T/n \ll p_T$ can have order one effect on the jet spectrum in heavy ion collisions.  In terms of the jet function, assuming $H(E=p_T/x)\sim (x/ p_T)^n$, we can approximate \eqn{eq:factorization}
\beq\label{eq:factorization-steep}
\frac{\rmd \sigma_{\rm incl}}{\rmd p_T}  \approx  H(E=p_T) \int_0^{1}\rmd x\, x^{n-1}\,J(x,E=p_T/x,R,\mu=p_T R/x) \,.
\eeq
It follows that when $n\gg 1$ the integral over $x$ receives the largest contribution close to the threshold region $x\sim 1$, with the result of inducing potentially large $\ln n$ which may provide an explanation as to the observed quantitative importance of the small-$R$ resummation even at moderate values of $R$ \cite{Dasgupta:2016bnd,Kang:2016mcy}.

In Laplace representation, from \eqn{eq:jet-function} we can rewrite \eqn{eq:factorization-steep} as
\beq\label{eq:jet-function-2}
J\left(x,\frac{p_T}{x}\right) = \frac{p_T}{x} \int \frac{\rmd \nu}{2\pi i} \, \rme^{ \frac{(1-x)}{x}p_T \nu}Q_{\nu}\left(\frac{p_T}{x}\right)  \, .
\eeq
It follows that,
\beq\label{eq:factorization-steep-LT}
\int_0^{1}\rmd x\, x^{n-1}\,J\left(x,\frac{p_T}{x}\right) =p_T \int \frac{\rmd \nu}{2\pi i} \,  \int_0^{1}\rmd x\, x^{n-2} \rme^{ \frac{(1-x)}{x} p_T \nu}Q_{\nu}\left(\frac{p_T}{x}\right)  \,.
\eeq
To gain some analytic insight we can look for a solution in the strong quenching case, which consists in linearizing  \eqn{eq:thermal-fp} around the thermal fixed point  that corresponds to ``total'' energy loss:
\beq
Q_\nu(E,\mu) = \rme^{-\nu E} (1+ q_\nu(E,\mu))\,,
\eeq
where we assume $q_\nu(E,\mu)\ll 1$.
To linear order in $q$, \eqn{eq:non-lin-eq-coll-mu} yields
\begin{align}\label{eq:strong-q}
\mu\frac{\del }{\del {\mu}}q_\nu(E,\mu) &\simeq \frac{\alpha_s}{\pi} \int_0^1 \rmd z \,  p_{gg}(z) \left[ q_\nu(zE,\mu)  +  q_\nu((1-z) E,\mu) - q_\nu(E,\mu)\right] \,.\nn
\end{align}

We recognize the DGLAP equation which can be solved in Mellin space \cite{Mehtar-Tani:2024mvl}
\beq
q_\nu(E,\mu) =  \int_{c-i\infty}^{c+i\infty}\frac{\rmd N}{2\pi i } E^{N-1} \exp\left[ \frac{\alpha_s}{\pi} \gamma(N) \ln \frac{\mu}{\mu_0} \right] q_{\nu,N}(\mu_0)\,.
\eeq
This is the general solution of the RG evolution down to some medium scale. One very important consequence is this limit is the cancellation of the $ \mu\sim ER$ dependence due to vacuum evolution and its replacement by $E\theta_c$.

Inserting this solution for strong quenching in \eqn{eq:factorization-steep-LT} 
\beq\label{eq:factorization-steep-LT-2}
 &&\int_0^{1}\rmd x\, x^{n-1}\,J\left(x,\frac{p_T}{x}\right) =\nn
 &&\quad \frac{1}{n-1} p_T\, \delta(p_T)+p_T   \,\,\rme^{ \frac{\alpha_s}{\pi} \gamma(n) \ln \frac{\mu}{\mu_0} } \int_0^1\rmd y y^{n-2} \int \frac{\rmd \nu}{2\pi i} \, \, \rme^{-\nu p_T} q_\nu(p_T/y,\mu_0) \,,\nn
\eeq
where we have integrated over $x$ and $N$.

Expression the result as function of the energy loss distribution and neglecting the irrelevant  $\delta(p_T)$ term, the  cross-section reads
\begin{align}\label{eq:factorization-steep}
\frac{\rmd \sigma_{\rm incl}}{\rmd p_T}  
&\approx p_T \, H(E=p_T,\mu=p_TR)\,  \,\,\rme^{ \frac{\alpha_s}{\pi} \gamma(n) \ln \frac{\mu}{\mu_0} } \int_0^1\rmd y \, y^{n-2}  \, \, S_{\rm loss}((1-y)p_T/y, p_T/y,\mu_0) \,
\end{align}
where we have used \eqn{eq:eloss-LT}.

Now, changing variable to $\epsilon=(1-y)p_T$ and expanding the above expression around $n\to \infty$ limit. We have
\beq
\gamma(n)  \to  - 2N_c \left( \ln n +\gamma_E-\frac{11}{12}\right) +\cO(1/n)\,.
\eeq
with $\gamma_E \approx 0.5772$ the Euler constant, and
\beq
p_T \int_0^1\rmd y \, y^{n-2}  \, \, S_{\rm loss}((1-y)p_T, p_T/y) &=& \int_0^{+\infty}\rmd \epsilon \, \left(\frac{p_T}{p_T+\epsilon}\right)^{n-2}  \, \, S_{\rm loss}\left(\epsilon, p_T+\epsilon\right) \nn
&= & \int_0^{+\infty}\rmd \epsilon \, \rme^{(n-2) \frac{\epsilon}{p_T} } \, \, S_{\rm loss}\left(\epsilon, p_T\right)  + \cO(\epsilon/p_T)+\cO(n \epsilon^2/p_T^2)\nn
&\simeq& Q_{\nu=p_T/n}(p_T)\,.
\eeq
As a result, we obtain for the large $n$ limit of the inclusive jet production cross-section:
\begin{align}\label{eq:factorization-steep-res}
\frac{\rmd \sigma_{\rm med}}{\rmd p_T}  \approx  \, H(p_T,\mu)\,  \,\,\exp\left(- \frac{2\alpha_s N_c}{\pi} \ln n \ln \frac{\mu}{\mu_0} \right) Q_{\nu=p_T/n}(p_T,\mu_0)\ \,
\end{align}
where we have assumed that $\mu\sim p_TR$ and $\mu_0\sim p_T \theta_c$.

This simple asymptotic result exhibits two main features of jet quenching:
\begin{enumerate}
\item In addition to the leading parton energy loss there is a Sudakov suppression due to the quenching of micro-jets within the jet. As result, if the a jet is measured it is composed of a single hard subjet of angular size $\theta_c$.
\item A second important property is the cancelation of the $R$ dependence generated by the DGLAP evolution. Of course, there subsides an $R$ dependence in the medium-induced cascade as part of the quenching factor.
\end{enumerate}
These effects were already discussed in \cite{Mehtar-Tani:2017web}.
In the leading logarithmic approximation (LLA), the vacuum cross-section can be expressed as:

\begin{align}\label{eq
} \frac{\rmd \sigma^{\rm vac}}{\rmd p_T} \approx H(p_T, \mu = p_T R)\,. \end{align}

Asymptotically, in the strong quenching limit, the nuclear modification factor is expected to behave approximately as:

\beq R_{AA} = \frac{\rmd \sigma^{\rm vac} / \rmd p_T}{\rmd \sigma^{\rm med} / \rmd p_T} \approx \rme^{-\frac{2\alpha_s N_c}{\pi} \ln(n) \ln(R/\theta_c)} \, Q_{\nu=p_T/n}(p_T, R)\,, \eeq
where $Q_{\nu=p_T/n}(p_T, R)$ encapsulates the quenching factor for a single color charge. 

In order to observe the Sudakov suppression factor in experimental data, it is necessary to impose an infrared (IR) cutoff to eliminate soft particles. These soft particles are responsible for the residual $R$-dependence in the medium-induced component, which leads to an enhancement of $R_{AA}$ due to energy recovery via soft emissions at large angles. With such a cutoff applied, we predict a systematic decrease in the nuclear modification factor $R_{AA}$ as the jet cone size $R$ increases.

In the next chapter, we will conclude this review by applying our approach to phenomenology. Specifically, we will solve the non-linear evolution equations numerically for a realistic quark-gluon plasma and compare the results with jet production data. Additionally, we will evaluate the theory and model uncertainties, examining the sensitivity of the framework to both the perturbative sector—where improvements can be made through higher-order corrections—and the non-perturbative sector, which relies on modeling.

%% file: phenomenology.tex
\chapter{Phenomenology} \label{chap:phenomenology}
Although this report primarily focuses on the theory of jet quenching, which has been continuously developed over the past few decades, we dedicate this final chapter to discussing its phenomenological applications. The aim is to bridge this formal theoretical framework with experimental observations, which ultimately represent the goal of this field. Indeed, these theoretical advancements have been largely motivated by the need to understand and interpret the wealth of data produced by ongoing heavy-ion jet programs at RHIC and LHC.

The objectives of the phenomenology of jets in HIC are multifaceted. One of the primary goals of studying fully reconstructed jets is to leverage the multidimensional nature of jet observables to probe the transport properties of non-equilibrium QCD matter. In particular, these studies aim to uncover the fundamental mechanisms governing energy transport from high-energy jets to soft plasma particles. Another significant potential discovery from these experiments is the QCD analog of the Landau-Pomeranchuk-Migdal (LPM) effect, previously observed in QED, which could be inferred from the production of medium-induced gluon jets. Additionally, the large separation of scales between the jet transverse momentum, $p_T$, on the order of 100–1000 GeV, and the plasma temperature, typically less than 1 GeV, provides a unique opportunity to explore a variety of emergent QCD phenomena. These include turbulent energy flow, QCD color decoherence, and anomalous diffusion in transverse momentum space, some of which are presented in this report.

At this stage, we pose the following question: {\bf to what extent is this first-principles-based approach to jet quenching predictive and amenable to theoretical improvements? }In other words, are theoretical uncertainties under control, and how significantly do non-perturbative processes—those not calculable from first principles and thus requiring modeling—affect high-$p_T$ jet observables?

While jet physics in smaller systems is not impacted by the presence of a soft QCD background, similar issues arise, such as the effects of hadronization, which are known to skew parton-level substructure observables like jet mass. The difference lies in the magnitude of these non-perturbative effects in heavy-ion collisions. Among the most prominent effects is the medium response, i.e., the back-reaction of the plasma on the jet itself, which alters the soft sector. Furthermore, the presence of a QCD medium may modify the hadronization mechanism, reshuffling color reconnections among final-state jet partons and leading to a modified hadronization pattern.

Nevertheless, there is hope that certain infrared and collinear (IRC) safe observables may be less sensitive to soft physics. In such cases, these observables could be mapped to perturbative partonic processes, which can be computed using perturbation theory combined with resummation techniques. 

\section{The jet nuclear modification factor as a substructure observable }

Perhaps the most iconic observable of jet quenching is the nuclear modification factor, defined as the ratio of measured per-event inclusive jet yield in heavy-ion (AA) collisions  $p_T$ in heavy-ion collisions, such as Pb-Pb collisions at the LHC at $\sqrt{s}=5.5$ TeV and Au-Au collisions at RHIC at $\sqrt{s}=200$ GeV, to the inclusive cross-section in proton-proton (pp) collisions properly normalized. The so-called $R_{_{AA}}$ serves as a measure of the modifications jets experience in the quark-gluon plasma (QGP):
\begin{equation}
R_{_{AA}} = \frac{1}{\langle T_{_{AA}} \rangle}  \frac{ \rmd^2N / \rmd p_T \rmd\eta}{\rmd^2\sigma_{pp}/ \rmd p_T d\eta}, 
\end{equation}
where $R_{AA}$ represents the ratio of the measured per-event inclusive jet yield in heavy-ion (AA) collisions to the inclusive cross-section in proton-proton (pp) collisions, scaled by the nuclear overlap $\langle T_{AA} \rangle$ for a given centrality class \cite{d'Enterria:2021ycp}. 

By definition, if final-state or initial-state interactions due to the differing nature of the colliding systems are negligible, $R_{_{AA}}$ is expected to be close to unity. This behavior is indeed observed in jet production in proton-nucleus (p-A) collisions, leading to the conclusion that no QGP is formed in these systems. However, this interpretation has been challenged by the striking flow-like behavior observed in high-multiplicity p-A events \cite{Weller:2017tsr}. 

Another example of $R_{_{AA}}$ close to unity is provided by direct photons. Due to their electromagnetic nature, direct photons interact weakly with the plasma and thus escape the medium without significant modification.

In contrast, QCD jets exhibit strong suppression in heavy-ion collisions. The nuclear modification factor for jets is typically around $0.5$ over a broad $p_T$ range, from approximately 50 GeV up to nearly the TeV scale. This suppression provides compelling evidence for substantial final-state interactions, where the color charges of the jet interact with and lose energy to an extended deconfined QCD medium.

The nuclear modification factor has been extensively studied but has recently taken a backseat to more differential observables, such as jet substructure observables, such as the energy-energy correlators \cite{Andres:2022ovj,Andres:2023xwr,Barata:2023zqg,Barata:2023bhh} and groomed jet observables \cite{CMS:2017qlm,Song:2023sxb,ALICE:2022hyz,CMS:2018fof,Caucal:2019uvr,Mehtar-Tani:2016aco,Caucal:2021bae}, which are expected to offer greater discriminating power among various model calculations. Notably, these models, despite incorporating vastly different physics—from fixed-order QCD computations and ad-hoc modifications of the vacuum parton cascade to strong-coupling approaches where jets lose energy to a strongly coupled plasma—have all succeeded in describing $R_{AA}$.

Recently, however, $R_{AA}$ has experienced a revival with the introduction of a new dimension: the jet opening angle $R$ \cite{CMS:2021vui,ALICE:2023waz}. This parameter complements existing variables such as jet $p_T$, collision centrality, and collision energy, offering new insights into the modification of jets in heavy-ion collisions.

While the inclusive jet cross-section in $pp$ collisions is largely insensitive to jet substructure—due to probability conservation, which ensures that only the total sum of the $p_T$ of hadrons making up the jet contributes to the cross-section regardless of substructure fluctuations—the situation is markedly different in heavy-ion collisions. In this case, the quark-gluon plasma acts as a \textit{filter}, suppressing different jet substructure configurations with varying quenching weights, which in turn influences the inclusive cross-section. This interplay is captured by the non-linear DGLAP equation \eqref{eq:non-lin-eq-coll}.

A remarkable consequence of the medium's resolution of the inner jet structure is that varying the cone size $R$ acts as a knob, effectively tuning the quenching factors for different jet substructure fluctuations. This phenomenon was strikingly demonstrated when existing models, tuned to a single value of $R$, struggled to accurately predict how the jet cross-section changes as a function of the cone size.

The framework presented in this report was used to compute the jet cross-section in Ref.~\cite{Mehtar-Tani:2021fud} by solving the non-linear DGLAP equation \eqn{eq:non-lin-eq-coll} in the large $n$ limit, which allows to perform a kind of saddle point approximation of the Laplace transform $\nu$, with an initial condition based on the exponential form of the quenching factor. Additional approximation were employed such as a step function to model the effect of the medium-induced shower below the scale $\omega_s$. Below this scale the radiation gluon energy is assumed to thermalize.  

The perturbative part of the calculation is based on a leading-logarithmic approximation, with the associated uncertainty estimated by varying the coherence angle $\theta_c$, which enters the leading collinear logarithms, $\ln(R/\theta_c)$. These logarithms are resummed to all orders by \eqref{eq:non-lin-eq-coll}. The medium response is modeled in a minimalistic manner through the non-perturbative component in \eqref{eq:lin-eq-jet-sol}, with the threshold energy set to $\omega_{\rm th} = \omega_s$. Soft gluons with $\omega < \omega_s$ cascade quasi-instantaneously to the thermal scale~\cite{Blaizot:2013hx} and are effectively treated within hydrodynamics. Their emission rate is, therefore, unaffected by transverse momentum broadening. Assuming that the distribution of these soft gluons becomes approximately uniform over the solid angle around the jet, we allow for the possibility that a fraction of this energy reenters the jet cone. This is accounted for by modifying the energy term as $\omega \to \omega (1-(R/R_{\rm rec})^2)$, where the recovery angle $R_{\rm rec}$ is treated as a free parameter.

The medium-induced gluon radiation spectrum has been computed up to next-to-leading order (NLO) (cf.~\eqref{eq:ioe-0} and \eqref{eq:nlo}) within the improved opacity expansion (IOE) in the soft limit~\cite{Mehtar-Tani:2019tvy,Mehtar-Tani:2019ygg,Barata:2020sav}. This formalism unifies the BDMPS approach with the GLV/higher-twist framework~\cite{Gyulassy:2000fs,Guo:2000nz}, providing an essential tool for phenomenological studies~\cite{Feal:2019xfl}. The IOE method, introduced in Chap.~\ref{chap:formalism}, has been shown to achieve remarkable accuracy when compared with exact numerical solutions~\cite{Andres:2020kfg}.

In our framework, the medium coupling $g_\med$, the only free parameter that determines energy loss, is to be extracted from the comparison to experimental data.

\begin{figure}
\centering
\includegraphics[width=0.4\columnwidth]{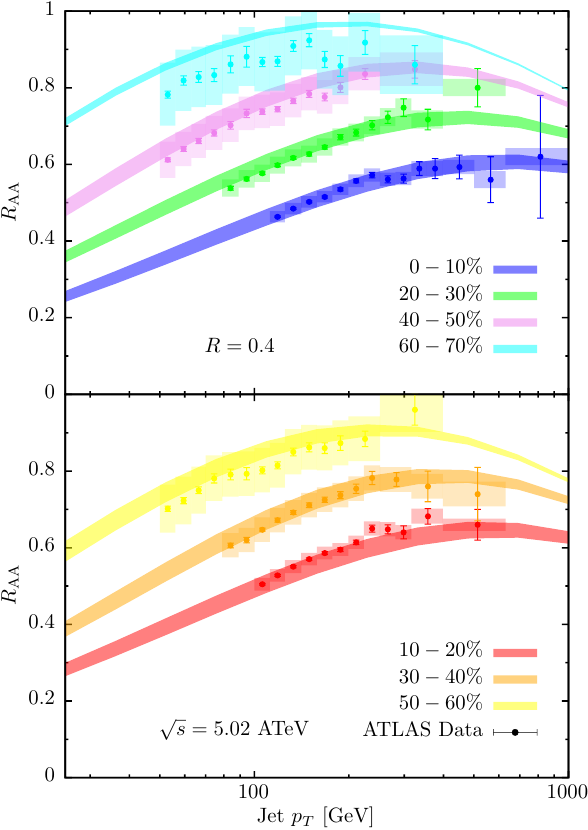}
\caption{\label{fig:jetraa} Calculation of inclusive jet  $R_{_{AA}}$ in PbPb collisions at $\sqrt{s}= 5.02$ TeV, compared to ATLAS data \cite{ATLAS:2018gwx}, for different centralities.}\label{fig:raa_atlas}
\end{figure}

Additional fluctuations on the path and medium density explored by the jet, which vary event by event, are taken care of by embedding our framework into a realistic heavy-ion environment in which the medium is described by the explosion of a liquid droplet of deconfined QCD matter.
Event-by-event in-medium path fluctuations of a jet through the QGP have been taken into account by embedding our framework into a realistic heavy ion environment as simulated in the VISHNU hydrodynamical model \cite{Shen:2014vra}. The value of $g_\med$ is thus constrained by the experimental data to be within the range $g_\med \in\lbrace 2.2, \, 2.3 \rbrace$.

The extracted parameters yield an average value $\langle \hat q_0 \rangle \simeq 0.41$ GeV$^2$/fm in 0-10\% central PbPb collisions. However, the logarithmic corrections to the bare medium parameters, resulting in $Q^2=14.2$ GeV$^2$ for the factorization scale and $\hat q =2.46$ GeV$^2$/fm, that is well within the perturbative regime.
We performed numerical computation and compare to the ATLAS data for $R=0.4$ are presented in Fig.~\ref{fig:raa_atlas} as function of centrality. 

\section{Theoretical uncertainties \label{sec:theo-uncertainties}}
We now discuss the sensitivity of our results to the various assumptions made in the setup, aiming to identify the primary sources of uncertainty. Our main findings are summarized in Tab.~\ref{tab:summary} for moderate cone sizes $0.2 \leq R \leq 0.6$.  First, the inclusion of the higher-twist radiative spectrum in the IOE has a mild impact on this observable, as such emissions predominantly occur at small angles. However, it provides an improved description at high-$p_T$. As expected, notable bias effects are observed through the strong sensitivity to the power of the steeply falling spectrum $n$, underscoring the importance of higher-order terms in the large-$n$ expansion. These terms can be systematically calculated, as outlined in Chap.~\ref{chap:jeteloss}.

More importantly, comparing the effects of varying the hard phase space (via $\theta_c$) with those of the parameters governing the behavior and recovery of soft gluons (via $\omega_s$ and $R_{\rm rec}$), we find that increased precision in the perturbative sector is necessary before the sensitivity to non-perturbative effects becomes dominant. For instance, going beyond leading-logarithmic accuracy to compute $\theta_c$ will be critical for rigorously studying the interesting and pronounced centrality dependence of this critical angle. 

The importance of the recovery parameter $R_{\rm rec}$ has been explored between two limiting scenarios: $R_{\rm rec} = 1$, representing almost complete energy recovery for large-$R$ jets, and $R_{\rm rec} = \infty$, corresponding to no energy recovery. Surprisingly, we observe relatively little sensitivity to this parameter at moderate cone sizes. In contrast, the sensitivity to $R_{\rm rec}$ becomes the dominant source of uncertainty only for large-$R$ jets, i.e., $R \approx 1$.

This analysis demonstrates that, for relatively small cone angles $R \sim 0.2 - 0.4$, the observable is much less sensitive to non-perturbative effects—beyond the inherent dependence on the quenching parameter—than to perturbative effects, such as the logarithmic phase space of resolved collinear vacuum splittings. This hints at the possibility of improving the accuracy of theoretical predictions by computing higher-order corrections.

In Tab.~\ref{tab:thetac} we extract the average value of the critical angle $\theta_c$, which governs the physics of decohernce, as function of centrality. These values will guide future higher order theoretical calculations. Indeed, for the these values of $\theta_c$ the $\ln R/\theta_c$ might not be large enough to necessitae resummations to all orders. A fullfledged fixed order calculation can provide a more accurate description beyond the leading logarithmic accuracy.

\begin{table}[t!]
\begin{center}
\begin{tabular}{p{0.2\columnwidth} | p{0.20\columnwidth} | p{0.15\columnwidth} }
Parameter & Variation & Effect  \\
\hline
$\theta_c$ & $[\theta_c/2, \, 2\theta_c]$ & $\lesssim 20\%$   \\
IOE & LO/NLO& $\sim 2\%$  \\
$n$ & $\pm 1$ & $\sim 10\%$   \\
$R_{\rm rec}$ & $[1, \, \infty]$ & $\lesssim 10\%$  \\
$\omega_s$ & $[\omega_s/2,\, 2\omega_s]$ & $\lesssim 8\%$    \\
\end{tabular}
\end{center}
\caption{Summary of the effect of relative change of $R_{AA}$ from varying key parameters for cone sizes $R=0.2-0.6$, see text for further details.}\label{tab:summary}
\end{table}%

\begin{table}[t!]
    \centering
   \begin{tabular}{c|c|c}
    & \multicolumn{2}{c}{$\theta_c$} \\
    \hline
     Centrality & RHIC & LHC \\ \hline 
        0-5\% & 0.13 & 0.09 \\
        5-10\% & 0.15 & 0.10 \\
        10-20\% & 0.17 & 0.12 \\
        20-30\% & 0.22 & 0.15 \\
        30-40\% & 0.27 & 0.19 \\
        40-50\% & 0.35 & 0.24 \\
        50-60\% & 0.45 & 0.32 \\
        60-70\% & 0.58 & 0.41 \\
    \end{tabular}
    \caption{Fitted values for $\theta_c$, using Eq.~\eqref{eq:coh-angle}, both at RHIC (AuAu collisions at $\sqrt{s}=0.2$ ATeV) and LHC (PbPb collisions at $\sqrt{s}=5.02$ TeV), for different centrality classes.}
    \label{tab:thetac}
\end{table}

\begin{figure*}[t!]
    \includegraphics[width=1\textwidth]{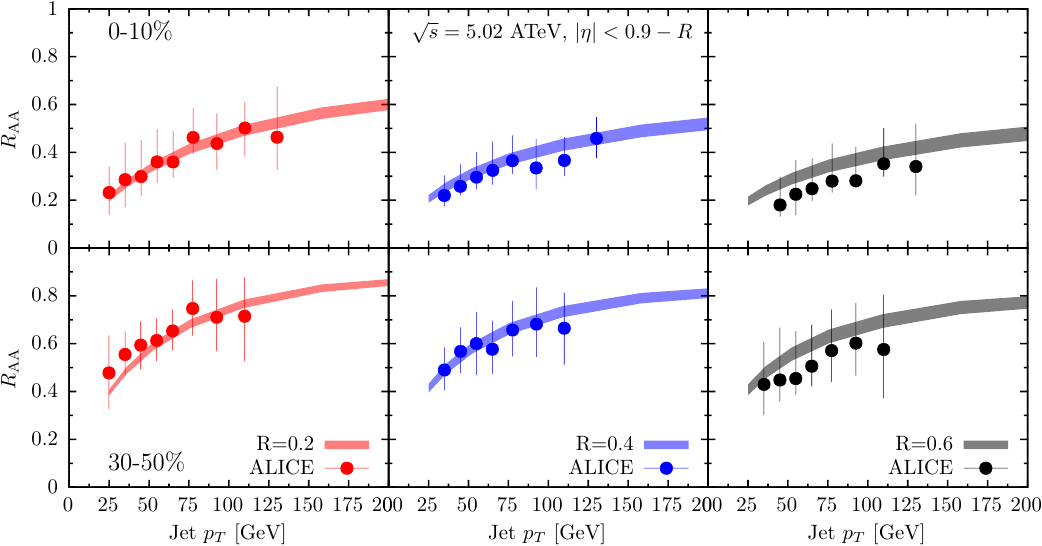}
    \caption{Comparison of theoretical predictions of $R_{\rm AA}$ against ALICE data~\cite{ALICE:2023waz} for PbPb collisions at $\sqrt{s}=5.02$ TeV, for the 0-10\% (30-50\%) centrality class in the upper (lower) row. Different columns display different values of jet-cone $R$.}
    \label{fig:dataraaalice}
\end{figure*}

We conclude this chapter with a prediction for the $R$-dependence of $R_{_{AA}}$, which was released prior to the publication of ALICE data~\cite{ALICE:2023waz}. The ALICE Collaboration compiled all theoretical predictions, including ours, as shown in Fig.~\ref{fig:dataraaalice}. The excellent agreement with data across a wide range of centralities, jet $p_T$, and, most notably, jet cone sizes $R$, serves as an encouraging validation of our leading-logarithmic approach. This success provides a solid foundation for future studies, particularly on jet substructure observables, which still lack systematic theoretical computations. Recently, this framework has been successfully applied to the study of jet $v_2$ \cite{Mehtar-Tani:2024jtd} and extended to smaller collision systems~\cite{Pablos:2025cli}. 

The framework developed here, while primarily applied to inclusive jet spectra, provides a flexible basis for extending phenomenological studies to more differential observables. In particular, jet substructure measurements — such as groomed momentum sharing, jet mass, and energy–energy correlators — offer enhanced sensitivity to the dynamics of medium-induced radiation, color coherence, and parton cascades. Applying this approach to such observables would enable a more differential characterization of jet quenching and open new avenues for connecting theory with precision data. We anticipate that future developments along these lines will play a central role in consolidating our understanding of QCD dynamics in dense matter and in fully exploiting the experimental programs at RHIC and the LHC.

%% file: conclusion.tex
\chapter{Conclusion and perspectives} \label{chap:conclusion}

Jet quenching theory has been evolving for over three decades, driven by groundbreaking theoretical ideas, beginning with Bjorken's early work and later expanded upon by Gyulassy and Wang. The advent of high-energy colliders such as RHIC and the LHC, where the hottest matter in the history of the universe is recreated -- albeit for an instant -- in ultrarelativistic heavy-ion collisions, has presented theorists with incredible challenges. These challenges involve constructing robust theoretical tools grounded in QCD to interpret the vast array of data on jet observables. The task is arduous, as it requires a framework that not only incorporates the well-understood physics of jets in vacuum, as studied in smaller systems like proton-proton or electron-positron collisions, but also addresses the non-linear interactions between jets and the QGP. Compounding the complexity is the emergence of intermediate scales between the jet scales $p_T$ and $p_T R$, and the plasma temperature $T$. These scales, arising from multiple interactions and coherence effects intrinsic to quantum systems, are characterized by non-linear dynamics and necessitate sophisticated theoretical approaches.

High-density effects are strikingly evident in the strong suppression of fully reconstructed jets observed at the LHC. Since the early works of Baier, Dokshitzer, Mueller, Peigné, and Schiff (BDMPS), it has been clear that higher-order effects play a critical role. Building on these pioneering efforts, I, along with my collaborators in Saclay and Santiago de Compostela, have sought to tackle the challenges of jet quenching. Our goal has been to develop a systematic perturbative framework that enables the precise computation of jet observables at the desired accuracy. 

In this concluding chapter, I will summarize the main findings presented in this work and place them in the context of the ongoing experimental programs at RHIC and the LHC. I will then take a step back to reflect on the challenges that lie ahead and the opportunities this field offers for broadening our understanding of QCD at high energy and high density.

This report is structured around three central themes. The first part is dedicated to introducing the physics of medium-induced radiation. after introducing the basic ideas and the formalism, I have reported on a recently developed technique that allows to have a unified description of the medium-induced spectrum, such as the Armesto-Salgado-Wiedemann (ASW) \cite{Salgado:2003gb, Armesto:2004vz}, the Gyulassy-Levai-Vitev \cite{Gyulassy:2000er} and the higher-twist \cite{Wang:2001ifa} approximations. This approach is based on an expansion around the harmonic oscillator that is used to describe multiple soft gluon exchanges and and account for the rare hard interactions perturbaitvely \cite{Mehtar-Tani:2019tvy,Mehtar-Tani:2019ygg,Barata:2020sav,Barata:2021wuf}. This analytic technique It agrees remarkably well with exact numerical simulations \cite{Andres:2020vxs} and constitutes one of the main recent advances toward higher precision jet phenomenlogy.

While progress has been made in understanding leading-order radiative processes, much less was known about how to translate these findings to multi-partonic systems that constitute fully developed jets. Our exploration of higher-order effects led to the construction of a probabilistic picture for medium-induced cascades, revealing emergent features absent at leading order. One notable feature is the self-similar energy transfer from hard modes to the plasma's temperature scale. This energy transport, reminiscent of wave turbulence, occurs down to the infrared scale without accumulation, a behavior that hints at an all-order mechanism of energy loss potentially at play in Pb-Pb collisions at the LHC \cite{Blaizot:2014ula}.

Another key result at higher orders is the renormalization of the jet quenching parameter, $\hat{q}$, which obeys a non-linear renormalization equation exhibiting saturation effects akin to those in the Balitsky-Kovchegov equation \cite{Balitsky:1995ub,Kovchegov:1999yj} -- a cornerstone in high-energy QCD for understanding parton density saturation at small Bjorken-$x$. One consequence of this result is that the medium’s interaction with the soft background is mediated by a gluon ladder with decreasing virtuality. Another significant implication is that resumming leading logarithms yields an anomalous scaling of $\hat{q}$, which modifies the transverse momentum diffusion of high-energy partons. At leading order, the diffusion process exhibits normal behavior, with the mean squared displacement growing linearly, $\langle k_\perp^2 \rangle \sim L$. However, at higher orders, this turns into a superdiffusive process, $\langle k_\perp^2 \rangle \sim L^\gamma$, where $\gamma > 1$. This anomalous diffusion is characterized by heavy Lévy tails, marking a significant departure from Rutherford scattering. These findings highlight the intricate dynamics of jet-plasma interactions and provide a foundation for future explorations of QCD in extreme conditions.

Of course, due to the complexity of the problem, certain assumptions were necessary to achieve analytical results. One such assumption is the locality of correlations in the plasma, which justifies the use of the independent multiple scattering approximation. It may be possible, however, to leverage the presence of emergent perturbative scales to construct an effective field theory that systematically factorizes long-distance and short-distance physics in the plasma. In such a framework, the multiple independent interaction approximation would emerge as a limiting case \cite{Mehtar-Tani:2024smp}. The inclusive properties of the turbulent gluon cascade have been extensively studied over the past decade \cite{Blaizot:2014ula,Blaizot:2014rla,Caucal:2019uvr,Mehtar-Tani:2018zba,Schlichting:2020lef,Mehtar-Tani:2022zwf,Soudi:2024yfy}, with substantial progress made at NLO \cite{Arnold:2020uzm,Arnold:2021pin,Arnold:2022fku,Arnold:2022mby,Arnold:2023qwi,Arnold:2024whj,Arnold:2024bph} including the resulting renormalization of the jet quenching parameter \cite{Blaizot:2014bha,Caucal:2022mpp,Ghiglieri:2022gyv,Caucal:2022fhc,Arnold:2021pin,Arnold:2021mow,Caucal:2021lgf,Blaizot:2019muz,Iancu:2018trm,Mehtar-Tani:2017ypq,Wu:2014nca,Iancu:2014sha,Liou:2013qya,Iancu:2014kga}.

The final part of this manuscript is dedicated to addressing the central problem of this work: the computation of jet quenching observables for phenomenological applications. We presented recently developed non-linear QCD evolution equations that describe the renormalization group (RG) evolution of jet functions, which encode the effects of both the early DGLAP collinear parton cascade and the medium-induced parton shower~\cite{Mehtar-Tani:2017web,Mehtar-Tani:2024mvl}. 

The interplay between these two parton cascades is governed by: (i) the medium scale, which sets the lower bound for DGLAP evolution, and (ii) color decoherence, introduced in Chap.~\ref{chap:decoherence}. The latter arises from interference effects~\cite{Mehtar-Tani:2010ebp,Mehtar-Tani:2011hma,Mehtar-Tani:2012mfa,Mehtar-Tani:2017ypq,Casalderrey-Solana:2011ule,Mehtar-Tani:2017ypq}, which enforce the property of color transparency for unresolved jet fluctuations. This decoherence is characterized by a critical angle $\theta_c$, which is typically smaller than the jet cone size.

Chap.~\ref{chap:phenomenology}, in particular, serves as a road test for the theoretical framework, which, in many respects, relies on asymptotic limits and thus has limited accuracy. On a positive note, recent experimental data provide multi-dimensional measurements that help constrain theoretical models. For example, recent ALICE results on the $R$)-dependence of the jet nuclear modification factor reveal that many model predictions -- often based on Monte Carlo event generators -- fail to accurately capture the observed $R$-dependence. This highlights the critical importance of theoretical frameworks that incorporate the correct underlying physics. While these frameworks may not yet achieve the precision of experimental data, attaining semi-quantitative agreement is a vital validation step and provides guidance for future higher-order calculations. 

In addition to the developments reported here, several noteworthy theoretical milestones have recently been achieved. These include investigations of jet quenching in flowing and inhomogeneous media~\cite{Andres:2022ndd,Sadofyev:2021ohn,Barata:2022krd,Barata:2023qds,Kuzmin:2023hko}, as well as extensive studies of jet substructure -- both analytic and Monte Carlo based -- that have revealed striking qualitative features of medium-modified jets~\cite{Chien:2016led,Mehtar-Tani:2016aco,Caucal:2021bae,Caucal:2019uvr,Barata:2023bhh} and have been confronted with experimental measurements~\cite{ALICE:2021aqk,ALICE:2022hyz,Song:2023sxb}. Promising new directions are also emerging, such as proposals to identify medium-induced hard processes through observables like the enhancement of charm production inside jets in heavy-ion relative to proton–proton collisions~\cite{Attems:2022otp,Attems:2022ubu}. Another particularly active line of research is the use of energy–energy correlators (EECs) as precision probes of QGP dynamics~\cite{Andres:2022ovj,Andres:2023ymw,Andres:2023xwr,Andres:2024xvk,Barata:2025fzd,Singh:2024vwb,Barata:2023bhh}, alongside the rapidly growing body of work on medium response and jet–medium interactions~\cite{Li:2010ts,Casalderrey-Solana:2016jvj,KunnawalkamElayavalli:2017hxo,Milhano:2017nzm,Tachibana:2017syd,Yang:2021qtl}. Together, these developments define vibrant research frontiers that complement the framework discussed in this review and deserve dedicated surveys in their own right.

In summary, advancing the field requires the development of precision theoretical tools to enable systematic analytic computations of jet observables in heavy-ion collisions, particularly in the perturbative sector. As alluded to above, a promising direction lies in the use of Effective Field Theory (EFT) approaches, which facilitate a robust separation of perturbative and non-perturbative contributions through novel factorization formulas. A significant step in this direction was recently achieved in~\cite{Mehtar-Tani:2024smp,Vaidya:2020cyi,Mehtar-Tani:2025xxd}. 

Additionally, we anticipate that a new generation of Monte Carlo (MC) event generators will incorporate these theoretical advancements. The aim is to reduce the wide variability in current modeling practices and enhance the reliability and predictive power of MC event generators.

Finally, an important avenue for future research is the application of the approach outlined in this report to substructure observables, such as angularities, energy-energy correlators, and groomed jet observables. These studies have the potential to provide valuable insights into the uncertainties in the theoretical description of various sectors, particularly the non-perturbative effects dominated by the soft particle background. This background has long been a major bottleneck in achieving a comprehensive understanding of the mechanisms governing jet quenching physics.

%% file: jqt-phys-rep.bbl
\providecommand{\href}[2]{#2}\begingroup\raggedright\begin{thebibliography}{100}

\bibitem{Gross:1973id}
D.~J. Gross and F.~Wilczek, \emph{Ultraviolet behavior of non-abelian gauge
  theories}, \href{http://dx.doi.org/10.1103/PhysRevLett.30.1343}{\emph{Phys.
  Rev. Lett.} {\bfseries 30} (1973) 1343--1346}.

\bibitem{Politzer:1973fx}
H.~D. Politzer, \emph{Reliable perturbative results for strong interactions?},
  \href{http://dx.doi.org/10.1103/PhysRevLett.30.1346}{\emph{Phys. Rev. Lett.}
  {\bfseries 30} (1973) 1346--1349}.

\bibitem{Fritzsch:1973pi}
H.~Fritzsch, M.~Gell-Mann and H.~Leutwyler, \emph{Advantages of the color octet
  gluon picture},
  \href{http://dx.doi.org/10.1016/0370-2693(73)90625-4}{\emph{Phys. Lett. B}
  {\bfseries 47} (1973) 365--368}.

\bibitem{Brandelik:1979bd}
R.~Brandelik and others (TASSO~Collaboration), \emph{Evidence for planar events
  in e+ e- annihilation at high energies},
  \href{http://dx.doi.org/10.1016/0370-2693(79)90861-6}{\emph{Phys. Lett. B}
  {\bfseries 86} (1979) 243--249}.

\bibitem{Sterman:1977wj}
G.~F. Sterman and S.~Weinberg, \emph{Jets from quantum chromodynamics},
  \href{http://dx.doi.org/10.1103/PhysRevLett.39.1436}{\emph{Phys. Rev. Lett.}
  {\bfseries 39} (1977) 1436--1439}.

\bibitem{ATLAS:2012yve}
G.~Aad and others (ATLAS~Collaboration), \emph{Observation of a new particle in
  the search for the standard model higgs boson with the atlas detector at the
  lhc}, \href{http://dx.doi.org/10.1016/j.physletb.2012.08.020}{\emph{Phys.
  Lett. B} {\bfseries 716} (2012) 1--29}.

\bibitem{CMS:2012qbp}
{\scshape CMS} collaboration, S.~Chatrchyan et~al., \emph{{Observation of a New
  Boson at a Mass of 125 GeV with the CMS Experiment at the LHC}},
  \href{http://dx.doi.org/10.1016/j.physletb.2012.08.021}{\emph{Phys. Lett. B}
  {\bfseries 716} (2012) 30--61},
  [\href{https://arxiv.org/abs/1207.7235}{{\ttfamily 1207.7235}}].

\bibitem{Gribov:1972ri}
V.~N. Gribov and L.~N. Lipatov, \emph{Deep inelastic e p scattering in
  perturbation theory}, {\emph{Sov. J. Nucl. Phys.} {\bfseries 15} (1972)
  438--450}.

\bibitem{Altarelli:1977zs}
G.~Altarelli and G.~Parisi, \emph{{Asymptotic Freedom in Parton Language}},
  \href{http://dx.doi.org/10.1016/0550-3213(77)90384-4}{\emph{Nucl. Phys. B}
  {\bfseries 126} (1977) 298--318}.

\bibitem{Dokshitzer:1977sg}
Y.~L. Dokshitzer, \emph{Calculation of the structure functions for deep
  inelastic scattering and e+ e- annihilation by perturbation theory in quantum
  chromodynamics}, {\emph{Sov. Phys. JETP} {\bfseries 46} (1977) 641--653}.

\bibitem{Aoki:2006we}
Y.~Aoki et~al., \emph{The order of the quantum chromodynamics transition
  predicted by the standard model of particle physics},
  \href{http://dx.doi.org/10.1038/nature05120}{\emph{Nature} {\bfseries 443}
  (2006) 675--678}.

\bibitem{Bazavov:2014pvz}
A.~Bazavov and others (HotQCD~Collaboration), \emph{Equation of state in
  (2+1)-flavor qcd},
  \href{http://dx.doi.org/10.1103/PhysRevD.90.094503}{\emph{Phys. Rev. D}
  {\bfseries 90} (2014) 094503}.

\bibitem{Kuti:1980gh}
J.~Kuti, J.~Polonyi and K.~Szlachanyi, \emph{Monte carlo study of su(2) gauge
  theory at finite temperature},
  \href{http://dx.doi.org/10.1016/0370-2693(81)90982-8}{\emph{Phys. Lett. B}
  {\bfseries 98} (1981) 199--204}.

\bibitem{Bjorken:1982tu}
J.~D. Bjorken, \emph{{Energy Loss of Energetic Partons in Quark - Gluon Plasma:
  Possible Extinction of High p(t) Jets in Hadron - Hadron Collisions}}, .

\bibitem{Arsene:2004fa}
I.~Arsene and others (BRAHMS~Collaboration), \emph{Quark-gluon plasma and color
  glass condensate at rhic? the perspective from the brahms experiment},
  \href{http://dx.doi.org/10.1016/j.nuclphysa.2005.02.130}{\emph{Nucl. Phys. A}
  {\bfseries 757} (2005) 1--27}.

\bibitem{Back:2004je}
B.~B. Back and others (PHOBOS~Collaboration), \emph{The phobos perspective on
  discoveries at rhic},
  \href{http://dx.doi.org/10.1016/j.nuclphysa.2005.03.084}{\emph{Nucl. Phys. A}
  {\bfseries 757} (2005) 28--101}.

\bibitem{Adams:2005dq}
J.~Adams and others (STAR~Collaboration), \emph{Experimental and theoretical
  challenges in the search for the quark-gluon plasma: The star collaboration's
  critical assessment of the evidence from rhic collisions},
  \href{http://dx.doi.org/10.1016/j.nuclphysa.2005.03.085}{\emph{Nucl. Phys. A}
  {\bfseries 757} (2005) 102--183}.

\bibitem{Adcox:2004mh}
K.~Adcox and others (PHENIX~Collaboration), \emph{Formation of dense partonic
  matter in relativistic nucleus-nucleus collisions at rhic: Experimental
  evaluation by the phenix collaboration},
  \href{http://dx.doi.org/10.1016/j.nuclphysa.2005.03.086}{\emph{Nucl. Phys. A}
  {\bfseries 757} (2005) 184--283}.

\bibitem{Shuryak:2004cy}
E.~Shuryak, \emph{{What RHIC experiments and theory tell us about properties of
  quark-gluon plasma?}},
  \href{http://dx.doi.org/10.1016/j.nuclphysa.2004.10.022}{\emph{Nucl. Phys. A}
  {\bfseries 750} (2005) 64--83},
  [\href{https://arxiv.org/abs/hep-ph/0405066}{{\ttfamily hep-ph/0405066}}].

\bibitem{Chatrchyan:2011sx}
S.~Chatrchyan and others (CMS~Collaboration), \emph{Observation and studies of
  jet quenching in pbpb collisions at $\sqrt{s_{NN}} = 2.76$ tev},
  \href{http://dx.doi.org/10.1103/PhysRevC.84.024906}{\emph{Phys. Rev. C}
  {\bfseries 84} (2011) 024906}.

\bibitem{Aad:2010bu}
G.~Aad and others (ATLAS~Collaboration), \emph{Observation of a
  centrality-dependent dijet asymmetry in lead-lead collisions at
  $\sqrt{s_{NN}} = 2.76$ tev with the atlas detector at the lhc},
  \href{http://dx.doi.org/10.1103/PhysRevLett.105.252303}{\emph{Phys. Rev.
  Lett.} {\bfseries 105} (2010) 252303}.

\bibitem{CMS:2021vui}
{\scshape CMS} collaboration, A.~M. Sirunyan et~al., \emph{{First measurement
  of large area jet transverse momentum spectra in heavy-ion collisions}},
  \href{http://dx.doi.org/10.1007/JHEP05(2021)284}{\emph{JHEP} {\bfseries 05}
  (2021) 284}, [\href{https://arxiv.org/abs/2102.13080}{{\ttfamily
  2102.13080}}].

\bibitem{ALICE:2023waz}
{\scshape ALICE} collaboration, S.~Acharya et~al., \emph{{Measurement of the
  radius dependence of charged-particle jet suppression in Pb\textendash{}Pb
  collisions at sNN=5.02TeV}},
  \href{http://dx.doi.org/10.1016/j.physletb.2023.138412}{\emph{Phys. Lett. B}
  {\bfseries 849} (2024) 138412},
  [\href{https://arxiv.org/abs/2303.00592}{{\ttfamily 2303.00592}}].

\bibitem{Connors:2017ptx}
M.~Connors, C.~Nattrass, R.~Reed and S.~Salur, \emph{Jet measurements in heavy
  ion physics},
  \href{http://dx.doi.org/10.1103/RevModPhys.90.025005}{\emph{Rev. Mod. Phys.}
  {\bfseries 90} (2018) 025005}.

\bibitem{Mehtar-Tani:2021fud}
Y.~Mehtar-Tani, D.~Pablos and K.~Tywoniuk, \emph{{Cone-Size Dependence of Jet
  Suppression in Heavy-Ion Collisions}},
  \href{http://dx.doi.org/10.1103/PhysRevLett.127.252301}{\emph{Phys. Rev.
  Lett.} {\bfseries 127} (2021) 252301},
  [\href{https://arxiv.org/abs/2101.01742}{{\ttfamily 2101.01742}}].

\bibitem{Larkoski:2014wba}
A.~J. Larkoski, S.~Marzani, G.~Soyez and J.~Thaler, \emph{{Soft Drop}},
  \href{http://dx.doi.org/10.1007/JHEP05(2014)146}{\emph{JHEP} {\bfseries 05}
  (2014) 146}, [\href{https://arxiv.org/abs/1402.2657}{{\ttfamily 1402.2657}}].

\bibitem{CMS:2017qlm}
{\scshape CMS} collaboration, A.~M. Sirunyan et~al., \emph{{Measurement of the
  Splitting Function in $pp$ and Pb-Pb Collisions at $\sqrt{s_{_{\mathrm{NN}}}}
  =$ 5.02 TeV}},
  \href{http://dx.doi.org/10.1103/PhysRevLett.120.142302}{\emph{Phys. Rev.
  Lett.} {\bfseries 120} (2018) 142302},
  [\href{https://arxiv.org/abs/1708.09429}{{\ttfamily 1708.09429}}].

\bibitem{Chien:2016led}
Y.-T. Chien and I.~Vitev, \emph{{Probing the Hardest Branching within Jets in
  Heavy-Ion Collisions}},
  \href{http://dx.doi.org/10.1103/PhysRevLett.119.112301}{\emph{Phys. Rev.
  Lett.} {\bfseries 119} (2017) 112301},
  [\href{https://arxiv.org/abs/1608.07283}{{\ttfamily 1608.07283}}].

\bibitem{Mehtar-Tani:2016aco}
Y.~Mehtar-Tani and K.~Tywoniuk, \emph{{Groomed jets in heavy-ion collisions:
  sensitivity to medium-induced bremsstrahlung}},
  \href{http://dx.doi.org/10.1007/JHEP04(2017)125}{\emph{JHEP} {\bfseries 04}
  (2017) 125}, [\href{https://arxiv.org/abs/1610.08930}{{\ttfamily
  1610.08930}}].

\bibitem{Caucal:2021bae}
P.~Caucal, A.~Soto-Ontoso and A.~Takacs, \emph{{Dynamical Grooming meets LHC
  data}}, \href{http://dx.doi.org/10.1007/JHEP07(2021)020}{\emph{JHEP}
  {\bfseries 07} (2021) 020},
  [\href{https://arxiv.org/abs/2103.06566}{{\ttfamily 2103.06566}}].

\bibitem{Caucal:2019uvr}
P.~Caucal, E.~Iancu and G.~Soyez, \emph{{Deciphering the $z_g$ distribution in
  ultrarelativistic heavy ion collisions}},
  \href{http://dx.doi.org/10.1007/JHEP10(2019)273}{\emph{JHEP} {\bfseries 10}
  (2019) 273}, [\href{https://arxiv.org/abs/1907.04866}{{\ttfamily
  1907.04866}}].

\bibitem{Barata:2023bhh}
J.~a. Barata, P.~Caucal, A.~Soto-Ontoso and R.~Szafron, \emph{{Advancing the
  understanding of energy-energy correlators in heavy-ion collisions}},
  \href{https://arxiv.org/abs/2312.12527}{{\ttfamily 2312.12527}}.

\bibitem{ALICE:2021aqk}
{\scshape ALICE} collaboration, S.~Acharya et~al., \emph{{Direct observation of
  the dead-cone effect in quantum chromodynamics}},
  \href{http://dx.doi.org/10.1038/s41586-022-04572-w}{\emph{Nature} {\bfseries
  605} (2022) 440--446}, [\href{https://arxiv.org/abs/2106.05713}{{\ttfamily
  2106.05713}}].

\bibitem{ALICE:2022hyz}
{\scshape ALICE} collaboration, S.~Acharya et~al., \emph{{Measurements of the
  groomed jet radius and momentum splitting fraction with the soft drop and
  dynamical grooming algorithms in pp collisions at $ \sqrt{s} $ = 5.02 TeV}},
  \href{http://dx.doi.org/10.1007/JHEP05(2023)244}{\emph{JHEP} {\bfseries 05}
  (2023) 244}, [\href{https://arxiv.org/abs/2204.10246}{{\ttfamily
  2204.10246}}].

\bibitem{Song:2023sxb}
{\scshape STAR} collaboration, Y.~Song, \emph{{Measurement of CollinearDrop jet
  mass and its correlation with SoftDrop groomed jet substructure observables
  in $\sqrt{s}=200$ GeV $pp$ collisions by STAR}},
  \href{https://arxiv.org/abs/2307.07718}{{\ttfamily 2307.07718}}.

\bibitem{Braaten:1991we}
E.~Braaten and M.~H. Thoma, \emph{Energy loss of a heavy quark in the
  quark-gluon plasma},
  \href{http://dx.doi.org/10.1103/PhysRevD.44.2625}{\emph{Phys. Rev. D}
  {\bfseries 44} (1991) 2625--2630}.

\bibitem{Gyulassy:1994ew}
M.~Gyulassy, X.-N. Wang and M.~Plümer, \emph{Radiative energy loss and p(t)
  broadening of high energy partons in quark gluon plasma},
  \href{http://dx.doi.org/10.1103/PhysRevD.51.3236}{\emph{Phys. Rev. D}
  {\bfseries 51} (1995) 3236--3240}.

\bibitem{Baier:1994bd}
R.~Baier, Y.~L. Dokshitzer, S.~Peigne and D.~Schiff, \emph{{Induced gluon
  radiation in a QCD medium}},
  \href{http://dx.doi.org/10.1016/0370-2693(94)01617-L}{\emph{Phys. Lett. B}
  {\bfseries 345} (1995) 277--286},
  [\href{https://arxiv.org/abs/hep-ph/9411409}{{\ttfamily hep-ph/9411409}}].

\bibitem{Baier:1996kr}
R.~Baier, Y.~L. Dokshitzer, A.~H. Mueller, S.~Peigne and D.~Schiff,
  \emph{{Radiative energy loss of high-energy quarks and gluons in a finite
  volume quark - gluon plasma}},
  \href{http://dx.doi.org/10.1016/S0550-3213(96)00553-6}{\emph{Nucl. Phys. B}
  {\bfseries 483} (1997) 291--320},
  [\href{https://arxiv.org/abs/hep-ph/9607355}{{\ttfamily hep-ph/9607355}}].

\bibitem{Wang:1991xy}
X.-N. Wang and M.~Gyulassy, \emph{{Gluon shadowing and jet quenching in A + A
  collisions at s**(1/2) = 200-GeV}},
  \href{http://dx.doi.org/10.1103/PhysRevLett.68.1480}{\emph{Phys. Rev. Lett.}
  {\bfseries 68} (1992) 1480--1483}.

\bibitem{Gyulassy:1990ye}
M.~Gyulassy and M.~Plumer, \emph{{Jet Quenching in Dense Matter}},
  \href{http://dx.doi.org/10.1016/0370-2693(90)91409-5}{\emph{Phys. Lett. B}
  {\bfseries 243} (1990) 432--438}.

\bibitem{Wang:1992qdg}
X.-N. Wang and M.~Gyulassy, \emph{{Gluon shadowing and jet quenching in A + A
  collisions at s**(1/2) = 200-GeV}},
  \href{http://dx.doi.org/10.1103/PhysRevLett.68.1480}{\emph{Phys. Rev. Lett.}
  {\bfseries 68} (1992) 1480--1483}.

\bibitem{Gyulassy:2000fs}
M.~Gyulassy, P.~Levai and I.~Vitev, \emph{{NonAbelian energy loss at finite
  opacity}}, \href{http://dx.doi.org/10.1103/PhysRevLett.85.5535}{\emph{Phys.
  Rev. Lett.} {\bfseries 85} (2000) 5535--5538},
  [\href{https://arxiv.org/abs/nucl-th/0005032}{{\ttfamily nucl-th/0005032}}].

\bibitem{Zakharov:1996fv}
B.~G. Zakharov, \emph{{Fully quantum treatment of the Landau-Pomeranchuk-Migdal
  effect in QED and QCD}}, \href{http://dx.doi.org/10.1134/1.567126}{\emph{JETP
  Lett.} {\bfseries 63} (1996) 952--957},
  [\href{https://arxiv.org/abs/hep-ph/9607440}{{\ttfamily hep-ph/9607440}}].

\bibitem{Zakharov:1997uu}
B.~G. Zakharov, \emph{{Radiative energy loss of high-energy quarks in finite
  size nuclear matter and quark - gluon plasma}},
  \href{http://dx.doi.org/10.1134/1.567389}{\emph{JETP Lett.} {\bfseries 65}
  (1997) 615--620}, [\href{https://arxiv.org/abs/hep-ph/9704255}{{\ttfamily
  hep-ph/9704255}}].

\bibitem{Gyulassy:2000er}
M.~Gyulassy, P.~Levai and I.~Vitev, \emph{{Reaction operator approach to
  nonAbelian energy loss}},
  \href{http://dx.doi.org/10.1016/S0550-3213(00)00652-0}{\emph{Nucl. Phys. B}
  {\bfseries 594} (2001) 371--419},
  [\href{https://arxiv.org/abs/nucl-th/0006010}{{\ttfamily nucl-th/0006010}}].

\bibitem{Wiedemann:1999fq}
U.~A. Wiedemann and M.~Gyulassy, \emph{{Transverse momentum dependence of the
  Landau-Pomeranchuk-Migdal effect}},
  \href{http://dx.doi.org/10.1016/S0550-3213(99)00458-7}{\emph{Nucl. Phys. B}
  {\bfseries 560} (1999) 345--382},
  [\href{https://arxiv.org/abs/hep-ph/9906257}{{\ttfamily hep-ph/9906257}}].

\bibitem{Baier:2001yt}
R.~Baier, Y.~L. Dokshitzer, A.~H. Mueller and D.~Schiff, \emph{{Quenching of
  hadron spectra in media}},
  \href{http://dx.doi.org/10.1088/1126-6708/2001/09/033}{\emph{JHEP} {\bfseries
  09} (2001) 033}, [\href{https://arxiv.org/abs/hep-ph/0106347}{{\ttfamily
  hep-ph/0106347}}].

\bibitem{Wiedemann:2000za}
U.~A. Wiedemann, \emph{{Gluon radiation off hard quarks in a nuclear
  environment: Opacity expansion}},
  \href{http://dx.doi.org/10.1016/S0550-3213(00)00457-0}{\emph{Nucl. Phys. B}
  {\bfseries 588} (2000) 303--344},
  [\href{https://arxiv.org/abs/hep-ph/0005129}{{\ttfamily hep-ph/0005129}}].

\bibitem{Arnold:2002ja}
P.~B. Arnold, G.~D. Moore and L.~G. Yaffe, \emph{{Photon and gluon emission in
  relativistic plasmas}},
  \href{http://dx.doi.org/10.1088/1126-6708/2002/06/030}{\emph{JHEP} {\bfseries
  06} (2002) 030}, [\href{https://arxiv.org/abs/hep-ph/0204343}{{\ttfamily
  hep-ph/0204343}}].

\bibitem{Salgado:2003gb}
C.~A. Salgado and U.~A. Wiedemann, \emph{{Calculating quenching weights}},
  \href{http://dx.doi.org/10.1103/PhysRevD.68.014008}{\emph{Phys. Rev. D}
  {\bfseries 68} (2003) 014008},
  [\href{https://arxiv.org/abs/hep-ph/0302184}{{\ttfamily hep-ph/0302184}}].

\bibitem{Landau:1953gr}
L.~D. Landau and I.~Pomeranchuk, \emph{{Electron cascade process at very
  high-energies}}, {\emph{Dokl. Akad. Nauk Ser. Fiz.} {\bfseries 92} (1953)
  735--738}.

\bibitem{Migdal:1956tc}
A.~B. Migdal, \emph{{Bremsstrahlung and pair production in condensed media at
  high-energies}},
  \href{http://dx.doi.org/10.1103/PhysRev.103.1811}{\emph{Phys. Rev.}
  {\bfseries 103} (1956) 1811--1820}.

\bibitem{Zapp:2013vla}
K.~C. Zapp, \emph{Jewel 2.0.0 - directions for use},
  \href{http://dx.doi.org/10.1140/epjc/s10052-014-2762-1}{\emph{Eur. Phys. J.
  C} {\bfseries 74} (2014) 2762}.

\bibitem{Lokhtin:2008xi}
I.~P. Lokhtin, L.~V. Malinina, S.~V. Petrushanko, A.~M. Snigirev, I.~Arsene and
  K.~Tywoniuk, \emph{Heavy ion event generator hydjet++ (hydrodynamics plus
  jets)}, \href{http://dx.doi.org/10.1016/j.cpc.2008.11.015}{\emph{Comput.
  Phys. Commun.} {\bfseries 180} (2009) 779--799}.

\bibitem{Caucal:2018dla}
P.~Caucal, E.~Iancu, A.~H. Mueller and G.~Soyez, \emph{{Vacuum-like jet
  fragmentation in a dense QCD medium}},
  \href{http://dx.doi.org/10.1103/PhysRevLett.120.232001}{\emph{Phys. Rev.
  Lett.} {\bfseries 120} (2018) 232001},
  [\href{https://arxiv.org/abs/1801.09703}{{\ttfamily 1801.09703}}].

\bibitem{Apolinario:2019amp}
L.~Apolinário, \emph{Jets in qcd matter: Monte carlo approaches},
  \href{http://dx.doi.org/10.1051/epjconf/201919902002}{\emph{EPJ Web Conf.}
  {\bfseries 199} (2019) 02002}.

\bibitem{JETSCAPE:2017eso}
S.~Cao and others (JETSCAPE~Collaboration), \emph{Multistage monte carlo
  simulation of jet modification in a static medium},
  \href{http://dx.doi.org/10.1103/PhysRevC.96.024909}{\emph{Phys. Rev. C}
  {\bfseries 96} (2017) 024909},
  [\href{https://arxiv.org/abs/1705.00050}{{\ttfamily 1705.00050}}].

\bibitem{He:2015pra}
Y.~He, T.~Luo, X.-N. Wang and Y.~Zhu, \emph{Linear boltzmann transport for jet
  propagation in the quark-gluon plasma: Elastic processes and medium recoil},
  \href{http://dx.doi.org/10.1103/PhysRevC.91.054908}{\emph{Phys. Rev. C}
  {\bfseries 91} (2015) 054908},
  [\href{https://arxiv.org/abs/1503.03313}{{\ttfamily 1503.03313}}].

\bibitem{Cao:2017zih}
S.~Cao and A.~Majumder, \emph{Nuclear modification of single hadron and jet
  production within a virtuality-ordered (matter) parton shower},
  \href{https://arxiv.org/abs/1712.10055}{{\ttfamily 1712.10055}}.

\bibitem{Cao:2016gvr}
S.~Cao, T.~Luo, G.-Y. Qin and X.-N. Wang, \emph{Linearized boltzmann transport
  model for jet propagation in the quark-gluon plasma: Heavy quark evolution},
  \href{http://dx.doi.org/10.1103/PhysRevC.94.014909}{\emph{Phys. Rev. C}
  {\bfseries 94} (2016) 014909},
  [\href{https://arxiv.org/abs/1605.06447}{{\ttfamily 1605.06447}}].

\bibitem{Park:2018fuo}
C.~Park, S.~Jeon and C.~Gale, \emph{Jet modification with medium recoil in
  quark-gluon plasma},  \href{https://arxiv.org/abs/1807.06550}{{\ttfamily
  1807.06550}}.

\bibitem{Schenke:2009gb}
B.~Schenke, C.~Gale and S.~Jeon, \emph{Martini: An event generator for
  relativistic heavy-ion collisions},
  \href{http://dx.doi.org/10.1103/PhysRevC.80.054913}{\emph{Phys. Rev. C}
  {\bfseries 80} (2009) 054913},
  [\href{https://arxiv.org/abs/0909.2037}{{\ttfamily 0909.2037}}].

\bibitem{Armesto:2009fj}
N.~Armesto, L.~Cunqueiro and C.~A. Salgado, \emph{Q-pythia: A medium-modified
  implementation of final state radiation},
  \href{http://dx.doi.org/10.1140/epjc/s10052-009-1133-9}{\emph{Eur. Phys. J.
  C} {\bfseries 63} (2009) 679--690},
  [\href{https://arxiv.org/abs/0907.1014}{{\ttfamily 0907.1014}}].

\bibitem{Gossiaux:2018}
P.~B. Gossiaux, J.~Aichelin, M.~Bluhm, T.~Gousset, R.~Katz, V.~Marin et~al.,
  \emph{A modified-boltzmann approach for modeling the hot qcd medium},
  \href{http://dx.doi.org/10.1103/PhysRevC.100.064911}{\emph{Phys. Rev. C}
  {\bfseries 100} (2019) 064911}.

\bibitem{Catani:1991hj}
S.~Catani, Y.~L. Dokshitzer, M.~H. Seymour and B.~R. Webber,
  \emph{Longitudinally invariant $k_t$ clustering algorithms for hadron hadron
  collisions},
  \href{http://dx.doi.org/10.1016/0550-3213(93)90166-M}{\emph{Nucl. Phys. B}
  {\bfseries 406} (1993) 187--224},
  [\href{https://arxiv.org/abs/hep-ph/9302288}{{\ttfamily hep-ph/9302288}}].

\bibitem{Dokshitzer:1997in}
Y.~L. Dokshitzer, G.~D. Leder, S.~Moretti and B.~R. Webber, \emph{Better jet
  clustering algorithms},
  \href{http://dx.doi.org/10.1088/1126-6708/1997/08/001}{\emph{JHEP} {\bfseries
  08} (1997) 001}, [\href{https://arxiv.org/abs/hep-ph/9707323}{{\ttfamily
  hep-ph/9707323}}].

\bibitem{Cacciari:2008gp}
M.~Cacciari, G.~P. Salam and G.~Soyez, \emph{The anti-$k_t$ jet clustering
  algorithm},
  \href{http://dx.doi.org/10.1088/1126-6708/2008/04/063}{\emph{JHEP} {\bfseries
  04} (2008) 063}, [\href{https://arxiv.org/abs/0802.1189}{{\ttfamily
  0802.1189}}].

\bibitem{Mehtar-Tani:2011hma}
Y.~Mehtar-Tani, C.~A. Salgado and K.~Tywoniuk, \emph{{Jets in QCD Media: From
  Color Coherence to Decoherence}},
  \href{http://dx.doi.org/10.1016/j.physletb.2011.12.042}{\emph{Phys. Lett. B}
  {\bfseries 707} (2012) 156--159},
  [\href{https://arxiv.org/abs/1102.4317}{{\ttfamily 1102.4317}}].

\bibitem{Mehtar-Tani:2012mfa}
Y.~Mehtar-Tani, C.~A. Salgado and K.~Tywoniuk, \emph{{The Radiation pattern of
  a QCD antenna in a dense medium}},
  \href{http://dx.doi.org/10.1007/JHEP10(2012)197}{\emph{JHEP} {\bfseries 10}
  (2012) 197}, [\href{https://arxiv.org/abs/1205.5739}{{\ttfamily 1205.5739}}].

\bibitem{Mehtar-Tani:2017ypq}
Y.~Mehtar-Tani and K.~Tywoniuk, \emph{{Radiative energy loss of neighboring
  subjets}},
  \href{http://dx.doi.org/10.1016/j.nuclphysa.2018.09.041}{\emph{Nucl. Phys. A}
  {\bfseries 979} (2018) 165--203},
  [\href{https://arxiv.org/abs/1706.06047}{{\ttfamily 1706.06047}}].

\bibitem{Casalderrey-Solana:2011ule}
J.~Casalderrey-Solana and E.~Iancu, \emph{{Interference effects in
  medium-induced gluon radiation}},
  \href{http://dx.doi.org/10.1007/JHEP08(2011)015}{\emph{JHEP} {\bfseries 08}
  (2011) 015}, [\href{https://arxiv.org/abs/1105.1760}{{\ttfamily 1105.1760}}].

\bibitem{Kovner:2003zj}
A.~Kovner and U.~A. Wiedemann, \emph{{Gluon radiation and parton energy loss}},
   \href{https://arxiv.org/abs/hep-ph/0304151}{{\ttfamily hep-ph/0304151}}.

\bibitem{Balitsky:2001gj}
I.~Balitsky, \emph{{High-energy QCD and Wilson lines}},
  \href{https://arxiv.org/abs/hep-ph/0101042}{{\ttfamily hep-ph/0101042}}.

\bibitem{Blaizot:2012fh}
J.-P. Blaizot, F.~Dominguez, E.~Iancu and Y.~Mehtar-Tani, \emph{{Medium-induced
  gluon branching}},
  \href{http://dx.doi.org/10.1007/JHEP01(2013)143}{\emph{JHEP} {\bfseries 01}
  (2013) 143}, [\href{https://arxiv.org/abs/1209.4585}{{\ttfamily 1209.4585}}].

\bibitem{Blaizot:2013vha}
J.-P. Blaizot, F.~Dominguez, E.~Iancu and Y.~Mehtar-Tani, \emph{{Probabilistic
  picture for medium-induced jet evolution}},
  \href{http://dx.doi.org/10.1007/JHEP06(2014)075}{\emph{JHEP} {\bfseries 06}
  (2014) 075}, [\href{https://arxiv.org/abs/1311.5823}{{\ttfamily 1311.5823}}].

\bibitem{Blaizot:2013hx}
J.-P. Blaizot, E.~Iancu and Y.~Mehtar-Tani, \emph{{Medium-induced QCD cascade:
  democratic branching and wave turbulence}},
  \href{http://dx.doi.org/10.1103/PhysRevLett.111.052001}{\emph{Phys. Rev.
  Lett.} {\bfseries 111} (2013) 052001},
  [\href{https://arxiv.org/abs/1301.6102}{{\ttfamily 1301.6102}}].

\bibitem{Blaizot:2014rla}
J.-P. Blaizot, L.~Fister and Y.~Mehtar-Tani, \emph{{Angular distribution of
  medium-induced QCD cascades}},
  \href{http://dx.doi.org/10.1016/j.nuclphysa.2015.03.014}{\emph{Nucl. Phys. A}
  {\bfseries 940} (2015) 67--88},
  [\href{https://arxiv.org/abs/1409.6202}{{\ttfamily 1409.6202}}].

\bibitem{Arnold:2023qwi}
P.~Arnold, O.~Elgedawy and S.~Iqbal, \emph{{Landau-Pomeranchuk-Migdal effect in
  sequential bremsstrahlung: Gluon shower development}},
  \href{http://dx.doi.org/10.1103/PhysRevD.108.074015}{\emph{Phys. Rev. D}
  {\bfseries 108} (2023) 074015},
  [\href{https://arxiv.org/abs/2302.10215}{{\ttfamily 2302.10215}}].

\bibitem{Blaizot:2014ula}
J.-P. Blaizot, Y.~Mehtar-Tani and M.~A.~C. Torres, \emph{{Angular structure of
  the in-medium QCD cascade}},
  \href{http://dx.doi.org/10.1103/PhysRevLett.114.222002}{\emph{Phys. Rev.
  Lett.} {\bfseries 114} (2015) 222002},
  [\href{https://arxiv.org/abs/1407.0326}{{\ttfamily 1407.0326}}].

\bibitem{Mehtar-Tani:2018zba}
Y.~Mehtar-Tani and S.~Schlichting, \emph{{Universal quark to gluon ratio in
  medium-induced parton cascade}},
  \href{http://dx.doi.org/10.1007/JHEP09(2018)144}{\emph{JHEP} {\bfseries 09}
  (2018) 144}, [\href{https://arxiv.org/abs/1807.06181}{{\ttfamily
  1807.06181}}].

\bibitem{Schlichting:2020lef}
S.~Schlichting and I.~Soudi, \emph{{Medium-induced fragmentation and
  equilibration of highly energetic partons}},
  \href{http://dx.doi.org/10.1007/JHEP07(2021)077}{\emph{JHEP} {\bfseries 07}
  (2021) 077}, [\href{https://arxiv.org/abs/2008.04928}{{\ttfamily
  2008.04928}}].

\bibitem{Mehtar-Tani:2022zwf}
Y.~Mehtar-Tani, S.~Schlichting and I.~Soudi, \emph{{Jet thermalization in QCD
  kinetic theory}},
  \href{http://dx.doi.org/10.1007/JHEP05(2023)091}{\emph{JHEP} {\bfseries 05}
  (2023) 091}, [\href{https://arxiv.org/abs/2209.10569}{{\ttfamily
  2209.10569}}].

\bibitem{Soudi:2024yfy}
I.~Soudi, \emph{{Medium-induced parton splitting in expanding QCD matter}},
  \href{https://arxiv.org/abs/2409.04806}{{\ttfamily 2409.04806}}.

\bibitem{Arnold:2020uzm}
P.~Arnold, T.~Gorda and S.~Iqbal, \emph{{The LPM effect in sequential
  bremsstrahlung: nearly complete results for QCD}},
  \href{http://dx.doi.org/10.1007/JHEP11(2020)053}{\emph{JHEP} {\bfseries 11}
  (2020) 053}, [\href{https://arxiv.org/abs/2007.15018}{{\ttfamily
  2007.15018}}].

\bibitem{Arnold:2021pin}
P.~Arnold, T.~Gorda and S.~Iqbal, \emph{{The LPM effect in sequential
  bremsstrahlung: analytic results for sub-leading (single) logarithms}},
  \href{http://dx.doi.org/10.1007/JHEP04(2022)085}{\emph{JHEP} {\bfseries 04}
  (2022) 085}, [\href{https://arxiv.org/abs/2112.05161}{{\ttfamily
  2112.05161}}].

\bibitem{Arnold:2022fku}
P.~Arnold, T.~Gorda and S.~Iqbal, \emph{{The LPM effect in sequential
  bremsstrahlung: incorporation of
  \textquotedblleft{}instantaneous\textquotedblright{} interactions for QCD}},
  \href{http://dx.doi.org/10.1007/JHEP11(2022)130}{\emph{JHEP} {\bfseries 11}
  (2022) 130}, [\href{https://arxiv.org/abs/2209.03971}{{\ttfamily
  2209.03971}}].

\bibitem{Arnold:2022mby}
P.~Arnold, O.~Elgedawy and S.~Iqbal, \emph{{Are Gluon Showers inside a
  Quark-Gluon Plasma Strongly Coupled? A Theorist\textquoteright{}s Test}},
  \href{http://dx.doi.org/10.1103/PhysRevLett.131.162302}{\emph{Phys. Rev.
  Lett.} {\bfseries 131} (2023) 162302},
  [\href{https://arxiv.org/abs/2212.08086}{{\ttfamily 2212.08086}}].

\bibitem{Arnold:2024whj}
P.~Arnold, O.~Elgedawy and S.~Iqbal, \emph{{Are in-medium quark-gluon showers
  strongly coupled? Results in the large-$N_f$ limit}},
  \href{https://arxiv.org/abs/2408.07129}{{\ttfamily 2408.07129}}.

\bibitem{Arnold:2024bph}
P.~Arnold, O.~Elgedawy and S.~Iqbal, \emph{{Strongly vs. weakly coupled
  in-medium showers: energy stopping in large-N$_{f}$ QED}},
  \href{http://dx.doi.org/10.1007/JHEP09(2024)131}{\emph{JHEP} {\bfseries 09}
  (2024) 131}, [\href{https://arxiv.org/abs/2404.19008}{{\ttfamily
  2404.19008}}].

\bibitem{Blaizot:2014bha}
J.-P. Blaizot and Y.~Mehtar-Tani, \emph{{Renormalization of the jet-quenching
  parameter}},
  \href{http://dx.doi.org/10.1016/j.nuclphysa.2014.05.018}{\emph{Nucl. Phys. A}
  {\bfseries 929} (2014) 202--229},
  [\href{https://arxiv.org/abs/1403.2323}{{\ttfamily 1403.2323}}].

\bibitem{Caucal:2022mpp}
P.~Caucal and Y.~Mehtar-Tani, \emph{{Transverse momentum broadening in large
  media from NLL BFKL to all orders in pQCD}},
  \href{http://dx.doi.org/10.1103/PhysRevD.108.014008}{\emph{Phys. Rev. D}
  {\bfseries 108} (2023) 014008},
  [\href{https://arxiv.org/abs/2209.08900}{{\ttfamily 2209.08900}}].

\bibitem{Ghiglieri:2022gyv}
J.~Ghiglieri and E.~Weitz, \emph{{Classical vs quantum corrections to jet
  broadening in a weakly-coupled Quark-Gluon Plasma}},
  \href{http://dx.doi.org/10.1007/JHEP11(2022)068}{\emph{JHEP} {\bfseries 11}
  (2022) 068}, [\href{https://arxiv.org/abs/2207.08842}{{\ttfamily
  2207.08842}}].

\bibitem{Caucal:2022fhc}
P.~Caucal and Y.~Mehtar-Tani, \emph{{Universality aspects of quantum
  corrections to transverse momentum broadening in QCD media}},
  \href{http://dx.doi.org/10.1007/JHEP09(2022)023}{\emph{JHEP} {\bfseries 09}
  (2022) 023}, [\href{https://arxiv.org/abs/2203.09407}{{\ttfamily
  2203.09407}}].

\bibitem{Arnold:2021mow}
P.~Arnold, \emph{{Universality (beyond leading log) of soft radiative
  corrections to $ \hat{q} $ in p$_{t}$ broadening and energy loss}},
  \href{http://dx.doi.org/10.1007/JHEP03(2022)134}{\emph{JHEP} {\bfseries 03}
  (2022) 134}, [\href{https://arxiv.org/abs/2111.05348}{{\ttfamily
  2111.05348}}].

\bibitem{Caucal:2021lgf}
P.~Caucal and Y.~Mehtar-Tani, \emph{{Anomalous diffusion in QCD matter}},
  \href{http://dx.doi.org/10.1103/PhysRevD.106.L051501}{\emph{Phys. Rev. D}
  {\bfseries 106} (2022) L051501},
  [\href{https://arxiv.org/abs/2109.12041}{{\ttfamily 2109.12041}}].

\bibitem{Blaizot:2019muz}
J.-P. Blaizot and F.~Dominguez, \emph{{Radiative corrections to the jet
  quenching parameter in dilute and dense media}},
  \href{http://dx.doi.org/10.1103/PhysRevD.99.054005}{\emph{Phys. Rev. D}
  {\bfseries 99} (2019) 054005},
  [\href{https://arxiv.org/abs/1901.01448}{{\ttfamily 1901.01448}}].

\bibitem{Iancu:2018trm}
E.~Iancu, P.~Taels and B.~Wu, \emph{{Jet quenching parameter in an expanding
  QCD plasma}},
  \href{http://dx.doi.org/10.1016/j.physletb.2018.10.007}{\emph{Phys. Lett. B}
  {\bfseries 786} (2018) 288--295},
  [\href{https://arxiv.org/abs/1806.07177}{{\ttfamily 1806.07177}}].

\bibitem{Wu:2014nca}
B.~Wu, \emph{{Radiative energy loss and radiative $p_{\bot}$-broadening of
  high-energy partons in QCD matter}},
  \href{http://dx.doi.org/10.1007/JHEP12(2014)081}{\emph{JHEP} {\bfseries 12}
  (2014) 081}, [\href{https://arxiv.org/abs/1408.5459}{{\ttfamily 1408.5459}}].

\bibitem{Iancu:2014sha}
E.~Iancu and D.~N. Triantafyllopoulos, \emph{{Running coupling effects in the
  evolution of jet quenching}},
  \href{http://dx.doi.org/10.1103/PhysRevD.90.074002}{\emph{Phys. Rev. D}
  {\bfseries 90} (2014) 074002},
  [\href{https://arxiv.org/abs/1405.3525}{{\ttfamily 1405.3525}}].

\bibitem{Liou:2013qya}
T.~Liou, A.~H. Mueller and B.~Wu, \emph{{Radiative $p_\bot$-broadening of
  high-energy quarks and gluons in QCD matter}},
  \href{http://dx.doi.org/10.1016/j.nuclphysa.2013.08.005}{\emph{Nucl. Phys. A}
  {\bfseries 916} (2013) 102--125},
  [\href{https://arxiv.org/abs/1304.7677}{{\ttfamily 1304.7677}}].

\bibitem{Iancu:2014kga}
E.~Iancu, \emph{{The non-linear evolution of jet quenching}},
  \href{http://dx.doi.org/10.1007/JHEP10(2014)095}{\emph{JHEP} {\bfseries 10}
  (2014) 095}, [\href{https://arxiv.org/abs/1403.1996}{{\ttfamily 1403.1996}}].

\bibitem{Mehtar-Tani:2024mvl}
Y.~Mehtar-Tani, \emph{{Non-linear dynamics of jet quenching}},
  \href{http://dx.doi.org/10.1007/JHEP04(2025)163}{\emph{JHEP} {\bfseries 04}
  (2025) 163}, [\href{https://arxiv.org/abs/2411.11992}{{\ttfamily
  2411.11992}}].

\bibitem{Banfi:2002hw}
A.~Banfi, G.~Marchesini and G.~Smye, \emph{{Away from jet energy flow}},
  \href{http://dx.doi.org/10.1088/1126-6708/2002/08/006}{\emph{JHEP} {\bfseries
  08} (2002) 006}, [\href{https://arxiv.org/abs/hep-ph/0206076}{{\ttfamily
  hep-ph/0206076}}].

\bibitem{Caucal:2021cfb}
P.~Caucal, A.~Soto-Ontoso and A.~Takacs, \emph{{Dynamically groomed jet radius
  in heavy-ion collisions}},
  \href{http://dx.doi.org/10.1103/PhysRevD.105.114046}{\emph{Phys. Rev. D}
  {\bfseries 105} (2022) 114046},
  [\href{https://arxiv.org/abs/2111.14768}{{\ttfamily 2111.14768}}].

\bibitem{Andres:2022ovj}
C.~Andres, F.~Dominguez, R.~Kunnawalkam~Elayavalli, J.~Holguin, C.~Marquet and
  I.~Moult, \emph{{Resolving the Scales of the Quark-Gluon Plasma with Energy
  Correlators}},
  \href{http://dx.doi.org/10.1103/PhysRevLett.130.262301}{\emph{Phys. Rev.
  Lett.} {\bfseries 130} (2023) 262301},
  [\href{https://arxiv.org/abs/2209.11236}{{\ttfamily 2209.11236}}].

\bibitem{Andres:2023ymw}
C.~Andres, F.~Dominguez, J.~Holguin, C.~Marquet and I.~Moult, \emph{{Seeing
  beauty in the quark-gluon plasma with energy correlators}},
  \href{http://dx.doi.org/10.1103/PhysRevD.110.L031503}{\emph{Phys. Rev. D}
  {\bfseries 110} (2024) L031503},
  [\href{https://arxiv.org/abs/2307.15110}{{\ttfamily 2307.15110}}].

\bibitem{Andres:2023xwr}
C.~Andres, F.~Dominguez, J.~Holguin, C.~Marquet and I.~Moult, \emph{{A coherent
  view of the quark-gluon plasma from energy correlators}},
  \href{http://dx.doi.org/10.1007/JHEP09(2023)088}{\emph{JHEP} {\bfseries 09}
  (2023) 088}, [\href{https://arxiv.org/abs/2303.03413}{{\ttfamily
  2303.03413}}].

\bibitem{Andres:2024xvk}
C.~Andres, F.~Dominguez, J.~Holguin, C.~Marquet and I.~Moult, \emph{{Simple
  Scaling Laws for Energy Correlators in Nuclear Matter}},
  \href{https://arxiv.org/abs/2411.15298}{{\ttfamily 2411.15298}}.

\bibitem{Barata:2025fzd}
J.~Barata, I.~Moult, A.~V. Sadofyev and J.~M. Silva, \emph{{Dissecting Jet
  Modification in the QGP with Multi-Point Energy Correlators}},
  \href{https://arxiv.org/abs/2503.13603}{{\ttfamily 2503.13603}}.

\bibitem{Singh:2024vwb}
B.~Singh and V.~Vaidya, \emph{{Factorization for energy-energy correlator in
  heavy ion collision}},
  \href{http://dx.doi.org/10.1007/JHEP06(2025)071}{\emph{JHEP} {\bfseries 06}
  (2025) 071}, [\href{https://arxiv.org/abs/2408.02753}{{\ttfamily
  2408.02753}}].

\bibitem{Li:2010ts}
H.~Li, F.~Liu, G.-l. Ma, X.-N. Wang and Y.~Zhu, \emph{{Mach cone induced by
  $\gamma$-triggered jets in high-energy heavy-ion collisions}},
  \href{http://dx.doi.org/10.1103/PhysRevLett.106.012301}{\emph{Phys. Rev.
  Lett.} {\bfseries 106} (2011) 012301},
  [\href{https://arxiv.org/abs/1006.2893}{{\ttfamily 1006.2893}}].

\bibitem{Casalderrey-Solana:2016jvj}
J.~Casalderrey-Solana, D.~Gulhan, G.~Milhano, D.~Pablos and K.~Rajagopal,
  \emph{{Angular Structure of Jet Quenching Within a Hybrid Strong/Weak
  Coupling Model}},
  \href{http://dx.doi.org/10.1007/JHEP03(2017)135}{\emph{JHEP} {\bfseries 03}
  (2017) 135}, [\href{https://arxiv.org/abs/1609.05842}{{\ttfamily
  1609.05842}}].

\bibitem{KunnawalkamElayavalli:2017hxo}
R.~Kunnawalkam~Elayavalli and K.~C. Zapp, \emph{{Medium response in JEWEL and
  its impact on jet shape observables in heavy ion collisions}},
  \href{http://dx.doi.org/10.1007/JHEP07(2017)141}{\emph{JHEP} {\bfseries 07}
  (2017) 141}, [\href{https://arxiv.org/abs/1707.01539}{{\ttfamily
  1707.01539}}].

\bibitem{Milhano:2017nzm}
G.~Milhano, U.~A. Wiedemann and K.~C. Zapp, \emph{{Sensitivity of jet
  substructure to jet-induced medium response}},
  \href{http://dx.doi.org/10.1016/j.physletb.2018.01.029}{\emph{Phys. Lett. B}
  {\bfseries 779} (2018) 409--413},
  [\href{https://arxiv.org/abs/1707.04142}{{\ttfamily 1707.04142}}].

\bibitem{Tachibana:2017syd}
Y.~Tachibana, N.-B. Chang and G.-Y. Qin, \emph{{Full jet in quark-gluon plasma
  with hydrodynamic medium response}},
  \href{http://dx.doi.org/10.1103/PhysRevC.95.044909}{\emph{Phys. Rev. C}
  {\bfseries 95} (2017) 044909},
  [\href{https://arxiv.org/abs/1701.07951}{{\ttfamily 1701.07951}}].

\bibitem{Yang:2021qtl}
Z.~Yang, W.~Chen, Y.~He, W.~Ke, L.~Pang and X.-N. Wang, \emph{{Search for the
  Elusive Jet-Induced Diffusion Wake in $Z/\gamma$-Jets with 2D Jet Tomography
  in High-Energy Heavy-Ion Collisions}},
  \href{http://dx.doi.org/10.1103/PhysRevLett.127.082301}{\emph{Phys. Rev.
  Lett.} {\bfseries 127} (2021) 082301},
  [\href{https://arxiv.org/abs/2101.05422}{{\ttfamily 2101.05422}}].

\bibitem{Gyulassy:1993hr}
M.~Gyulassy and X.-N. Wang, \emph{Multiple collisions and induced gluon
  bremsstrahlung in qcd},
  \href{http://dx.doi.org/10.1016/0550-3213(94)90079-5}{\emph{Nucl. Phys. B}
  {\bfseries 420} (1994) 583--614}.

\bibitem{Wang:1994fx}
X.-N. Wang, M.~Gyulassy and M.~Plumer, \emph{{The LPM effect in QCD and
  radiative energy loss in a quark gluon plasma}},
  \href{http://dx.doi.org/10.1103/PhysRevD.51.3436}{\emph{Phys. Rev. D}
  {\bfseries 51} (1995) 3436--3446},
  [\href{https://arxiv.org/abs/hep-ph/9408344}{{\ttfamily hep-ph/9408344}}].

\bibitem{Baier:1996sk}
R.~Baier, Y.~L. Dokshitzer, A.~H. Mueller, S.~Peigne and D.~Schiff,
  \emph{{Radiative energy loss and p(T) broadening of high-energy partons in
  nuclei}}, \href{http://dx.doi.org/10.1016/S0550-3213(96)00581-0}{\emph{Nucl.
  Phys. B} {\bfseries 484} (1997) 265--282},
  [\href{https://arxiv.org/abs/hep-ph/9608322}{{\ttfamily hep-ph/9608322}}].

\bibitem{Carbonell:1998rj}
J.~Carbonell, B.~Desplanques, V.~A. Karmanov and J.~F. Mathiot,
  \emph{{Explicitly covariant light front dynamics and relativistic few body
  systems}}, \href{http://dx.doi.org/10.1016/S0370-1573(97)00089-6}{\emph{Phys.
  Rept.} {\bfseries 300} (1998) 215--347}.

\bibitem{Burkardt:1995ct}
M.~Burkardt, \emph{{Light front quantization}}, {\emph{Adv. Nucl. Phys.}
  {\bfseries 23} (1996) 1--74},
  [\href{https://arxiv.org/abs/hep-ph/9505259}{{\ttfamily hep-ph/9505259}}].

\bibitem{Brodsky:1997de}
S.~J. Brodsky, H.-C. Pauli and S.~S. Pinsky, \emph{{Quantum chromodynamics and
  other field theories on the light cone}},
  \href{http://dx.doi.org/10.1016/S0370-1573(97)00089-6}{\emph{Phys. Rept.}
  {\bfseries 301} (1998) 299--486},
  [\href{https://arxiv.org/abs/hep-ph/9705477}{{\ttfamily hep-ph/9705477}}].

\bibitem{Becher:2014oda}
T.~Becher, \emph{{Introduction to Soft-Collinear Effective Theory}},
  \href{https://arxiv.org/abs/1410.1892}{{\ttfamily 1410.1892}}.

\bibitem{Collins:2011zzd}
J.~C. Collins, \emph{{Foundations of Perturbative QCD}}, vol.~32 of
  \emph{Cambridge Monographs on Particle Physics, Nuclear Physics and
  Cosmology}.
\newblock Cambridge University Press, 2011,
  \href{http://dx.doi.org/10.1017/CBO9780511975592}{10.1017/CBO9780511975592}.

\bibitem{Blaizot:2008yb}
J.-P. Blaizot and Y.~Mehtar-Tani, \emph{{The Classical field created in early
  stages of high energy nucleus-nucleus collisions}},
  \href{http://dx.doi.org/10.1016/j.nuclphysa.2008.11.010}{\emph{Nucl. Phys. A}
  {\bfseries 818} (2009) 97--119},
  [\href{https://arxiv.org/abs/0806.1422}{{\ttfamily 0806.1422}}].

\bibitem{McLerran:1993ni}
L.~D. McLerran and R.~Venugopalan, \emph{{Computing quark and gluon
  distribution functions for very large nuclei}},
  \href{http://dx.doi.org/10.1103/PhysRevD.49.2233}{\emph{Phys. Rev. D}
  {\bfseries 49} (1994) 2233--2241},
  [\href{https://arxiv.org/abs/hep-ph/9309289}{{\ttfamily hep-ph/9309289}}].

\bibitem{McLerran:1993ka}
L.~D. McLerran and R.~Venugopalan, \emph{{Gluon distribution functions for very
  large nuclei at small transverse momentum}},
  \href{http://dx.doi.org/10.1103/PhysRevD.49.3352}{\emph{Phys. Rev. D}
  {\bfseries 49} (1994) 3352--3355},
  [\href{https://arxiv.org/abs/hep-ph/9311205}{{\ttfamily hep-ph/9311205}}].

\bibitem{Kovchegov:1998bi}
Y.~V. Kovchegov and A.~H. Mueller, \emph{{Gluon production in current nucleus
  and nucleon - nucleus collisions in a quasiclassical approximation}},
  \href{http://dx.doi.org/10.1016/S0550-3213(98)00384-8}{\emph{Nucl. Phys. B}
  {\bfseries 529} (1998) 451--479},
  [\href{https://arxiv.org/abs/hep-ph/9802440}{{\ttfamily hep-ph/9802440}}].

\bibitem{Mueller:2012bn}
A.~H. Mueller and S.~Munier, \emph{{$p_{\perp}$-broadening and production
  processes versus dipole/quadrupole amplitudes at next-to-leading order}},
  \href{http://dx.doi.org/10.1016/j.nuclphysa.2012.08.005}{\emph{Nucl. Phys. A}
  {\bfseries 893} (2012) 43--86},
  [\href{https://arxiv.org/abs/1206.1333}{{\ttfamily 1206.1333}}].

\bibitem{Nikolaev:1990ja}
N.~N. Nikolaev and B.~G. Zakharov, \emph{{Color transparency and scaling
  properties of nuclear shadowing in deep inelastic scattering}},
  \href{http://dx.doi.org/10.1007/BF01483577}{\emph{Z. Phys. C} {\bfseries 49}
  (1991) 607--618}.

\bibitem{Moliere:1948zz}
G.~Moliere, \emph{{Theory of the scattering of fast charged particles. 2.
  Repeated and multiple scattering}}, {\emph{Z. Naturforsch. A} {\bfseries 3}
  (1948) 78--97}.

\bibitem{Bethe:1953va}
H.~A. Bethe, \emph{{Moliere's theory of multiple scattering}},
  \href{http://dx.doi.org/10.1103/PhysRev.89.1256}{\emph{Phys. Rev.} {\bfseries
  89} (1953) 1256--1266}.

\bibitem{Barata:2020rdn}
J.~a. Barata, Y.~Mehtar-Tani, A.~Soto-Ontoso and K.~Tywoniuk, \emph{{Revisiting
  transverse momentum broadening in dense QCD media}},
  \href{http://dx.doi.org/10.1103/PhysRevD.104.054047}{\emph{Phys. Rev. D}
  {\bfseries 104} (2021) 054047},
  [\href{https://arxiv.org/abs/2009.13667}{{\ttfamily 2009.13667}}].

\bibitem{Mehtar-Tani:2024smp}
Y.~Mehtar-Tani, F.~Ringer, B.~Singh and V.~Vaidya, \emph{{Factorization for jet
  production in heavy-ion collisions}},
  \href{https://arxiv.org/abs/2409.05957}{{\ttfamily 2409.05957}}.

\bibitem{Mehtar-Tani:2025xxd}
Y.~Mehtar-Tani, F.~Ringer, B.~Singh and V.~Vaidya, \emph{{Open quantum system
  approach to inclusive jet production in heavy-ion collisions}},
  \href{https://arxiv.org/abs/2504.00101}{{\ttfamily 2504.00101}}.

\bibitem{Mehtar-Tani:2019tvy}
Y.~Mehtar-Tani, \emph{{Gluon bremsstrahlung in finite media beyond multiple
  soft scattering approximation}},
  \href{http://dx.doi.org/10.1007/JHEP07(2019)057}{\emph{JHEP} {\bfseries 07}
  (2019) 057}, [\href{https://arxiv.org/abs/1903.00506}{{\ttfamily
  1903.00506}}].

\bibitem{Mehtar-Tani:2019ygg}
Y.~Mehtar-Tani and K.~Tywoniuk, \emph{{Improved opacity expansion for
  medium-induced parton splitting}},
  \href{http://dx.doi.org/10.1007/JHEP06(2020)187}{\emph{JHEP} {\bfseries 06}
  (2020) 187}, [\href{https://arxiv.org/abs/1910.02032}{{\ttfamily
  1910.02032}}].

\bibitem{Barata:2020sav}
J.~a. Barata and Y.~Mehtar-Tani, \emph{{Improved opacity expansion at NNLO for
  medium induced gluon radiation}},
  \href{http://dx.doi.org/10.1007/JHEP10(2020)176}{\emph{JHEP} {\bfseries 10}
  (2020) 176}, [\href{https://arxiv.org/abs/2004.02323}{{\ttfamily
  2004.02323}}].

\bibitem{Barata:2021wuf}
J.~a. Barata, Y.~Mehtar-Tani, A.~Soto-Ontoso and K.~Tywoniuk,
  \emph{{Medium-induced radiative kernel with the Improved Opacity Expansion}},
  \href{http://dx.doi.org/10.1007/JHEP09(2021)153}{\emph{JHEP} {\bfseries 09}
  (2021) 153}, [\href{https://arxiv.org/abs/2106.07402}{{\ttfamily
  2106.07402}}].

\bibitem{Andres:2020vxs}
C.~Andres, L.~Apolin\'ario and F.~Dominguez, \emph{{Medium-induced gluon
  radiation with full resummation of multiple scatterings for realistic
  parton-medium interactions}},
  \href{http://dx.doi.org/10.1007/JHEP07(2020)114}{\emph{JHEP} {\bfseries 07}
  (2020) 114}, [\href{https://arxiv.org/abs/2002.01517}{{\ttfamily
  2002.01517}}].

\bibitem{Baier:1998kq}
R.~Baier, Y.~L. Dokshitzer, A.~H. Mueller and D.~Schiff, \emph{{Medium induced
  radiative energy loss: Equivalence between the BDMPS and Zakharov
  formalisms}},
  \href{http://dx.doi.org/10.1016/S0550-3213(98)00546-X}{\emph{Nucl. Phys. B}
  {\bfseries 531} (1998) 403--425},
  [\href{https://arxiv.org/abs/hep-ph/9804212}{{\ttfamily hep-ph/9804212}}].

\bibitem{Caron-Huot:2010qjx}
S.~Caron-Huot and C.~Gale, \emph{{Finite-size effects on the radiative energy
  loss of a fast parton in hot and dense strongly interacting matter}},
  \href{http://dx.doi.org/10.1103/PhysRevC.82.064902}{\emph{Phys. Rev. C}
  {\bfseries 82} (2010) 064902},
  [\href{https://arxiv.org/abs/1006.2379}{{\ttfamily 1006.2379}}].

\bibitem{Apolinario:2014csa}
L.~Apolin{\'a}rio, N.~Armesto, J.~G. Milhano and C.~A. Salgado,
  \emph{{Medium-induced gluon radiation and colour decoherence beyond the soft
  approximation}}, \href{http://dx.doi.org/10.1007/JHEP02(2015)119}{\emph{JHEP}
  {\bfseries 02} (2015) 119},
  [\href{https://arxiv.org/abs/1407.0599}{{\ttfamily 1407.0599}}].

\bibitem{Caucal:2020zcz}
P.~Caucal, \emph{{Jet evolution in a dense QCD medium}}.
\newblock PhD thesis, Saclay, 9, 2020.
\newblock \href{https://arxiv.org/abs/2010.02874}{{\ttfamily 2010.02874}}.

\bibitem{Arnold:2009ik}
P.~B. Arnold, S.~Cantrell and W.~Xiao, \emph{{Stopping distance for high energy
  jets in weakly-coupled quark-gluon plasmas}},
  \href{http://dx.doi.org/10.1103/PhysRevD.81.045017}{\emph{Phys. Rev. D}
  {\bfseries 81} (2010) 045017},
  [\href{https://arxiv.org/abs/0912.3862}{{\ttfamily 0912.3862}}].

\bibitem{Fister:2014zxa}
L.~Fister and E.~Iancu, \emph{{Medium-induced jet evolution: wave turbulence
  and energy loss}},
  \href{http://dx.doi.org/10.1007/JHEP03(2015)082}{\emph{JHEP} {\bfseries 03}
  (2015) 082}, [\href{https://arxiv.org/abs/1409.2010}{{\ttfamily 1409.2010}}].

\bibitem{Baier:2000sb}
R.~Baier, A.~H. Mueller, D.~Schiff and D.~T. Son, \emph{{'Bottom up'
  thermalization in heavy ion collisions}},
  \href{http://dx.doi.org/10.1016/S0370-2693(01)00191-5}{\emph{Phys. Lett. B}
  {\bfseries 502} (2001) 51--58},
  [\href{https://arxiv.org/abs/hep-ph/0009237}{{\ttfamily hep-ph/0009237}}].

\bibitem{Arnold:2002zm}
P.~B. Arnold, G.~D. Moore and L.~G. Yaffe, \emph{{Effective kinetic theory for
  high temperature gauge theories}},
  \href{http://dx.doi.org/10.1088/1126-6708/2003/01/030}{\emph{JHEP} {\bfseries
  01} (2003) 030}, [\href{https://arxiv.org/abs/hep-ph/0209353}{{\ttfamily
  hep-ph/0209353}}].

\bibitem{Jeon:2003gi}
S.~Jeon and G.~D. Moore, \emph{{Energy loss of leading partons in a thermal QCD
  medium}}, \href{http://dx.doi.org/10.1103/PhysRevC.71.034901}{\emph{Phys.
  Rev. C} {\bfseries 71} (2005) 034901},
  [\href{https://arxiv.org/abs/hep-ph/0309332}{{\ttfamily hep-ph/0309332}}].

\bibitem{Blaizot:2015jea}
J.-P. Blaizot and Y.~Mehtar-Tani, \emph{{Energy flow along the medium-induced
  parton cascade}},
  \href{http://dx.doi.org/10.1016/j.aop.2016.01.002}{\emph{Annals Phys.}
  {\bfseries 368} (2016) 148--176},
  [\href{https://arxiv.org/abs/1501.03443}{{\ttfamily 1501.03443}}].

\bibitem{Nazarenko:2011zz}
S.~Nazarenko, \emph{{Wave Turbulence}}, vol.~825 of \emph{Lecture Notes in
  Physics}.
\newblock Springer, 2011,
  \href{http://dx.doi.org/10.1007/978-3-642-15942-5}{10.1007/978-3-642-15942-5}.

\bibitem{Kang:2013raa}
Z.-B. Kang, E.~Wang, X.-N. Wang and H.~Xing, \emph{{Next-to-Leading Order QCD
  Factorization for Semi-Inclusive Deep Inelastic Scattering at Twist 4}},
  \href{http://dx.doi.org/10.1103/PhysRevLett.112.102001}{\emph{Phys. Rev.
  Lett.} {\bfseries 112} (2014) 102001},
  [\href{https://arxiv.org/abs/1310.6759}{{\ttfamily 1310.6759}}].

\bibitem{Mehtar-Tani:2013pia}
Y.~Mehtar-Tani, J.~G. Milhano and K.~Tywoniuk, \emph{{Jet physics in heavy-ion
  collisions}}, \href{http://dx.doi.org/10.1142/S0217751X13400137}{\emph{Int.
  J. Mod. Phys. A} {\bfseries 28} (2013) 1340013},
  [\href{https://arxiv.org/abs/1302.2579}{{\ttfamily 1302.2579}}].

\bibitem{Hatta:2007cs}
Y.~Hatta, E.~Iancu and A.~H. Mueller, \emph{{Deep inelastic scattering off a
  N=4 SYM plasma at strong coupling}},
  \href{http://dx.doi.org/10.1088/1126-6708/2008/01/063}{\emph{JHEP} {\bfseries
  01} (2008) 063}, [\href{https://arxiv.org/abs/0710.5297}{{\ttfamily
  0710.5297}}].

\bibitem{Chesler:2008uy}
P.~M. Chesler, K.~Jensen, A.~Karch and L.~G. Yaffe, \emph{{Light quark energy
  loss in strongly-coupled N = 4 supersymmetric Yang-Mills plasma}},
  \href{http://dx.doi.org/10.1103/PhysRevD.79.125015}{\emph{Phys. Rev. D}
  {\bfseries 79} (2009) 125015},
  [\href{https://arxiv.org/abs/0810.1985}{{\ttfamily 0810.1985}}].

\bibitem{JET:2013cls}
{\scshape JET} collaboration, K.~M. Burke et~al., \emph{{Extracting the jet
  transport coefficient from jet quenching in high-energy heavy-ion
  collisions}}, \href{http://dx.doi.org/10.1103/PhysRevC.90.014909}{\emph{Phys.
  Rev. C} {\bfseries 90} (2014) 014909},
  [\href{https://arxiv.org/abs/1312.5003}{{\ttfamily 1312.5003}}].

\bibitem{Munier:2003vc}
S.~Munier and R.~B. Peschanski, \emph{{Geometric scaling as traveling waves}},
  \href{http://dx.doi.org/10.1103/PhysRevLett.91.232001}{\emph{Phys. Rev.
  Lett.} {\bfseries 91} (2003) 232001},
  [\href{https://arxiv.org/abs/hep-ph/0309177}{{\ttfamily hep-ph/0309177}}].

\bibitem{Kovchegov:2012mbw}
Y.~V. Kovchegov and E.~Levin, \emph{{Quantum Chromodynamics at High Energy}},
  vol.~33.
\newblock Oxford University Press, 2013,
  \href{http://dx.doi.org/10.1017/9781009291446}{10.1017/9781009291446}.

\bibitem{Kowalski:2003hm}
H.~Kowalski and D.~Teaney, \emph{{An Impact parameter dipole saturation
  model}}, \href{http://dx.doi.org/10.1103/PhysRevD.68.114005}{\emph{Phys. Rev.
  D} {\bfseries 68} (2003) 114005},
  [\href{https://arxiv.org/abs/hep-ph/0304189}{{\ttfamily hep-ph/0304189}}].

\bibitem{Lappi:2011ju}
T.~Lappi, \emph{{Gluon spectrum in the glasma from JIMWLK evolution}},
  \href{http://dx.doi.org/10.1016/j.physletb.2011.08.011}{\emph{Phys. Lett. B}
  {\bfseries 703} (2011) 325--330},
  [\href{https://arxiv.org/abs/1105.5511}{{\ttfamily 1105.5511}}].

\bibitem{Stasto:2000er}
A.~M. Stasto, K.~J. Golec-Biernat and J.~Kwiecinski, \emph{{Geometric scaling
  for the total gamma* p cross-section in the low x region}},
  \href{http://dx.doi.org/10.1103/PhysRevLett.86.596}{\emph{Phys. Rev. Lett.}
  {\bfseries 86} (2001) 596--599},
  [\href{https://arxiv.org/abs/hep-ph/0007192}{{\ttfamily hep-ph/0007192}}].

\bibitem{Iancu:2002tr}
E.~Iancu, K.~Itakura and L.~McLerran, \emph{{Geometric scaling above the
  saturation scale}},
  \href{http://dx.doi.org/10.1016/S0375-9474(02)01010-2}{\emph{Nucl. Phys. A}
  {\bfseries 708} (2002) 327--352},
  [\href{https://arxiv.org/abs/hep-ph/0203137}{{\ttfamily hep-ph/0203137}}].

\bibitem{Kwiecinski:2002ep}
J.~Kwiecinski and A.~M. Stasto, \emph{{Geometric scaling and QCD evolution}},
  \href{http://dx.doi.org/10.1103/PhysRevD.66.014013}{\emph{Phys. Rev. D}
  {\bfseries 66} (2002) 014013},
  [\href{https://arxiv.org/abs/hep-ph/0203030}{{\ttfamily hep-ph/0203030}}].

\bibitem{Brunet:1997zz}
E.~Brunet and B.~Derrida, \emph{{Shift in the velocity of a front due to a
  cutoff}}, \href{http://dx.doi.org/10.1103/PhysRevE.56.2597}{\emph{Phys. Rev.
  E} {\bfseries 56} (1997) 2597--2604},
  [\href{https://arxiv.org/abs/cond-mat/0005362}{{\ttfamily
  cond-mat/0005362}}].

\bibitem{Munier:2003sj}
S.~Munier and R.~B. Peschanski, \emph{{Traveling wave fronts and the transition
  to saturation}},
  \href{http://dx.doi.org/10.1103/PhysRevD.69.034008}{\emph{Phys. Rev. D}
  {\bfseries 69} (2004) 034008},
  [\href{https://arxiv.org/abs/hep-ph/0310357}{{\ttfamily hep-ph/0310357}}].

\bibitem{Balitsky:1995ub}
I.~Balitsky, \emph{{Operator expansion for high-energy scattering}},
  \href{http://dx.doi.org/10.1016/0550-3213(95)00638-9}{\emph{Nucl. Phys. B}
  {\bfseries 463} (1996) 99--160},
  [\href{https://arxiv.org/abs/hep-ph/9509348}{{\ttfamily hep-ph/9509348}}].

\bibitem{Kovchegov:1999yj}
Y.~V. Kovchegov, \emph{{Unitarization of the BFKL Pomeron on a nucleus}},
  \href{http://dx.doi.org/10.1103/PhysRevD.60.034008}{\emph{Phys. Rev. D}
  {\bfseries 60} (1999) 034008},
  [\href{https://arxiv.org/abs/hep-ph/9901281}{{\ttfamily hep-ph/9901281}}].

\bibitem{DEramo:2012uzl}
F.~D'Eramo, M.~Lekaveckas, H.~Liu and K.~Rajagopal, \emph{{Momentum Broadening
  in Weakly Coupled Quark-Gluon Plasma (with a view to finding the
  quasiparticles within liquid quark-gluon plasma)}},
  \href{http://dx.doi.org/10.1007/JHEP05(2013)031}{\emph{JHEP} {\bfseries 05}
  (2013) 031}, [\href{https://arxiv.org/abs/1211.1922}{{\ttfamily 1211.1922}}].

\bibitem{Mehtar-Tani:2010ebp}
Y.~Mehtar-Tani, C.~A. Salgado and K.~Tywoniuk, \emph{{Anti-angular ordering of
  gluon radiation in QCD media}},
  \href{http://dx.doi.org/10.1103/PhysRevLett.106.122002}{\emph{Phys. Rev.
  Lett.} {\bfseries 106} (2011) 122002},
  [\href{https://arxiv.org/abs/1009.2965}{{\ttfamily 1009.2965}}].

\bibitem{Casalderrey-Solana:2012evi}
J.~Casalderrey-Solana, Y.~Mehtar-Tani, C.~A. Salgado and K.~Tywoniuk,
  \emph{{New picture of jet quenching dictated by color coherence}},
  \href{http://dx.doi.org/10.1016/j.physletb.2013.07.046}{\emph{Phys. Lett. B}
  {\bfseries 725} (2013) 357--360},
  [\href{https://arxiv.org/abs/1210.7765}{{\ttfamily 1210.7765}}].

\bibitem{Dokshitzer:1991wu}
Y.~L. Dokshitzer, V.~A. Khoze, A.~H. Mueller and S.~I. Troian, \emph{{Basics of
  perturbative QCD}}.
\newblock 1991.

\bibitem{Mueller:1981ex}
A.~H. Mueller, \emph{{On the Multiplicity of Hadrons in QCD Jets}},
  \href{http://dx.doi.org/10.1016/0370-2693(81)90581-5}{\emph{Phys. Lett. B}
  {\bfseries 104} (1981) 161--164}.

\bibitem{Bassetto:1982ma}
A.~Bassetto, M.~Ciafaloni, G.~Marchesini and A.~H. Mueller, \emph{{Jet
  Multiplicity and Soft Gluon Factorization}},
  \href{http://dx.doi.org/10.1016/0550-3213(82)90161-4}{\emph{Nucl. Phys. B}
  {\bfseries 207} (1982) 189--204}.

\bibitem{Dokshitzer:1982ia}
Y.~L. Dokshitzer, V.~S. Fadin and V.~A. Khoze, \emph{{On the Sensitivity of the
  Inclusive Distributions in Parton Jets to the Coherence Effects in QCD Gluon
  Cascades}}, \href{http://dx.doi.org/10.1007/BF01571703}{\emph{Z. Phys. C}
  {\bfseries 18} (1983) 37}.

\bibitem{Mehtar-Tani:2017web}
Y.~Mehtar-Tani and K.~Tywoniuk, \emph{{Sudakov suppression of jets in QCD
  media}}, \href{http://dx.doi.org/10.1103/PhysRevD.98.051501}{\emph{Phys. Rev.
  D} {\bfseries 98} (2018) 051501},
  [\href{https://arxiv.org/abs/1707.07361}{{\ttfamily 1707.07361}}].

\bibitem{Casalderrey-Solana:2015bww}
J.~Casalderrey-Solana, D.~Pablos and K.~Tywoniuk, \emph{{Two-gluon emission and
  interference in a thin QCD medium: insights into jet formation}},
  \href{http://dx.doi.org/10.1007/JHEP11(2016)174}{\emph{JHEP} {\bfseries 11}
  (2016) 174}, [\href{https://arxiv.org/abs/1512.07561}{{\ttfamily
  1512.07561}}].

\bibitem{Dominguez:2019ges}
F.~Dom{\'\i}nguez, J.~G. Milhano, C.~A. Salgado, K.~Tywoniuk and V.~Vila,
  \emph{{Mapping collinear in-medium parton splittings}},
  \href{http://dx.doi.org/10.1140/epjc/s10052-019-7563-0}{\emph{Eur. Phys. J.
  C} {\bfseries 80} (2020) 11},
  [\href{https://arxiv.org/abs/1907.03653}{{\ttfamily 1907.03653}}].

\bibitem{Abreu:2024wka}
S.~Abreu, X.~M. L\'opez, G.~Milhano and A.~Soto-Ontoso, \emph{{A generalized
  picture of colour decoherence in dense QCD media}},
  \href{https://arxiv.org/abs/2410.24135}{{\ttfamily 2410.24135}}.

\bibitem{Dominguez:2012ad}
F.~Dominguez, C.~Marquet, A.~M. Stasto and B.-W. Xiao, \emph{{Universality of
  multiparticle production in QCD at high energies}},
  \href{http://dx.doi.org/10.1103/PhysRevD.87.034007}{\emph{Phys. Rev. D}
  {\bfseries 87} (2013) 034007},
  [\href{https://arxiv.org/abs/1210.1141}{{\ttfamily 1210.1141}}].

\bibitem{Mehtar-Tani:2011vlz}
Y.~Mehtar-Tani and K.~Tywoniuk, \emph{{Jet coherence in QCD media: the antenna
  radiation spectrum}},
  \href{http://dx.doi.org/10.1007/JHEP01(2013)031}{\emph{JHEP} {\bfseries 01}
  (2013) 031}, [\href{https://arxiv.org/abs/1105.1346}{{\ttfamily 1105.1346}}].

\bibitem{Kang:2016mcy}
Z.-B. Kang, F.~Ringer and I.~Vitev, \emph{{The semi-inclusive jet function in
  SCET and small radius resummation for inclusive jet production}},
  \href{http://dx.doi.org/10.1007/JHEP10(2016)125}{\emph{JHEP} {\bfseries 10}
  (2016) 125}, [\href{https://arxiv.org/abs/1606.06732}{{\ttfamily
  1606.06732}}].

\bibitem{Dasgupta:2014yra}
M.~Dasgupta, F.~Dreyer, G.~P. Salam and G.~Soyez, \emph{{Small-radius jets to
  all orders in QCD}},
  \href{http://dx.doi.org/10.1007/JHEP04(2015)039}{\emph{JHEP} {\bfseries 04}
  (2015) 039}, [\href{https://arxiv.org/abs/1411.5182}{{\ttfamily 1411.5182}}].

\bibitem{Dasgupta:2001sh}
M.~Dasgupta and G.~P. Salam, \emph{{Resummation of nonglobal QCD observables}},
  \href{http://dx.doi.org/10.1016/S0370-2693(01)00725-0}{\emph{Phys. Lett. B}
  {\bfseries 512} (2001) 323--330},
  [\href{https://arxiv.org/abs/hep-ph/0104277}{{\ttfamily hep-ph/0104277}}].

\bibitem{Andres:2023jao}
C.~Andres, L.~Apolin\'ario, F.~Dominguez and M.~G. Martinez, \emph{{In-medium
  gluon radiation spectrum with all-order resummation of multiple scatterings
  in longitudinally evolving media}},
  \href{https://arxiv.org/abs/2307.06226}{{\ttfamily 2307.06226}}.

\bibitem{Moore:2021jwe}
G.~D. Moore, S.~Schlichting, N.~Schlusser and I.~Soudi, \emph{{Non-perturbative
  determination of collisional broadening and medium induced radiation in QCD
  plasmas}}, \href{http://dx.doi.org/10.1007/JHEP10(2021)059}{\emph{JHEP}
  {\bfseries 10} (2021) 059},
  [\href{https://arxiv.org/abs/2105.01679}{{\ttfamily 2105.01679}}].

\bibitem{Schlichting:2021idr}
S.~Schlichting and I.~Soudi, \emph{{Splitting rates in QCD plasmas from a
  nonperturbative determination of the momentum broadening kernel
  C(q\ensuremath{\perp})}},
  \href{http://dx.doi.org/10.1103/PhysRevD.105.076002}{\emph{Phys. Rev. D}
  {\bfseries 105} (2022) 076002},
  [\href{https://arxiv.org/abs/2111.13731}{{\ttfamily 2111.13731}}].

\bibitem{Adhya:2019qse}
S.~P. Adhya, C.~A. Salgado, M.~Spousta and K.~Tywoniuk, \emph{{Medium-induced
  cascade in expanding media}},
  \href{http://dx.doi.org/10.1007/JHEP07(2020)150}{\emph{JHEP} {\bfseries 07}
  (2020) 150}, [\href{https://arxiv.org/abs/1911.12193}{{\ttfamily
  1911.12193}}].

\bibitem{Caucal:2020uic}
P.~Caucal, E.~Iancu and G.~Soyez, \emph{{Jet radiation in a longitudinally
  expanding medium}},
  \href{http://dx.doi.org/10.1007/JHEP04(2021)209}{\emph{JHEP} {\bfseries 04}
  (2021) 209}, [\href{https://arxiv.org/abs/2012.01457}{{\ttfamily
  2012.01457}}].

\bibitem{Catani:1990rr}
S.~Catani, B.~R. Webber and G.~Marchesini, \emph{{QCD coherent branching and
  semiinclusive processes at large x}},
  \href{http://dx.doi.org/10.1016/0550-3213(91)90390-J}{\emph{Nucl. Phys. B}
  {\bfseries 349} (1991) 635--654}.

\bibitem{Blaizot:2015lma}
J.~P. Blaizot and Y.~Mehtar-Tani, \emph{{Jet Structure in Heavy Ion
  Collisions}}, \href{http://dx.doi.org/10.1142/S021830131530012X}{\emph{Int.
  J. Mod. Phys. E} {\bfseries 24} (2015) 1530012},
  [\href{https://arxiv.org/abs/1503.05958}{{\ttfamily 1503.05958}}].

\bibitem{Chen:2022muj}
H.~Chen, M.~Jaarsma, Y.~Li, I.~Moult, W.~J. Waalewijn and H.~X. Zhu,
  \emph{{Collinear Parton Dynamics Beyond DGLAP}},
  \href{https://arxiv.org/abs/2210.10061}{{\ttfamily 2210.10061}}.

\bibitem{Chang:2013rca}
H.-M. Chang, M.~Procura, J.~Thaler and W.~J. Waalewijn, \emph{{Calculating
  Track-Based Observables for the LHC}},
  \href{http://dx.doi.org/10.1103/PhysRevLett.111.102002}{\emph{Phys. Rev.
  Lett.} {\bfseries 111} (2013) 102002},
  [\href{https://arxiv.org/abs/1303.6637}{{\ttfamily 1303.6637}}].

\bibitem{Barata:2024bmx}
J.~a. Barata, I.~Moult and J.~a.~M. Silva, \emph{{Tracking Energy Loss in Heavy
  Ion Collisions}},  \href{https://arxiv.org/abs/2409.18174}{{\ttfamily
  2409.18174}}.

\bibitem{Lee:2024tzc}
K.~Lee, I.~Moult and X.~Zhang, \emph{{Revisiting Single Inclusive Jet
  Production: Small-$R$ Resummation at Next-to-Leading Logarithm}},
  \href{https://arxiv.org/abs/2410.01902}{{\ttfamily 2410.01902}}.

\bibitem{vanBeekveld:2024jnx}
M.~van Beekveld, M.~Dasgupta, B.~K. El-Menoufi, J.~Helliwell, A.~Karlberg and
  P.~F. Monni, \emph{{Two-loop anomalous dimensions for small-R jet versus
  hadronic fragmentation functions}},
  \href{http://dx.doi.org/10.1007/JHEP07(2024)239}{\emph{JHEP} {\bfseries 07}
  (2024) 239}, [\href{https://arxiv.org/abs/2402.05170}{{\ttfamily
  2402.05170}}].

\bibitem{Chen:2017zte}
W.~Chen, S.~Cao, T.~Luo, L.-G. Pang and X.-N. Wang, \emph{{Effects of
  jet-induced medium excitation in $\gamma$-hadron correlation in A+A
  collisions}},
  \href{http://dx.doi.org/10.1016/j.physletb.2017.12.015}{\emph{Phys. Lett. B}
  {\bfseries 777} (2018) 86--90},
  [\href{https://arxiv.org/abs/1704.03648}{{\ttfamily 1704.03648}}].

\bibitem{Park:2018acg}
C.~Park, S.~Jeon and C.~Gale, \emph{{Jet modification with medium recoil in
  quark-gluon plasma}},
  \href{http://dx.doi.org/10.1016/j.nuclphysa.2018.10.057}{\emph{Nucl. Phys. A}
  {\bfseries 982} (2019) 643--646},
  [\href{https://arxiv.org/abs/1807.06550}{{\ttfamily 1807.06550}}].

\bibitem{Dasgupta:2016bnd}
M.~Dasgupta, F.~A. Dreyer, G.~P. Salam and G.~Soyez, \emph{{Inclusive jet
  spectrum for small-radius jets}},
  \href{http://dx.doi.org/10.1007/JHEP06(2016)057}{\emph{JHEP} {\bfseries 06}
  (2016) 057}, [\href{https://arxiv.org/abs/1602.01110}{{\ttfamily
  1602.01110}}].

\bibitem{d'Enterria:2021ycp}
D.~d'Enterria and C.~Loizides, \emph{{Progress in the Glauber model at collider
  energies}},
  \href{http://dx.doi.org/10.1146/annurev-nucl-102419-060007}{\emph{Annu. Rev.
  Nucl. Part. Sci.} {\bfseries 71} (2021) 315--344},
  [\href{https://arxiv.org/abs/2011.14909}{{\ttfamily 2011.14909}}].

\bibitem{Weller:2017tsr}
R.~D. Weller and P.~Romatschke, \emph{{One fluid to rule them all: viscous
  hydrodynamic description of event-by-event central p+p, p+Pb and Pb+Pb
  collisions at $\sqrt{s}=5.02$ TeV}},
  \href{http://dx.doi.org/10.1016/j.physletb.2017.09.077}{\emph{Phys. Lett. B}
  {\bfseries 774} (2017) 351--356},
  [\href{https://arxiv.org/abs/1701.07145}{{\ttfamily 1701.07145}}].

\bibitem{Barata:2023zqg}
J.~a. Barata, J.~G. Milhano and A.~V. Sadofyev, \emph{{Picturing QCD jets in
  anisotropic matter: from jet shapes to energy energy correlators}},
  \href{http://dx.doi.org/10.1140/epjc/s10052-024-12514-1}{\emph{Eur. Phys. J.
  C} {\bfseries 84} (2024) 174},
  [\href{https://arxiv.org/abs/2308.01294}{{\ttfamily 2308.01294}}].

\bibitem{CMS:2018fof}
{\scshape CMS} collaboration, A.~M. Sirunyan et~al., \emph{{Measurement of the
  groomed jet mass in PbPb and pp collisions at $ \sqrt{s_{\mathrm{NN}}}=5.02 $
  TeV}}, \href{http://dx.doi.org/10.1007/JHEP10(2018)161}{\emph{JHEP}
  {\bfseries 10} (2018) 161},
  [\href{https://arxiv.org/abs/1805.05145}{{\ttfamily 1805.05145}}].

\bibitem{Guo:2000nz}
X.-f. Guo and X.-N. Wang, \emph{{Multiple scattering, parton energy loss and
  modified fragmentation functions in deeply inelastic e A scattering}},
  \href{http://dx.doi.org/10.1103/PhysRevLett.85.3591}{\emph{Phys. Rev. Lett.}
  {\bfseries 85} (2000) 3591--3594},
  [\href{https://arxiv.org/abs/hep-ph/0005044}{{\ttfamily hep-ph/0005044}}].

\bibitem{Feal:2019xfl}
X.~Feal, C.~A. Salgado and R.~A. Vazquez, \emph{{Jet quenching test of the QCD
  matter created at RHIC and the LHC needs opacity-resummed medium induced
  radiation}},
  \href{http://dx.doi.org/10.1016/j.physletb.2021.136251}{\emph{Phys. Lett. B}
  {\bfseries 816} (2021) 136251},
  [\href{https://arxiv.org/abs/1911.01309}{{\ttfamily 1911.01309}}].

\bibitem{Andres:2020kfg}
C.~Andres, F.~Dominguez and M.~Gonzalez~Martinez, \emph{{From soft to hard
  radiation: the role of multiple scatterings in medium-induced gluon
  emissions}}, \href{http://dx.doi.org/10.1007/JHEP03(2021)102}{\emph{JHEP}
  {\bfseries 03} (2021) 102},
  [\href{https://arxiv.org/abs/2011.06522}{{\ttfamily 2011.06522}}].

\bibitem{ATLAS:2018gwx}
{\scshape ATLAS} collaboration, M.~Aaboud et~al., \emph{{Measurement of the
  nuclear modification factor for inclusive jets in Pb+Pb collisions at
  $\sqrt{s_\mathrm{NN}}=5.02$ TeV with the ATLAS detector}},
  \href{http://dx.doi.org/10.1016/j.physletb.2018.10.076}{\emph{Phys. Lett. B}
  {\bfseries 790} (2019) 108--128},
  [\href{https://arxiv.org/abs/1805.05635}{{\ttfamily 1805.05635}}].

\bibitem{Shen:2014vra}
C.~Shen, Z.~Qiu, H.~Song, J.~Bernhard, S.~Bass and U.~Heinz, \emph{{The
  iEBE-VISHNU code package for relativistic heavy-ion collisions}},
  \href{http://dx.doi.org/10.1016/j.cpc.2015.08.039}{\emph{Comput. Phys.
  Commun.} {\bfseries 199} (2016) 61--85},
  [\href{https://arxiv.org/abs/1409.8164}{{\ttfamily 1409.8164}}].

\bibitem{Mehtar-Tani:2024jtd}
Y.~Mehtar-Tani, D.~Pablos and K.~Tywoniuk, \emph{{Jet suppression and azimuthal
  anisotropy from RHIC to LHC}},
  \href{http://dx.doi.org/10.1103/PhysRevD.110.014009}{\emph{Phys. Rev. D}
  {\bfseries 110} (2024) 014009},
  [\href{https://arxiv.org/abs/2402.07869}{{\ttfamily 2402.07869}}].

\bibitem{Pablos:2025cli}
D.~Pablos and A.~Takacs, \emph{{Bayesian Constraints on Pre-Equilibrium Jet
  Quenching and Predictions for Oxygen Collisions}},
  \href{https://arxiv.org/abs/2509.19430}{{\ttfamily 2509.19430}}.

\bibitem{Armesto:2004vz}
N.~Armesto, C.~A. Salgado and U.~A. Wiedemann, \emph{{Low-p(T) collective flow
  induces high-p(T) jet quenching}},
  \href{http://dx.doi.org/10.1103/PhysRevC.72.064910}{\emph{Phys. Rev. C}
  {\bfseries 72} (2005) 064910},
  [\href{https://arxiv.org/abs/hep-ph/0411341}{{\ttfamily hep-ph/0411341}}].

\bibitem{Wang:2001ifa}
X.-N. Wang and X.-f. Guo, \emph{{Multiple parton scattering in nuclei: Parton
  energy loss}},
  \href{http://dx.doi.org/10.1016/S0375-9474(01)01130-7}{\emph{Nucl. Phys. A}
  {\bfseries 696} (2001) 788--832},
  [\href{https://arxiv.org/abs/hep-ph/0102230}{{\ttfamily hep-ph/0102230}}].

\bibitem{Andres:2022ndd}
C.~Andres, F.~Dominguez, A.~V. Sadofyev and C.~A. Salgado, \emph{{Jet
  broadening in flowing matter: Resummation}},
  \href{http://dx.doi.org/10.1103/PhysRevD.106.074023}{\emph{Phys. Rev. D}
  {\bfseries 106} (2022) 074023},
  [\href{https://arxiv.org/abs/2207.07141}{{\ttfamily 2207.07141}}].

\bibitem{Sadofyev:2021ohn}
A.~V. Sadofyev, M.~D. Sievert and I.~Vitev, \emph{{Ab~initio coupling of jets
  to collective flow in the opacity expansion approach}},
  \href{http://dx.doi.org/10.1103/PhysRevD.104.094044}{\emph{Phys. Rev. D}
  {\bfseries 104} (2021) 094044},
  [\href{https://arxiv.org/abs/2104.09513}{{\ttfamily 2104.09513}}].

\bibitem{Barata:2022krd}
J.~a. Barata, A.~V. Sadofyev and C.~A. Salgado, \emph{{Jet broadening in dense
  inhomogeneous matter}},
  \href{http://dx.doi.org/10.1103/PhysRevD.105.114010}{\emph{Phys. Rev. D}
  {\bfseries 105} (2022) 114010},
  [\href{https://arxiv.org/abs/2202.08847}{{\ttfamily 2202.08847}}].

\bibitem{Barata:2023qds}
J.~a. Barata, X.~Mayo~L\'opez, A.~V. Sadofyev and C.~A. Salgado, \emph{{Medium
  induced gluon spectrum in dense inhomogeneous matter}},
  \href{http://dx.doi.org/10.1103/PhysRevD.108.034018}{\emph{Phys. Rev. D}
  {\bfseries 108} (2023) 034018},
  [\href{https://arxiv.org/abs/2304.03712}{{\ttfamily 2304.03712}}].

\bibitem{Kuzmin:2023hko}
M.~V. Kuzmin, X.~Mayo~L{\'o}pez, J.~Reiten and A.~V. Sadofyev, \emph{{Jet
  quenching in anisotropic flowing matter}},
  \href{http://dx.doi.org/10.1103/PhysRevD.109.014036}{\emph{Phys. Rev. D}
  {\bfseries 109} (2024) 014036},
  [\href{https://arxiv.org/abs/2309.00683}{{\ttfamily 2309.00683}}].

\bibitem{Attems:2022otp}
M.~Attems, J.~Brewer, G.~M. Innocenti, A.~Mazeliauskas, S.~Park, W.~van~der
  Schee et~al., \emph{{Medium-Enhanced cc\textasciimacron{} Radiation}},
  \href{http://dx.doi.org/10.1103/PhysRevLett.132.212301}{\emph{Phys. Rev.
  Lett.} {\bfseries 132} (2024) 212301},
  [\href{https://arxiv.org/abs/2209.13600}{{\ttfamily 2209.13600}}].

\bibitem{Attems:2022ubu}
M.~Attems, J.~Brewer, G.~M. Innocenti, A.~Mazeliauskas, S.~Park, W.~van~der
  Schee et~al., \emph{{The medium-modified $ g\to c\overline{c} $ splitting
  function in the BDMPS-Z formalism}},
  \href{http://dx.doi.org/10.1007/JHEP01(2023)080}{\emph{JHEP} {\bfseries 01}
  (2023) 080}, [\href{https://arxiv.org/abs/2203.11241}{{\ttfamily
  2203.11241}}].

\bibitem{Vaidya:2020cyi}
V.~Vaidya and X.~Yao, \emph{{Transverse momentum broadening of a jet in
  quark-gluon plasma: an open quantum system EFT}},
  \href{http://dx.doi.org/10.1007/JHEP10(2020)024}{\emph{JHEP} {\bfseries 10}
  (2020) 024}, [\href{https://arxiv.org/abs/2004.11403}{{\ttfamily
  2004.11403}}].

\end{thebibliography}\endgroup
